\documentclass[hyper]{JHEP3}

\input{epsf}
\usepackage{epsfig}
\usepackage{amssymb}
\usepackage{amsfonts}
\usepackage{amsbsy}
\usepackage[all]{xy}
\usepackage{amsmath}
\usepackage{enumerate}

\usepackage{amssymb,amscd}
\usepackage{mathrsfs}
\usepackage{amsmath,amsthm}
\usepackage{xspace}
\usepackage{url}

\def\ch{{\rm ch}}

\def\noself{non-selfintersecting}

\def\IC{\mathbb{C}}
\def\IH{\mathbb{H}}

\def\IP{\mathbb{P}}

\def\IR{{\mathbb{R}}}
\def\IS{{\mathbb{S}}}
\def\IZ{{\mathbb{Z}}}

\def\fg{\mathfrak{g}}

\def\CI {{\cal I}}
\def\CM {{\cal M}}
\def\tCM{\widetilde{\cal M}}
\def\CN{{\cal N}}
\def\CK {{\cal K}}
\def\CN {{\cal N}}
\def\CR {{\cal R}}

\def\CF {{\cal F}}
\def\CJ {{\cal J}}
\def\CP {{\cal P }}
\def\CL {{\cal L}}

\def\CW {{\cal W}}
\def\CX {{\cal X}}
\def\CY {{\cal Y}}
\def\tCY {\widetilde{\cal Y}}
\def\CO {{\cal O}}
\def\CZ {{\cal Z}}
\def\CE {{\cal E}}
\def\CG {{\cal G}}
\def\CH {{\cal H}}
\def\CB {{\cal B}}
\def\CS {{\cal S}}
\def\CA{{\cal A}}
\def\CK{{\cal K}}
\def\CQ{{\cal Q}}

\def\CZ{{\cal Z}}

\def\half{\frac{1}{2}}

\def\hW{\widehat W}

\renewcommand{\Im}{{\rm Im }}
\renewcommand{\Re}{{\rm Re }}

\def\one{{\hbox{ 1\kern-.8mm l}}}

\def\vol{{\rm vol\,}}
\def\p{\partial}

\def\be{\bar{e}}

\def\half{\frac{1}{2}}

\def\hk{hyperk\"ahler\xspace}

\newcommand{\qbinom}[2]{\genfrac{[}{]}{0pt}{}{#1}{#2}}
\newcommand{\ti}[1]{\textit{#1}}

\def\fg{\mathfrak{g}}

\def\lieg{\mathfrak{g}}
\def\liet{\mathfrak{t}}
\def\p{\partial}

\def\IC{\mathbb{C}}

\def\IZ{{\mathbb{Z}}}
\def\IR{{\mathbb{R}}}
\def\IP{\mathbb{P}}

\def\CI {{\cal I}}
\def\CM {{\cal M}}
\def\tCM {\widetilde{\cal M}}
\def\CN{{\cal N}}
\def\CK {{\cal K}}
\def\CN {{\cal N}}
\def\CR {{\cal R}}

\def\CF {{\cal F}}
\def\CJ {{\cal J}}
\def\CP {{\cal P }}
\def\CL {{\cal L}}

\def\CW {{\cal W}}
\def\CX {{\cal X}}
\def\CY {{\cal Y}}
\def\CO {{\cal O}}
\def\CZ {{\cal Z}}
\def\CE {{\cal E}}
\def\CG {{\cal G}}
\def\CH {{\cal H}}
\def\CB {{\cal B}}
\def\CS {{\cal S}}
\def\CA{{\cal A}}
\def\CK{{\cal K}}
\def\CQ{{\cal Q}}

\def\CZ{{\cal Z}}

\def\half{\frac{1}{2}}
\def\ttheta{\tilde\theta}

\def\be{\begin{equation}
}

\def\ee{\end{equation}}

\newcommand{\inprod}[1]{\langle#1\rangle}
\newcommand\N{{\cal N}}

\newcommand\inn{{\mathrm{in}}}
\newcommand\out{{\mathrm{out}}}

\newcommand\eps{\epsilon}

\newcommand\fro{{\overline{\underline{\Omega}}}}

\newcommand\WKB{{\mathrm{WKB}}}

\newcommand\BPS{{\mathrm{BPS}}}

\newcommand\ts{\tilde{s}}

\newcommand{\bS}{\mathbf S}

\DeclareMathOperator{\Tr}{{Tr}}
\DeclareMathOperator{\Hom}{{Hom}}
\DeclareMathOperator{\Hol}{{Hol}}

\DeclareMathOperator{\Edges}{{Edges}}

\newcommand{\insfig}[2]{\begin{figure}[htbp] \centering \includegraphics[scale=0.275]{figures/#1-crop.pdf} \caption{#2} \label{fig:#1} \end{figure}}

\title{Framed BPS States}

\author{Davide Gaiotto$^1$, Gregory W. Moore$^2$, Andrew Neitzke$^3$\\
$^1$ School of Natural Sciences, Institute for Advanced Study, \\
Princeton, NJ 08540, USA\\
 $^2$ NHETC and Department of Physics and Astronomy,
Rutgers University,\\
Piscataway, NJ 08855--0849, USA\\
$^3$ Department of Mathematics, University of Texas at Austin,\\ Austin, TX 78712, USA\\
\\
{\tt dgaiotto@ias.edu, gmoore@physics.rutgers.edu, neitzke@math.utexas.edu} }

\abstract{We consider a class of line operators in $d=4, \CN=2$
supersymmetric field theories which leave four supersymmetries
unbroken.  Such line operators support a new class of BPS states which
we call ``framed BPS states.''  These include
halo bound states similar to those of $d=4, \CN=2$ supergravity,
where (ordinary) BPS particles are loosely bound to the line operator.
Using this construction, we give a new proof
of the Kontsevich-Soibelman wall-crossing formula for the ordinary BPS
particles, by reducing it to the semiprimitive wall-crossing formula.
After reducing on $S^1$, the expansion of the vevs of the line operators in
the IR provides a new physical interpretation of the ``Darboux coordinates''
on the moduli space $\CM$ of the theory.
Moreover, we introduce a ``protected spin character'' which keeps track
of the spin degrees of freedom of the framed BPS states. We show that the
generating functions of protected spin characters admit a multiplication
which defines a deformation of the algebra of holomorphic functions on   $\CM$.
As an illustration of these ideas, we consider the six-dimensional
$(2,0)$ field theory of $A_1$ type compactified on a Riemann surface $C$.  Here we show
(extending previous results) that
line operators are classified by certain laminations on a suitably decorated
version of $C$, and we compute the
spectrum of framed BPS states in several explicit examples. Finally we
indicate some interesting connections to the theory of cluster algebras.}

\begin{document}


\bibliographystyle{utphys}

\section{Introduction and Summary}

This paper continues an investigation \cite{Gaiotto:2008cd,Gaiotto:2009hg} into
the properties of moduli spaces $\CM$ naturally associated to $\CN=2$
supersymmetric quantum field theories in four dimensions
\cite{Seiberg:1994rs,Seiberg:1996nz}. These moduli spaces are
the moduli spaces of vacua of such theories on $\IR^3 \times S^1$. They can be
given the structure of completely integrable systems and have been the subject of
much intense research.  The focus of \cite{Gaiotto:2008cd,Gaiotto:2009hg}
was on a collection of functions on $\CM$, the so-called ``Darboux coordinates'' which
are both useful and interesting. They are useful because they provide a neat
way to compute \hk\ metrics on $\CM$, because they provide a framework within which
one can prove the Kontsevich-Soibelman wall-crossing formula \cite{ks1}, and because
they can be used to construct scattering amplitudes in $\CN=4$ theories at
strong coupling \cite{Alday:2009yn,Alday:2009dv}. They are interesting because
they have many beautiful asymptotic and analytic properties, linking them to several other
subjects including the Thermodynamic Bethe ansatz of integrable systems theory
and the Fock-Goncharov coordinates of (higher) Teichm\"uller theory.

In the rest of this introduction we give a brief expository description of the
four main results of this paper.

In view of the central importance of the Darboux coordinates on $\CM$ it is
desirable to give them a more direct physical interpretation. One of the
primary goals of this paper is to provide just  such an interpretation: They are vevs
of line operators in the $\CN=2$ theory. This is our first main result.
 This interpretation is encapsulated
in equation \eqref{eq:trh-large-r} of Section \ref{sec:linehol}:
\begin{equation} \label{eq:intro-eq}
\inprod{L_\zeta} = \sum_\gamma \fro(L_\zeta, \gamma) \CY_\gamma.
\end{equation}
The left side of this equation is the vacuum expectation value of a supersymmetric line operator $L_\zeta$.
This operator sits at a single point in space, stretches along the time direction,
and preserves four Poincar\'e supercharges specified by an angle $\zeta$ --- the
same four supersymmetries preserved by a BPS particle of phase $\zeta$. (Section
\ref{sec:Susy-BPS-line} goes over this definition in detail and summarizes some
basic aspects of the theory of supersymmetric line operators.)  Supersymmetric
't Hooft-Wilson operators provide examples of possible $L_\zeta$, but there are others, as we will see.
The right side of \eqref{eq:intro-eq} is an expansion in the Darboux coordinates $\CY_\gamma$.
Heuristically speaking, $\CY_{\gamma}$
can be thought of as the vev of a line operator in the low energy IR theory, obtained by inserting
an infinitely heavy dyon, which carries electromagnetic charge $\gamma$ and central charge of
phase $\zeta$.  In this sense, \eqref{eq:intro-eq} expresses how UV line operators $L_\zeta$ decompose
into IR line operators.

The coefficients $\fro(L_\zeta, \gamma)$
of the expansion \eqref{eq:intro-eq} are integers, and in fact define
a new kind of BPS index, counting a new kind of BPS state which we call a
\emph{framed BPS state}. These states are introduced in Section
 \ref{sec:Hilbert-Halo-WC}. Briefly, the   Hilbert space $\CH_L$ of the theory in the presence of
the line operator $L$ is a representation of the unbroken part of the $\N=2$ supersymmetry.
It is still graded by electromagnetic charge $\gamma$, and in the sector
labeled by $\gamma$, the energy satisfies a modified BPS bound
\begin{equation}
E\geq  - {\rm Re} (Z_\gamma/\zeta).
\end{equation}
Framed BPS states are states in $\CH_L$ which saturate this bound.
To reduce confusion, we will call the
BPS states of the original $\CN=2$ theory, in the absence of line operators,
the ``vanilla BPS states.''

We can refine the BPS index to account for the spin information
of the framed BPS states. This very interesting index, defined in \eqref{eq:Def-Framed-PSC} and
\eqref{eq:PSC-one} of Section \ref{subsec:BPS-Deg-PSC}, is called a \emph{protected spin character} (PSC).
A PSC can be defined for both framed and vanilla BPS states.  The PSC is a function of a variable $y$
and is denoted $\fro(L_\zeta,\gamma;y)$. Its specialization to $y=-1$ gives the BPS index.
The second and third main results of this paper revolve around
the properties of the generating functions of framed protected spin characters (such as \eqref{eq:form-gen}):
\begin{equation}\label{eq:intro-eq-2}
F( L_\zeta ) := \sum_{\gamma} \fro(  L_\zeta,
\gamma;y) X_\gamma.
\end{equation}
Here the formal variables $X_{\gamma}$ in the generating function
are meant to be thought of as elements of a
noncommutative Heisenberg algebra, and hence satisfy
\begin{equation}
X_{\gamma} X_{\gamma'} = y^{\langle \gamma, \gamma' \rangle} X_{\gamma+\gamma'}
\end{equation}
where $\langle \gamma, \gamma' \rangle$ is the usual Dirac-Schwinger-Zwanziger
antisymmetric product of charges.  The second main result of this paper is that the
algebra of the generating functions \eqref{eq:intro-eq-2} gives a
noncommutative deformation of the algebra of functions on $\CM$, where $y$ is the
deformation parameter.
 This multiplication law is justified on physical grounds in Section
\ref{subsec:Deformed-Ring}.  The resulting noncommutative algebras are illustrated
in many concrete examples in Section \ref{sec:Formal-line-operators}. In Section
\ref{sec:Examples-of-M} we describe the commutative ring of Darboux coordinates in a way suitable
to make clear that the algebras of Section \ref{sec:Formal-line-operators} are indeed
noncommutative deformations.  In one important class of examples (the $A_1$ theories)
this algebra is related to the algebra of
quantum geodesic length operators in quantum Teichm\"uller theory,
as described in Section \ref{sec:Quantum-Holonomy}.

The main difference between the  PSC and the ``refined BPS index,''
which has been the subject of many recent investigations
\cite{Gukov:2004hz,Diaconescu:2007bf,Dimofte:2009bv,Dimofte:2009tm}, is
that the PSC is, on \emph{a priori} grounds, an index. Nevertheless, in the
examples we have examined, the two quantities agree. This agreement is nontrivial
and implies the absence of certain representations of supersymmetry among both the
framed and vanilla BPS states. This surprising agreement is formalized as a ``strong positivity
conjecture'' in Section \ref{subsec:BPS-Deg-PSC}. The truth of this conjecture would have far-reaching
consequences. We show in Section \ref{sec:Formal-line-operators} that strong positivity
implies that the ring of line operators and their vevs can be computed essentially by
formal algebraic means.  Given the geometrical interpretations of PSC's this is a highly nontrivial
result.  Moreover, the strong positivity properties of line operators are reminiscent of some
deep mathematics, including the canonical bases of quantum groups of Lusztig and Kashiwara,
the Laurent phenomenon of cluster algebras of Fomin and Zelevinsky, and the universal Laurent polynomials of
Fock and Goncharov.  We have not attempted to make these connections very precise,
and we leave that to future work.

The framed BPS states undergo wall-crossing analogous to that of the vanilla BPS states.
The analog of walls of marginal stability are the BPS walls defined in \eqref{eq:Def-BPS-Walls}.
Near these walls some of the framed BPS states have a very simple physical description
which is closely analogous to the ``halo'' configurations of multi-centered supergravity solutions
which were discovered by Frederik Denef. The wall-crossing mechanism is essentially identical
to that described in \cite{Denef:2007vg} in which entire Fock spaces of halo configurations
are created or annihilated as parameters cross a BPS wall. The third main result of this paper is a
formula describing the  wall-crossing of the generating functions \eqref{eq:intro-eq-2}.
It is essentially given by conjugation with quantum dilogarithms (the precise
formula is stated in \eqref{eq:Spin-KS-tmn-S} and
\eqref{eq:Spin-KS-ii}).

A corollary of this framed wall-crossing formula is a simple physical derivation of the
so-called ``motivic wall-crossing formula'' of Kontsevich and Soibelman for the \ti{vanilla} PSC's,
described in Section \ref{subsec:Motivic-WCF}.
Roughly, the idea is to study the spectrum of the vanilla BPS states indirectly through their effects on the
framed ones.  (This is related to our earlier derivation of the wall-crossing formula
\cite{Gaiotto:2008cd}, which also involved studying vanilla BPS states indirectly through their effects on
something else --- in that case, on the effective Lagrangian of a dimensionally reduced theory.)
In this way we show that the motivic KSWCF follows directly from the
simple physics of halo Fock spaces.

Finally, in the remainder of the paper we demonstrate our fourth main result:  a description
 of how to compute framed BPS indices and framed PSC's
in an interesting class of examples of $\CN=2$ theories, namely the class $\CS$ of theories
obtained by a partially twisted compactification of the $d=6$ nonabelian $(2,0)$ theories
on a Riemann surface $C$ of genus $g$ with $n$ punctures
\cite{Witten:1997sc,Gaiotto:2009we,Gaiotto:2009hg}. After a brief review of $d=6$
 $(2,0)$ theories in Section \ref{sec:Six-Dim-View} we use the
  surface operators of the $d=6$ theory
to  arrive at the key result \eqref{eq:L-5d-Holon} which makes computations possible. This
result states that one can
identify $\CM$ with a moduli space of flat connections on $C$ and express the
quantum expectation value of a line operator as a classical holonomy of the flat connection.
In Section \ref{sec:Examples} we
spell out many examples drawn from the class of ``$A_1$ theories,'' theories
in $\CS$ based on the $d=6$ theories with $\lieg = su(2)$.  Along the way, in Section \ref{sec:A1},
we extend the classification of line operators for the conformal $A_1$ theories
(proposed in \cite{Drukker:2009tz}) to asymptotically free theories.
The upshot is that isotopy classes of
closed curves on $C$ are replaced by objects known as ``laminations''
(see Section \ref{subsec:Lams-Def} for the definition).  We thereby make contact with
some very interesting work of Fock and Goncharov \cite{MR2349682,MR2567745}.
Once this connection is made, computations of framed BPS indices are in principle
straightforward and can be reduced to an essentially well-known algorithm --- which we call the \emph{traffic rule}
algorithm --- described in detail in Appendix \ref{app:traffic}.

The results of the fairly straightforward traffic-rule computations lead to some physically
surprising and mathematically nontrivial results. For example, the Wilson line amplitude in the
fundamental representation for the $SU(2)$ $\CN_f=0$ theory is given by a three-term
expression \eqref{eq:wilson-vev-nf0}
of the form
\begin{equation}
\CY_{\gamma_e} + \CY_{-\gamma_e} + \CY_{\gamma_m},
\end{equation}
where $\gamma_e$ is half the charge of the $W$-boson and $\gamma_m$ is a mutually nonlocal charge.
The first two terms are expected from semiclassical reasoning but the third one is a
suprise. It is exponentially small in the weak coupling domain. One can go on in this
vein with several examples, as we do in Section \ref{sec:Examples}. We mention here only
one more example: In
the $SU(2)$ $\CN_f=4$ theory at strong coupling the Wilson line supports $7$ framed BPS hypermultiplets,
as revealed by the simple computation  \eqref{eq:Nf4-Wilson}. It would probably be somewhat
nontrivial to reproduce this result by studying the geometry of monopole moduli spaces.

In Sections \ref{sec:Cluster-Algebras} and \ref{sec:Tropic} we briefly explore some
relations to interesting mathematics, and show that these relations have useful
physical applications. In Section \ref{sec:Cluster-Algebras} we show that the algebra
of formal line operators is an example of a quantum cluster algebra.  This viewpoint
proves useful in constructing the formal line operators for the $\CN=2^*$ theory
and suggests refinements of some of the positivity conjectures in the cluster algebra
literature. We introduce the concept of a ``cluster $\CN=2$ theory.''
In Section \ref{sec:Tropic} we similarly introduce a notion of a ``tropical $\CN=2$
 theory,'' and  we introduce ``tropical labels'' for line operators.
These should be thought of as defining a kind of UV to IR mapping of the enumeration
of (simple) line operators. In these two sections we have just scratched the surface of what
is perhaps a very deep connection.

Some technical arguments and some conventions are relegated to the appendices.

Despite its (regrettable) length, the present paper leaves many pretty stones unturned. We point out a few
open problems and future directions for research in Section \ref{sec:Open-Problems}.

\section{BPS line operators}\label{sec:Susy-BPS-line}

\subsection{Definition of line operators of type $\zeta$ }

In this paper we will follow the definition of line operators
advocated in \cite{Kapustin:2005py,Kapustin:2006pk}. We consider a
quantum field theory defined by relevant perturbations from an
ultraviolet fixed point theory, and a line operator is a conformally
invariant boundary condition for that uv theory on $AdS_2 \times
S^2$.

In this paper we focus on theories with $d=4, \CN=2$ supersymmetry.
Thus the uv fixed point has $su(2,2\vert 2)$ superconformal
symmetry. (Our conventions are spelled out in Appendix
\ref{app:Susy-Conventions}.) We will usually focus on line operators  in $\IR^{1,3}$
which are located at a spatial origin $x^1 = x^2 =x^3 =0$ and extend in the
time direction.  In this case the
unbroken subalgebra of the conformal algebra $so(2,4)$ is generated
by $D,P_0,K_0$, (these  generate a copy of $sl(2,\IR)$), and spatial
rotations $M_{ij}$ (these generate a copy of $so(3)$). Thus, the uv
boundary condition defining our line operator preserves
$sl(2,\IR)\oplus so(3) \cong so(4^*)$, the Lie algebra of isometries of
$AdS_2 \times S^2$. In addition we choose to study operators
preserving the $su(2)_R$ $R$-symmetry of the superconformal algebra.
We also choose to preserve half the supersymmetry, and hence we want
to study line operators preserving the superalgebra $osp(4^*\vert
2)$.  There are, in fact many ways of choosing the odd generators of
this subalgebra, and the existence of this family will be of some
importance in what follows. There is an   involution $I_\zeta$ of
the superalgebra induced by conjugating the  spatial reflection $x^i
\to - x^i$ by a $U(1)_R$ transformation of the superconformal group.
On the Poincar\'e supersymmetries this involution acts by
\begin{equation}
\begin{split}
Q_\alpha^A & \rightarrow \zeta^{-1}  \epsilon^{AB}
\sigma^0_{\alpha\dot \beta}\bar
Q^{\dot \beta}_{~ B} \\
\bar Q^{\dot \alpha}_{~ A} & \rightarrow  \zeta  \epsilon_{AB}
\bar\sigma^{0 \dot \alpha\beta} Q_\beta^{~B}
\end{split}
\end{equation}
with a similar action on the special conformal supersymmetries
$S_\alpha^{~A}, \bar S^{\dot \alpha A}$. Here $\zeta$ is a phase
  resulting from the $U(1)_R$ rotation. The
fixed subalgebra of this automorphism will be denoted $osp(4^*\vert
2)_\zeta$ and is isomorphic to  $osp(4^*\vert 2)$ for all $\vert \zeta \vert =1 $.
 Indeed, all such subalgebras are rotated into each other by a $U(1)_R$ transformation.
We will denote line operators preserving the above  superalgebra
by $L(\zeta ; \cdots)$, or $L_\zeta( \cdots ) $,  where $\cdots$ will
refer to other parameters defining a specific line operator such as
those discussed in Section \ref{subsec:LabelsForLines} below.
As we will see from examples in Sections \ref{sec:Formal-line-operators}
and \ref{sec:Examples}, the line operators are in general only single-valued on the universal
cover of the $\zeta$  circle.

We will be studying field theories which are perturbed from their
superconformal fixed points either by moving on the Coulomb branch
or by turning on masses. In the IR description of the theory there
is $d=4, \CN=2$ Poincar\'e supersymmetry. The involution $I_\zeta$
also acts on this superalgebra with the additional rule that
$I_\zeta: Z \to \zeta^2 \bar Z$, where $Z$ is the central charge
operator.   It will be convenient for us to choose a squareroot of
$\zeta^{-1}$, call it $\xi$, and define the fixed supercharges under
the involution by
\begin{equation}\label{eq:Q-zeta-susy}
\CR_{\alpha}^{~A} = \xi^{-1} Q_{\alpha}^{~A} + \xi \sigma^0_{\alpha
\dot \beta } \bar Q^{\dot \beta A}
\end{equation}

The theory in the presence of a line operator $L$ extended along
the time direction has a Hilbert space denoted $\CH_L$. The $\CR_{\alpha}^{~A}$
are operators on this Hilbert space.

\subsection{Semiring of line operators}\label{subsubsec:RingStructure}

An interesting observation \cite{Kapustin:2006hi}, developed in
further detail in \cite{Kapustin:2007wm}, is that supersymmetric
line operators in ${\CN=2}$ gauge theories form a semiring,
with a rather intricate structure, which depends in detail on the
matter content of the theory.

To define the sum we   say that the correlation function
of the sum  $L+L'$ of two line operators is simply the
sum of the correlation functions of $L$ and $L'$.
The Hilbert space of the theory in the presence of the sum of two operators
is simply the sum of two superselection sectors, the Hilbert spaces associated with the two operators:
\begin{equation}\label{eq:sum-line}
\CH_{L+L'}:= \CH_L \oplus \CH_{L'}.
\end{equation}
We define a \emph{simple line operator} to be a line operator which is
not a sum of two other line operators.

 Moreover, one can define a product of operators $LL'$ by considering
the path integral with the insertion of two line operators.
The line operators we are considering have the special property
that their correlation functions are independent of the distance of separation.
To see this,  note that the
supersymmetries $T_{\alpha}^A = \xi^{-1} Q_\alpha^A -  \xi
\sigma^0_{\alpha \dot \beta } \bar Q^{\dot \beta A}$ which are odd
under the involution $I_\zeta$ satisfy the commutation relation
\begin{equation}
[\CR_\alpha^A, T_\beta^B ] = 2 \epsilon_{\alpha\beta} \epsilon^{AB}
(\bar   Z - \zeta^{-2} Z) + 8 \zeta^{-1} \epsilon^{AB}
\sigma^{0m}_{\alpha\beta} P_m.
\end{equation}
Taking the symmetric part in $\alpha\beta$ we can write
the spatial translation operators $P_i$ as a $\CR_\alpha^A$
commutator, and therefore correlation functions of line operators of
type $I_\zeta$ all aligned with the $x^0$ axis will be independent
of space coordinates.
We can let these operators approach one
another and then, by locality, the product should be equivalent to some other line
defect. Thus,  the operator product
expansion should define a ring multiplication. We will determine it for
several interesting examples below.
It is interesting to relate the Hilbert space associated to the product of two line operators
to the Hilbert spaces associated to the two line operators. This is possible, but surprisingly subtle.
It will lead us to define an interesting non-commutative deformation of the product
of line operators in Section \ref{subsec:Deformed-Ring}.

\subsection{Labels for line operators in Lagrangian gauge theories}\label{subsec:LabelsForLines}

Many $d=4, \CN=2$ theories admit a Lagrangian description, in terms of
vectormultiplets and hypermultiplets. For such theories we can be more specific about
the definition of the line operators we are interested in following
\cite{Kapustin:2005py} and \cite{Kapustin:2006hi}
(notice that our $\zeta$ is $t^{-1}$ in the latter reference).

We begin by reviewing some Lie algebra theory.
Let us first begin with  a simple Lie algebra  $\lieg$ and define some
standard lattices. Choose a Cartan subalgebra $\liet \subset \lieg$.
Then there is a canonical set of roots $\Phi(\lieg)\subset \liet^*$ which
arise from diagonalizing  the adjoint action of $\liet$
on $\lieg$. For each root $\alpha \in \Phi(\lieg)$ there is a copy
of $sl(2)_\alpha \subset \lieg$ and we can canonically define
the corresponding \emph{coroot}
$H_{\alpha}\subset \liet$ which generates a Cartan subalgebra
of $sl(2)_\alpha$ and is normalized so that
 $\langle \beta , H_{\alpha} \rangle = 2 (\beta,\alpha)/(\alpha,\alpha)$.
(On the left we write the canonical pairing of $\liet^*$ with $\liet$, while
 on the right we have used a Cartan-Killing form, but the normalization drops out
 since we assume $\lieg$ is simple.)
The roots and coroots  generate lattices $\Lambda_r\subset
\liet^*$ and $\Lambda_{cr}\subset \liet$, respectively.
There are two useful ways to think about the coroot lattice.
  By Lie's theorem there is a unique simply connected
and connected compact Lie group $\tilde G$ whose lie algebra is
$\lieg$.  Elements of the coroot lattice can be identified with
homomorphisms of $U(1)$ into
$\tilde T$, the maximal torus of $\tilde G$ with Lie algebra $\liet$.
Alternatively, $\Lambda_{cr}$ is the kernel
of $\exp[ 2\pi \cdot]: \liet \to \tilde T$.

Using the
perfect pairing of $\liet$ with $\liet^*$ (and \emph{not} the Killing
form) we then have canonically
defined weight and magnetic weight lattices:
\begin{equation}
\begin{split}
\Lambda_{wt} & := \Lambda_{cr}^* \subset \liet^* \\
\Lambda_{mw} & := \Lambda_{r}^* \subset \liet.
\end{split}
\end{equation}
 A standard result states that the center $Z(\tilde G)$ is
isomorphic to the quotients
\begin{equation}
Z(\tilde G) \cong \Lambda_{mw}/\Lambda_{cr} \cong
\Lambda_{wt}/\Lambda_r.
\end{equation}
Finally, let  $\CW$ be the Weyl group of the Lie algebra $\lieg$.
This is isomorphic to $N(\tilde T)/\tilde T$ where
$N(\tilde T)$ is the normalizer of
$\tilde T$ in $\tilde G$.

Now, our theories will have matter fields in general and therefore we
should consider a general compact connected  simple Lie group with Lie
algebra $\lieg$. Again by Lie's theorem, such a group must be of the
form  $G\cong \tilde G/\CZ$, where $\CZ\subset Z(\tilde G)$ is
a subgroup of the center. Thus, $Z(G) \cong Z(\tilde G)/\CZ$.
 Since the center is fixed by the action of
the Weyl group   we can find a $\CW$-invariant lattice $\Lambda_G$,
where
\begin{equation}
 \Lambda_{cr}\subset \Lambda_G \subset \Lambda_{mw}
 \end{equation}
so that $\CZ \cong  \Lambda_G/\Lambda_{cr} $. Indeed can identify
$\Lambda_G$ with the kernel of the exponential map for the group $G$:
\begin{equation}
\Lambda_G = \{ P \in \liet \vert \exp(2\pi P) = 1_G \}
\end{equation}
That is $\Lambda_G \cong {\rm Hom}(U(1), T)$, where $T= \tilde
T/\CZ$ is a maximal torus of $G$. Dually, the character group of
$G$ is ${\rm Hom}(T, U(1))$, and is canonically dual to ${\rm
Hom}(U(1),T)$ because there is a perfect pairing
\begin{equation}
{\rm Hom}(U(1),T) \times {\rm Hom}(T,U(1)) \rightarrow {\rm
Hom}(U(1),U(1))\cong \IZ
\end{equation}
 Thus the character group is
isomorphic to $\Lambda_G^*$, and for this reason $\Lambda_G$ is sometimes
called the cocharacter lattice of $G$.  $\Lambda_G^*$  can be identified with the weight
lattice $\Lambda_{wt}(G)$. This lattice may also be viewed as the
weights in representations of $\tilde G$ which transform trivially
under $\CZ$. This completes our review of Lie algebra theory.

As we have said, with explicit field multiplets we can construct some
explicit line operators. The most obvious line operators we can consider are
the  supersymmetric Wilson lines \cite{Baulieu:1997nj,Maldacena:1998im,Rey:1998ik}
along a path $p$ along the time direction, at some fixed $\vec x$,
in a representation $\rho_\CR: G\to {\rm End}(V)$ of the gauge group:
\begin{equation}\label{eq:Susy-Wilson-Zeta}
L_\zeta(\vec x ;\CR) =
\rho_{\CR} P\exp \oint_{p} \left( \frac{\varphi}{2\zeta } - i A
- \frac{\zeta \bar \varphi}{2 }  \right).
\end{equation}
In addition, we can construct 't Hooft operators. Recall that these are
defined by imposing boundary conditions on
the fields in the infinitesimal neighborhood of the line operator. In our case
 we require the limiting values of the fields on small linking spheres $S^2$
  to correspond to   a supersymmetric magnetic monopole preserving the Poincar\'e
  subalgebra of $osp(4^*\vert 2)_\zeta$.
Recall that spherically symmetric magnetic monopole in a theory with gauge group $G$ is
specified by a transition function for a $G$-bundle on  $S^2$ (which determines the
principal bundle on the physical space $\IR^4-\{p\}$). Such $G$-bundles are
 identified with $\Lambda_G
\cong {\rm Hom}(U(1),T)$. Indeed, for any such homomorphism we can construct
a solution preserving the Poincar\'e subalgebra of  $osp(4^*\vert 2)_\zeta$, by
slightly modifying the construction of Kapustin. (See
Appendix \ref{app:Fixed-Point-Equations}
equation \eqref{eq:Embed-fix-sol}). Thus, each homomorphism defines an 't Hooft line
operator $L_\zeta (p;v)$ where $v \in \Lambda_G$. \footnote{
Here it is important to
understand $\zeta$ to be valued in the covering space of $U(1)_R$. This is not obvious
from the definition we have reviewed. However, quantum anomalies relate a change in
the phase $\zeta$ to a shift in the $\theta$-angle. The Witten effect for line operators
\cite{Kapustin:2005py,Henningson:2006hp}
shows that the 't Hooft operator cannot be a single-valued function of $\zeta$.  }

As is well-known, one can combine the above two constructions
and define Wilson-'t Hooft operators $L_\zeta(p;v,\CR)$ where $\CR$ is a
representation of the commutant of $v$. As shown in \cite{Kapustin:2005py} the  set of such
Wilson-'t Hooft operators is \emph{ a priori}  labeled by elements of
 $\Lambda_{mw} \times \Lambda_{wt}$.
However, there are locality conditions on collections of compatible
line operators.  For example,
in the presence of an 't Hooft operator labeled by $(P,0)$
a Wilson line operator labeled by $(0,Q)$ will only make sense
for true (not projective) representations of the structure group
of the gauge bundle on $\IR^3- \{0 \} $. More generally, we expect
that  consistent sets of simple line operators can be constructed from
maximal  sublattices $\CL\subset \Lambda_{mw} \times  \Lambda_{wt}$
which   satisfy a Dirac-like quantization condition:
\begin{equation}\label{eq:Line-Op-Locality}
\forall (P,Q), (P',Q')\in \CL : \qquad \qquad \langle P, Q' \rangle - \langle P' , Q \rangle \in \IZ
\end{equation}
Since operators   related by Weyl transformations are gauge-equivalent we should
require that $\CL$ is $\CW$-invariant, and in this case the gauge isomorphism
classes of the line operators are in one-one correspondence with $\bar\CL=\CL/\CW$.
  A natural example of
 such a consistent set is
\begin{equation}\label{eq:Wilson-tHooft-B}
\bar\CL_G := (\Lambda_G \times \Lambda_G^*)/\CW
\end{equation}
At the two extremes, for $G=\tilde G$ we have $(\Lambda_{cr}\times \Lambda_{wt})/\CW$
while for the adjoint group $G = G_{\rm ad}:= \tilde G/Z(\tilde G)$
we have $(\Lambda_{mw} \times \Lambda_{r})/\CW$.

The most general maximal
mutually local sublattice
$\CL$ sits in an exact sequence
\begin{equation}\label{eq:label-seq}
0 \rightarrow \Lambda_G^* \rightarrow \CL \rightarrow \Lambda_G  \rightarrow 0
\end{equation}
In order to see this consider first the projection $(P,Q) \to P$. This defines a sublattice of
$\Lambda_{mw}$ which must be $\Lambda_G$ for some $G$. Then since
$\CL$ is maximal and local, the kernel of the map must be $(0,Q)$ where
$Q \in \Lambda_G^*$.  While the sequence \eqref{eq:label-seq} is split it is
not naturally split --- reflecting the possibility of the Witten effect ---
so if we consider local systems (e.g. over the coupling constant
space or the  $\zeta$ plane $\IC^*$) then \eqref{eq:label-seq} can be nontrivial.

A more general mutual locality condition on line operators can be stated
which should apply to nonlagrangian theories.
The Hilbert space in the presence of the line operator $\CH_L$ should
be a representation of the spatial rotation group $SU(2)$. If we consider
two     line operators $L_1,L_2$
at $\vec x_1, \vec x_2\in \IR^3$ then the rotation group is broken to
the group $U(1)_{\vec x_1 \vec x_2}$ which double-covers the group of rotations about the
$\vec x_1 \vec x_2$ axis. The Hilbert space $\CH_{L_1 L_2}$ associated
to the product of the line operators is certainly a representation
of the universal cover of $U(1)_{\vec x_1 \vec x_2}$. Our mutual locality
condition is the statement that $\CH_{L_1 L_2}$ is
 a true representation of $U(1)_{\vec x_1 \vec x_2}$.
In $\CN=2$ gauge theories with a Lagrangian description we can recover
the previous statement of the mutual locality condition by moving far onto
the Coulomb branch so that the fields sourced by the line operators can be
put in the Cartan subalgebra. Then we can apply the classical computation
of the electromagnetic field in the presence of a pair of dyons to derive
\eqref{eq:Line-Op-Locality}.

\subsubsection{Example:  $A_1$ theories}\label{subsubsec:A1-labels}

In this section we will discuss line operator labels in $A_1$ theories.
These are the most general Lagrangian $\CN=2$ SCFTs built from $su(2)$ gauge groups.
They are members of a larger class of theories denoted
$\CS$ in \cite{Gaiotto:2009hg}, to which we refer for background and
notation. These $d=4, \CN=2$ theories are defined by
considering a partially twisted $ADE$ $(0,2)$ six dimensional theory on a Riemann surface $C$
which is decorated with punctures.  The remarkable S-duality properties of these theories were
elucidated in \cite{Gaiotto:2009we}. A typical theory in the class $\CS$ admits several S-dual descriptions,
as a non-Abelian gauge theory weakly coupled to non-Lagrangian matter theories. We expect
half-BPS line operators of the Wilson-'t Hooft type to exist for the non-Abelian gauge groups,
but we do not know at this point how to classify the line operators which are available in the
non-Lagrangian matter theories.
The $A_1$ theories are a useful example, as they share many of the interesting properties of
the theories of the $\CS$ class, but have a Lagrangian description in all S-duality frames.

A warmup example is an $su(2)$ gauge theory with  ${\cal N}=4$ supersymmetry \cite{Kapustin:2006pk}.
There is a single coroot $H$ and $\Lambda_{cr} = H \IZ$, and there is a single
root $\alpha$ so $\Lambda_r = \alpha \IZ$. The duality pairing is
$\langle \alpha, H\rangle = 2$ and hence $\Lambda_{wt} = \half \alpha \IZ$
and $\Lambda_{mw} = \half H \IZ$.  Electric charge and magnetic charge
are traditionally defined by choosing isomorphisms $\Lambda_{wt} \cong \IZ$
and $\Lambda_{mw} \cong \IZ$ respectively, so we identify
$(P,Q) = (\frac{p}{2} H, \frac{q}{2}\alpha)$ with a pair of integers $(p,q)\in \IZ\times \IZ$.
The consistency condition \eqref{eq:Line-Op-Locality} becomes
$pq'-p'q= 0 \mod 2$.
  With this isomorphism
understood, if  $G = SU(2)$ then
$\Lambda_G \times \Lambda_G^* = 2\IZ \times \IZ$, i.e. we have even magnetic charge
and any integral electric charge whereas if $G=SO(3)$ we have
$\Lambda_G \times \Lambda_G^* =  \IZ \times 2\IZ$, i.e. any integral magnetic charge,
but even electric charge. The two are related by $S$-duality. However, there is a third
consistent set of line operators, which is not of the form $\CL_G$, and is obtained by
shifting the theta angle of the $SO(3)$ theory by a half-period. \footnote{If we normalize the
theta angle for the $SO(3)$ theory by $\frac{\theta}{8 \pi^2} \int_M {\Tr}_{2} F\wedge F$ then
the periodicity of $\theta$ on a general 4-manifold $M$ is $8 \pi$. However, on
spin $4$-manifolds it is $4\pi$. Of course, $\IR^4$ is spin.
In this sense we are shifting by a half-period.} This consists of the
set of $(p,q)$ such that $p=q \mod 2$.

Now let us return to the $A_1$ theory on a curve $C$ of genus $g$ with $n$ punctures.
Recall that a weak-coupling
limit is specified by choosing a pants decomposition associated with separating curves
$c_i$, $i=1,\dots, 3g-3+n$. The Lie algebra of the gauge group is $\lieg = su(2)^{\oplus 3g-3 + n}$,
and there are hypermultiplets $\Phi_{i_1i_2i_3}$ associated with each trinion bounded by
a triplet of curves $(c_{i_1}, c_{i_2},c_{i_3})$. These hypermultiplets are in the
representation $2_{i_1}\otimes 2_{i_2} \otimes 2_{i_3}$ with a reality condition
(for a total of $16$ independent real scalar fields) \cite{Gaiotto:2009we}.

First of all, let us note that there are several possible gauge groups associated with a given
weak coupling cusp.
Let $\tilde G= SU(2)^{3g-3+n}$. Then allowed gauge groups are of the form $\tilde G/\CZ$ where
$\CZ$ is any subgroup of the center $Z(\tilde G)$ which acts trivially on all the $2g-2+n$ hypermultiplets.
We can express this more concretely by associating a number $\epsilon(c_i)\in \IZ_2$
for each $i$ thus fixing an isomorphism $Z(\tilde G) \cong \IZ_2^{3g-3+n}$.
Then   the subgroup $\CZ_{\rm max}$ which acts trivially on the trinions is
defined by $2g-2+n$ conditions on the vectors $\vec \epsilon$. There is one relation between
these relations and hence   $\CZ_{\rm max}$ is generated by $g$ elements.
Fix some isomorphism $\CZ_{\rm max} \cong \IZ_2^g$. We denote $G_{\rm max}:= \tilde G/\CZ_{\rm max}$.
The allowed gauge groups for the $A_1$ theory at the cusp defined by the
separating curves $c_i$ are
in one-one correspondence with subgroups   $\CZ \subset \CZ_{\rm max}$.
(Incidentally, the number of such subgroups is the Galois number
\begin{equation}
\CG_g = 1+ \sum_{k=1}^g \frac{ (2^g-1)(2^{g-1}-1) \cdots (2^{g-k+1}-1)}{(2^k-1)(2^{k-1}-1)\cdots (2-1)}.
\end{equation}
For large $g$ we have $  \CG_g \sim 2^{\frac{1}{4}g^2 }$, so there can be quite a large
variety of choices.)

Now, for  $\lieg = su(2)^{\oplus 3g-3+n}$ we extend the isomorphism described above for $su(2)$
to an isomorphism $\Lambda_{mw} \times \Lambda_{wt} \cong \IZ^{3g-3+n} \times \IZ^{3g-3+n}$. Consistent
sets of line operators will be constructed from maximal $\CW$-invariant
collections of vectors $(\vec p, \vec q) \subset \Lambda_{G_{\rm max}} \times \Lambda_{wt}$
satisfying $\vec p\cdot \vec q' - \vec p' \cdot \vec q = 0 \mod 2$ for all pairs $(\vec p, \vec q)$
and $(\vec p', \vec q')$.
The vector $\vec p$ specifies an $SO(3)^{3g-3+n}$ bundle on $\IR^3-\{ 0 \} $ and the restriction
$\vec p \in \Lambda_{G_{\rm max}}$ arises because the hypermultiplet fields $\Phi_{abc}$ must
be sections of the associated bundle in the representation $2_a \otimes 2_b \otimes 2_c$.
We fix a fundamental domain for the action of the Weyl group by taking $p_i \geq 0$ and
$q_i \geq 0$ if $p_i=0$.

A beautiful
observation of Drukker, Morrison, and Okuda  \cite{Drukker:2009tz} is that the classification
of simple line operators in $A_1$ theories
 is closely related to the Dehn-Thurston classification of isotopy classes of
\noself\ curves on $C$. Recall that for such a curve $c$
 we can define $p_i = c \# c_i$ and $q_i = {\rm Twist}_{c_i}(c)$.
Here $\# $ is the homotopy intersection number, defined as the positive sum of intersections,
minimized over isotopy classes. For details of the twist see \cite{Drukker:2009tz}.
(In what follows $c$ sometimes denotes a curve, and sometimes denotes its isotopy class,
depending on context.)  It turns out
rather beautifully that $p_a + p_b + p_c = 0 \mod 2$ for all trinions and this is precisely
the condition that $\vec p \in \Lambda_{G_{\rm max}}$.  We can therefore
 define a  one-one and onto map, which we denote $\tau_{DMO}$, from the set of all  isotopy classes of
 \noself\ paths on $C$
 to    $(\Lambda_{G_{\rm max}} \times \Lambda_{wt})/\CW$.

Now, what consistent sets of line operators, $\CL$,  can we define in these theories?
Under the correspondence
$\tau_{DMO}$
the condition \eqref{eq:Line-Op-Locality} is equivalent to the condition
\begin{equation}\label{eq:max-even}
c \# c' = 0 \mod 2.
\end{equation}
We will call a set $\CP$ of (isotopy classes of) paths \emph{even} if
\eqref{eq:max-even} holds for all pairs in $\CP$, and \emph{maximal even}
if $\CP$ is not properly contained
in any larger even set.  Thus, a consistent set of line operators $\CL$ defines
a maximal even set of paths on $C$. In particular, if we choose $G = \tilde G/\CZ$
for some subgroup $\CZ \subset \CZ_{\rm max}$, then $(\Lambda_G \times\Lambda_G^*)/\CW$
determines a maximal even set of paths which we can denote $\CP_G$.
Conversely, given $\CP$ we can project it to the magnetic charges
and reconstruct an exact sequence as in \eqref{eq:label-seq}.

The $S$-duality group, i.e. the modular group, will change the gauge group, as in the
$\CN=4$ example mentioned above. It is possible to give an explicit formula for how
$\CZ$ changes in terms of mod-two homology. This complements the discussion of the
$S$-duality action on the line operators in \cite{Drukker:2009tz}.

\section{Hilbert spaces, halos, and wall-crossing}\label{sec:Hilbert-Halo-WC}

\subsection{Framed BPS States} \label{sec:framed-BPS-states}

So far we have only discussed UV aspects of the $\CN=2$ theory.
We are now going to pass to the IR.  We need a little notation.
We denote the Coulomb branch as $\CB$, its singular divisor as
$\CB^{sing}$, and the lattice of vanilla electromagnetic and flavor charges
as $\Gamma$.
\footnote{As is well known  $\Gamma$ undergoes monodromy if $u$ is continued
along nontrivial paths in $\CB^*:=\CB - \CB^{sing}$. Thus one should speak
of $\Gamma_u$. In mathematical terms $\Gamma$ is a fibration of lattices
over $\CB^*$ and defines a ``local system,'' i.e. it has a flat connection.
One can work on the universal cover $\widehat\CB$ of
$\CB-\CB^{sing}$ where $\Gamma$ may be trivialized, or one
can work on the base, bearing in mind that there is nontrivial monodromy.
This is a matter of taste which we leave to the reader. For further background
see  \cite{Gaiotto:2008cd,Gaiotto:2009hg} and references therein.}

We now study the Hilbert space of the theory on
$\IR^{1,3}$, with vacuum at infinity labeled by $u\in \CB$, and with a simple line operator
$L_\zeta$ inserted at the origin $x^i=0$.  This Hilbert space, which we
denote by $\CH_{u,L,\zeta}$, is a representation
of the Poincar\'e sub-superalgebra of  $osp(4^*\vert 2)_\zeta$.
In addition it is graded by the charge:
\begin{equation}\label{eq:Grade-Hilbert}
\CH_{u,L,\zeta} = \bigoplus_{\gamma \in \Gamma_L}
\CH_{u,L,\zeta,\gamma}.
\end{equation}
Here $\Gamma_L$ is the Poisson lattice of
electromagnetic and flavor charges in the presence of $L$, and is a
torsor for the lattice $\Gamma$ of vanilla charges.  That is, it is of the form
\begin{equation}\label{eq:Lattice-Shift}
\Gamma_L = \Gamma + \gamma_L
\end{equation}
where $\gamma_L \in \Gamma \otimes \IR$, and $\langle \gamma_L, \gamma\rangle \in \IZ$
for all $\gamma \in \Gamma$.
In terms of a path integral formulation, the direct sum over $\Gamma$ in \eqref{eq:Grade-Hilbert}
becomes a sum over topological sectors.

The need for a sum over $\gamma$ in \eqref{eq:Grade-Hilbert}
can be seen easily in a weakly coupled pure
$SU(N)$ theory, far out on the Coulomb branch so that the gauge symmetry
is strongly broken to $U(1)^{N-1}$. Consider for example the Wilson line operator
\eqref{eq:Susy-Wilson-Zeta}.
In the naive classical limit the vev of this operator
would be a sum of vevs corresponding to supersymmetric Wilson
lines of the IR theory, labeled by the weights of the representation $\CR$.
\footnote{Perhaps surprisingly,
in concrete examples we will see that this naive answer is missing something:
there are extra contributions to the sum over weights.  A simple example
is the Wilson operator of pure $SU(2)$ gauge theory in the
fundamental representation, for which we give the answer in
\eqref{eq:wilson-vev-nf0} below. There are two terms corresponding to the
weights of the fundamental representation, but there is also a third term. We
will attribute that third term to an interesting boundstate of a particle with
the line operator.  }
This example also makes clear the physical origin of the shift in  \eqref{eq:Lattice-Shift}.
Consider the example of a fundamental Wilson line operator in a
pure $SU(2)$ gauge theory:  all vanilla BPS states   carry even electric charge
(in conventions where a W-boson has charge $2$), but a state in the presence of a
fundamental Wilson loop should be able to carry an odd electric charge.
The freedom to add vanilla BPS particles to the system makes it obvious that
$\Gamma_L$ should be a torsor for $\Gamma$.

The mutual locality condition on line operators shows that the $\Gamma_L$ for
a consistent set of line operators $L \in \CL$ should all lie in a common lattice $\Gamma_{\CL}$,
such that
\begin{equation}\label{eq:Consist-LO-Latt}
\Gamma \subset \Gamma_{\CL},
\end{equation}
with an integral antisymmetric form on $\Gamma_{\CL}$ restricting to the standard one on $\Gamma$.

Each of the sectors
$\CH_{u,L,\zeta,\gamma}$ is a representation of the Poincar\'e
sub-superalgebra of $osp(4^*\vert 2)_\zeta$.
When we quantize with time slices of constant $x^0$, the supersymmetry
operators satisfy the Hermiticity conditions
\begin{equation}
\begin{split}
(\CR_{1}^{~1})^\dagger & = - \CR_{2}^{~2}, \\
(\CR_{1}^{~2})^\dagger & =   \CR_{2}^{~1}.
\end{split}
\end{equation}
Moreover, in the theory deformed from its superconformal point, there will be sectors of the
Hilbert space with nonzero central charges for the $\CN = 2$ Poincar\'e
supersymmetry.  The operators $\CR_{\alpha}^{~A}$ satisfy the algebra
\begin{equation}
\{ \CR_{\alpha}^{~A} , \CR_{\beta}^{~B} \} = 4 \left( E + {\rm
Re}(Z/\zeta) \right) \epsilon_{\alpha\beta} \epsilon^{AB}
\end{equation}
where $E=P^0$ is the energy operator. Combining this with the Hermiticity
conditions gives the BPS bound
\begin{equation} \label{eq:bps-bound}
E + \Re \, (Z / \zeta) \geq 0,
\end{equation}
where $Z = Z_\gamma(u)$ is the standard central charge associated
with the vacuum $u\in \CB$ in the sector $\gamma$.

Now we are ready for one of the main definitions of this paper:  we define a
\emph{framed BPS state} to be one which saturates this BPS bound, i.e.
$E = -\Re \, (Z / \zeta)$. (The motivation for the name comes from the notion
of framed quiver representations.)
We can specialize
\eqref{eq:Grade-Hilbert} to the subspace consisting of framed BPS states:
\begin{equation}\label{eq:Grade-BPS-Hilbert}
\CH_{u,L,\zeta}^{\rm BPS} = \bigoplus_{\gamma \in \Gamma_L}
\CH_{u,L,\zeta,\gamma}^{\rm BPS}.
\end{equation}

Note that \eqref{eq:bps-bound}
differs from the standard BPS bound $E \geq \vert Z \vert$.
There is a nice heuristic for understanding \eqref{eq:bps-bound},
using an IR version of the standard interpretation of a Wilson-'t Hooft
operator as the insertion of an infinitely heavy dyon.
Extend the lattice $\Gamma$ by one extra
flavor charge $\gamma_f$, and consider a very heavy particle, of charge $\gamma_f - \gamma$
and central charge $Z = \zeta M  -  Z_\gamma$, where $M>0$.  The renormalized BPS
bound in the limit $M \to +\infty$ is just
\begin{equation}\label{eq:heuristic-bound}
E \geq  \lim_{M\to +\infty} \left(\vert \zeta M  -  Z_\gamma\vert - M \right)= - {\rm \Re}(Z_\gamma/\zeta).
\end{equation}

BPS particles
of total charge $\gamma$ have energy bounded below by $\vert Z_\gamma(u)\vert$.
However, we will argue below
that in the presence of line operators
there can be interesting quantum states
with energy levels below $\vert Z_\gamma(u)\vert$, but above the bound
$-\Re \, (Z / \zeta)$. Viewed from far away these look like a ``core particle''
of charge $\gamma_c$, located at the position of the line operator,
 interacting with  one (or more)  ``halo particle(s)'' of charge $\gamma_h$ such
that $\gamma_c + \gamma_h = \gamma$. States in which the halo particle is not bound -- i.e.
does not have its wavefunction essentially confined to a finite region of space --
have energies which form a continuum starting at
$-\Re \, (Z_{\gamma_c}(u) / \zeta)+ \vert Z_{\gamma_h}(u)\vert$.
In addition there are bound states
 analogous to the multi-centered boundstates of supergravity
\cite{Denef:2000nb,Denef:2000ar}.
For these states   the wavefunction of the
halo particle is essentially confined to a compact region of space.
Such states have a discrete energy spectrum.
As long as there is no halo charge which
saturates the bound $-\Re \, (Z_{\gamma_h}(u) / \zeta) = \vert Z_{\gamma_h}(u)\vert$,
there is a nonzero energy gap between the framed BPS states
and the continuum states.
In Figure \ref{fig:spectrum} we show a schematic picture of the energy spectrum.

\insfig{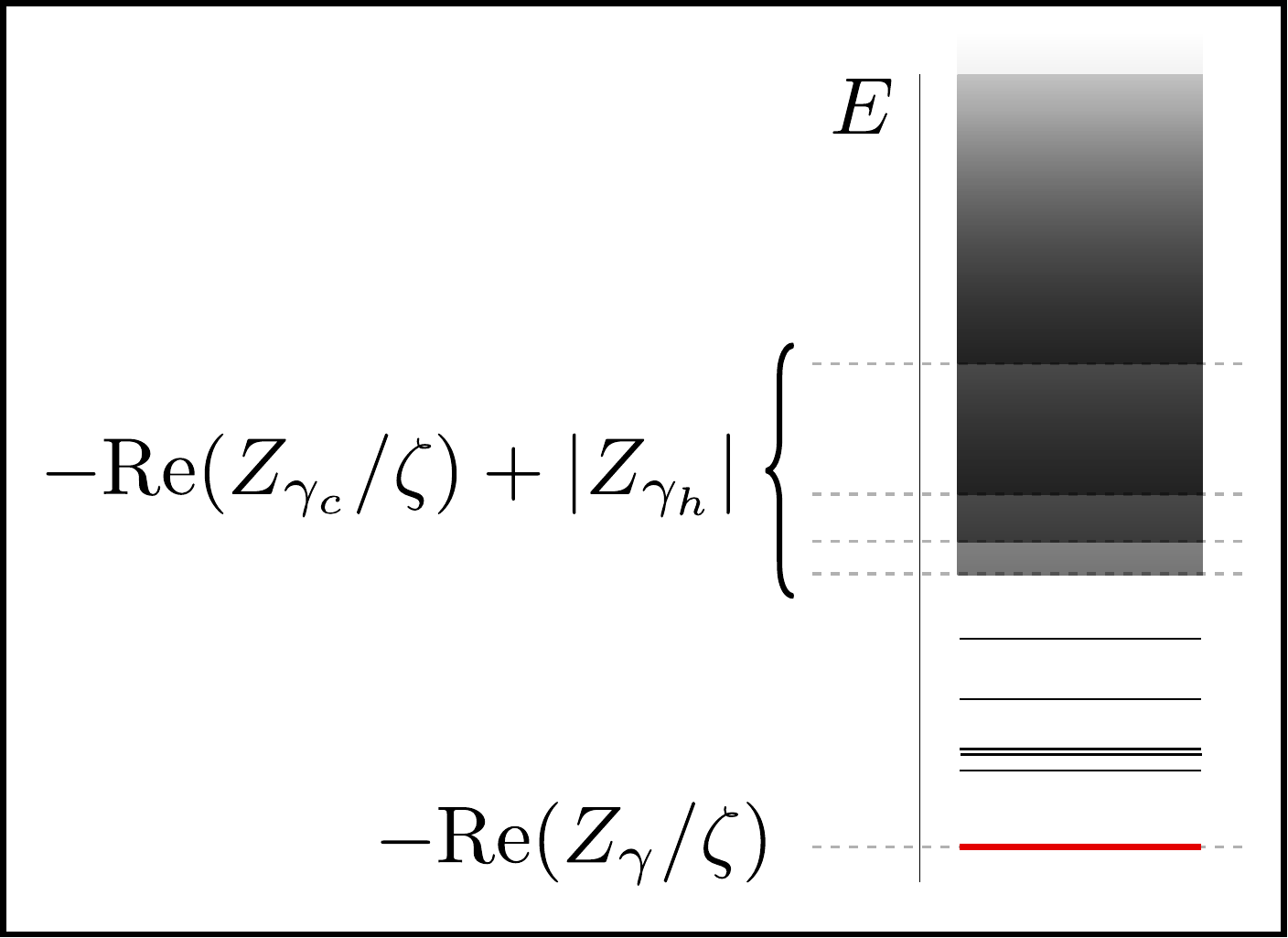}{A schematic picture of the spectrum of energies of states with charge $\gamma$, in the theory
with the line operator $L_\zeta$ inserted.
At the bottom we see the framed BPS states.  Next there are discrete excited states bound to the line operator.
Finally there are various continua of unbound states, corresponding to different possible decompositions of $\gamma$
as $\gamma_c + \gamma_h$.  The framed BPS states are safely separated from the continua except when
there is a $\gamma_h$ with $-\Re \, (Z_{\gamma_h}(u) / \zeta)= \vert Z_{\gamma_h}(u)\vert$.}

The condition $-\Re \, (Z_\gamma(u) / \zeta)= \vert Z_{\gamma}(u)\vert$ defines the \ti{BPS walls}:\footnote{In
\cite{Gaiotto:2008cd,Gaiotto:2009hg} we considered ``BPS rays,'' which are just the projection of the BPS walls
to the $\zeta$-sphere at fixed $u$.}
\begin{equation}\label{eq:Def-BPS-Walls}
\hW(\gamma ):=\{ (u,\zeta) \vert Z_{\gamma}(u)/\zeta \in
\IR_- \}\subset \widetilde \CB \times \widetilde\IC^*.
\end{equation}
As we vary parameters in the space $\tilde \CB \times \IC^*$, framed BPS states can
mix with the continuum whenever we hit one of the BPS walls.

\subsection{Framed BPS degeneracies and the Protected Spin Characters}\label{subsec:BPS-Deg-PSC}

The framed BPS states transform in representations of the Lie algebra
of spatial rotations $so(3)$.  At fixed grading $\gamma$ they are finite dimensional
and we can define a simple   index
 counting the framed BPS degeneracies defined by
\begin{equation}
\fro(u,L_\zeta, \gamma) := {\rm Tr}_{\CH_{u,L,\zeta,\gamma}^{\rm
BPS}} (-1)^{(2J_3)}
\end{equation}
In fact, we can define a more refined quantity which counts
framed BPS states.  We would like to define a quantity
which keeps track of the spin information. The simple spin character ${\rm Tr}
y^{2J_3}$ would naively do the job, but it might jump
unpredictably if several BPS states pair up into an unprotected
representation of the SUSY algebra.  To get a more robust index we recall
that framed BPS states transform under
$su(2)_R$ as well, and one of the four supercharges $\CR_\alpha^A$
is a singlet under the diagonal combination of the
spin $so(3)$ and $su(2)_R$, namely,
\begin{equation}
\CQ := \epsilon_{\alpha A} \CR_{\alpha}^A.
\end{equation}
So we can define what we will call a \emph{(framed) Protected Spin Character},
abbreviated PSC:\footnote{We thank Juan Maldacena for
suggesting this improved index.}
\begin{equation}\label{eq:Def-Framed-PSC}
\fro(u,L_\zeta, \gamma;y) := {\rm Tr}_{\CH_{u,L,\zeta,\gamma}^{\rm
BPS}} (-1)^{2J_3} (-y)^{2\CJ_3} = {\rm Tr}_{\CH_{u,L,\zeta,\gamma}^{\rm
BPS}} y^{2J_3} (-y)^{2I_3}
\end{equation}
where $I_3$ is an R-symmetry
generator and
\begin{equation}
\CJ_3 := J_3 + I_3.
\end{equation}
The PSC is an index because $\CQ$ is a singlet under $\CJ_a$,
anticommutes with $(-1)^F$, and is invertible on long representations
of the algebra of the $\CR_{\alpha}^A$.
Usually we will shorten the notation and leave the
dependence on $u$ implicit.

Specializing to $y=-1$ we recover
the framed BPS degeneracies $\fro(u,L_\zeta, \gamma)$.  Thanks
to the rigidity of small representations of supersymmetry, they
are invariant under deformations of parameters provided no
states enter or leave the Hilbert space.  However, there will be
wall-crossing phenomena analogous to the usual wall-crossing of
BPS states in the absence of line operators.  We investigate
this in Section \ref{subsec:Halos}.

A PSC can also be defined  for the usual vanilla BPS states.
The definition of the vanilla PSC is based on the claim that the quantity
\begin{equation}\label{eq:PSC-one}
{\rm Tr}_{\CH}(2 J_3) (-1)^{2J_3} (-y)^{2\CJ_3}
\end{equation}
vanishes on long representations $\CH$ of the $\CN=2$ algebra.  To prove this
consider the $so(3)\oplus su(2)_R$ character ${\rm Tr} x_1^{2J_3} x_2^{2I_3}$.  We obtain \eqref{eq:PSC-one}
from it by
\begin{equation}\label{eq:Inermed-PSC}
{\rm Tr}(2 J_3) (-1)^{2J_3} (-y)^{2\CJ_3} = x_1{\p \over \p x_1}\left({\rm Tr} x_1^{2J_3} x_2^{2I_3}\right)\vert_{x_1=-x_2 = y}
\end{equation}
Now, in the representation theory of the little superalgebra of a
massive particle the long
representations, considered as representations of $so(3)\oplus su(2)_R$,
have the form
\begin{equation}
\rho_{hh}\otimes \rho_{hh}\otimes \mathfrak{h}_{\ell}
\end{equation}
where $\mathfrak{h}_{\ell}$ is an arbitrary finite dimensional representation
of $so(3)\oplus su(2)_R$ and $\rho_{hh}$ is the \emph{half-hypermultiplet}
representation $\rho_{hh}\cong (\half;0) \oplus (0;\half)$.
On the other hand, the short
representations of the little superalgebra, considered as
representations of  $so(3)\oplus su(2)_R$,
have the form
\begin{equation}
\rho_{hh} \otimes \mathfrak{h}_{s}
\end{equation}
where $\mathfrak{h}_{s}$ is an arbitrary finite dimensional representation
of  $so(3)\oplus su(2)_R$. Now observe that
\begin{equation}
{\rm Tr}_{\rho_{hh}}  x_1^{2J_3} x_2^{2I_3} = x_1 + x_1^{-1} + x_2 + x_2^{-1}
\end{equation}
and hence \eqref{eq:Inermed-PSC} vanishes on long representations, but is
nonvanishing on short representations. In fact, its value on short
representations is just
\begin{equation}
(y- y^{-1}) {\rm Tr}_{\mathfrak{h}_s} y^{2J_3} (-y)^{2I_3} .
\end{equation}
Therefore, we can define
an (unframed) Protected Spin Character $\Omega(u,\gamma;y)$ by
\begin{equation}
(y- y^{-1}) \Omega(u,\gamma;y) := {\rm Tr}(2 J_3) (-1)^F (-y)^{2\CJ_3}.
\end{equation}
i.e.
\begin{equation}
\Omega(u,\gamma;y) = {\rm Tr}_{\mathfrak{h}_s} y^{2J_3} (-y)^{2I_3} .
\end{equation}
Note that if the isotypical decomposition of   $\mathfrak{h}_s$ as
an $su(2)_R$ representation contains only the singlet then
 the Protected Spin Character {\it coincides}
with the spin character of $\mathfrak{h}_s$. Thus, we immediately see that
a standard hypermultiplet, which  has   $\mathfrak{h}_s = (0;0)$, has PSC $\Omega(u,\gamma;y)=1$.
A standard W-boson, on the other hand,  has $\mathfrak{h}_s= (\half;0)$ and therefore
gives $\Omega(u,\gamma;y)=y+1/y$.
In fact, in all the examples of $\CN=2$ field theories with a good UV limit
of which we are aware, the BPS particles have representations
whose decompositions under $su(2)_R$ only contain singlets.
We therefore call BPS particles where $\mathfrak{h}_s$ has
nontrivial representations of $su(2)_R$ \emph{exotic BPS particles}.  Similarly, we call framed BPS states which transform
nontrivially under $su(2)_R$ \emph{exotic framed BPS states}.  For nonexotic BPS states
our PSC  is the same as the  ``refined index''
of \cite{Gukov:2004hz,Dimofte:2009bv,Dimofte:2009tm} up to a (convention-dependent) sign.

We now state a set  of ``positivity conjectures''
for both vanilla  and framed BPS states. The strongest of these is
the ``no exotics conjecture'' mentioned above: The isotypical decomposition
of $\mathfrak{h}_s$ contains only $su(2)_R$ singlets. That is,
 there are no exotic particles at smooth points on the
Coulomb branch.\footnote{On Higgs or hybrid branches there are complications in formulating
such a statement.} Note that by wall-crossing the
existence of exotic BPS particles would surely lead to exotic framed BPS states.
 A weaker conjecture is the  ``strong positivity conjecture'' which states that
 $\Omega(u,\gamma;y)$ (and its framed counterpart) is a linear combination
 of $su(2)$ characters $\chi_n(y)$ with \emph{nonnegative integral} coefficients.
 The ``weak positivity conjecture'' merely requires $\Omega(u,\gamma;y)$ to be
 positive integer at $y=+1$.

The no-exotics conjecture implies the strong positivity conjecture, but
the converse need not hold. A sufficient condition
for the strong positivity conjecture is that all isotypical components
of $\mathfrak{h}_s$ have integral $su(2)_R$ spin. This condition is,
however, not necessary.

We will see in  Section \ref{sec:Formal-line-operators} that the
strong positivity conjecture (even in a weakened form)
has far reaching, beautiful consequences for the framed and standard BPS spectrum. There is strong evidence it is true for all theories in the $A_1$ class, in a very non-trivial fashion. Another fact which appears to be true in all our examples is that for a fixed line operator the
spaces $\CH_{u,L,\zeta,\gamma}^{\rm BPS}$ are only nonvanishing for a finite number of charges $\gamma$.
That is, there are only a finite number of nonvanishing framed BPS invariants for a fixed line operator.
We have no simple field-theoretical argument
to justify this statement, but we suspect that this finiteness property might be universally true in field theory.
However, it must be admitted that all our evidence is based on the $A_1$ class of theories. This is one
reason why it is important to understand better the higher rank theories in the class $\CS$, a project to which
we hope to return.
\footnote{Recently (August 2012) it has been shown by Diaconescu et. al. that
for pure $SU(K)$ gauge theories the no-exotics conjecture is in fact true.}

A final observation is in order.  Spin characters have been studied recently in the literature, but always
in the context of field theory rather than supergravity.  Our construction suggests one reason why
the field theory case is preferred:  in supergravity
there is no $SU(2)_R$ symmetry available, hence no way of constructing a protected spin character.

\subsection{Halos and halo Fock spaces}\label{subsec:Halos}

We now describe a particularly interesting class of framed BPS states, namely bound
states between a ``core'' supported in a small neighborhood of
the line operator and a surrounding ``halo'' of BPS particles.
Such configurations are closely related to Denef's halo solutions of $\CN=2$
supergravity
\cite{Denef:2000nb,Denef:2000ar,Denef:2002ru}.
In \cite{Denef:2007vg} this phenomenon was used to
derive the semi-primitive wall-crossing formula.
As we will see shortly, halos are important for the framed version of the wall-crossing
problem as well.

We begin by working classically, in the simple case of a BPS core of charge $\gamma_c$, bound to a
single halo particle of charge $\gamma_h$.  A variant of an
argument from \cite{Denef:2000nb} shows the following.\footnote{We
give a few details of the derivation in Appendix
\ref{app:Low-Energy-FxdPoint}.} The energy of a halo particle probing the
IR background associated with the core charge $\gamma_c$ is
\begin{equation}\label{eq:Halo-Particle-Energy}
E_{halo} = \vert Z_{\gamma_h}(u(r))\vert \left(1 + \cos(\alpha_h -
\alpha_\zeta)\right) -  {\rm Re}(Z_{\gamma_h}(u)/\zeta).
\end{equation}
Recall that $u \in \CB$ stands for the value of the vacuum moduli at
infinity. The moduli detected by the probe particle at a distance $r$
from the line operator depends on $r$, and we denote
$Z_{\gamma_h}(u(r)) =\vert Z_{\gamma_h}(u(r))\vert e^{i
\alpha_h(r)}$.  On the other hand $\zeta$ is independent of $r$, and we
write its phase as  $\zeta = e^{i \alpha_\zeta}$.

The halo particle minimizes its energy at some $r$ with
$e^{i\alpha_h(r)} / \zeta = -1$.  This gives
a close analog of Denef's formula for the boundstate
radius of BPS black holes \cite{Denef:2000nb}:\footnote{Indeed,
using the heuristic explained near equation
\eqref{eq:heuristic-bound} with central charge $\zeta M - Z_{\gamma_c}$
and charge $\gamma_f + \gamma_c$, Denef's formula reduces to \eqref{eq:halo-rad}
in the $M\to +\infty$ limit.}
\begin{equation}\label{eq:halo-rad}
r_{\rm halo} = \frac{\langle \gamma_h, \gamma_c\rangle}{2 {\rm
Im}(Z_{\gamma_h}(u)/\zeta)}.
\end{equation}
This result has several important consequences.
First, it gives an analog of the Denef stability condition:
the halo configuration only exists for $(u,\zeta)$ such that $r_{\rm halo}$ given in \eqref{eq:halo-rad} is positive.
Second, note that $r_{\rm halo} \to \infty$ when $ Z_{\gamma_h}(u) = -
\vert Z_{\gamma_h}(u)\vert \zeta $ --- in other words, when $(u, \zeta)$ lie on
the BPS wall $\hW(\gamma_h)$.  Unless we are very near this wall, $r_{\rm
halo}$ is much smaller than the natural cutoff for the validity of
the IR description.  So the picture is that as one approaches the
BPS wall $\hW(\gamma_h)$, halo configurations built from of particles of charge
(a multiple of) $\gamma_h$ reach a size which justifies treating them purely
within the IR theory.  These haloes grow to infinite size when we reach the wall,
and disappear on the other side.

What are the \ti{quantum} states associated with these haloes?
Fortunately it is easy to quantize the classical
halo configurations:  being mutually BPS, the halo particles
do not interact with one another, and simply generate a $\IZ_2$-graded Fock space
of quantum states.
To describe this mathematically, let $\CH'(\gamma_h;u)$ be the space of vanilla BPS states of charge $\gamma$,
with a half-hypermultiplet factored out.  It is a
representation of spatial $so(3)$ and $su(2)_R$.  We
define a $\IZ_2$ grading  by $-(-1)^{2J_3}$.
The usual BPS
degeneracy $\Omega(\gamma_h;u)$ is minus the superdimension of
$\CH'(\gamma_h;u)$ with respect to this grading.  Letting
$J_{\gamma_c,\gamma_h}= \half (\vert \langle \gamma_c,
\gamma_h\rangle \vert-1)$,
introduce the representation $(J_{\gamma_c,\gamma_h})$ of spatial $so(3)$,
of dimension $\vert \langle \gamma_c, \gamma_h\rangle \vert$.
This accounts for the spin of the
electromagnetic field from the halo-core interaction.  We
build a $\IZ_2$-graded Fock space on the finite-dimensional
vector space $(J_{\gamma_c,\gamma_h})\otimes \CH'(\gamma_h;u)$,
considering the $\IZ_2$ grading of $(J_{\gamma_c,\gamma_h})$ to
be \emph{even}.\footnote{
Given a finite dimensional $\IZ_2$-graded vector space $V= V^0 \oplus V^1$
of superdimension $(n_0\vert n_1)$,
the associated $\IZ_2$-graded Fock space is the symmetric algebra on $V^0$ tensor the
anti-symmetric algebra on $V^1$.  More prosaically, we choose a basis $\alpha_i$, $i=1,\dots, n_0$,
and $\gamma_s$, $s=1, \dots, n_1$, then represent the bosonic Heisenberg algebra with $\alpha_i$
corresponding to creation operators, and the fermionic Clifford algebra with $\gamma_s$ corresponding
to creation operators.  The
unusual-looking $\IZ_2$ gradings have a beautiful
physical explanation \cite{Denef:2002ru}, ultimately related to the fact that the
magnetic field forces the spin of the halo particle to point inwards.}
Concretely, if we
define a set of integers $a_{m, \gamma_h}$ by
\begin{equation}
\Omega(u,\gamma_h;-z) =  {\Tr}_{\CH'(\gamma_h;u)}(-z)^{2J_3} z^{2 I_3} =
\sum_{-M_{\gamma_h}}^{M_{\gamma_h}} a_{m, \gamma_h} z^m
\end{equation}
where $M_{\gamma_h} \ge 0$ is twice the maximal $\CJ_3$
of a halo particle, our Fock space is generated by $\vert
a_{m,\gamma_h}\vert $ creation operators of $2 \CJ_3$ eigenvalue
$m+m'$, for each $m'$ of the form $m' = -2J_{\gamma_c,\gamma_h}, -2J_{\gamma_c,\gamma_h} + 2,
\dots,2J_{\gamma_c,\gamma_h}-2, 2J_{\gamma_c,\gamma_h}$. The
oscillators are fermionic for $m$ even (i.e.
$a_{m,\gamma_h}>0$) and bosonic for $m$ odd (i.e.
$a_{m,\gamma_h}<0$). Note that $\Omega(\gamma_h;u) = \sum_m
a_{m, \gamma_h}$. If $\langle\gamma_c,\gamma_h \rangle=0$ then
we consider $(J_{\gamma_c,\gamma_h})$ to be the zero vector space,
and no halos form. Of course, $a_{m, \gamma_h}$ is a piecewise
continuous integer function of $u$, but we usually suppress the
dependence in the notation.

\subsection{Framed wall-crossing}\label{subsec:Halo-Wall-Crossing}

In Section \ref{subsec:BPS-Deg-PSC} we defined the protected spin character $\fro(L,\gamma;u, \zeta)$
which ``counts'' framed BPS states.  But as we noted in Section \ref{sec:framed-BPS-states}, when
$(\zeta, u)$ cross a BPS wall, the framed BPS
states can mix with the continuum.  Hence the standard arguments
for the invariance of $\fro(L,\gamma;u, \zeta)$ break down.
This is the framed version of the
``wall-crossing'' problem.
Fortunately, it is easier than the
full-fledged wall-crossing problem for the vanilla BPS states:  the states which disappear
from the framed BPS spectrum are just the simple halo states which we have described in Section
\ref{subsec:Halos}.  So we can completely describe the jump of $\fro(L,\gamma;u, \zeta)$ as we cross the wall.

Consider a path $(u_t,\zeta_t)\in \widehat \CB \times \widehat{U(1)}$ which
crosses a wall $\widehat{W}(\gamma_h)$.  As noted above, when we reach the wall
$r_{\rm halo}$ goes to infinity.  Hence an entire Fock
space of halo boundstates, constructed from halo particles whose
charge is positively proportional to $\gamma_h$, either appears or
disappears from the BPS spectrum.
Now let us describe the effect of this
on the (framed) Protected Spin Characters.
Let $\{x_{\gamma}\}$ be a basis for the group algebra of
$\Gamma$, so that
\begin{equation} \label{eq:untwisted}
x_{\gamma} x_{\gamma'} = x_{\gamma +
\gamma'}.
\end{equation}
Then form the generating functional
\begin{equation}\label{eq:gen}
F(u, L, \zeta, \{x_\gamma\}; y) := \sum_{\gamma} \fro(u, L, \zeta,
\gamma; y) x_\gamma.
\end{equation}
(In order to keep the notation from getting too heavy we will
often suppress some of the variables when they can be safely
understood as implicit.)
Each creation operator of type $m, m', \gamma_h$ described above
contributes a factor
\begin{equation}
(1 + (-1)^{m} y^{m+m'} x_{\gamma_h} )^{  a_{m,\gamma_h}}
\end{equation}
to the product representation of the trace over the Fock space.
Define $\Gamma_h := \IZ \gamma_h$, and for $\gamma_c \in \Gamma_{\CL}$ let
\begin{equation}
F_{\bar \gamma_c} := \sum_{\gamma'\in \Gamma_h} \fro(u, L, \zeta,
\gamma_c +\gamma'; y) x_{\gamma_c+\gamma'}.
\end{equation}
(The sum only depends on the projection $\bar\gamma_c\in \Gamma_L/\Gamma_h$.)
This is the piece of $F$ corresponding to states with charge of type $\gamma_c + \ell \gamma_h$.
Let $F_{\bar\gamma_c}^\pm$ denote this generating function on the side of the
wall with ${\rm Im}(Z_{\gamma_h}(u)/\zeta)>0$ and with ${\rm Im}(Z_{\gamma_h}(u)/\zeta)<0$,
respectively. Then we have
\begin{equation} \label{eq:transform}
F^\pm_{\bar\gamma_c}  = F^\mp_{\bar\gamma_c}  \prod_{\gamma_h}
\prod_{m=-M_{\gamma_h}}^{M_{\gamma_h}} \prod_{m'= - 2J_{\gamma_c,
\gamma_h}}^{2J_{\gamma_c, \gamma_h}} (1 + (-1)^{m } y^{m+m'}
x_{\gamma_h} )^{  a_{m,\gamma_h}}.
\end{equation}
Here we have slightly abused notation:  the first product over
$\gamma_h$ means the product over all halo charges giving the same
wall $\hW(\gamma_h)$. Whether we choose $F^+$ or $F^-$ on the
LHS (i.e. whether we gain or lose a Fock space) depends
on the direction in which the wall is
crossed and the sign of $\langle \gamma_c,\gamma_h\rangle$.
(See \eqref{eq:spin-ks-tmn-ii} below for a more precise statement.)

This transformation   resembles the effect of a ``coordinate transformation''
of the formal variables
$x_\gamma$ which multiplies each $x_{\gamma_c}$ by a rational function of $x_{\gamma_h}$.
Unfortunately, this interpretation is not quite compatible with the multiplication rule
\eqref{eq:untwisted}.  It does work out well if $y=1$, as we now show.
In this case \eqref{eq:transform} specializes to
\begin{equation} \label{eq:tspec}
F^\pm_{\bar\gamma_c}  = F^\mp_{\bar\gamma_c}  \prod_{\gamma_h}
\prod_{m=-M_{\gamma_h}}^{M_{\gamma_h}} (1 + (-1)^{m}
x_{\gamma_h} )^{  \lvert \langle \gamma_h, \gamma_c\rangle \rvert a_{m,\gamma_h} }.
\end{equation}
The ``coordinate'' transformation
\begin{equation}\label{eq:y-one-prod}
x_\gamma \to x_\gamma  \prod_{\gamma_h}
\prod_{m=-M_{\gamma_h}}^{M_{\gamma_h}}  (1 + (-1)^{m } x_{\gamma_h} )^{\langle \gamma_h, \gamma\rangle a_{m,\gamma_h}}
\end{equation}
precisely reproduces \eqref{eq:tspec} as we cross the wall from the side
${\rm Im}(Z_{\gamma_h}(u)/\zeta) < 0$ to the side ${\rm Im}(Z_{\gamma_h}(u)/\zeta) > 0$.
Moreover, it is compatible with the product law \eqref{eq:untwisted}.

If $y=-1$, the factor $y^{m'}$ in \eqref{eq:transform} coincides with $- (-1)^{\langle \gamma_h, \gamma_c\rangle }$.
There is a simple trick to tame these extra signs:  modify the multiplication rule from
\eqref{eq:untwisted} to
\begin{equation} \label{eq:twisted-mult}
\hat x_{\gamma} \hat x_{\gamma'} = (-1)^{\langle \gamma, \gamma'\rangle } \hat x_{\gamma +\gamma'}.
\end{equation}
Then, using the relation
$\hat x_{\gamma_c + n \gamma_h} = \hat x_{\gamma_c} \left( (-1)^{\inprod{\gamma_h, \gamma_c}} \hat x_{\gamma_h} \right)^n$,
we get an analog of \eqref{eq:tspec},
\begin{equation} \label{eq:transform2}
F^\pm_{\gamma_c}  = F^\mp_{\gamma_c}  \prod_{\gamma_h}
\prod_{m=-M_{\gamma_h}}^{M_{\gamma_h}} (1 -
\hat x_{\gamma_h} )^{ \lvert \inprod{\gamma_h, \gamma_c} \rvert a_{m,\gamma_h}}.
\end{equation}
The transformation rule
\begin{equation}\label{eq:y-min-one-gen}
\hat x_\gamma \to \hat x_\gamma  \prod_{\gamma_h}
(1 - \hat x_{\gamma_h} )^{\inprod{\gamma_h, \gamma} \Omega(\gamma_h)}
\end{equation}
applied to the $y=-1$ generating functional correctly reproduces \eqref{eq:transform2},
and is compatible with the
twisted multiplication rule \eqref{eq:twisted-mult}.

To state this more carefully, if $\zeta_{\rm cw}$ is on the clockwise side of the
BPS wall $\hW(\gamma_h)$ and $\zeta_{\rm ccw}$ is on the counterclockwise side, then, suppressing
all other irrelevant indices, the generating function $F$ obeys
\begin{equation}\label{eq:Careful-rule}
F\left(\zeta_{\rm cw}, \hat x_\gamma\right) = F\left(\zeta_{\rm ccw} , \hat x_\gamma (1- \hat x_{\gamma_h})^{-\langle \gamma, \gamma_h\rangle \Omega(\gamma_h)} \right).
\end{equation}

The transformation \eqref{eq:y-min-one-gen} used above
is exactly the symplectomorphism introduced in
\cite{ks1}:  defining $\CK_{\gamma_h}$ by
\begin{equation}\label{eq:KS-sympo}
\CK_{\gamma_h}(\hat x_\gamma) := \hat x_\gamma (1-\hat x_{\gamma_h})^{\langle
\gamma, \gamma_h \rangle },
\end{equation}
\eqref{eq:y-min-one-gen} is simply
$\hat x_\gamma \to \CK_{\gamma_h}^{-\Omega(\gamma_h)}(\hat x_\gamma)$.

\subsubsection{Noncommuting variables and quantum dilogarithms}

For general $y$, we cannot interpret \eqref{eq:transform} in terms of a change of variables,
at least not \ti{commuting} variables.  But suppose we introduce
formal variables $X_\gamma$ satisfying the relation
\begin{equation}\label{eq:Heis-Alg}
X_\gamma X_{\gamma'} = y^{\inprod{\gamma, \gamma'}} X_{\gamma +
\gamma'},
\end{equation}
and again consider the generating function
\begin{equation}\label{eq:form-gen}
F(u,L,\zeta,\{X_\gamma\};y) := \sum_{\gamma} \fro(u, L, \zeta,
\gamma;y) X_\gamma,
\end{equation}
now as a function of these \ti{noncommuting} variables.
In this case, as we will now see, the effect of adding or subtracting halo boundstates
is nicely summarized by a certain transformation of the $X_\gamma$.  This transformation
is implemented by conjugation with the quantum dilogarithm of Faddeev and Kashaev \cite{Faddeev:1993rs}.
(See \cite{Teschner:2005bz}, Appendix A, for a useful summary of the various
properties and sobriquets enjoyed by this function.)

First we need a mathematical lemma.
Suppose $\langle \gamma_c, \gamma_h \rangle = n$ is not zero.  Let us define
Laurent polynomials in $y$ by expanding the commutative
variables in the group algebra:
\begin{equation}
x_{\gamma_c} (1 + y^{n-1} x_{\gamma_h})(1 + y^{n-3}
x_{\gamma_h})\cdots (1 + y^{3-n} x_{\gamma_h})(1 + y^{1-n}
x_{\gamma_h}) = \sum_{j=0}^{\vert n\vert}  P^{(n)}_j(y) x_{\gamma_c
+ j \gamma_h}
\end{equation}
Physically it is obvious that  $P^{(n)}_j(y)$ is just the
character of the $j^{th}$ antisymmetric product
$\Lambda^j \rho_{\vert n \vert }$ where $\rho_{N}$
is the $N$-dimensional irreducible representation of $SU(2)$.
\footnote{Incidentally, it is amusing and will be useful
to note that there is a ``$q$-binomial theorem''
that also identifies (take $n>0$):
\begin{equation}
P^{(n)}_j(y) = \frac{ [n]_y ! }{ [j]_y ! [n-j]_y !}:=\qbinom{n}{j}
\end{equation}
where
\begin{equation}
[n]_y := \frac{y^{n} - y^{-n}}{y - y^{-1}} .
\end{equation}
}
Our lemma states that the
\emph{same} Laurent polynomials appear when expanding the
noncommutative expression
\begin{equation}
X_{\gamma_c} \Phi_n( X_{\gamma_h}) = \sum_{j=0}^{\vert n \vert}
P^{(n)}_j(y) X_{\gamma_c + j \gamma_h}
\end{equation}
where
\begin{equation}
\Phi_n(\xi) := \begin{cases}  \prod_{s=1}^{n} (1 + y^{-(2s-1)} \xi )
& n >0 \\
 1 & n=0 \\
 \prod_{s=1}^{\vert n\vert } (1 + y^{(2s-1)} \xi ) & n< 0 \\
 \end{cases}
 \end{equation}
The proof of this lemma is straightforward and will be omitted.

Using this lemma we can state that
the effect of a wall-crossing is to transform \emph{all} the
noncommutative variables in \eqref{eq:form-gen} by
\begin{equation}\label{eq:Spin-KS-tmn}
X_\gamma \to X_\gamma \prod_{m=-M_{\gamma_h}}^{M_{\gamma_h}}
\Phi_{\langle \gamma, \gamma_h\rangle} ( (-1)^m y^m
X_{\gamma_h})^{\epsilon a_{m,\gamma_h}}
\end{equation}
In this transformation law we are crossing the wall $\widehat{W}(\gamma_h)$.  The sign $\epsilon=+1$ if we cross from the side
where $ \langle \gamma_h, \gamma \rangle \ {\rm
Im}(Z_{\gamma_h}(u)/\zeta) < 0$  to the side where $ \langle
\gamma_h, \gamma \rangle \ {\rm Im}(Z_{\gamma_h}(u)/\zeta) > 0$
since in this case we \emph{gain} a Fockspace of halo particles
around the core charge $\gamma$. The sign $\epsilon=-1$ if we cross
the wall in the other direction since in this case we \emph{lose} a
Fockspace of halo particle boundstates. If we denote by $F^+, F^-$ the
formal generating function \eqref{eq:form-gen} on the   side of the wall
with $ \ {\rm Im}(Z_{\gamma_h}(u)/\zeta) > 0$ and $<0$, respectively
then (suppressing all other irrelevant variables):
\begin{equation}\label{eq:spin-ks-tmn-ii}
F^+(X_\gamma)  = F^-\left( X_\gamma \left( \prod_{m=-M_{\gamma_h}}^{M_{\gamma_h}}
\Phi_{\langle \gamma, \gamma_h\rangle} ( (-1)^m y^m
X_{\gamma_h})^{a_{m,\gamma_h}}\right)^{{\rm sign}(\langle \gamma_h,
\gamma\rangle ) }\right)
\end{equation}

The transformation \eqref{eq:Spin-KS-tmn} is elegantly summarized by
conjugation with the quantum dilogarithm. Define
\begin{equation}
\Phi(X) := \prod_{k=1}^\infty ( 1 + y^{2k-1} X)^{-1}
\end{equation}
Then the transformations \eqref{eq:Spin-KS-tmn} and \eqref{eq:spin-ks-tmn-ii}
are simply equivalent to   the rule:
\begin{equation}\label{eq:Spin-KS-tmn-S}
F^+(X_\gamma) =   S_{\gamma_h} F^-(X_\gamma ) S_{\gamma_h}^{-1}
\end{equation}
where
\begin{equation}\label{eq:Spin-KS-ii}
S_{\gamma_h} =  \prod_{\gamma_h} \prod_{m=-M_{\gamma_h}}^{M_{\gamma_h}} \Phi (
(-1)^m y^m X_{\gamma_h})^{a_{m,\gamma_h}}
\end{equation}
where once again the first product means we take the product over all
parallel charges to $\gamma_h$.
Accordingly, the generating function of framed degeneracies
\eqref{eq:form-gen} transforms by conjugation with $S_{\gamma_h}$.

\subsection{The motivic wall-crossing formula}\label{subsec:Motivic-WCF}

Let $\widehat\CB$ be the  universal cover of the
smooth part of the Coulomb branch $\CB$ and  consider
\begin{equation}\label{eq:Chamber-Space}
\Xi: =  \widehat \CB \times \widehat \IC^* - \cup_{\gamma: \exists u: \Omega(\gamma;u) \not=0 }  \widehat{W}(\gamma)
\end{equation}
This space is divided into chambers - the connected components of $\Xi$.     Let us
label these chambers by an index denoted $c$.
Suppose we consider two chambers $c_1,c_2$ of $\Xi$. Consider a path  $\CP$
in $\widehat \CB \times \widehat \IC^*$
 connecting these chambers. The generating functions
  \eqref{eq:form-gen} for a line operator
$L$ in the two chambers
 will be related by some composition of transformations of the form \eqref{eq:Spin-KS-tmn-S}:
 \footnote{Observe that here
  we have snuck in the assumption that the generating functions can be continued
  from the unit circle $\vert \zeta \vert =1$ to $\IC^*$. That assumption is not
  essential in the present argument, nor in the discussions of Sections \ref{sec:Formal-line-operators} and \ref{sec:Cluster-Algebras}. However it will be important in Section \ref{sec:linehol}.}
\begin{equation}\label{eq:Path-Transf}
F(L;c_1) = S(\CP)  F(L;c_2)
\end{equation}
where $S(\CP)$ is the path ordered product of transformations associated to the
walls $\widehat{W}(\gamma_h^i)$ crossed by the $\CP$ on going from $c_1$ to $c_2$
\begin{equation}
S(\CP)= \prod_{\gamma_h^i} {\rm Ad}(S_{\gamma_h^i} ).
\end{equation}
We would like to conclude that the product of transformations
 $\prod_{\gamma_h^i} Ad(S_{\gamma_h^i} ) $  is independent of the path $\CP$ joining
 $c_1$ to $c_2$. This will be true provided the theory has enough line operators $L$ that knowing
the action of the    transformations \eqref{eq:Path-Transf}   is strong enough
to   determine completely $S(\CP)$.
We do not have a totally general reason why this works (indeed, we do not have a totally
general reason why there should be any line operators at all).  However, we will see in
examples below that there are always enough line operators; indeed, in the examples
there is a 1-1 correspondence between line operators and electromagnetic charges at any
point of the Coulomb branch, and one can recover any $X_\gamma$ as a linear combination of the functions
$F(L,\{X_\gamma\})$. See Section \ref{sec:Tropic} below for further discussion.

\insfig{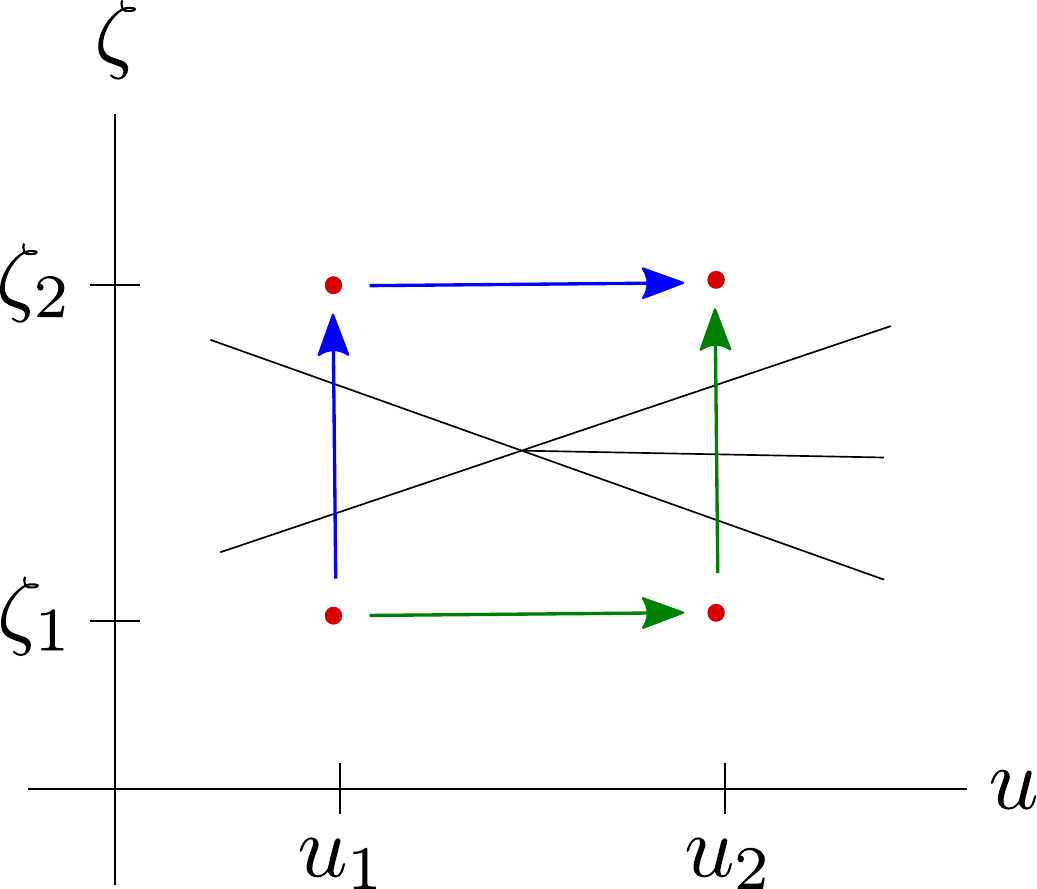}{Two paths from $(\zeta_1, u_1)$ to $(\zeta_2, u_2)$.  The BPS rays appear as codimension-1 loci
in the joint $(\zeta, u)$ space.  Each BPS wall $\widehat{W}(\gamma)$  has a coordinate transformation attached.
The walls can coalesce and split as shown, and hence the two paths in general cross different  sets of BPS walls.
Nevertheless,  the composition of the corresponding
coordinate transformations must be independent of the path.}

The path-independence of $S(\CP)$ is actually a
version of the ``motivic wall-crossing formula'' of Kontsevich and Soibelman \cite{ks1}.
To see this consider  two paths  joining  $(\zeta_1, u_1)$ to $(\zeta_2,
u_2)$ in $\widehat\CB \times \widehat \IC^*$ as illustrated in Figure \ref{fig:zeta-u-wall},
and let us compare generating functions $F(L, c_1)$ and $F(L,c_2)$ for the two chambers $c_1,c_2$ in $\Xi$
containing $(\zeta_1, u_1)$ to $(\zeta_2,u_2)$, respectively.
Suppose we go along a path $(\zeta_1, u_1) \to (\zeta_2,
u_1) \to (\zeta_2, u_2)$; and we assume that no $\arg Z_\gamma(u)$
enters or leaves the interval $(\arg \zeta_1, \arg \zeta_2)$ as $u$
varies along the path from $u_1$ to $u_2$. Then all the functions
$F(L,c_1)$ get transformed by the ordered product
\begin{equation}
{\bS}(\zeta_1, \zeta_2;u_1) = \prod_{\gamma: \arg Z(\gamma) \in (\zeta_1, \zeta_2)} {\rm Ad}(S_\gamma).
\end{equation}
On the other hand, if we take the path which follows  $(\zeta_1, u_1) \to (\zeta_1, u_2) \to
(\zeta_2, u_2)$ then they get transformed instead by ${\bS}(\zeta_1,
\zeta_2;u_2)$. Thus we have, for all line operators $L$:
\begin{equation}\label{eq:SF-SF}
{\bS}(\zeta_1, \zeta_2;u_1)F(L,c_1)={\bS}(\zeta_1, \zeta_2;u_2)F(L,c_1).
\end{equation}
We would like to conclude from \eqref{eq:SF-SF} that the two
transformations are indeed equal:
\begin{equation} \label{eq:s-equal}
{\bS}(\zeta_1, \zeta_2;u_1) = {\bS}(\zeta_1, \zeta_2;u_2).
\end{equation}
As we discussed above, we must make an assumption that there exist
sufficiently many line operators $L$ that we can conclude
\eqref{eq:s-equal}.

One slight difference from the standard discussion is that we here only obtain
an equality of products of quantum dilogs in the ``adjoint representation.'' Nevertheless,
following the discussion from equation \eqref{eq:y-one-prod} to
\eqref{eq:y-min-one-gen} we learn that the specialization to $y=-1$
gives the equality \eqref{eq:s-equal} where now
\begin{equation}\label{eq:Symplectic-S}
{\bS}(\zeta_1, \zeta_2;u_1) = \prod_{\gamma: \arg Z(\gamma) \in (\zeta_1, \zeta_2)} \CK_\gamma^{\Omega(\gamma_\BPS; u_1)}.
\end{equation}
The equation \eqref{eq:s-equal} with ${\bS}$ given by
\eqref{eq:Symplectic-S}  is the form of the  Kontsevich-Soibelman WCF
which was discussed in \cite{Gaiotto:2008cd,Gaiotto:2009hg}.  Knowing $\bS(\zeta_1, \zeta_2;u)$
for all $\zeta_{1,2}$ is equivalent to knowing the set of BPS indices
at $u$.  Determining the full BPS spectrum requires more information, namely
the protected spin characters.

We end this section by commenting on some related literature.
Wall-crossing formulas involving noncommutative ($q$-deformed) quantum dilogarithms
appeared in the work of Kontsevich and Soibelman \cite{ks1}.  The classical limit ($q = -1$) was also discussed there, and
yielded directly a WCF for the second helicity supertrace in $\N=2$ theories, proven physically in \cite{Gaiotto:2008cd}.
However, initially it was not completely clear how to extend this success to the $q$-deformed context.
In \cite{Dimofte:2009bv} it was proposed that
one should identify $q$ with the parameter $y$ counting the spins of BPS states;
in \cite{Dimofte:2009tm} this proposal was sharpened to a precise wall-crossing formula,
which was shown to work in several examples.
Nevertheless a general physical proof was still missing.
A novel argument was given in \cite{Cecotti:2009uf}, where the commutation relations
\eqref{eq:Heis-Alg} and the wall-crossing formula
were connected to the A model open topological string and hence to Chern-Simons theory.
Here we have given a direct proof using general notions from four-dimensional
gauge theory.

\subsection{The deformed ring of line operators}\label{subsec:Deformed-Ring}

As we mentioned in Section \ref{subsubsec:RingStructure} the parallel supersymmetric line operators
preserving $osp(4^*\vert 2)_\zeta$ at different points $\vec x_i\in \IR^3$
can be multiplied and re-expanded as sums of line operators. This operation
defines a ring, which should be commutative, since there is no natural ordering
of points in $\IR^3$, and associative, since they are operators. It is natural
to wonder how this ring structure is related to the framed protected spin characters
and framed BPS degeneracies. It turns out that the generating functions $F(L)$ defined in
equation \eqref{eq:form-gen}
are well suited to discussing this question. In this section we are going to see that they can be
used to define a very interesting noncommutative deformation of the ring of line operators.

Suppose we put one line operator $L$ at $\vec x=0$ and another $L'$ at $\vec x' = (0,0,z)$
displaced along the $z$-axis. Note that rotations generated by $J_3$ are symmetries of
this configuration but the $so(3)$ symmetry has been broken to $so(2)$. If $z\not=0$ and
$L$ and $L'$ are mutually local in the sense explained in Section \ref{subsec:LabelsForLines}
the Hilbert space $\CH_{L_{\vec x} L'_{\vec x'}}$ is a representation of $U(1)_{\vec x \vec x'}\subset SU(2)$.
If $L$ and $L'$ are simple then the   Hilbert space $\CH_{LL'}$ is also
graded by $\Gamma + \gamma_L + \gamma_{L'}$,
and the summands in the grading satisfy
\begin{equation}\label{eq:LL-grade}
\CH_{L_{\vec x} L'_{\vec x'},\gamma_0}^{\rm BPS} = \bigoplus_{\gamma+ \gamma'=\gamma_0} \CH_{L,\gamma}^{\rm BPS}
 \otimes \CH_{L',\gamma'}^{\rm BPS}
\otimes N_{\gamma,\gamma'}.
\end{equation}
Here $N_{\gamma,\gamma'}$ is a one-dimensional representation of $U(1)_{\vec x \vec x'}$. It results
from the electromagnetic fields excited by the pair of dyons. Its contribution to the protected spin
character is therefore\footnote{This should be contrasted with what happens for the primitive
wall-crossing formula for vanilla BPS states. In that case the electromagnetic field has $SU(2)$ spin
$2j = \vert \langle \gamma, \gamma' \rangle \vert -1$. The extra $-1$ arises from the alignment of the
center of mass degrees of freedom of the two separate constituents.}
\begin{equation}
{\Tr}_{N_{\gamma,\gamma'}} y^{2J_3} = y^{\langle \gamma, \gamma' \rangle}.
\end{equation}
Recalling the definition (\ref{eq:form-gen}) of the generating functions $F(L)$ and the
Heisenberg relation \eqref{eq:Heis-Alg} we see
that equation \eqref{eq:LL-grade} can be used to define a \emph{noncommutative} product
of line operators:
\begin{equation}\label{eq:Non-Comm-Line-Prod}
F(L\circ_y L') := F(L) F(L').
\end{equation}
In general $F(L)F(L') \not= F(L') F(L)$, as
is clear from examples discussed below.

In Section \ref{subsubsec:RingStructure} we described a commutative product on line operators. It might
therefore seem surprising to find that there is a noncommutative deformation
of this product.  The distinction is whether one first takes the OPE of
operators $\lim_{z\to 0} LL' $, to produce a sum of
line operators, or instead first takes the trace on the Hilbert space.
These procedures do not commute:
\begin{equation}
\lim_{z\to 0} F(LL')\not= F(\lim_{z\to 0} LL').
\end{equation}

 In order to understand better the relation between the commutative and
 noncommutative products, let us suppose that there is a
 collection of simple objects $L_i$ which generate the set of all line operators.
 For example, any of the collections $\CL$ of line operators discussed in Section
 \ref{subsec:LabelsForLines} would do.  The simple objects
 have the property that any line operator can be written as $\sum c_i L_i$ where $c_i$
 are positive integers. In order to define the ring structure it suffices to
 compute the fusion coefficients
 \begin{equation}\label{eq:simp-mult}
 \lim_{z\to 0} L_i L_j = \sum_k c_{ij}^k L_k.
 \end{equation}
 However, the analogous decomposition of Hilbert spaces takes the form
 \begin{equation}
 \CH_{L_i L_j}= \oplus_k  N_{ij}^k \otimes \CH_{L_k}
 \end{equation}
 This is an equality of $U(1)_{\vec x \vec x'}$ representations. The surprise
 (at least for the authors) is that in some examples it turns out that
  $N_{ij}^k$ can be   a nontrivial
 representation of $U(1)_{\vec x \vec x'}$. We introduce
\begin{equation}
c_{ij}^k(y) := {\Tr}_{N_{ij}^k} y^{2J_3}
\end{equation}
This is a Laurent polynomial in $y$ with positive integral coefficients and
satisfies $c_{ij}^k(1) = c_{ij}^k$. In these terms the commutative ring
of line operators \eqref{eq:simp-mult} is deformed to
\begin{equation}
F(L_i\circ_y L_j) = \sum_{k} c_{ij}^k(y) F(L_k)
\end{equation}
Because the $\CN=2$ theory is parity invariant we have
$c_{ij}^k(y)= c_{ji}^k(1/y)$.  Note the product $\circ_y$ is therefore commutative
when $y^2=1$.

In order to understand the need for $N_{ij}^k$ to be a non-trivial representation of $SO(2)$, we need to be a little cautious about the
identification of the correct quantum number $J_3$ (and hence $F = 2 J_3 \, \mathrm{mod} \, 2$) for the system of two separate line operators. First of all, in order for $J_3$  even to be defined, we need the two operators to be brought together along the $z$-axis. Furthermore, while for a single line operator the definition of $J_3$ is unambiguous (constant shifts would not be compatible with the nonabelian rotation group), for a system of two defects the unbroken rotation group is abelian, and
overall constant shifts may appear when comparing the $J_3$ which appear in the
definition of the framed PSCs for $L L'$ and for the individual line operators $L_i$.
This shift ambiguity is captured by the $SO(2)$ grading of the multiplicities $N_{ij}^k$.

Notice that our product formulae pass a very stringent test, consistency
with wall-crossing: as long as the products are executed with the rule
$X_\gamma X_{\gamma'} = y^{\langle \gamma,\gamma' \rangle} X_{\gamma+\gamma'}$,
the action of the wall-crossing transformations commutes with the product.

\section{Formal line operators and their remarkable wall-crossing properties}\label{sec:Formal-line-operators}

The combination of the halo wall-crossing picture and the strong positivity conjectures
is very powerful.
Strong positivity does not leave space for cancellations in the
spin character. The coefficients of $y^m X_{\gamma}$ are actual dimensions of Hilbert spaces,
graded by the IR charges and $so(3)$ spin. As we vary $\zeta$,
the parameters and vacuum expectation values of the theory, the
halo wall-crossing concretely adds or removes subspaces of these Hilbert spaces, but the
dimensions must always remain non-negative.
This is a very powerful constraint, and in this subsection we would like to explore its consequences for
some concrete $\CN=2$ theories with simple, well understood vanilla BPS spectra.

\subsection{Chambers and the generating function}\label{subsec:Chambers}

In this section we make some formal definitions which summarize the
behavior of the generating functions \eqref{eq:form-gen}.

Introduce the noncommutative ring
$\IZ[y,y^{-1}, X_{\gamma}]/I$
where the ideal $I$ is generated by the relations
$y X_\gamma = X_\gamma y$ and
$X_\gamma X_{\gamma'} = y^{\langle \gamma, \gamma'\rangle } X_{\gamma+\gamma'}$.
Now recall the space $\Xi$ defined in \eqref{eq:Chamber-Space}. As mentioned there, $\Xi$
is divided into chambers.  We will label the chambers by $c$, where $c$ varies over some
index set.
We define a \emph{strongly positive formal line operator} to be a collection of elements $F(c) \in
 \IZ[y,y^{-1}, X_{\gamma}]/I$
such that

\begin{enumerate}

\item  Across walls $\widehat{W}(\gamma)$  between chambers $c^+, c^-$,
\begin{equation}
F(c^+)  =  S_{\gamma}  F(c^-)  S_{\gamma}^{-1},
\end{equation}
where $c^+$, $c^-$  is the chamber  where $ \Im(Z_\gamma/\zeta)  $ is positive,
negative respectively and $S_\gamma$ is defined in \eqref{eq:Spin-KS-ii}.

\item   In each chamber $c$ of $\Xi$,

\begin{equation}\label{eq:finite-positive-sum}
F(c)  =  \sum_{\gamma}  P_\gamma^c(y) X_\gamma
\end{equation}
where $P_\gamma^c(y)$ is the character of some (true, not virtual)
representation of $SU(2)$.

\end{enumerate}

The second item above defines the ``strongly positive'' condition. We note that the definition
we have given is very closely related to the \emph{universal Laurent polynomials} introduced
by Fock and Goncharov in their study of cluster ensembles \cite{MR2567745}.

If the strong positivity conjecture holds, the generating function
$F(L)$ of a line operator \eqref{eq:form-gen} is a strongly positive formal line operator.
We denote formal line operators as $F$. As for actual line operators, when considering families
of line operators, $F$  is subject to monodromies in the UV parameters. Of course, if we
define chambers on $\CB \times \IC^*$ rather than its universal cover, then we will have monodromy
in this space too.

In the next subsections we will examine some examples of formal line operators in
some simple $\CN=2$ theories.  In these cases we will find some
important simplifications in the chamber structure.
For example, we will only consider walls involving a single hypermultiplet.  The full
quantum dilogarithm technology of Section \ref{subsec:Halo-Wall-Crossing} is a bit of overkill
in this case ---  we can just apply \eqref{eq:transform} directly to remove or add haloes as necessary.
The wall-crossing rule for a single hypermultiplet
in passing \emph{from} a region with ${\rm Im}(Z_{\gamma_h}/\zeta) < 0$ \emph{to} a region with ${\rm Im}(Z_{\gamma_h}/\zeta) > 0$ comes out to be simply
\begin{equation}\label{eq:explicit-dilog-conj}
X_{\gamma_c} \to \begin{cases} \sum_{j=0}^N \ch \left( \Lambda^j \rho_N \right) X_{\gamma_c + j \gamma_h} & \langle \gamma_h,\gamma_c\rangle = N > 0 \\
\sum_{j=0}^\infty (-1)^j \ch \left( S^j \rho_N \right) X_{\gamma_c + j \gamma_h} & \langle \gamma_h,\gamma_c\rangle = - N < 0 \\
\end{cases}
\end{equation}
Here $\rho_N$ is the $N$-dimensional representation of $SU(2)$. The first line
of \eqref{eq:explicit-dilog-conj} clearly preserves positivity.
However, the second line shows that it is highly nontrivial to maintain positivity.
Note that in the first case the sign of the inner product is such that we move into a region
which supports a halo of $\gamma_h$ particles around the core $\gamma_c$, while in the second
case we move into a region which does not support such halo configurations.
The rule for transforming \emph{from} a region with ${\rm Im}(Z_{\gamma_h}/\zeta) > 0$
\emph{to} a region with  ${\rm Im}(Z_{\gamma_h}/\zeta) < 0$ similarly works out to be
\begin{equation}\label{eq:explicit-dilog-conj-ii}
X_{\gamma_c} \to \begin{cases} \sum_{j=0}^\infty (-1)^j \ch \left( S^j \rho_N \right) X_{\gamma_c + j \gamma_h} & \langle \gamma_h,\gamma_c\rangle = N > 0 \\
\sum_{j=0}^N \left( \ch \Lambda^j \rho_N \right) X_{\gamma_c + j \gamma_h}  & \langle \gamma_h,\gamma_c\rangle = - N < 0 \\
\end{cases}
\end{equation}
Again, positivity is manifestly preserved when we move into a region that supports
halo configurations, but otherwise it is not.
Of course, applying \eqref{eq:explicit-dilog-conj} followed by \eqref{eq:explicit-dilog-conj-ii}
must give the identity operator.  This implies the somewhat nontrivial identity
\begin{equation}\label{eq:Char-Ident}
\sum_{j+k=\ell} (-1)^j \ch\left( S^j \rho_N \right) \ch \left( \Lambda^k \rho_N \right) = \delta_{\ell,0},
\end{equation}
for integers $\ell \geq 0$. This is how
halos are ``removed'' when using \eqref{eq:explicit-dilog-conj} to pass from a region
supporting halo configurations to one which does not.

We would like to make one final remark on equations \eqref{eq:explicit-dilog-conj} and
\eqref{eq:explicit-dilog-conj-ii}. Note that when the expression is finite it is manifestly strongly
positive, but when it is infinite it is not.  This suggests that finiteness
of the expansion in $X_\gamma$ is correlated with strong positivity.  In Sections
\ref{subsec:Formal-line-ops-U1} and \ref{sec:formaln3} we will construct strongly positive
formal line operators in some simple field theories.  However, in Section \ref{subsec:formal-SU2} we will
treat a more complicated field theory where it is
difficult for us to prove strong positivity of the formal line operators we construct.
Nevertheless, motivated by all the explicit examples
and the above observation, we would like to conjecture that
\emph{if a formal line operator is strongly positive in a single chamber and has a
finite Darboux expansion in all chambers then it is in fact strongly positive. }

\subsection{Formal line operators and $U(1)$ gauge theory}\label{subsec:Formal-line-ops-U1}

There is a convenient toy model which captures key aspects of the behavior of
formal line operators:  a $U(1)$ gauge theory with a single hypermultiplet of
electric charge $1$.  This theory is not well defined in the UV, but it does capture
the behavior of well defined theories near singularities of the Coulomb branch $\CB$.

In this example $\CB^*$ is the punctured disc, with $\Gamma$ a rank two local system.
The fiber of $\Gamma$ is generated by $\gamma_1, \gamma_2$, with $\langle \gamma_1, \gamma_2 \rangle = +1$,
and the counterclockwise monodromy around the origin acts by
\begin{equation}
\gamma_1 \to \gamma_1 + \gamma_2, \qquad \gamma_2 \to \gamma_2.
\end{equation}
We denote $p \gamma_1 + q \gamma_2$ by $(p,q)$.

We can take $Z_{\gamma_2}(u) = u$.  We will not need an explicit
formula for $Z_{\gamma_1}(u)$, because the only nonzero
vanilla BPS degeneracies are $\Omega(\pm \gamma_2) = +1$.  There is no
wall-crossing for the vanilla BPS degeneracies.

We expect to be able to find line operators corresponding to Wilson -'t Hooft line operators of various
electric and magnetic charges. Because of the one-loop beta function due to the
presence of the charged particle, we expect to see the effect of an anomalous R-symmetry: an R-symmetry transformation
requires a shift of the $\theta$ angle, which in turn induces, by Witten's effect, a shift of the electric charge of
magnetically charged objects. This results in a monodromy of line operator labels under rotation of $\zeta$ by $2 \pi$.
There will be a beautiful interplay between this monodromy and wall-crossing.

The universal cover $\widehat\CB \times \widehat\IC^*$ is $\IC \times \IC$ with coordinates
$(\log u, \log \zeta)$.   Given our vanilla BPS spectrum, the walls are
\begin{align}
\hW(\gamma_2) &= \amalg_{n \in \IZ} \{ (\log u, \log \zeta) :  \arg u - \arg \zeta = (2n+1)\pi \}, \\
\hW(-\gamma_2) &= \amalg_{n \in \IZ} \{ (\log u, \log \zeta) :  \arg u - \arg \zeta = 2n \pi \},
\end{align}
and hence the chambers can be taken to be
\begin{equation}
c_n: = \{ (\log u, \log \zeta) : n \pi < \arg u - \arg \zeta < (n+1)\pi \}.
\end{equation}

Let us now try to construct a formal line operator $F^{(n)}_{p,q}$, by declaring its
value in chamber $c_n$ to be $F^{(n)}_{p,q}(c_n) = X_{p,q}$. This is the naive generating function for
a Wilson -'t Hooft line operator of electric charge $q$ and magnetic charge $p$.
The transformation laws \eqref{eq:explicit-dilog-conj} and \eqref{eq:explicit-dilog-conj-ii} then determine
$F^{(n)}_{p,q}$ in all other chambers, and we may ask:  do they obey the strong
positivity constraint?

First of all, if $p=0$ there is
no transformation and hence $F^{(n)}_{0,q}(c) = X_{0,q}$ for all chambers $c$.  This certainly obeys
the constraint of formal positivity.  Since they are actually independent of $n$ we
denote these formal line operators simply by $F_{0,q}$. These are Wilson line operators of charge $q$,
and have no interesting wall-crossing because the vanilla particles carry electric charge only.

Now, note that in the stability condition
 $\langle \gamma_h, \gamma_c \rangle {\rm Im} Z_{\gamma_h}/\zeta >0  $  we need only consider
 $\gamma_h = \pm \gamma_2$, and in fact the condition is the same for these two charges.
 If $\gamma_c = (p,q)$ then the stability condition becomes
 $-p {\rm Im} Z_{\gamma_2}/\zeta >0$.  Thus for $p>0$ there can be halo configurations in $c_n$
 for $n$ odd and not for $n$ even, and if $p<0$ there can be halo configurations in $c_n$ for
 $n$ even and not for $n$ odd. See Figures \ref{fig:chamber-u1-even} and \ref{fig:chamber-u1-odd}.

\insfig{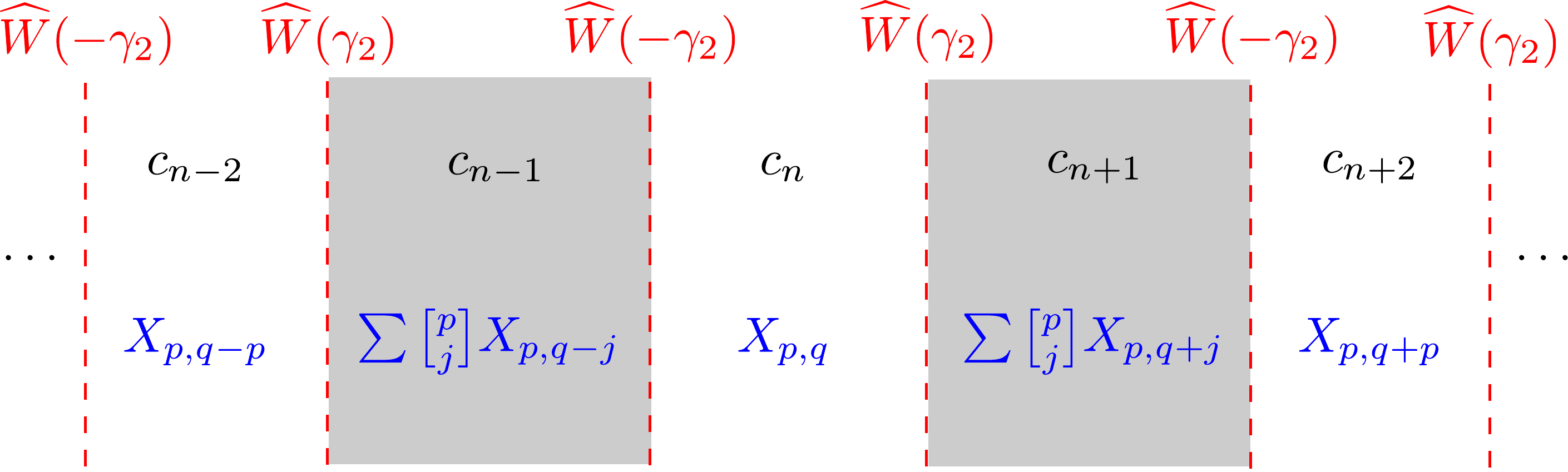}{Chambers and strongly positive formal line operators in the very simplest
case, a $U(1)$ gauge theory with a single BPS hypermultiplet. Here $n=0 \mod 2$ and $p>0$. Regions
which are stable to halo formation are shaded.}
\insfig{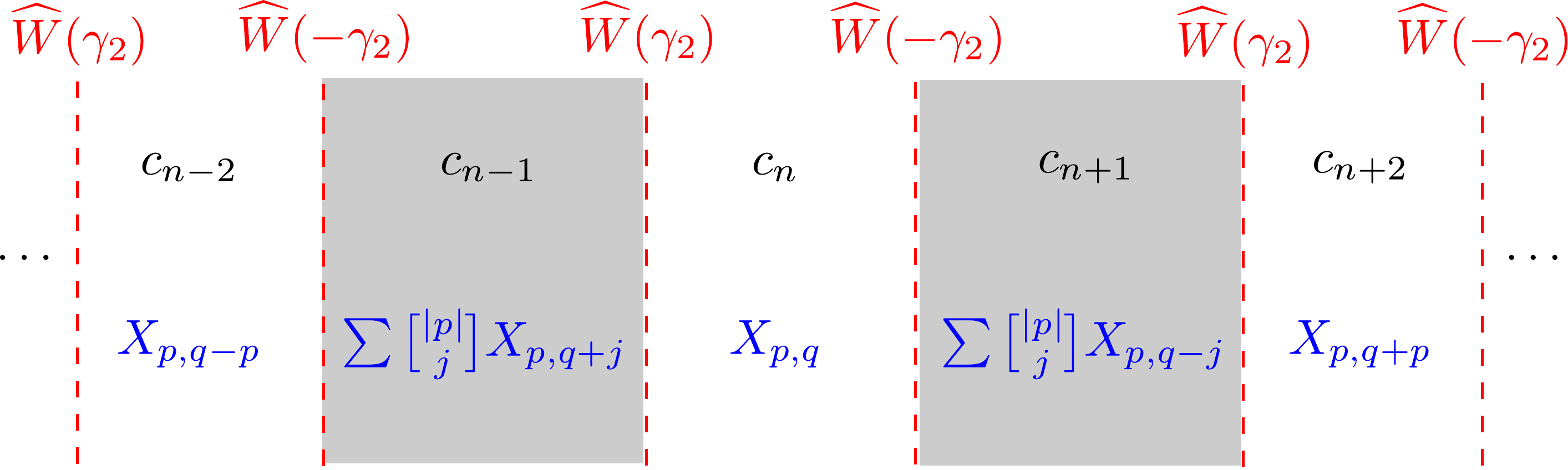}{Chambers and strongly positive formal line operators in the very simplest
case, a $U(1)$ gauge theory with a single BPS hypermultiplet. Here $n=1 \mod 2$ and $p<0$. Regions
which are stable to halo formation are shaded.}

In analyzing the behavior of $F^{(n)}_{p,q}$ there are four cases to consider: $n$
can be even or odd, and $p$ can be positive or negative.

\begin{enumerate}

\item   $n=0 \mod 2$, $p>0$. In this case we cross a wall of type $\hW(\gamma_2)$ in going from
$c_n$ to $c_{n+1}$. We apply the second line of \eqref{eq:explicit-dilog-conj-ii} and find that
\begin{equation}\label{eq:n-even-p-pos}
F^{(n)}_{p,q}(c_{n+1}) = \sum_{j=0}^p \qbinom{p}{j} X_{p,q+j}.
\end{equation}
Similarly, in passing from $c_n$ to $c_{n-1}$ we pass through a wall of type $\hW(-\gamma_2)$.
Now we apply the first line of \eqref{eq:explicit-dilog-conj} and obtain
\begin{equation}
F^{(n)}_{p,q}(c_{n-1}) = \sum_{j=0}^p \qbinom{p}{j}  X_{p,q-j}.
\end{equation}
When we continue from $c_{n+1}$ to $c_{n+2}$ we could apply our formal rules and
use the identity \eqref{eq:Char-Ident}, but it is simpler to work backwards, recalling
that the stability condition
does not allow halos in the chamber $c_{n+2}$.  Therefore, in passing from $c_{n+2}$ to $c_{n+1}$
through the wall $\widehat{W}(-\gamma_2)$, we must be creating a halo of particles of charge
$-\gamma_2$.  This operation would indeed produce \eqref{eq:n-even-p-pos}, if and only if
\begin{equation}\label{eq:u1-monod}
F^{(n)}_{p,q}(c_{n+2} ) = X_{p, q+p}.
\end{equation}
In a similar way, continuing to chamber $c_{n-2}$ we find $F^{(n)}_{p,q}(c_{n-2} ) = X_{p, q-p}$.
The pattern clearly continues in both directions and thus $F^{(n)}_{p,q}$ is a
formal line operator. Moreover, this equation shows that $F^{(n)}_{p,q} = F^{(n+2)}_{p,q+p}$, so without loss of generality we can
reduce these operators to a single set defined by the value in the chamber $c_0$.
The shift of electric charge as $n \to n+2$ has two independent interpretations:
if we interpret it as the result of a monodromy of the $\zeta$ variable, it is the monodromy of UV labels
$F_{p,q} \to  F_{p,q+p}$, while if we interpret it as the result of a monodromy in the $u$ plane
it expresses the monodromy of the IR charge lattice $\Gamma$ around the singular locus $u=0$.

Finally, this example illustrates that in expanding $F$ in terms of $X_{\gamma}$,
there is no invariant distinction between the summands representing cores and those representing haloes.
Indeed, in crossing $\widehat{W}(\gamma_2)$ from $c_n$ to $c_{n+1}$
the core charge is $(p, q)$, but when crossing $\widehat{W}(-\gamma_2)$
from $c_{n+2}$ to $c_{n+1}$, the core charge is $(p, q+p)$.  This is as expected from our physical picture:
the distinction between cores and haloes is sharp only when we are very close to a wall.

\item $n=0~\mod 2$, $p<0$. Now the stability condition allows halos in the chambers $c_n$
and not in $c_{n\pm 1}$, so we should expect trouble since when we try to continue to the
adjacent chambers $c_{n \pm 1}$ we cannot have halo configurations. Indeed, now crossing the wall $\widehat{W}(\gamma)$
into chamber $c_{n+1}$ we must apply line 1 of \eqref{eq:explicit-dilog-conj-ii}.  Acting
with this transformation on $X_{p,q}$ clearly gives an expression which
violates strong positivity.  Thus, $F^{(n)}_{p,q}$ does not give a formal line operator.

\item $n=1 ~ \mod 2$, $p>0$. This is similar to the case of $(n=0~\mod 2, p<0)$ and does not
give a formal line operator.

\item $n=1~ \mod 2$, $p<0$. This is similar to the case of $(n=0~\mod 2, p>0)$.  It does
give a formal line operator, with
\begin{equation}\label{eq:n-odd-p-neg}
F^{(n)}_{p,q}(c_{n+1}) = \sum_{j=0}^{\vert p\vert} \qbinom{ \vert p \vert }{ j} X_{p,q-j},
\end{equation}
and $F^{(n)}_{p,q}(c_{n+2})=X_{p,q+p}$.  Again $F^{(n)}_{p,q} = F^{(n+2)}_{p,q+p}$,
so we can restrict without loss of generality to $F^{(1)}_{p,q}$.

\end{enumerate}

In summary, because $F^{(n)}_{p,q} = F^{(n+2)}_{p,q+p}$ it suffices to
consider the chambers $n=0$ and $n=1$ in order to construct
those strongly positive formal line operators which reduce to a monomial in some chamber.
The resulting operators
are in 1-1 correspondence with pairs of integers $(p,q)$.  Hence we may simply
denote them by $F_{p,q}$. If $p=0$ then $F_{0,q}= X_{0,q}$ in all chambers. If $p>0$
then $F_{p,q} = X_{p,q}$ in chamber $c_0$ and if $p<0$ then $F_{p,q} = X_{p,q}$ in
chamber $c_1$.  These are clearly simple line operators
  and in fact they are the only simple line operators: given
 any formal line operator we can split it into a piece with positive magnetic charges, a piece with no magnetic charges and a piece with
 negative magnetic charges only. Clearly the piece with positive magnetic charges can be decomposed into simple pieces just by
 decomposing it into monomials in a chamber with even $n$. A similar statement holds true for the piece with negative magnetic charge and odd $n$,
 and the piece with no magnetic charge and any $n$.
  The formal line operators $F_{p,q}$ should be identified with the protected spin characters of
'tHooft-Wilson loops in the $U(1)$ theory.

Even in this simplest example the algebra of formal line
operators is already somewhat non-trivial, although simple enough
to write down explicitly.  Indeed,
$F_{\pm 1, 0}$, $F_{0,\pm 1}$ already generate a ring.
 One readily works out the relation  $F_{1,0} F_{-1,0} = 1 + y F_{0,1}$
 by hand, but higher multiplication involves cumbersome manipulation
 of $q$-binomial coefficients.   The ring relations can be easily worked out
 as follows. First, it is very easy to check that
 \begin{equation}
 F_{p,q} F_{r,s} = y^{ps-qr} F_{p+r, q+s}
 \end{equation}
 holds if $p\geq 0, r\geq 0$ or if $p\leq 0, r\leq 0$ because we can evaluate
 in chambers $c_0$ or $c_1$ where the operator is a monomial.  The case where $p,r$ have
 opposite sign is more nontrivial.  Suppose for definiteness that $p\geq 0, r\leq 0$. Then, by
 definition:
 \begin{equation}
 F_{p,q} F_{r,s} = \begin{cases} X_{p,q} \left( \sum_{j=0}^{\vert r\vert} \qbinom{\vert r\vert }{ j} X_{r,s+j}
 \right) & {\rm in } ~~ c_0 \\
 \left( \sum_{j=0}^p \qbinom{p }{ j} X_{p,q+j} \right) X_{r,s} & {\rm in } ~~ c_1 \\
 \end{cases}
 \end{equation}
 and hence carrying out the multiplications we have:
 \begin{equation}
 F_{p,q} F_{r,s} = \begin{cases} y^{ps-qr} \sum_{j=0}^{\vert r\vert}
 \qbinom{\vert r\vert }{ j} y^{p j} X_{p+r,q+s+j} & {\rm in }  ~~ c_0 \\
 y^{ps-qr}  \sum_{j=0}^p \qbinom{p }{ j}  y^{-r j} X_{p+r,q+s+j} & {\rm in }  ~~ c_1 \\
 \end{cases}
 \end{equation}
Now, to identify what line operator the RHS corresponds to we should view the result in the chamber
in which the line operator with magnetic charge $p+r$ is a monomial. Thus we obtain the product on
line operators:
 \begin{equation}\label{eq:U1-ring-rel}
 F_{p,q} F_{r,s} = \begin{cases} y^{ps-qr} \sum_{j=0}^{\vert r\vert}
 \qbinom{\vert r\vert }{ j} y^{p j} F_{p+r,q+s+j} &  p+r \geq 0 \\
 y^{ps-qr}  \sum_{j=0}^p \qbinom{p }{ j} y^{-r j} F_{p+r,q+s+j} & p+r \leq 0 \\
 \end{cases}
 \end{equation}
We can easily obtain the other case where $p\leq 0, r \geq 0$ by recalling that $F(L_1) F(L_2)$
is the same as $F(L_2) F(L_1)$ with the replacement $y \to 1/y$.
A brute force
check of \eqref{eq:U1-ring-rel} implies   nontrivial  identities on $y$-binomials (
the $y$-deformed Pascal relations),
\begin{equation}
y^j \qbinom{p}{j} = \qbinom{p-1}{j} + y^p \qbinom{p-1}{j-1}.
\end{equation}

\subsubsection{Higher charges}

A natural modification of this construction is to consider a $U(1)$ theory coupled to an  hypermultiplet
of electric charge $Q$. That is, we now consider the BPS spectrum
 \begin{equation}\label{eq:Q-BPS-spec}
 \Omega(\gamma) = \begin{cases} 1 & \gamma = \pm Q \gamma_2, \\  0 & {\rm else.} \\ \end{cases}
 \end{equation}
We can get the formal line operators immediately by a small trick:
if we rescale the electric and magnetic charge lattices of the theory by an opposite factor of $Q$, we go back to
the $Q=1$ case.  The algebra relations above apply to $F'_{p',q'}$  where
$F'_{p',q'} := F_{Q^{-1} p', Q q'} $. Honest line operators must have an integral number of charges.
Thus, we now obtain the relation
\begin{equation}\label{eq:Q-U1-algebra}
F_{1,0} F_{-1,0} = \sum_{j=0}^Q \qbinom{Q }{ j} y^{Qj} F_{0,Qj}.
\end{equation}

\subsection{Formal line operators and $N=3$ Argyres-Douglas theory} \label{sec:formaln3}

We now describe the formal line operators for the simplest example of a theory with wall-crossing.
This is the $N=3$ Argyres-Douglas theory. See Section 9.4 of \cite{Gaiotto:2009hg} for further details about this
theory. Briefly, $\CB= \IC - \{ u_+, u_-\}$. The universal cover is already somewhat complicated,
so we will work at definite regions on the $u$-plane and consider the chambers in the (universal
cover of the) $\zeta$-plane. The best description of the local system is that
$\Gamma = H_1( \Sigma_u, \IZ)^-$ where $\Sigma_u$ is the fiber of a family of Riemann surfaces
over $\CB$ defined by
\begin{equation}
\lambda^2 = (z^3 - 3 \Lambda^2  z + u) (dz)^2.
\end{equation}
where $z\in \IC$.
The fibers $\Gamma_u$ are rank two and will be taken to have generators $\gamma_1, \gamma_2$ with
$\langle \gamma_1, \gamma_2 \rangle = 1$. Vectors $p\gamma_1 + q\gamma_2$ will sometimes
be denoted $(p,q)$. There are two singular points at $u=u_\pm = \pm 2 \Lambda^3$ around
which $\Gamma$ has monodromy. (See equation (9.30) of \cite{Gaiotto:2009hg}.) The spectrum of
BPS states was determined in \cite{Gaiotto:2009hg}.  The base $\CB$ should be divided into
  three regions $\CB_s, \CB_{w}^\pm$ on which $\Gamma$ can be trivialized. $\CB_{s}$ is a region
around $u=0$ and $\CB_{w}^\pm$ are two regions around $u=\infty$. There is a connected wall passing through
$u_\pm$ which separates $\CB_s$ from $\CB_{w}^\pm$. We have
\begin{equation}
\Omega(\gamma;u) = \begin{cases} 1 & \gamma = \pm \gamma_1, \pm \gamma_2 \\
0 & {\rm else} \\ \end{cases}
\end{equation}
for $u \in \CB_s$ and wall crossing then determines
\begin{equation}
\Omega(\gamma;u) = \begin{cases} 1 & \gamma = \pm \gamma_1, \pm \gamma_2, \pm (\gamma_1 + \gamma_2) \\
0 & {\rm else} \\ \end{cases}
\end{equation}
for $u \in \CB_w^+$. The spectrum in $\CB_w^-$ follows from monodromy or wall-crossing.

\insfig{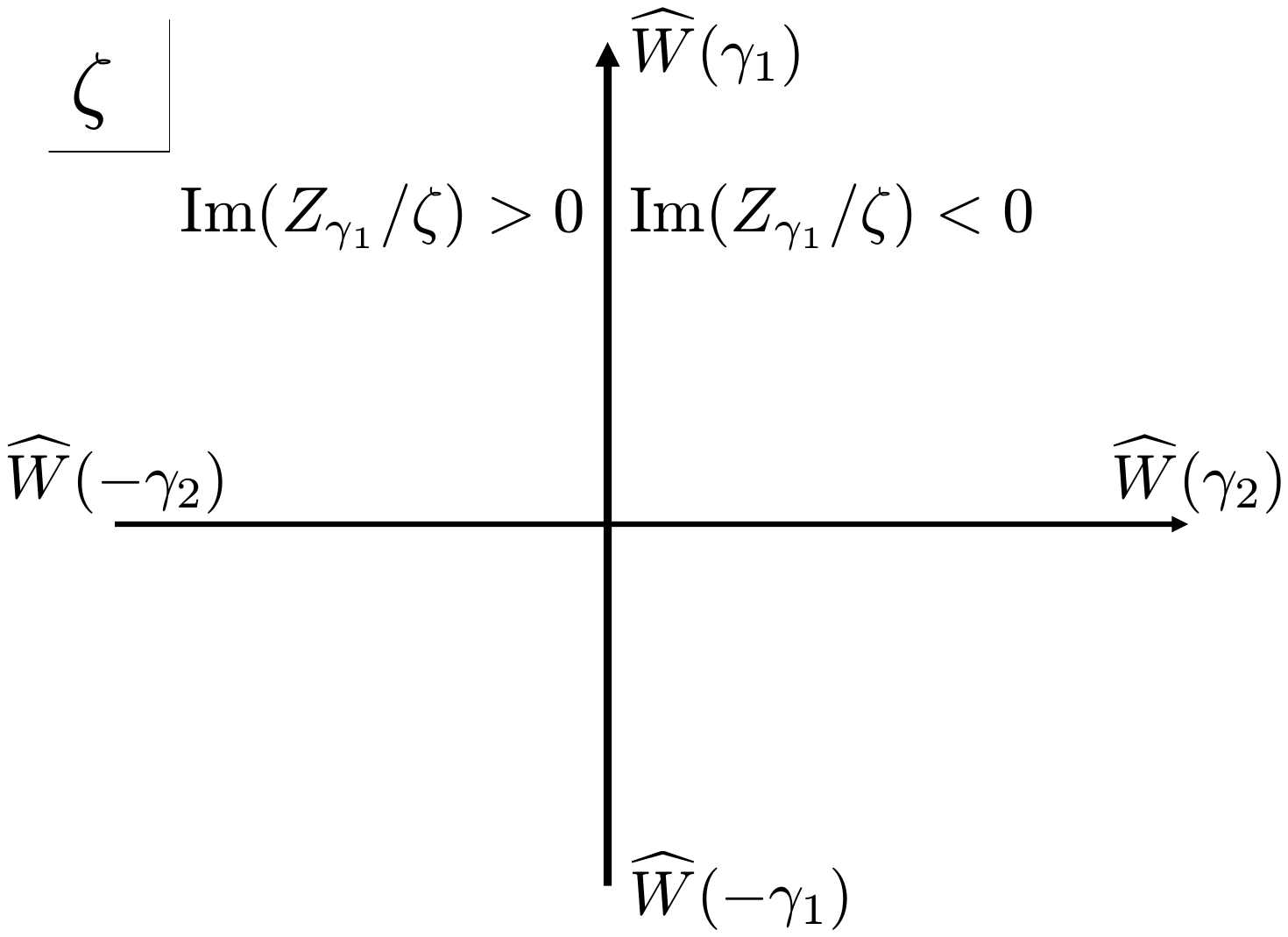}{BPS walls for the $N=3$ Argyres-Douglas theory at $u=0$ with $\Lambda>0$ projected
to the $\zeta$-plane.}

A suitable choice of cycles of $\Sigma_u$ at $u=0$ gives $Z_{\gamma_1} = - i \Lambda^{5/2} K$ and $Z_{\gamma_2} = -K\Lambda^{5/2}$
with $K>0$ a (computable) constant.
  The projection of the BPS walls for $u=0$ into the $\zeta$ plane then
appear as in figure \ref{fig:ADN3-FL-1}. Note that in crossing each wall $\widehat{W}(\gamma_h)$
we pass from the region with ${\rm Im}(Z_{\gamma_h}/\zeta)>0$ to the region with ${\rm Im}(Z_{\gamma_h}/\zeta)<0$
when traveling in the \emph{clockwise} direction. Thus when transforming across walls in the clockwise direction
we should apply the rule \eqref{eq:explicit-dilog-conj-ii} while when transforming in the counterclockwise
directions we should apply the rule \eqref{eq:explicit-dilog-conj}.

Let us follow the strategy of Section \ref{subsec:Formal-line-ops-U1}
 and try to find formal line operators which reduce to a
monomial $X_{p,q}$ in some chamber in the $\zeta$-plane.  Introduce
chambers $c_n$ where $\frac{\pi}{2} (1-n) < \arg\zeta < \frac{\pi}{2}(2-n) $.
The first obstruction comes from single wall-crossings in the clockwise or
counterclockwise directions. This leads to the result that $X_{p,q}$ can only
satisfy formal positivity if:

\begin{enumerate}

\item $p\geq 0, q\geq 0 $ in $c_n$ for $n=1\mod 4$

\item $p\leq 0, q\geq 0 $ in $c_n$ for $n=2\mod 4$

\item $p\leq 0, q\leq 0 $ in $c_n$ for $n=3\mod 4$

\item $p\geq 0, q\leq 0 $ in $c_n$ for $n=0\mod 4$

\end{enumerate}

We will now show that this is the only obstruction.

\insfig{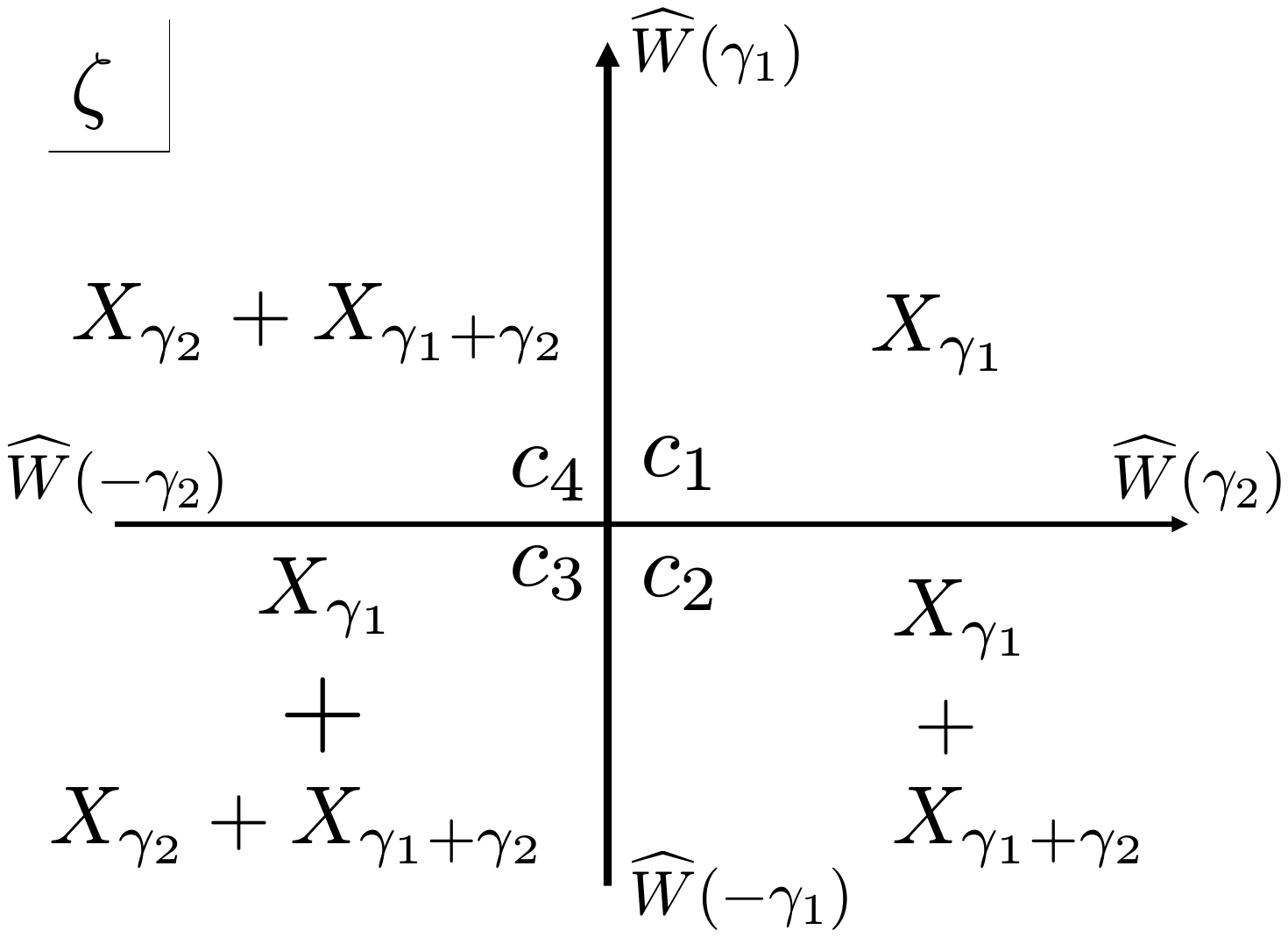}{The first sheet of the 5-sheeted cover contains four chambers.}

\insfig{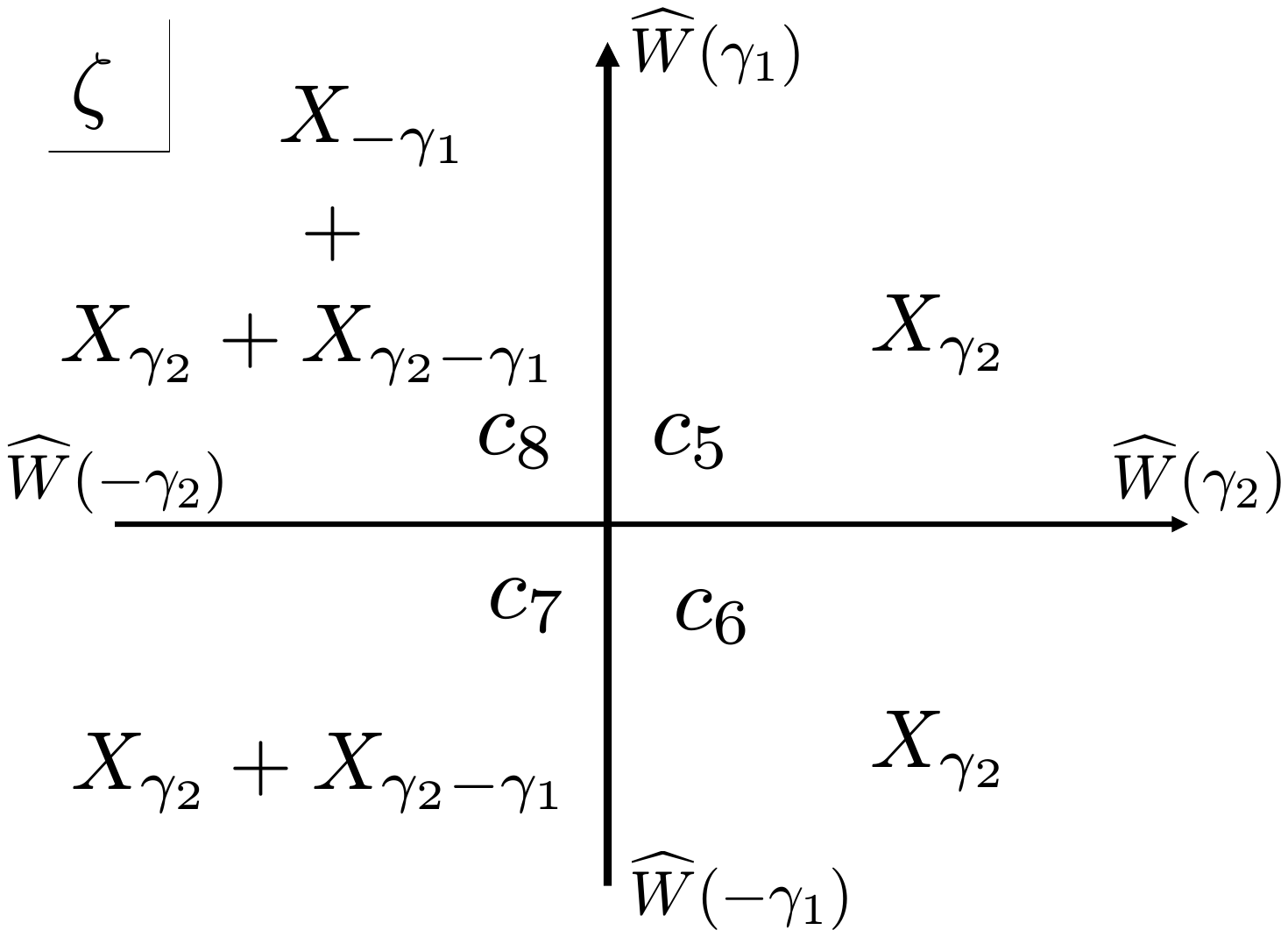}{Continuing the analytic continuation in the clockwise direction to the second sheet.}

\insfig{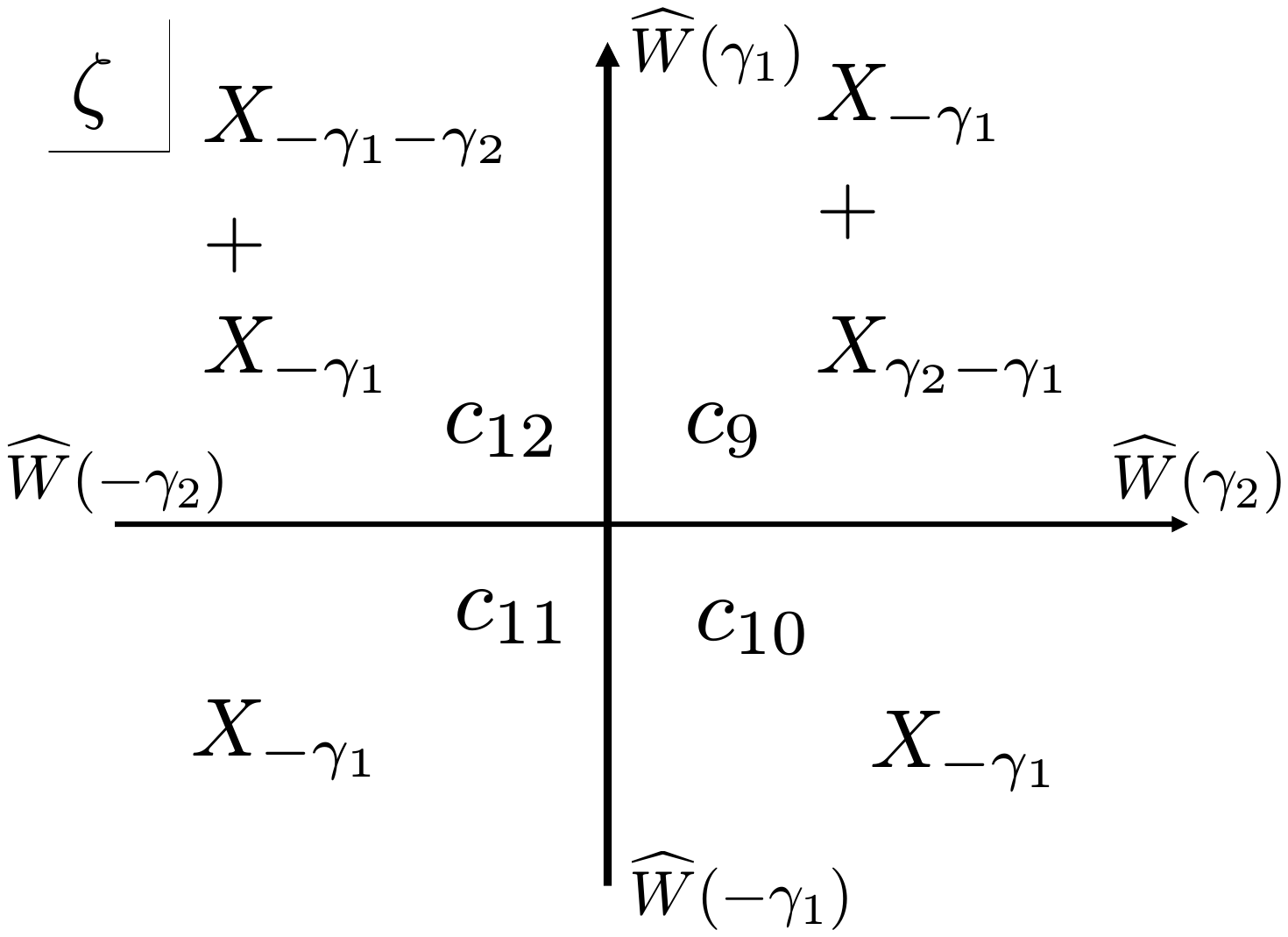}{Continuing the analytic continuation in the clockwise direction to the third sheet.}

\insfig{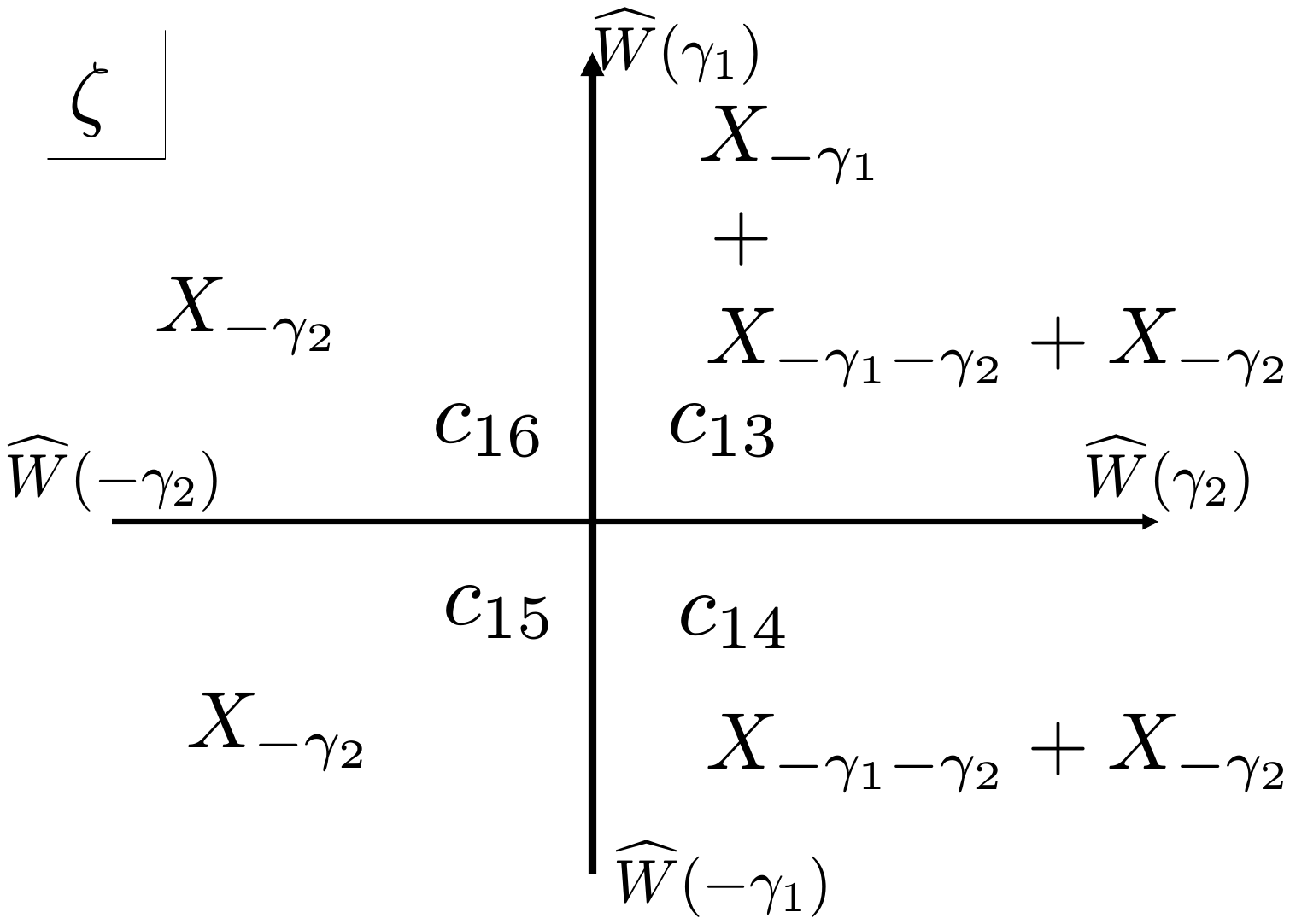}{Continuing the analytic continuation in the clockwise direction to the fourth sheet.}

\insfig{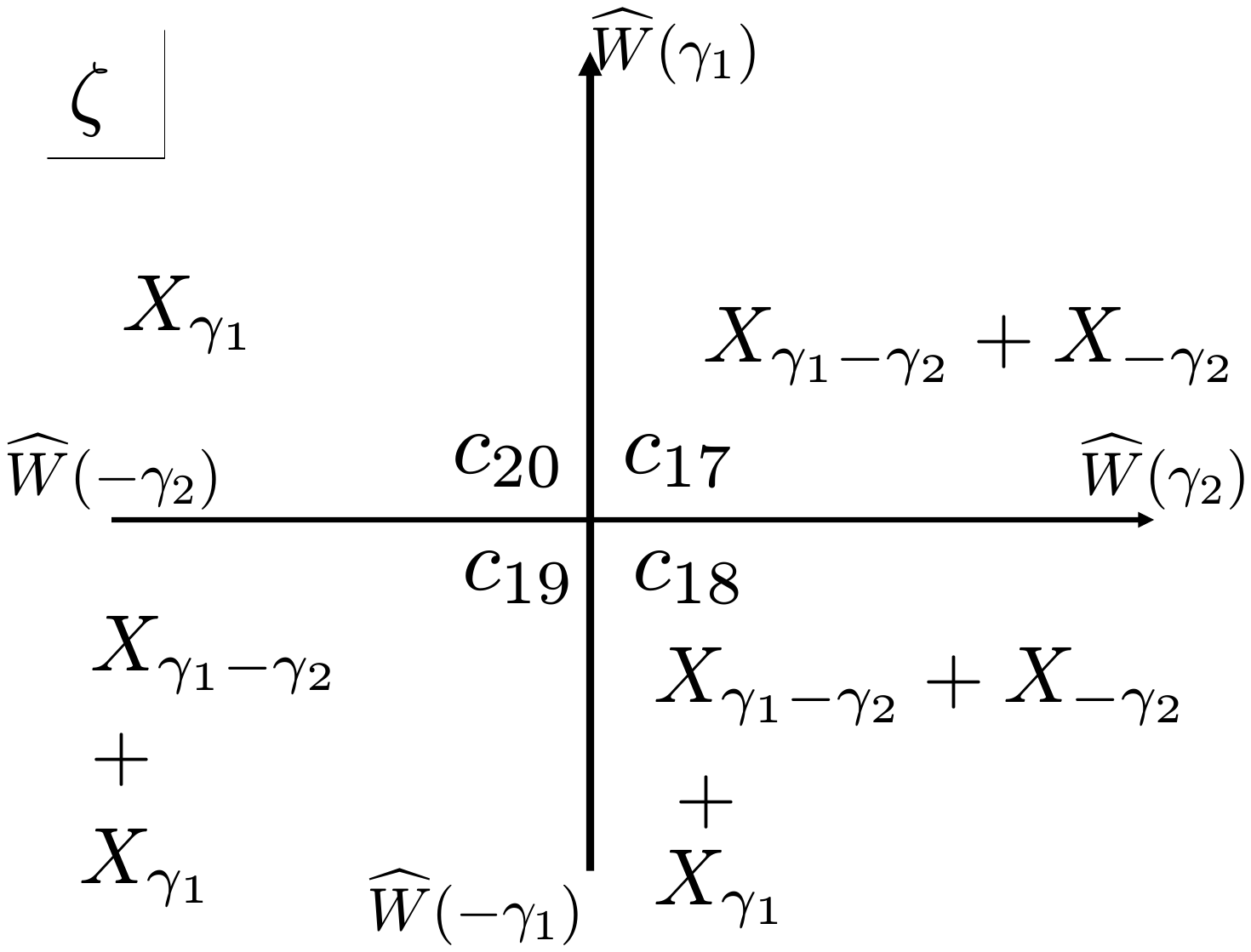}{Continuing the analytic continuation in the clockwise direction to the fifth sheet.
Clockwise wall crossing from chamber $c_{20}$ produces $X_{\gamma_1}$ and therefore the whole
process repeats. Thus chamber $21$ on the universal cover
is identified with chamber $1$, and chamber numbers should be identified modulo $20$ when
working on the $5$-fold cover.}

For definiteness, let us start with $X_{\gamma_1}$ in a chamber $0 < \arg\zeta < \frac{\pi}{2}$
 in the universal cover $\widehat{\IC^*}$. Wall crossing in the clockwise direction
 generates the expressions shown in figure \ref{fig:ADN3-FL-2}. We observe some of the same phenomena
 we saw in the $U(1)$ example. Crossing from $c_1$ to $c_2$ we gain a halo with core charge $\gamma_1$.
 Crossing from $c_2$ to $c_3$ we gain another halo, but this time the core charge is $\gamma_1+\gamma_2$
 and the halo charge is $-\gamma_1$.
 Crossing from $c_3$ to $c_4$ we lose a halo particle, with the core charge again interpreted as
 $\gamma_1+\gamma_2$ but with halo particle of charge $-\gamma_2$. Now if we continue in the clockwise direction
 from this chamber
 we generate a new expression, $X_{\gamma_2}$. We have again removed a halo particle from a halo
 with core $\gamma_2$. Not surprisingly, we have found there is monodromy and we really must discuss
 the wall-crossing on the universal cover $\widehat{\IC^*}$. When doing so we find the  schematic pattern:
 \begin{equation}
 \cdots \to X \to X \to X+X \to X+X+X \to X+X \to X \to X \to X+X \to \cdots
 \end{equation}
 It turns out that the pattern is repeated with a period of $20=4\times 5$. Modding out the universal
 cover by this periodicity we obtain a $5$-fold cover of the $\zeta$ plane. Each sheet contains
 four chambers for a total of $20$ chambers  labeled $c_1, \dots, c_{20}$. The explicit expressions
 for the formal line operator in all $20$ chambers are illustrated in   figures \ref{fig:ADN3-FL-2}
 to \ref{fig:ADN3-FL-6}.

 \insfig{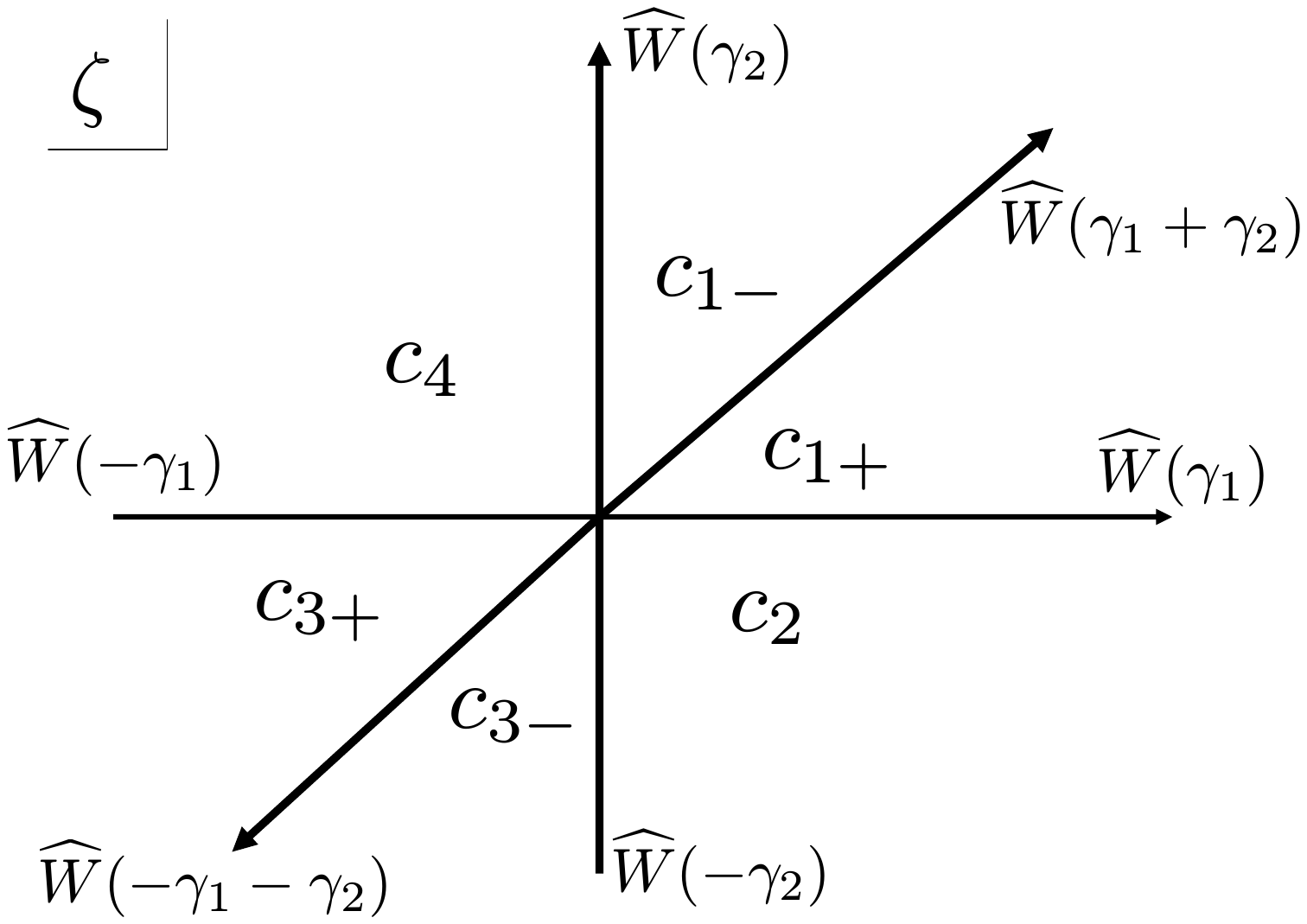}{Chamber structure in one of the weak coupling regions.  }

 Unfortunately, this does not completely construct the strongly positive line operator because
 we must check that it is strongly positive in the weak coupling region. Moving through the marginal stability wall into the region $\CB_w^+$ the walls for $\gamma_1$ and $\gamma_2$ merge and produce a third wall leading to the
 new chamber structure in the $\zeta$ plane shown in figure \ref{fig:ADN3-FL-weak}. Each of the odd chambers
  splits into two so there are $6$ chambers on each sheet for a total of $30$ chambers. Note that the expressions in
 the even numbered chambers cannot change from the strong coupling region. It is straightforward to check that
 wall crossing produces strongly positive elements in the odd chambers $c_{2n+1,\pm}$.

 Thus, we have completely constructed a strongly positive formal line operator.
 Using the monodromy operation we can in
 fact produce $5$ distinct formal line operators defined by
\begin{align}
\begin{split}
F_1(c_1) & =  X_{\gamma_1} \\
F_2(c_1) & =  X_{\gamma_2} \\
F_3(c_1) & = X_{-\gamma_1} + X_{\gamma_2-\gamma_1}\\
F_4(c_1) & = X_{-\gamma_1} + X_{-\gamma_1-\gamma_2}+X_{-\gamma_2}\\
F_5(c_1) & = X_{-\gamma_2} + X_{\gamma_1-\gamma_2} \\
\end{split}
\end{align}

Using these operators we can produce a number of other strongly positive operators as follows.
Consider $F_{p,q}:=y^{-pq} F_1^p F_2^q$ for $n,m\geq 0$. In the chamber $c_1$ with $u$
in the strong coupling regime
this has value $X_{(p,q)}$. We claim this generates a strongly positive formal line operator.
A straightforward attempt to check this rapidly becomes very difficult, but we can prove it using
the following trick. First note that wall-crossing straightforwardly produces strongly positive
expressions if we move two chambers in either the clockwise or counterclockwise direction. Indeed
we find the explicit expressions:
\begin{align}\label{eq:Fpq-chs-N3}
\begin{split}
F_{p,q}(c_{19}) & = \sum_{j=0}^q \sum_{k=0}^{p+j} \qbinom{q}{j}\qbinom{p+j}{k} X_{p+j,q-k}\\
F_{p,q}(c_{20}) & = \sum_{j=0}^q \qbinom{q}{j} X_{p+j,q}\\
F_{p,q}(c_1) & = X_{p,q} \\
F_{p,q}(c_2) & = \sum_{j=0}^p \qbinom{p}{j} X_{p,q+j}\\
F_{p,q}(c_3) & = \sum_{j=0}^p \sum_{k=0}^{q+j} \qbinom{p}{j}\qbinom{q+j}{k} X_{p-k,q+j}\\
\end{split}
\end{align}
Of course, since $q$-binomials are true spin characters so are their products. Thus
\eqref{eq:Fpq-chs-N3} satisfies strong positivity.
Now observe that $F_1$ becomes a monomial in the $8$ chambers $c_{20},c_{1}$, $c_5,c_6$, $c_{10},c_{11}$,
$c_{15},c_{16}$. Similarly, $F_2$ becomes a monomial in the $8$ chambers
  $c_1,c_2$, $c_{6},c_{7}$,
$c_{11},c_{12}$, $c_{16},c_{17}$.  Thus in chambers $c_1, c_6, c_{11},c_{16}$ the expression
$F_{p,q}$ will be a monomial. Explicitly we have
\begin{align}\label{eq:Fpq-mon-chs}
\begin{split}
F_{p,q}(c_6) & = X_{(-q,p)} \\
F_{p,q}(c_{11}) & = X_{(-q,-p)} \\
F_{p,q}(c_{16}) & = X_{(q,-p)}\\
\end{split}
\end{align}
Now, applying the two-step clockwise and anticlockwise transformations
analogous to \eqref{eq:Fpq-chs-N3} to each of the expressions in \eqref{eq:Fpq-mon-chs}
one checks that wall-crossing produces strongly positive expressions. In this way we can cover all
twenty chambers for the strong coupling region. An argument given at the end of Section
\ref{subsec:Seeds-from-N=2} implies that there is no obstruction to extending these
as strongly positive operators in the weak coupling region.

In a similar way we can generate strongly positive formal line operators from products of the
other line operators $F_i$.  Relabeling $F_{p,q}$ above
by $F_{p,q}^{(1)}$ we can define (with $p\geq 0, q\geq 0$ in all cases):

\begin{enumerate}

\item
$F_{p,q}^{(2)} = y^{-pq} F_2^p F_3^q $ which is a monomial in $c_2, c_7, c_{12}, c_{17}$

\item
$F_{p,q}^{(3)}= y^{-pq} F_3^p F_4^q $ which is a monomial in $c_3, c_8, c_{13}, c_{18}$

\item
$F_{p,q}^{(4)} = y^{-pq} F_4^p F_5^q $ which is a monomial in $c_4, c_9, c_{14}, c_{19}$

\item
$F_{p,q}^{(5)}= y^{-pq} F_5^p F_1^q $ which is a monomial in $c_5, c_{10}, c_{15}, c_{20}$

\end{enumerate}

Should we expect the $F_{p,q}^{(i)}$ to exhaust the list of \emph{simple} strongly positive formal line operators?
We can sketch a proof, which can be formalized with the tools presented in Section \ref{sec:Tropic}.
Consider the expansion of some formal line operator $F$ in chamber $c_1$ and look at terms of the form $a_{n_1, n_2} X_{n_1 \gamma_1 + n_2 \gamma_2}$
with $n_1 \geq 0$ and $n_2 \geq 0$. Is $\tilde F = F - \sum_{n_1>0, n_2>0}  a_{n_1, n_2} F^{(1)}_{n_1, n_2}$ still strongly positive?
It is positive in $c_1$, and has a finite number of terms in all other chambers, hence it must be positive there as well
(as the wall-crossing from a chamber to the next can only add infinitely many negative terms to a positive sum).
Next we can consider $\tilde F$ in chamber $c_2$ and subtract appropriate multiples of $F^{(2)}_{p,q}$, etc.
At the end of the process we are left with a decomposition of $F$ into a positive sum of $F^{(i)}_{p,q}$.

We can now explore the algebra generated by these operators. By computing in chamber $c_1$ it is straightforward
to compute
\begin{equation}
F_1 F_3 = 1 + y F_2
\end{equation}
Since the other operators are generated by monodromy we have
\begin{equation}\label{eq:Deformd-N3-rels}
F_{i-1} F_{i+1} = 1 + y F_i
\end{equation}
To get the multiplication in the other order we take $y \to 1/y$.

\subsection{Formal line operators in pure $SU(2)$ gauge theory}\label{subsec:formal-SU2}

We now turn to gauge theories based on a gauge group with Lie algebra $su(2)$ and
no flavors. At first sight these are deceptively similar to the $N=3$ Argyres-Douglas theory.
To describe the chamber structure and BPS spectrum we rely on
Section 10.1  of \cite{Gaiotto:2009hg}. Once again, $\CB= \IC - \{ u_+, u_-\}$, and the
best description of the charge lattice is that
$\Gamma = H_1( \Sigma_u, \IZ)^-$, where $\Sigma_u$ is now the fiber of a family of Riemann surfaces
over $\CB$ defined by
\begin{equation}
\lambda^2 = \left(\frac{\Lambda^2}{z^3 } + \frac{2u}{z^2} + \frac{\Lambda^2}{z} \right) (dz)^2.
\end{equation}
where $z\in \IC^*$.
The fibers $\Gamma_u$ are once again rank two and vectors $p\gamma_1 + q\gamma_2$ will sometimes
be denoted $(p,q)$.  One very important difference from the
$N=3$ Argyres-Douglas theory is that now
\begin{equation}\label{eq:SU2-innprod}
\langle \gamma_1, \gamma_2 \rangle = 2.
\end{equation}
There are two singular points at $u=u_\pm = \pm   \Lambda^2$ around
which $\Gamma$ has monodromy. (See (10.4), (10.5) of \cite{Gaiotto:2009hg}.)
The base $\CB$ should be divided into
three regions $\CB_s, \CB_{w}^\pm$ on which $\Gamma$ can be trivialized. $\CB_{s}$ is a region
around $u=0$ and $\CB_{w}^\pm$ are two regions around $u=\infty$. There is a connected wall passing through
$u_\pm$ which separates $\CB_s$ from $\CB_{w}^\pm$. We have
\begin{equation}
\Omega(\gamma;u) = \begin{cases} 1 & \gamma = \pm \gamma_1, \pm \gamma_2 \\
0 & {\rm else} \\ \end{cases}
\end{equation}
for $u \in \CB_s$.  The inner product \eqref{eq:SU2-innprod}
leads to a more complicated wall-crossing than in the $N=3$ theory:  we find
\begin{equation}\label{eq:SU2-weak-spec}
\Omega(\gamma;u) = \begin{cases} -2 & \gamma = \pm (\gamma_1 +  \gamma_2) \\
1 & \gamma = \pm [\gamma_1 + n(\gamma_1+\gamma_2) ], \quad n \geq 0 \\
1 & \gamma = \pm [\gamma_2 + n(\gamma_1+\gamma_2) ], \quad n \geq 0 \\
0 & {\rm else} \\ \end{cases}
\end{equation}
for $u \in \CB_w^+$. The spectrum in $\CB_w^-$ follows from monodromy or wall-crossing.

\insfig{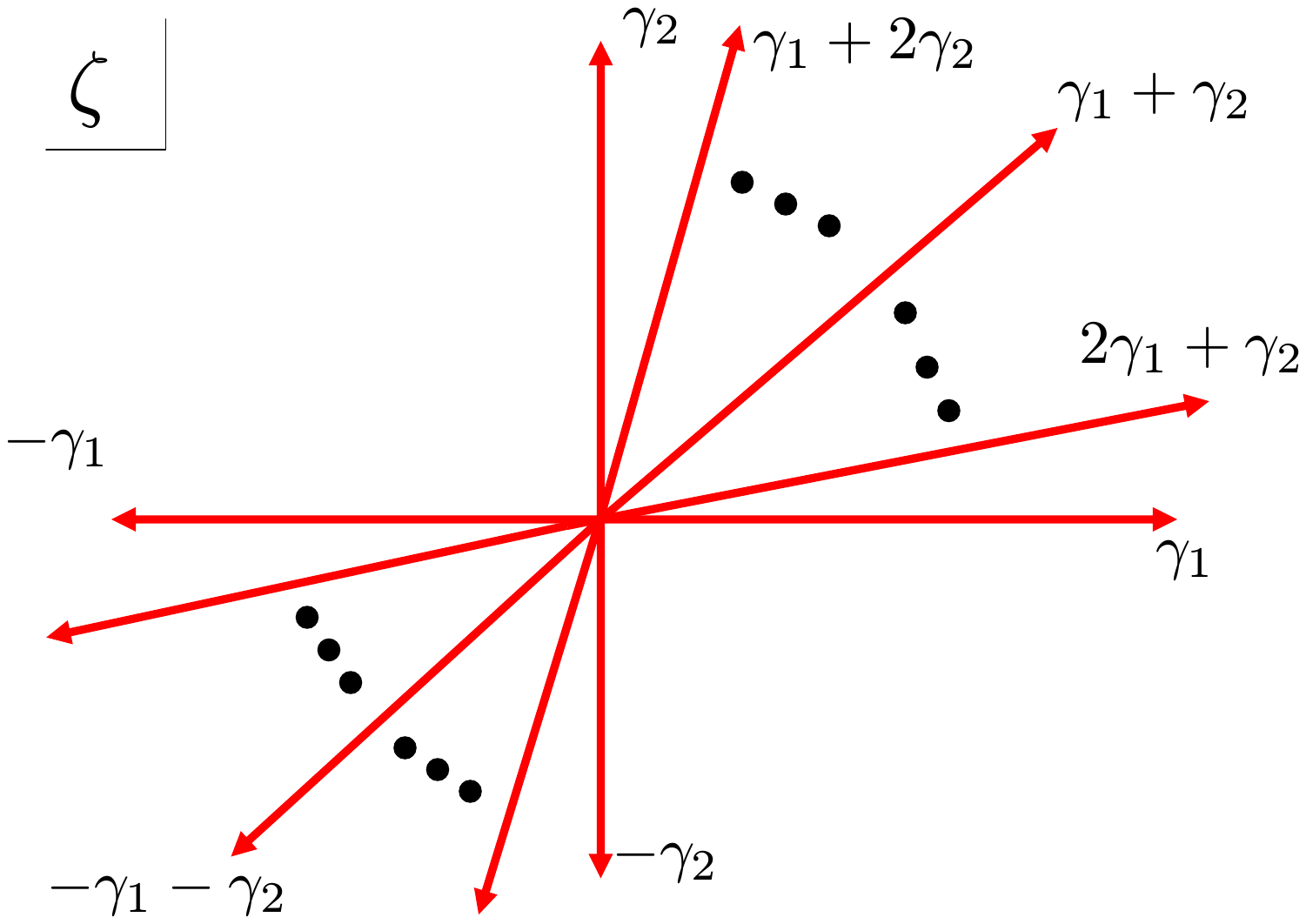}{BPS walls for the pure $SU(2)$ theory for $u$ in the weak coupling region, projected
to the $\zeta$-plane.  For clarity the $\widehat{W}$ has been dropped in labeling the
walls. In the first quadrant there are walls for charges $n\gamma_1 + (n+1)\gamma_2$
and $(n+1)\gamma_1 + n\gamma_2$ for $n\geq 0$ which
accumulate at the wall for $\gamma_1 + \gamma_2$ from either side. The wall crossing
on each of these walls is generated by a single hypermultiplet, except for the central walls
at $\pm (\gamma_1 +\gamma_2)$.}

Now one can compute that at $u \approx 0$ there is  a suitable choice of basis cycles $\gamma_1, \gamma_2$
satisfying \eqref{eq:SU2-innprod} so that the chamber structure in the $\zeta$ plane is identical to
that of Figure \ref{fig:ADN3-FL-1}. In the weak coupling, on the other hand, the situation is
dramatically different. For fixed $u \in \CB_w^+$ the projection of the chambers to the $\zeta$ plane
is shown in Figure \ref{fig:SU2-FL-weak}. One can choose a path to large $u$ to compare $\gamma_1, \gamma_2$
with the usual basis of magnetic and electric charges. We can thereby identify
$\gamma_1 = -\half H_\alpha +\alpha$ and $\gamma_2 =\half H_\alpha$, so that $\gamma_1+\gamma_2$ is the
charge of the $W$-boson.

Let us suppose $u \in \CB_s$.
As in the $N=3$ case it is easy to check which monomials $X_{(p,q)}$ remain consistent
with strong positivity after
single wall-crossings in both clockwise and counterclockwise directions. The result is again:

\begin{enumerate}

\item $p\geq 0, q\geq 0 $ in $c_n$ for $n=1\mod 4$

\item $p\leq 0, q\geq 0 $ in $c_n$ for $n=2\mod 4$

\item $p\leq 0, q\leq 0 $ in $c_n$ for $n=3\mod 4$

\item $p\geq 0, q\leq 0 $ in $c_n$ for $n=0\mod 4$

\end{enumerate}

In the remainder of this section we are going to show that under arbitrary wall-crossings
these monomials generate finite expansions in $X_{\gamma}$. According to the remark made at the end of
Section \ref{subsec:Chambers} this is conjecturally sufficient to prove strong positivity.
We will
give some plausibility arguments for this assumption below. Using the reasoning
at the end of Section  \ref{subsec:Seeds-from-N=2} we expect that
there is no obstruction to extending these
as strongly positive operators in the weak coupling region.

If we attempt to construct a strongly positive formal line operator starting from a
monomial we rapidly find somewhat complicated expressions, even in the strong coupling
regime. We will see that   there  is no finite monodromy in the $\zeta$ plane.
Let us begin by examining an example. It is convenient to introduce the
notation
\begin{equation}
[a,b]:= X_{\half(a \gamma_1 + b\gamma_2)},
\end{equation}
 and consider a formal line operator
defined by starting with   $F(c_1)=[1,0]$. We now start generating the values in the
other chambers by wall-crossing in  the clockwise direction. It will insightful to
organize the terms in triangular arrays. Doing this makes it easy to recognize the
annihilation of halos across walls. Moving in the clockwise direction we find that $F(c_2)$ is given by:

\begin{equation}\label{eq:SU2-sc-rot1}
\begin{array}{cc}
 &[1,2]\\
 &+ \\
 & [1,0]
\end{array}
\end{equation}
$F(c_3)$ is given by
\begin{equation}\label{eq:SU2-sc-rot2}
 \begin{array}{ccccc}
   [-3,2] & + & \rho_2 [-1,2] &  + & [1,2]\\
    &  & &  & +\\
    & &  & & [1,0]
 \end{array}
 \end{equation}
Here to save writing we identify the character of a representation
with the representation itself. Thus $\rho_N$ stands for $[N]$.
This proves useful as the expressions become lengthier.
Both the row and the column terms in \eqref{eq:SU2-sc-rot2} can be viewed as halo configurations
with $[1,2]$ serving as the core charge.
Next, crossing into chamber $c_4$ the vertical column is killed
because a halo is removed but two other vertical columns are produced due to
halos being created and so $F(c_4)$ is
\begin{equation}
\begin{array}{cccccc}
  &[-3,2] & + & \rho_2 [-1,2] & + & [1,2] \\
     &  + &   & + &   &   \\
 & \rho_3 [-3,0] &   & \rho_2 [-1,0] &  &  \\
  &+ &  &   &   &   \\
  & \rho_3 [-3,-2] &   &   &   &   \\
  &+ &   &  &   &   \\
  &[-3,-4] &   &   &   &
\end{array}
\end{equation}
Once again we see that the corner term $[-3,2]$ serves as a core particle for a
horizontal and vertical halo.
Now, moving into the next chamber across $\widehat{W(\gamma_1)}$ the horizontal halo in the first
row collapses to its core and   we get $F(c_5)$
\begin{equation}
\begin{array}{cccccccccc}
  &[-3,2] &   &   &   &  &   &   &   &   \\
 & + &   &   &   &  &   &   &   &   \\
 & \rho_3 [-3,0] &   & \rho_2 [-1,0] &   &   &   &   &   &   \\
 &  + &   &  + &   &   &   &   &   &   \\
 & \rho_3 [-3,-2] & + & (\rho_2+\rho_4)[-1,-2] & + & \rho_3[1,-2] &   &   &   &   \\
 &  + &   &  + &   &   &   &   &   &   \\
 & [-3,-4] & +& \rho_4 [-1,-4] & + & \Lambda^2 \rho_4 [1,-4] & + & \rho_4 [3,-4] & + & [5,-4]
\end{array}
\end{equation}
An interesting feature of this expression is that we must use the identity on
representations $\rho_2 \rho_3 = \rho_2 + \rho_4$. In order to recognize the
horizontal halo in the third row we use $\rho_2 \rho_3$ while to recognize the
two vertical halos in the second column we must use $\rho_2 + \rho_4$. As a final example,
continuing clockwise across $\widehat{W(\gamma_2)}$ we find $F(c_6)$ is
\begin{equation}
\begin{array}{ccccccccccc}
    &   &   &   &   &   &   &   &   &   &  [5,6] \\
    &   &   &   &   &   &   &   &   &   &  + \\
    &   &   &   &   &   &   &   &   &   &  \rho_5 [5,4] \\
    &   &   &   &   &   &   &   &   &   &  + \\
    &   &   &   &   &   &   &   & \rho_4[3,2]  &   &  \Lambda^3\rho_5 [5,2] \\
    &   &   &   &   &   &   &   & +  &   &  + \\
    &   &   &   &   &   &\rho_3 [1,0]   &   & \rho_4\rho_3 [3,0]  &   &  \Lambda^2\rho_5 [5,0] \\
    &   &   &   &   &   &+  &   & +  &   &  + \\
    &   &   &   & \rho_2 [-1,-2]  &   &(\rho_3+\Lambda^2\rho_4) [1,-2]   &   & \rho_4\rho_3 [3,-2]  &   & \rho_5 [5,-2] \\
        &   &   &   &   &   &+  &   & +  &   &  + \\
         &   & [-3,-4]  &  + & \rho_4 [-1,-4]  & +  & \Lambda^2\rho_4 [1,-4]   &  + & \rho_4 [3,-4]  & +  &  [5,-4]
\end{array}
\end{equation}
Now it is not at all obvious that moving into chamber $c_7$ across the wall $\widehat{W(-\gamma_1)}$
 will preserve strong positivity. The problematic
row is the second to last row. We must write $\rho_3 + \Lambda^2\rho_4 = \rho_2^2 + \rho_5$ and
$\rho_4 \rho_3 = \rho_5 \rho_2 + \rho_2 $ so that this row can be written as:
\begin{equation}
\rho_2 ([-1,-2] + \rho_2 [1,-2] + [3,-2]) + \rho_5 ([1,-2] + \rho_2 [3,-2] + [5,-2]) .
\end{equation}
In this form it is clear that the horizontal halos will collapse and strong positivity is preserved.

Proceeding in the counterclockwise direction we observe a similar pattern. Crossing from
$c_1$ to $c_0$ across $\widehat{W(\gamma_1)}$ gives simply $F(c_0)=[1,0]$ and then $F(c_{-1})$ is
\begin{equation}
\begin{array}{cc}
& [1,0] \\
& + \\
& [1,-2]
\end{array}
\end{equation}
passing to $F(c_{-2})$ we find
\begin{equation}
\begin{array}{cccccc}
  &  &   &   &   &  [1,0] \\
  &  &   &   &   & + \\
  & [-3,-2] & + & \rho_2 [-1,-2] & + & [1,-2]
\end{array}
\end{equation}
and so forth.

The general pattern is the following. Starting in a chamber
 of the type $c_n$ with $n=1 \mod 4$ we begin with a right triangle
   with vertices $[a,b], [a-2b,b], [a,b-2a]$, where $a,b$ satisfy
   some inequalities $a\leq 0, b\leq 0, a-2b \geq 0, b-2a\geq 0, \cdots$.
   Then the wall-crossings create and destroy halos forming a right triangular
   array in each chamber so that after four wall-crossings we return to a triangular array with
   $a' = - 3a +4 b, b' = -4 a + 5b$.  In our example above $a= 1-4k, b= -4k$ after
   $k$ clockwise turns. It seems very nontrivial to check directly the strong positivity
   of the interior lattice points of these triangles, but if the pattern of
   triangles continues this must be the case. That is, the main challenge in
   proving strong positivity is showing that the expressions remain polynomials
   in the $[a,b]$ upon wall-crossing. We will in fact give a rigorous argument
   that this finiteness is the case at the end of this section.

\insfig{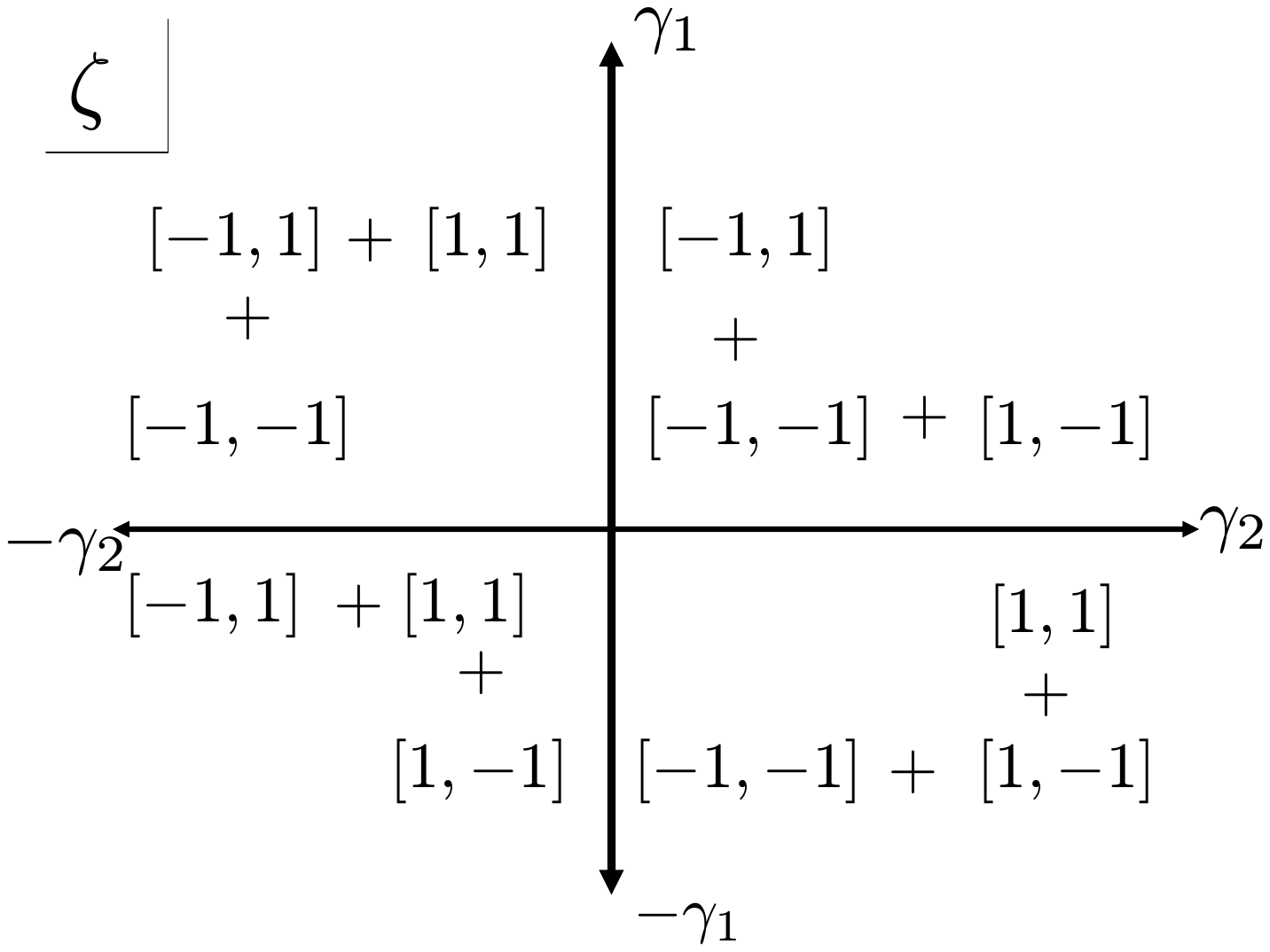}{The Wilson line operator at strong coupling. There is no monodromy
in the $\zeta$ plane for this expression. The corner charge serves as a core charge
for both horizontal and vertical halos. Upon crossing each wall one halo is destroyed
and one halo is created. }

While the above expressions become somewhat formidable, there is a remarkable
and special case of operators where the triangle has no monodromy in the $\zeta$
plane in the strong coupling region,
so $a' = -3a+4b = a $ and $b' = -4a + 5 b = b$,
that is, $a=b$ with $a,b<0$. The most basic case $a=b=-1$ will turn out to correspond to the Wilson
line in the fundamental representation in the physical theory.  Its  value in the four chambers
is shown in figure \ref{fig:SU2-FL-Wilson}. The expression for the Wilson operator
is sufficiently simple that we can also check strong positivity in the weak coupling
region. We begin with the expressions in $c_0$ and $c_2$ which are the same in the
strong and weak coupling domain $\CB_w^+$. Crossing from $c_0$ in the clockwise direction
we find expressions of the form $[1,1]+[2n-1,2n+1] + [-1,-1]$ next to the wall
for $n\gamma_1 + (n+1)\gamma_2$ and crossing this wall in the clockwise direction
one halo is destroyed while one is created to produce $[1,1]+[2n+1,2n+3] + [-1,-1]$.
Similarly, crossing counterclockwise from $c_2$ we obtain expressions
$[1,1]+[2n+1,2n-1] + [-1,-1]$ and crossing the wall for $(n+1)\gamma_1 + n\gamma_2$
we get $[1,1]+[2n+3,2n+1] + [-1,-1]$. Thus, we have explicitly constructed a
formal line operator corresponding to the Wilson loop.

The Wilson line operator, which we will call $\hat W$, can be used to generate a series of
formal line operators as follows. We define recursively
\begin{equation}\label{eq:rec-rels-1}
G_{2k} \hat W = y G_{2k+1} + 1 + y^{-1} G_{2k-1},
\end{equation}
\begin{equation}\label{eq:rec-rels-2}
G_{2k+1}\hat W = y G_{2k+2} + y^{-1} G_{2k}.
\end{equation}
We start the recursive procedure
with an operator which in chamber $c_1$ is $G_0 = [1,1]$. (Note that this is the
operator which is ``missing'' from the square formed by the vertices of the Wilson operator.)
Recall that $[a,b]:= X_{\half(a \gamma_1 + b\gamma_2)}$
so that $[a,b]\cdot [c,d] = y^{\half(ad-bc)} [a+c,b+d]$. Carrying out the multiplications
 we generate a series of
operators whose values in chamber $c_1$ are

\begin{equation}
\begin{split}
G_{-5}& = [ -2,-4] + \rho_4 [0,-4] + \Lambda^2 \rho_4 [2,-4] + \rho_4 [4,-4] + [ 6,-4] + \rho_2 ( [-2,-2] + \rho_2 [0,-2]+[2,-2]) + [-2,0]\\
G_{-4}& = [-1,-3] + \rho_3 [1,-3] + \rho_3 [3,-3] + [5,-3] + [-1,-1] + [1,-1] \\
G_{-3} & = [0,-2] + \rho_2 [2,-2] + [4,-2] \\
G_{-2} & = [1,-1] + [3,-1] \\
G_{-1} & = [2,0] \\
G_{0} & = [1,1] \\
G_1 & = [0,2] \\
G_2 & = [-1,1] + [-1,3] \\
G_3 & = [-2,0] + \rho_2 [-2,2] + [-2,4] \\
G_4 & = [-3,-1] + \rho_3 [-3,1] + \rho_3 [-3,3] + [-3,5] + [-1,-1] + [-1,1] \\
G_5 & = [-4,-2] + \rho_4 [-4,0] + \Lambda^2 \rho_4 [-4,2] + \rho_4 [-4,4] + [-4,6] + \rho_2 ( [-2,-2] + \rho_2 [-2,0]+[-2,2]) + [0,-2]
\end{split}
\end{equation}
(Again, if one organizes these in a $2\times 2$  grid they form triangular arrays.)

\insfig{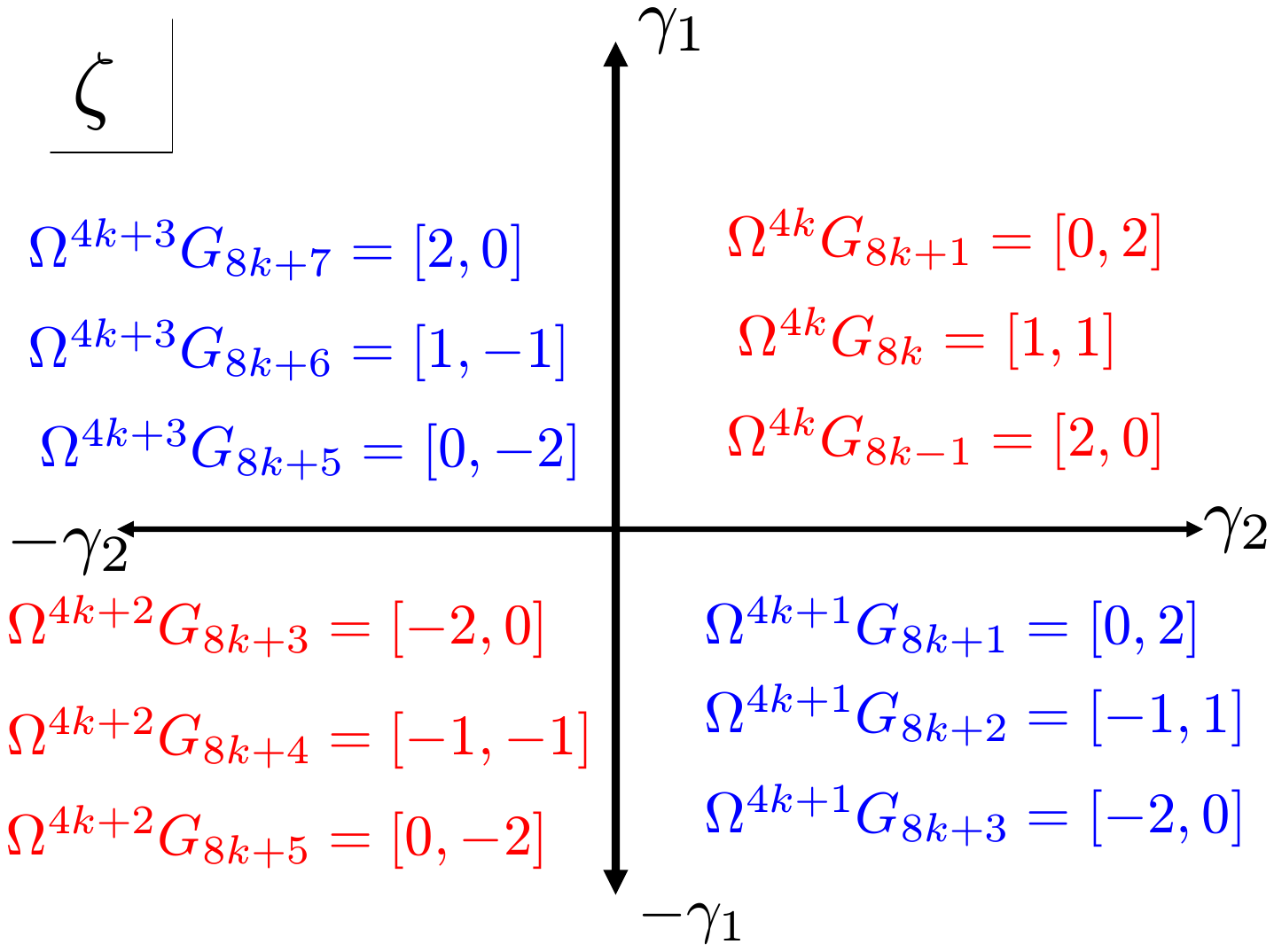}{This figure illustrates how wall-crossings turn the $G_n$ into monomials.
For example, if $n=8k$, then $4k$ wall-crossings in the clockwise direction from $c_1$ will turn $G_n$
into the monomial $[1,1]$.  }

%
%

Once again the expressions become rather complicated. However, they can be brought under control
since a sufficient number of wall-crossings brings them to monomial form. Let $\Omega$ denote the operation
of wall-crossing across a wall in the clockwise direction.
Then we claim that the wall-crossings turn $G_n$ into a monomial according to the pattern shown in figure
\ref{fig:SU2-FL-Monomials}. The pattern shown in this figure
can be proven by using induction together with two sets of
four very simple identities, which can be readily checked by hand:
\begin{equation}\label{eq:4fold-wilson-idents}
\begin{split}
c_{1 + 4\ell} \qquad  & [1,1]\hat W = y[0,2] + 1 + y^{-1} [2,0] \\
c_{2 + 4\ell} \qquad  & [-1,1]\hat W = y[-2,0] + 1 + y^{-1} [0,2] \\
c_{3 + 4\ell} \qquad  & [-1,-1]\hat W = y[0,-2] + 1 + y^{-1} [-2,0] \\
c_{4 + 4\ell} \qquad  & [1,-1]\hat W = y[2,0] + 1 + y^{-1} [0,-2] \\
\end{split}
\end{equation}
\begin{equation}
\begin{split}
c_{1 + 4\ell} \qquad  & [2,0]\hat W = y[1,1]  + y^{-1} \Omega[1,-1], \\
c_{2 + 4\ell} \qquad  & [0,2]\hat W = y[-1,1]  + y^{-1} \Omega[1,1], \\
c_{3 + 4\ell} \qquad  & [-2,0]\hat W = y[-1,-1]  + y^{-1} \Omega[-1,1], \\
c_{4 + 4\ell} \qquad  & [0,-2]\hat W = y[1,-1]  + y^{-1} \Omega[-1,-1].
\end{split}
\end{equation}
A number of highly nontrivial results follow from these identities. First, since $G_{2n+1}$
have a wall-crossing image by $\Omega^n$ which admits a squareroot the operator itself
admits a squareroot. We set $\hat V_n := \sqrt{G_{2n+1}}$. Thus we have for example
\begin{equation}
\begin{split}
\hat V_{-3} = \sqrt{G_{-5}} & = [-1,0] + [-1,-2] + \rho_2 [1,-2] + [3,-2],\\
\hat V_{-2} = \sqrt{G_{-3}} & = [0,-1] + [2,-1],\\
\hat V_{-1} = \sqrt{G_{-1}} & = [1,0],\\
\hat V_0 = \sqrt{G_1} & = [0,1],\\
\hat V_1 = \sqrt{G_3} & =    [-1,0] + [-1,2],\\
\hat V_2 = \sqrt{G_5} & =    [0,-1] + [-2,-1] + \rho_2 [-2,1] + [-2,3].
\end{split}
\end{equation}

By passing to a convenient chamber in which expressions become monomials (or
simple $\Omega$ images of monomials) we can easily
establish many useful ring relations. First of all, the
$G$'s are related to the $\hat V$'s by
\begin{equation}\label{eq:SU2-formring-1}
\begin{split}
G_{2n+1} & = \hat V_n^2 \\
G_{2n} & = y^{-1/2}
\hat V_{n-1} \hat V_n \\
\end{split}
\end{equation}
Next, the $\hat V$'s satisfy the relations:
\begin{equation}\label{eq:SU2-formring-2}
\hat V_{n-1} \hat V_{n+1} = 1+ y \hat V_n^2
\end{equation}
\begin{equation}\label{eq:SU2-formring-3}
  \hat V_{n} \hat W =  y^{1/2} \hat V_{n+1} + y^{-1/2} \hat V_{n-1}
\end{equation}
We will see later that it is sometimes useful to work with the subalgebra of the
$G$'s. Using the above method we find
\begin{equation}\label{eq:SU2-formring-4}
\begin{split}
G_{2k-1} G_{2k+1} & = y^2 G_{2k}^2 \\
G_{2k-1} G_{2k+3} & = (1+ y G_{2k+1})(1+y^3 G_{2k+1}) \\
\end{split}
\end{equation}
for the odd-indexed operators and
\begin{equation}\label{eq:SU2-formring-5}
G_{2k} G_{2k+2} = y G_{2k+1} (1+ y G_{2k+1})
\end{equation}
and so on for the even-indexed operators.

In Section \ref{subsec:SU2-Functions} below, we will see how the $y=1$ specialization
of these relations nicely reproduces the ring of functions on moduli spaces for
$SU(2)$ and $SO(3)$ gauge theories.  In Section \ref{subsec:SU2-Lams}, especially
in the equations \eqref{eq:chebyshev-expls} below, we will reproduce the
sequence of operators $\hat V_n$ at $y=+1$ from laminations.
In this way, the $\hat V_n$ above provide a computation of the protected spin
characters for the $SU(2)$ $N_f=0$ theory.

We are now in a position to give a compelling (to us)  argument that the $\hat V_n$ and $G_{2n}$
do indeed define
strongly positive formal line operators.  When strong positivity fails there is
typically an infinite series (see for examples \eqref{eq:explicit-dilog-conj},
\eqref{eq:explicit-dilog-conj-ii}). Thus, we will use finiteness as a surrogate
for strong positivity in our argument. We wish to consider the various wall-crossings
of $G_n$ and in particular it is not obvious from applying the rules \eqref{eq:explicit-dilog-conj}
and
\eqref{eq:explicit-dilog-conj-ii}) that the expression will be finite in some chamber $c_m$.
However, using the recursion relations we can always express $G_n$ as a finite polynomial
in $G_\ell, G_{\ell+1}, W, y^{\pm 1}$ where we choose $\ell$ so that $G_{\ell}$ and $G_{\ell+1}$
reduce to monomials in chamber $c_m$.  This shows that $G_n$ is finite in every chamber.
A similar argument applies to $\hat V_n$.

Thus, we conclude that $G_n$ and $\hat V_n$ are all strongly positive formal line operators.
Furthermore, any line operator generated from a monomial $[p,q]$ in some chamber can be written
as a product of appropriate $\hat V_n$'s. For example if we are in a chamber $c_n$ with $n=1 \mod 4$
then $p,q\geq 0$ and we can choose $k$ so that $[p,q] = y^{-pq/2} \hat V_{4k-1}^p \hat V_{4k}^q$.
It follows that all the wall-crossing images of $[p,q]$ are finite polynomials in $[a,b]$ with
coefficients in the representation ring of $su(2)$. Thus, given our main assumption above,
 $[p,q]$ does indeed generate a strongly
positive formal line operator.

As we mentioned before, these arguments apply for $u$ in the strong coupling region. We have
not attempted to prove strong positivity in the weak coupling region, but we expect it to hold
because of the relation to physical line operators described in Sections \ref{subsec:SU2-Functions}
and \ref{subsec:SU2-Lams}.

\subsection{Physical vs. formal line operators}\label{subsec:Physics-Formal}

In Section \ref{subsec:BPS-Deg-PSC} we stated four conjectures: the
strong and weak positivity conjectures
for framed and vanilla BPS states. In this section we state a further conjecture
which, if true, allows one to derive the full set of simple physical line operators
using purely algebraic means, once the vanilla BPS spectrum is known, even if the theory is non-Lagrangian.

The strong positivity conjecture for framed BPS states implies that the generating function
\eqref{eq:form-gen} of a physical line operator defines a strongly positive formal line operator. We would like to
conjecture that the converse is true, namely, that any strongly positive formal line operator is
the generating function of framed BPS degeneracies for some physical line operator.
The main piece of evidence for this conjecture is that, as we will see in Section
\ref{sec:Examples}, it is true in several examples.

There is a neat consistency check of this conjecture using the action of the monodromy group.
The specific UV labeling of a physical line operator $L$
must be independent of the IR data $u$, It may be subject to monodromies
only when transported along homotopically non-trivial paths in the gauge coupling moduli space.
It might  also be subject to monodromy when transported around the origin in the $\zeta$ plane,
although only in theories where $U(1)_R$ is anomalous. The reason is that
 in such theories a rotation of $\zeta$ can be traded
for a rotation of the gauge coupling scales, i.e. a shift of the theta angles, which can induce
a monodromy of the UV labels.
Formal line operators undergo mondromy transformations in the form of a product
of KS transformations for all the BPS walls crossed under the monodromy.
This is consistent with known properties of the KS
transformations in field theories \cite{Gaiotto:2008cd}:
the product of KS factors around singularities in the $u$ plane gives the identity (or more precisely the
appropriate monodromy transformation of the IR charge lattice) but
the product of all KS factors around the origin in the $\zeta$ plane does not.
So no formal line operator has monodromy transformations which
would be inappropriate for the generating function of a physical line operator

Unfortunately, as the above examples amply demonstrate, it is
rather hard to work directly with formal line operators, and we would like
to suggest that in fact an even stronger conjecture is true allowing us to work with
simpler objects.  We set $y=1$ and define a (possibly larger)
space of  collections $\CL(c)$, defined in the chambers of $\Xi$, where $\CL(c)$ is
a  positive integral linear combinations of $x_\gamma$ and across neighboring
chambers $\CL(c^\pm)$    are related by appropriate $y=1$ KS transformations.
We might call the elements of this space {\it formal laminations}.
The reason for the name ``lamination'' will become clear in Section \ref{sec:Examples}.
The space of formal laminations is easy to study, because it admits a natural product,
and is typically generated by a finite set of generators. Note that
if we expand the product in \eqref{eq:y-one-prod} only positive
integer coefficients appear in the expansion (for the appropriate sign of  $\gamma$).
Thus the positivity is nicely consistent with physical expectations.

Of course, formal line operators can be seen as formal laminations, by setting $y=1$.
It is not obvious, though, that the non-commutative product of two formal line operators will produce a linear combination
\begin{equation}
F_i \circ_y F_j = \sum_{k} c_{ij}^k(y) F_k
\end{equation}
of formal line operators. On the other hand, it is obvious that the set of formal laminations
is closed under multiplication.  In the $A_1$ examples we focus on, the space of formal laminations
appears to coincide with the space of {\it physical} line operators in the gauge theory. It is much
easier to construct formal laminations than formal line operators! Thus, it would be a very powerful
result if the space of formal
laminations already coincides with that of formal line operators. We would therefore suggest that this is a
question worth pursuing.

\section{Relation to cluster algebras}\label{sec:Cluster-Algebras}

The algebra of formal line operators is closely related to the
mathematical theory of \emph{cluster algebras}.  (The literature on cluster algebras
is by now vast; a couple of pointers are \cite{MR1887642,MR2132323,cluster-intro}
and \cite{cluster-portal}.)  In this section we sketch a little bit of that relation.
As we will discuss in Section \ref{sec:Open-Problems}, there are further things to understand.
In addition to making a connection with an interesting subject in mathematics,
we will see that this connection allows us to construct formal line operators
using only ``local'' rules in moduli space, i.e., local transformation
rules between chambers of $\Xi$, which can be given using only a partial knowledge
of the BPS spectrum. Moreover, we will also show how only partial knowledge of the
BPS spectrum can allow one, in principle, to construct the entire BPS spectrum.

\subsection{Basic definitions of cluster algebras}\label{subsec:Cluster-Defs}

The definition of a cluster algebra begins with a \emph{seed}, which is simply an
$m\times n$ matrix $B_{ij}$ with $m\geq n$ and integral entries so that the $n\times n$ upper block
is skew-symmetrizable. That is, there are positive integers $d_i$ so that  $B_{ij}(d_j)^{-1}$
for $1\leq i,j\leq n$ is antisymmetric:  $d_i B_{ij} = - d_j B_{ji}$. For each index $1\leq k \leq n$
one defines a \emph{seed mutation} along the direction $k$ to be a transformation $\mu_k: B\to B'$:
\begin{equation}\label{eq:seed-mtrx}
B'_{ij} = \begin{cases} - B_{ij} & {\rm If} ~~ i = k ~~ {\rm or}~~ j=k \\
B_{ij} + {\rm sgn}(B_{ik}) [B_{ik} B_{kj}]_+ & {\rm Else} \\
\end{cases}
\end{equation}
Here we have introduced the notation $[x]_+ = {\rm Max}\{ 0,x\}$ and
\begin{equation}
{\rm sgn}(x) = \begin{cases} 1 & x>0 \\  0 & x =0 \\ -1 & x < 0 \\ \end{cases}
\end{equation}
Cluster variables are $A_1, \dots, A_m$. For each $k$, $1\leq k \leq n$,
  cluster transformations are defined by
$\mu_k(A_i) = A_i$ for $i\not= k$ and
\begin{equation}
\mu_k(A_k) = \frac{ \prod_{j\vert B_{kj}>0} A_j^{B_{kj}} + \prod_{j\vert B_{kj}<0} A_j^{-B_{kj}} }{A_k}
\end{equation}
This recursive procedure defines a sequence of generators of an algebra known as a cluster algebra.
The variables $A_i$ for $n<i\leq m$ are called ``frozen variables.''
Cluster algebras turn out to have many beautiful and remarkable properties.

It was shown by Fock and Goncharov in \cite{MR2567745} that if we define
\begin{equation}
x_i = \prod_j A_j^{B_{ij} }
\end{equation}
then the corresponding transformation of the $x_i$ is $\mu_k: x_i \to x_i'$ where
\begin{equation}\label{eq:Mutation-x}
 x_i' = \begin{cases}  x_k^{-1} & i=k \\
x_i (1+ x_k^{-{\rm sgn}(B_{ik})})^{-B_{ik}} & i \not= k \\
\end{cases}
\end{equation}
It was stressed by Kontsevich and Soibelman \cite{ks1} that \eqref{eq:Mutation-x}
can be interpreted in terms of the symplectic transformations we call
 KS transformations.

\subsection{Cluster algebra structure  in $\CN=2$ theories }\label{subsec:Seeds-from-N=2}

In order to make a connection to $\CN=2$ theory we proceed as follows.

Quite generally, let us define a charge $\gamma \in \Gamma_u$ to be a  \emph{root} if $\Omega(\gamma;u)\not=0$.
Let $\CR(u)$ be the root system at fixed $u$.
We define a system of \emph{positive roots} to be a disjoint decomposition $\CR(u) = \CR^+(u)
\amalg \CR^-(u)$ where if $\gamma \in \CR^+(u)$ then $-\gamma \in \CR^-(u)$. We will only consider
systems of positive roots such that the associated BPS rays lies in a half-space in the $\zeta$ plane.
Conversely, a choice of a half-space in the $\zeta$ plane determines a decomposition of the root
system into positive and negative roots.
Given a system of positive roots we can define a  \emph{simple root} to be a positive root which is
not the sum of two other positive roots.  Given any point $(u,\zeta)$ we can canonically define
a system of positive roots by declaring those to be the roots  so that
${\rm Im}Z_{\gamma}/\zeta >0$. Equivalently, given a point $(u,\zeta)$ which is not on a BPS
 wall we can define a system of positive roots by taking the roots whose BPS rays
 lie in the half-plane on the counter-clockwise side of the line through $\zeta$. At fixed $u$ the point
 $(u,\zeta)$ sits in a chamber in the $\zeta$-plane bounded by two walls: One is a wall for a
 simple root on the counter-clockwise side and one is a wall for minus a simple root on the clockwise side,
 with respect to this canonical system of positive roots. If we vary $u$ at fixed $\zeta$ then $(u,\zeta)$ will not
 cross into another chamber of $\Xi$ unless a BPS ray for $\pm $ a simple root sweeps past $\zeta$.
 Thus, we can label the chambers in $\Xi$ by systems of
simple roots $\Delta$.

Now, for any chamber $c$ of $\Xi$  use the simple roots $\{\gamma_i\}$ to   define  a matrix
\begin{equation}
B_{ij} = \langle \gamma_i, \gamma_j\rangle   \Omega(\gamma_j;u)
\end{equation}
This is clearly skew-symmetrizable, provided the  $\Omega(\gamma_j;u)$ are all positive.
The frozen indices $n<i\leq m$ label generators of the
flavor sublattice.
We will now propose a transformation rule across BPS walls for the set of simple roots, which guarantees that seed mutations occur when $(u,\zeta)$ passes along a path from one
chamber to the next. As we have just observed it must cross a wall for $\pm$ a simple
root. The projection of the path in the $\zeta$ plane moves counter-clockwise
across a simple root $Z_{\gamma_k}$,  from a region with ${\rm Im}Z_{-\gamma_k}/\zeta <0 $ to a
region with ${\rm Im}Z_{-\gamma_k}/\zeta > 0 $. This crosses the wall $\widehat{W(-\gamma_k)}$.
Similarly, if the path moves   clockwise across the negative
of a simple root $Z_{-\gamma_k}$, from a region with ${\rm Im}Z_{\gamma_k}/\zeta >0 $ to a
region with ${\rm Im}Z_{\gamma_k}/\zeta < 0 $ it is crossing the wall $\widehat{W(\gamma_k)}$.

For a path which moves clockwise across $\widehat{W(\gamma_k)}$ the transformation
\begin{equation}\label{eq:charge-mutation}
\mu_{k,+}: \gamma_i \to \gamma_i' = \begin{cases} - \gamma_k & i=k \\   \gamma_i + \gamma_k [B_{ik}]_+  & i \not=k\\
\end{cases}
\end{equation}
produces the new system of simple roots in the new chamber $\mu_{k,+}(c)$. Similarly, for a path which moves
counter-clockwise rotation across $\widehat{W(-\gamma_k)}$, where $\gamma_k$ is a
simple root, the transformation
\begin{equation}
\mu_{k,-}: \gamma_i \to \gamma_i' = \begin{cases} - \gamma_k & i=k \\   \gamma_i - \gamma_k [B_{ik}]_-  & i \not=k\\
\end{cases}
\end{equation}
produces the new system of simple roots in the new chamber $\mu_{k,-}(c)$. Here we have introduced the notation
  $[x]_- := {\rm Min}\{x,0\} $. One can check that $\mu_{k,+}$ is a two-sided  inverse to $\mu_{k,-}$.

Now, one can check that, provided $\Omega(\gamma_j';u) = \Omega(\gamma_j;u)$ the matrix $B'_{ij}$ for the
new chamber is (for both  $\mu_{k,\pm}$) related to the matrix $B_{ij}$ in the old chamber
by precisely the seed mutation $\mu_k$ defined in equation \eqref{eq:seed-mtrx}.

Moreover, the Kontsevich-Soibelman transformation
\begin{equation}
\CK_{\gamma_0}^{-\Omega(\gamma_0;u)} x_\gamma = x_{\gamma} (1-\sigma(\gamma_0) x_{\gamma_0})^{-\langle \gamma,\gamma_0\rangle \Omega(\gamma_0)}
\end{equation}
coincides with the cluster transformation \eqref{eq:Mutation-x} if we have $\sigma(\gamma_0)=-1$,
 $x_i' = x'_{\gamma_i'}$ , $x_i = x_{\gamma_i}$, $\gamma_0 = \gamma_k$ and $\gamma_i' = \mu_{k,+}\gamma_i$.
Thus, for example, if we define $\hat x_{\gamma} = \sigma(\gamma) x_{\gamma}$ then the
framed BPS degeneracies can be computed using   cluster transformations
 induced by the simple root
mutation according to \eqref{eq:Careful-rule}.

This statement can be extended to the full $y \neq \pm 1$ wallcrossing
by using the theory of quantum cluster algebras, \cite{MR2233852}, but for simplicity we will look
at the classical case only.

We will say that an $\CN=2$ field theory has the \emph{cluster algebra property} if
 the transformations across chambers satisfy the above properties. That is,
 $\Omega(\gamma_i;u)>0$ for the simple roots and $\Omega(\mu_{k,\pm} \gamma_j;u) =
 \Omega(\gamma_j;u)$.
(To avoid confusion regarding this second condition, observe that the chamber walls
are \emph{not} walls of marginal stability, so the condition is well defined.)

One of the main
 advantages of having the cluster algebra property is that one can give an algorithm
 for constructing the BPS spectrum and the line operators in the theory.
To see this, consider first the BPS spectrum. Suppose we know the BPS degeneracies of the
simple roots in some chamber, and suppose we know the values of $Z_{\gamma_i}(u)$. These
will vary throughout the chamber, but will remain in a half-plane. Fix $u$ and choose a
value of $\zeta$ in this chamber (thus, ${\rm Im}Z_{\gamma_i}(u)/\zeta >0$). Now consider
moving $\zeta$ in the counterclockwise direction. It will hit a BPS line for some simple
root $\gamma_k$. We can now use the mutation $\mu_{k,-}$
to recompute the new basis of simple roots. We can proceed in this way by continuing
to rotate $\zeta$ counter-clockwise.
If one can rotate $\zeta$ by a full angle of $\pi$ then the entire BPS spectrum will have been
constructed. As we will see in two examples below, there can be obstructions to doing this.
Basically, all chambers have sets of simple roots which are hypermultiplets only,
and the BPS rays for particles of higher spin are surrounded by an infinite set of rays  for  infinitely many hypermultiplets, with the rays accumulating on the higher spin ray. Thus the higher spin rays are
 never associated to the boundary of a chamber. If one simply moves in the $\zeta$ plane,
the process will get bogged down at the first of such infinite sequence of hypermultiplets, unable to
reach the higher spin particle at the ``end'' of the sequence. On the other hand, in $A_1$ theories and possibly
in all theories in the  class ${\cal S}$, it appears that
all chambers can be connected by a finite number of steps, as long as one is allowed to move
along generic paths in the space of $(Z_\gamma, \zeta)$. Thus, a  possible strategy to compute the
spectrum would be to
use such paths to ``jump'' ahead of an obstruction, and then track back in order to control the infinite sequence of
hypermultiplets on both sides of the obstruction, and hence find the residual coordinate transformation {\it across} the obstruction.
This can be further decomposed into transformations $\CK_\gamma$
to read off the full spectrum of the theory.

As an example of the above procedure,
suppose that $\Gamma$ is two-dimensional, $\Omega(\gamma_1)=\Omega(\gamma_2)=1$,
and for some positive integer $n$
\begin{equation}
B_{ij} = \begin{pmatrix} 0 & n \\ -n & 0 \\ \end{pmatrix}.
\end{equation}
Begin with a chamber with $\zeta$ on the real axis so that ${\rm Im} Z_1 > 0 $ and
${\rm Im} Z_2 >0$.  If ${\rm arg}Z_1 > {\rm arg} Z_2$, then we make the series of transformations
$\mu_{2,-}, \mu_{1,-}, \mu_{2,-}, \cdots$.  The resulting pattern of simple roots has period
4 and is simply
\begin{equation}
\begin{split}
\gamma_1^{[1]}=   \gamma_1, & \quad   \gamma_2^{[1]}= - \gamma_2, \\
\gamma_1^{[2]}= - \gamma_1, & \quad \gamma_2^{[2]}= -\gamma_2, \\
\gamma_1^{[3]}=  - \gamma_1, & \quad \gamma_2^{[3]}=  \gamma_2.
\end{split}
\end{equation}
On the other hand, if  $\arg Z_2 > \arg Z_1$  then moving $\zeta$ counterclockwise one
encounters the sequence of mutations $\mu_{1,-}, \mu_{2,-}, \mu_{1,-},\cdots$. If the initial
chamber is labeled as $j=0$ then we obtain a sequence of chambers with
\begin{equation}
\begin{split}
\gamma_1^{[2j+1]} & = -\gamma_1^{[2j]}  , \\
\gamma_2^{[2j+1]} & = n\gamma_1^{[2j]}+ \gamma_2^{[2j]},
\end{split}
\end{equation}
\begin{equation}
\begin{split}
\gamma_1^{[2j+2]} & =\gamma_1^{[2j+1]}+ n\gamma_2^{[2j+1]},   \\
\gamma_2^{[2j+2]} & =  - \gamma_2^{[2j+1]}.
\end{split}
\end{equation}
This sequence of transformations has period $6$ for $n=1$. It was explicitly realized in
the weak coupling region of $N=3$ Argyres-Douglas theories in
Section \ref{sec:formaln3}. For $n=2$ the sequence does not have a finite periodicity and
$\mu_{2,-}\mu_{1,-}$ is transformation by the
matrix
\begin{equation}
1 + N = \begin{pmatrix} 3 & 2 \\ -2 & -1 \\ \end{pmatrix}
\end{equation}
Since $N^2 =0$ we can write immediately
\begin{equation}
\begin{split}
\gamma^{[2j]}_1 & = (2j+1)\gamma_1 + 2j  \gamma_2, \\
\gamma^{[2j]}_2 & =  -2j \gamma_1  -  (2j-1) \gamma_2.
\end{split}
\end{equation}
This was realized in the $SU(2)$ $N_f=0$ theories in Section \ref{subsec:formal-SU2}.
Note that the BPS rays accumulate on rays along $\pm (Z_1 + Z_2)$.
%
%
%
Thus one cannot continue beyond this limit through a full angle of $\pi$. However,
by continuing in the clockwise direction one finds a second limiting value
corresponding to the opposite ray, so that in fact the full spectrum is indeed
generated.

For the case $n>2$ an interesting phenomenon appears. The transformation matrix from
$2j$ to $2j+2$ is
\begin{equation}
\begin{pmatrix} n^2 -1 & n \\ -n &  -1 \\ \end{pmatrix} = S^{-1} \begin{pmatrix} \lambda_+ & 0 \\
0 & \lambda_- \\ \end{pmatrix} S
\end{equation}
where $1+\lambda_\pm = \half (n^2  \pm n \sqrt{n^2-4} ) $.  Moving in the counter-clockwise
direction the limiting slope of the BPS rays is
\begin{equation}\label{eq:slope-ccw}
\frac{ \xi {\rm Im}Z_1 +   {\rm Im}Z_2 }{\xi {\rm Re} Z_1 +   {\rm Re} Z_2}
\end{equation}
where $\xi = \frac{\lambda_+ +1 }{n} > 1$. Now if we consider rotating in the clockwise direction
we encounter  a sequence of mutations $\mu_{2,+}$, $\mu_{1,+}$, $\mu_{2,+},\dots$
leading to a sequence of simple roots
\begin{equation}
\begin{split}
\gamma_1^{[2j+1]} & =   \gamma_1^{[2j]} + n   \gamma_2^{[2j]} \\
\gamma_2^{[2j+1]} & =  -   \gamma_2^{[2j]} \\
\end{split}
\end{equation}
\begin{equation}
\begin{split}
\gamma_1^{[2j+2]} & =     -\gamma_1^{[2j+1]}    \\
\gamma_2^{[2j+2]} & = n \gamma_1^{[2j+1]} + \gamma_2^{[2j+1]}  \\
\end{split}
\end{equation}
The limiting slope is now
\begin{equation}\label{eq:slope-cw}
\frac{ {\rm Im}Z_1 +  \xi  {\rm Im}Z_2 }{  {\rm Re} Z_1 +  \xi {\rm Re} Z_2}
\end{equation}
and the ray points in the quadrant with $-\pi < {\rm arg}\zeta < -\pi/2$.
Since $\xi >1$ the two limiting BPS rays subtend an angle less than $\pi$.
There is therefore a ``gap region'' as shown in Figure \ref{fig:gap}
through which the above procedure cannot explore.
The spectrum of populated charges in that region has been computed \cite{gctv},
and the result makes it clear what the difficulty is: the phases of $Z_\gamma$ for
$\Omega(\gamma) \neq 0$ are {\it dense}
in that region, so that it is not really possible to speak about chambers.

 \insfig{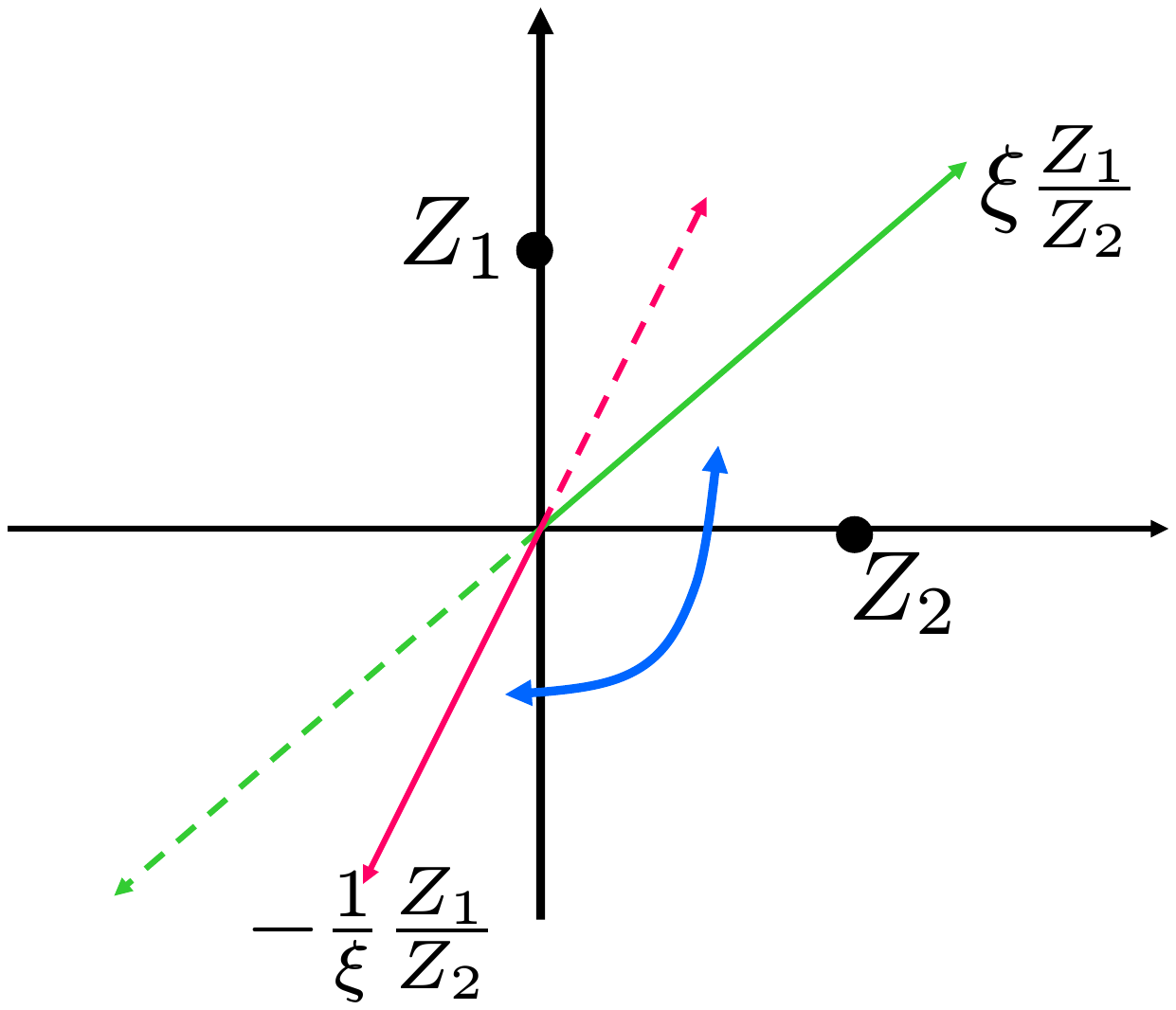}{When $n>2$ in the rank-2 example, we can only use cluster transformations at fixed $u$
 to determine the BPS spectrum of rays lying in an angular sector of angle less than $\pi$,
 indicated with the blue arrow. We have made a convenient choice of phases for
 $Z_1$ and $Z_2$.}

There are some immediate payoffs when  an $\CN=2$ theory has the cluster algebra
property.  The Laurent phenomena of \cite{MR1888840}
allow us to construct formal laminations.  The construction of formal
line operators involves some kind of ``quantum Laurent phenomenon'' which
does not appear to have been discussed in the mathematics literature (although it
is close to the ``universal Laurent polynomials'' of Fock and Goncharov.)
If our purpose is  just to check the strong positivity of a formal line operator,
we do not really care about the labeling of the charge lattice, hence we can
just express the generating function in terms of the $X_i$, forget about the $\gamma_i$
and make sure strong positivity is respected by all possible cluster transformations.
This can be used to streamline the derivations we gave in \ref{sec:formaln3}
 and \ref{subsec:formal-SU2} and extend them trivially to weak coupling regions, as the sequences of cluster transformations
 at strong and weak coupling are indistinguishable. For example,
 if we forget about the charge labeling, the $N=3$ Argyres-Douglas theory has only five distinct chambers in all,
 labeled by the five triangulations of a pentagon. The cluster transformations have period
 $5$ and the basic formal line operators
 simply undergo a periodic sequence $X\to X+X \to X+X+X \to X+X \to X \to X$.
 The long sequences of period $20=4 \times 5$ we encountered at strong coupling and
 $30 = 6 \times 5$ at weak coupling were simply the combination of the period $5$
 cluster transformations and period $4$ or $6$ relabeling  of charges.

We conjecture that theories in the class $\CS$ have the cluster algebra property.
This conjecture is motivated by the work of Fock and Goncharov on higher Teichm\"uller
theory \cite{MR2233852}. If our conjecture is true, then the algorithm described above
might provide a way to compute the BPS spectrum of higher rank $\CN=2$ gauge theories.
We think this is an important idea which should be pursued.
Given the above observations a natural question one can ask is whether every cluster
algebra arises from an $\CN=2$ theory. This too we leave to the future.

\subsection{Cluster algebra structure  in $A_1$ theories }\label{subsec:Seeds-from-N=2}

In this section we will show that indeed the  subclass of theories of $A_1$ type
enjoy the cluster algebra property. This can be proved by the relation between
the chambers and the WKB triangulations of $C$ explained in
\cite{Gaiotto:2009hg}.  Recall from Section 6 of that paper
that given $(u,\zeta)$ there is a canonical (WKB) triangulation
of $C$. It was shown in Section 7.8 of
\cite{Gaiotto:2009hg} that there is also an
associated system of simple roots $\{ \gamma_E \}_{E\in \CE(T)}$.
Moreover, each of these simple roots supports a single hypermultiplet with PSC equal to $1$.
When $(u,\zeta)$ vary the triangulation changes by isotopy unless there is a flip of an edge,
and this can only happen when $(u,\zeta)$ cross a BPS wall.  Thus, the walls of the chamber
are a subset of the walls $\widehat{W}(\pm \gamma_E)$ for $E\in \CE(T)$.  Generically, we
expect all the edges to contribute walls of the chamber. The transformation rules for the $\{ \gamma_E \}_{E\in \CE(T)}$
due to a flip of an edge $E_k$ are computed in Section 7.6 of \cite{Gaiotto:2009hg} , and agree with the $\mu_{k,\pm}$.

We can label the chambers by ideal triangulations of $C$ together with a topological
class of an ``adapted double cover.'' An adapted double cover of $C$ is one with one branch point
in each triangle of $C$ together with a choice of orientation of  the lifts of the edges $E$
so that these lifts are either all ingoing or all outgoing at the preimages of the singular points
of $C$.

\subsection{Formal line operators in $\CN=2^*$ $SU(2)$ gauge theory}\label{subsec:Formal-N2-star}

We now turn to the $\CN=2^*$ theory with gauge algebra $su(2)$. This is obtained from the
$\CN=4$ theory by giving a mass $m$ to an adjoint hypermultiplet. It is the $A_1$
theory where $C$ is a once-punctured torus.\footnote{More precisely, the $A_1$ theory where $C$ is a punctured torus is the $\N=2^*$
theory plus some decoupled fields, namely a doublet of half-hypermultiplets under an $SU(2)$ flavor symmetry associated to the puncture.}
The Seiberg-Witten
differential is given by
\begin{equation}
\lambda^2 = \left( m^2 \wp(z\vert \tau) + u\right) (dz)^2,
\end{equation}
where $z$ is a flat coordinate on an elliptic curve of modulus $\tau$, and
$\wp(z\vert \tau)$ is the Weierstrass function.
 The Seiberg-Witten curve is of genus two with two punctures.
The local system $\Gamma$ is of rank $3$ and can be taken
to have generators $\gamma_1, \gamma_2, \gamma_3$ with $\langle \gamma_i, \gamma_{i+1} \rangle =2$
for $i=1,2,3$, where the index $i$ is understood to be cyclic of order $3$. Note that
$\gamma_1 + \gamma_2 + \gamma_3$  generates the one-dimensional annihilator of $\langle \cdot, \cdot \rangle$.
This is the rank $1$ ``flavor lattice.''

The normalizable parameter
is $u$ and $\CB^* = \IC-\CS$, where $\CS$ consists of three singular points $u_i$,
where $\wp(z_i\vert \tau) = - u_i/m^2$
with $\wp'(z_i\vert\tau) =0$. The local system $\Gamma$ has monodromy around these three points.
The theory is known to have $SL(2,\IZ)$ symmetry, where
$PSL(2,\IZ)$ acts in the standard way on the coupling constant $\tau$.

The BPS spectrum $\Omega(\gamma;u)$ of the $\CN=2^*$ theory
is, unfortunately, not known explicitly.\footnote{We note that the spectrum
is implicitly
encoded in the ``spectrum generator'' of \cite{Gaiotto:2009hg}. This transformation
is easily written down, but its decomposition into Kontsevich-Soibelman factors
is not straightforward.}
Nevertheless, it is clear that the spectrum has some stark differences from the examples
we have thus far examined.  In particular, there
is no ``strong coupling region'' with a finite spectrum.
We expect that the  spectrum is always
infinite, and is controlled by a variable number of vector multiplets, each one accompanied by
an infinite cohort of hypermultiplets.
Thus the chamber structure is very complicated.  Here then is an example where the local
cluster transformation rules can be used very effectively.

In each chamber $[c]$ there are   three simple roots $\gamma_i[c]$; the chamber is
bounded by six walls $\widehat{W}(\pm \gamma_i[c])$.
$\sum_i \gamma_i[c]$ is pure flavor and we have
\emph{either}
$\langle \gamma_i[c],\gamma_{i+1}[c]\rangle=2$
\emph{or}
$\langle \gamma_i[c],\gamma_{i+1}[c]\rangle=-2$.
We call these chambers of positive and negative type, respectively.
The chamber is the set $\{ (u,\zeta): {\rm Im}Z_{\gamma_i[c]}(u)/\zeta > 0 \}$.
Note that a mutation $\mu_{i,\pm}$ takes a chamber of positive type to one of
negative type.

It is convenient
to define
\begin{equation}
[p,q,r]:= X_{\half (p \gamma_1 + q \gamma_2 + r \gamma_3)}.
\end{equation}
When we wish to emphasize the chamber we can write $[p,q,r](c)$.
We first observe that, in contrast to our previous examples, there are no monomials $[p,q,r]$ which
can generate formal line operators. In our chamber ${\rm Im} Z_i/\zeta >0$ for $i=1,2,3$. Examining
the rules \eqref{eq:explicit-dilog-conj}, \eqref{eq:explicit-dilog-conj-ii} we see that we must have
$\langle \gamma, \gamma_i[c] \rangle <0$ for $i=1,2,3$ where
$\gamma = \half (p\gamma_1[c]+q\gamma_2[c]+r\gamma_3[c])$. On the other hand,
$\gamma_1[c]+\gamma_2[c]+\gamma_3[c]$ is in the kernel of $\langle \cdot, \cdot \rangle$ and hence these
three inequalities can never be simultaneously satisfied.

Even though we cannot generate formal line operators from monomials, there are three nice
analogs of the Wilson line operator $W$ which proved so useful in Section \ref{subsec:formal-SU2}:
\begin{equation}\label{eq:star-formal-Wilson}
\begin{split}
W_1[c] & = [0,1,1] + [0, -1, -1] + [0,-1,1],\\
W_2[c] & = [1,0,1] + [-1,0, -1] + [1,0,-1],\\
W_3[c] & = [1,1,0] + [-1, -1, 0] + [-1,1,0].
\end{split}
\end{equation}
These are the unique combinations which transform well across all walls of the chamber if
it is of positive type with $\langle \gamma_i[c],\gamma_{i+1}[c]\rangle=+2$. For
chambers of negative type, $\langle \gamma_i[c],\gamma_{i+1}[c]\rangle=-2$, the three
Wilson operators are of the form
\begin{equation}
\begin{split}
W_1[c] & = [0,1,1] + [0, -1, -1] + [0,1,-1],\\
W_2[c] & = [1,0,1] + [-1,0, -1] + [-1,0,1],\\
W_3[c] & = [1,1,0] + [-1, -1, 0] + [1,-1,0].
\end{split}
\end{equation}
%

%
Let $\mu_{i,\pm}$ denote the transformation (mutation) across  wall $\widehat{W}(\pm \gamma_i[c])$.
The transformation of $W_i[c]$ by $\mu_{j,\pm}$ for $j\not=i$
follows the familiar pattern from the $SU(2)$ Wilson line: One 2-term (spin 1/2)
halo collapses and one term grows into
a 2-term halo, so that we continue to get a 3-term expression in the next chamber. That is, for $i \not=j$
we have the simple result
\begin{equation}
\mu_{i,\pm}(W_j[c]) = W_j[\mu_{i,\pm}(c)].
\end{equation}

On the other hand, if we mutate $W_i[c]$ by $\mu_{i,\pm}$ then two terms are unchanged,
while the third
term becomes a 3-term (spin 1) halo.
For example, using $\mu_{3,+}$ to transform to  chamber $c' = \mu_{3,+}(c)$ we find:
\begin{equation}\label{eq:mu-three-W3}
\begin{split}
\mu_{3,+} W_3[c] &=  [1,1,0](c) + [-1, -1, 0](c) + ([-1,1,0]+ [2] [-1,1,2] + [-1,1,4])(c)\\
& = [1,1,2](c') + [-1,-1,-2](c') + \left( [-1,1,2] + [2] [-1,1,0] + [-1,1,-2]\right)(c'). \\
\end{split}
\end{equation}

As in the case of $SU(2)$ $N_f=0$ the ring relations of the Wilson line operators are beautifully
connected to their transformation properties across chamber walls.
  For example
\begin{align}
W_1[c] W_2[c]&= y^{-\half}W_3[c]+y^{\half}(X_{\half \gamma_1[c]+\half \gamma_2[c]+\gamma_3[c]} +X_{\half \gamma_1[c]-\half \gamma_2[c]+\gamma_3[c]} +X_{-\half \gamma_1[c]-\half \gamma_2[c]-\gamma_3[c]} +\notag \\&+X_{\half \gamma_1[c]-\half \gamma_2[c]-\gamma_3[c]} + (y+y^{-1})X_{-\half \gamma_2[c]+\half\gamma_1[c]})
\end{align}
The generating function multiplied by $y^\half$ simplifies across either $\pm \gamma_3[c]$ walls: it loses a Fock space
and goes to a simple Wilson line, as we can recognize from \eqref{eq:mu-three-W3}.
%
%
%
%
The general relation for products of this type is:
\begin{equation}\label{eq:quant-skein-rel}
W_{i}[c]  W_{i+1}[c] = y^{-1/2} W_{i+2}[c]   + y^{1/2} \mu_{i+2,+} W_{i+2}[\mu_{i+2,-}(c)]
\end{equation}
For $W_{i+1}[c] W_{i}[c]$ we change $y\to 1/y$.

We expect that, by successively multiplying by Wilson lines we can recursively
generate the entire ring of formal line operators, as in the case of Section \ref{subsec:formal-SU2}.
%
%

Step by step we are uncovering a beautiful structure in the line operators of this theory.
In the second half of the paper we will develop more powerful tools to analyze
all $A_1$ theories. In the process, we will rediscover the formulae of this section.
For comparison with Sections \ref{subsec:N2-star-moduli}
and \ref{subsec:SU2-N2star-Examples} let's do one last calculation:
we note that $[2,2,2]$ and $[-2,-2,-2]$ are
central and that we can choose coefficients $a_{ijk}\in \IZ[y^{1/2},y^{-1/2}]$ suitably such that
\begin{equation}\label{eq:quant-mod-eq}
W_1^2 + W_2^2 + W_3^2 - a_{ijk} W_i W_j W_k = 1 + y^{-2} + B(y^2-y^{-2}) - \left(y^{-1} + (y-y^{-1})B\right)([2,2,2]+[-2,-2,-2])
\end{equation}
is a central operator. Here $B$ is an arbitrary constant.
The $y \to 1$ limit of \eqref{eq:quant-mod-eq} is \eqref{eq:moduli-equation}.


\section{Line operators and holomorphic functions on moduli spaces}\label{sec:linehol}

\subsection{Expansion in ``Darboux coordinates''}

We have been studying a $d=4$, $\CN=2$ theory, equipped with a collection $\CL$ of line operators.
We now consider the theory on a Euclidean spacetime
$\IR^3 \times S^1$, where the circle has radius $R$, and we take periodic
boundary conditions for the fermions.  At low energies, the theory is
described by a three-dimensional sigma model with a \hk target space $\CM_\CL$.
(In Section \ref{sec:moreonmoduli} below we will elaborate on how the moduli
spaces differ from one another for different $\CL$.)

This sigma model was the main object of study in
\cite{Gaiotto:2008cd}.  Let us quickly recall a few basic points to fix notation.
Topologically $\CM_\CL$ is a torus fibration over $\CB$.  The fiber over a point $u \in \CB$ is the torus of
electric and magnetic Wilson lines of the abelian gauge fields around $S^1$.
After choosing a quadratic refinement $\sigma: \Gamma_\CL \to \IZ_2$ of the pairing
$(-1)^{\inprod{\gamma, \gamma'}}$, the torus fiber can be identified with the character torus
$\Hom(\Gamma_\CL, \IR / 2 \pi \IZ)$.  We write $\theta$ for an element in this torus.
On $\CM_\CL$ there are complex ``Darboux coordinates'' $\CX_{\gamma}$
which were studied in detail in \cite{Gaiotto:2008cd,Gaiotto:2009hg}.\footnote{We refer
to \cite{Gaiotto:2008cd} for our conventions on \hk
manifolds, their complex structures, and the definitions of the
Darboux coordinates $\CX_{\gamma}$.  For a brief summary
see Section 2 of \cite{Gaiotto:2009hg}.}
In what follows it will be more convenient to consider instead
coordinates $\CY_\gamma$, which differ from $\CX_\gamma$ only by the quadratic refinement:
\begin{equation}\label{eq:Ygam-Xgam}
\CX_\gamma = \sigma(\gamma) \CY_\gamma,
\end{equation}
The $\CY_{\gamma}$ are actually more canonical than $\CX_\gamma$:  their
definition is independent of the choice of $\sigma$.  They obey a twisted product law
\begin{equation}\label{eq:Y-twisted}
 \CY_\gamma \CY_{\gamma'} = (-1)^{\inprod{\gamma,\gamma'}} \CY_{\gamma+\gamma'}.
\end{equation}

Now we can take the line operators $L_\zeta$ of the 4-d theory
to wrap the compactification $S^1$, thus defining corresponding loop operators.
In the low energy limit these become point operators in the 3-d theory.
The expectation value $\inprod{L_\zeta}$ of such a point operator
is
a function on the moduli space $\CM_\CL$.
As we explain in Appendix \ref{app:Holomorphy}, the fact that $L_\zeta$ preserves $osp(4^*\vert 2)_\zeta$
implies that the $\inprod{L_\zeta}$ are in fact \ti{holomorphic}
functions on $\CM_\CL$ in complex structure $J^{(\zeta)}$.
It follows that they can be expressed as functions of the $\CY_\gamma$.
We conjecture that
\begin{equation} \label{eq:trh-large-r}
\inprod{L_\zeta  } = \sum_\gamma \fro(L_\zeta, \gamma) \CY_\gamma.
\end{equation}

Now let us try to justify \eqref{eq:trh-large-r}.
We find the following argument compelling, although it falls short of a proof.
We begin by considering $\inprod{L_\zeta}$ as a trace over the
Hilbert space of the four-dimensional theory. Writing this trace explicitly
is slightly tricky:
\begin{equation}\label{eq:Lvev-trace}
\inprod{L_\zeta} = {\rm Tr }_{\CH_{u,L_\zeta}} (-1)^F e^{-2 \pi R H}  e^{i \theta\cdot \CQ}\sigma(\CQ).
\end{equation}
There are two subtleties here.  First, the torus
 fibers of $\CM_\CL$ define a local system with nontrivial monodromy. $\CQ$ is the
 charge operator measuring the IR charge in $\Gamma$, and is thus
 valued in a local system. The notation $\theta\cdot \CQ$ is the natural evaluation,
 which
 has no monodromy.  The second subtlety is that $\theta$
 defines boundary conditions for both electric \emph{and} magnetic Wilson lines.
 In order
 to implement such a boundary condition we should use a self-dual formalism.
 Now, it is well-known that to define the path integral of a self-dual abelian gauge
 theory it is necessary to introduce a quadratic refinement of a bilinear form
 on the relevant cohomology underlying the Dirac quantization of the  self-dual
 gauge theory \cite{Witten:1996hc,Freed:2000ta,Hopkins:2002rd,Belov:2006jd}. In the present case
 we should take the Dirac quantization condition to be valued in $H^2(M_4;\Gamma)$.
 In the case $M_4 = (\IR^3-\{\vec 0 \}) \times S^1 $, and using compactly
 supported cohomology, we are led to include the quadratic refinement $\sigma(\CQ)$
 in the trace. That is the
origin of the $\sigma(\CQ)$ in equation \eqref{eq:Lvev-trace}.

In the $R \to \infty$ limit the trace \eqref{eq:Lvev-trace} is projected to
the states with the lowest energy, namely the
framed BPS states. We can then use the infrared
description of the theory.  The leading
contribution we would expect from a framed BPS state with charge
$\gamma$ would be of the form
\begin{equation}
\sigma(\gamma) (-1)^{F} \exp(2 \pi R \, \Re (Z_\gamma / \zeta) + i \theta_\gamma).
\end{equation}
So for $R\to \infty$ we have
\begin{equation}\label{eq:Lzeta-asymp}
\inprod{L_\zeta} \sim \sum_\gamma \fro(L_\zeta, \gamma) \exp(2\pi R
\, \Re (Z_\gamma / \zeta) + i \ttheta_\gamma),
\end{equation}
where $e^{i \ttheta_{\gamma}}:= e^{i \theta_{\gamma}}\sigma(\gamma)$.
This expansion is much like the desired \eqref{eq:trh-large-r}, except that
instead of $\CY_\gamma$ we have $\exp(2\pi R \, \Re (Z_\gamma / \zeta) + i \ttheta_\gamma)$.
This indeed matches the $R \to \infty$ asymptotics of $\CY_\gamma$
as determined in \cite{Gaiotto:2008cd}.
Now how about the extension to finite $R$?
Let us provisionally write this expansion as
\begin{equation} \label{eq:trh-large-r-prov}
\inprod{L_\zeta  } = \sum_\gamma \fro(L_\zeta, \gamma) \CY'_\gamma,
\end{equation}
for some set of holomorphic functions on $\CM$ in complex structure $\zeta$.
In the IR theory the $\CY'_\gamma$
capture the effect of the insertion of a source of charge $\gamma$.
In particular, we expect all the detailed information about the UV line operator to be captured
by the $\fro(L_\zeta, \gamma)$, so that the $\CY'_\gamma$ do not depend on the
specific choice of UV line operator $L_\zeta$.
If we assume that there are sufficiently many line operators $L_\zeta$ then we can
in fact regard \eqref{eq:trh-large-r} as a linear transformation defining the $\CY'_\gamma$.
We would like to show that $\CY_{\gamma} = \CY'_\gamma$.

Recall from \cite{Gaiotto:2008cd,Gaiotto:2009hg} that the $\CX_{\gamma}$, and hence
$\CY_{\gamma}$, can be
uniquely characterized by a set of axioms expressing their holomorphy, asymptotic behavior for
$R\to \infty$ and $\zeta \to 0,\infty$, reality properties under $\zeta \to -1/\bar \zeta$, wall-crossing,
and multiplication.  Let us compare these properties for $\CY'_\gamma$ and $\CY_\gamma$.

First, concerning holomorphy, we
know that  $\inprod{L_\zeta}$ is holomorphic on $\CM^\zeta$, and it is holomorphic
in $\zeta\in \IC^*$ by assumption.
Next, as we have seen, the asymptotics of $\CY'_{\gamma}$ as $R \to \infty$
are those of $\CY_{\gamma}$, namely,
\begin{equation}\label{eq:X-asymps}
\CY'_\gamma \sim \CY_{\gamma}^{\rm sf} = \exp( \pi R
 \frac{Z_\gamma}{\zeta} + i \ttheta_\gamma + \pi R \zeta \bar Z_{\gamma} ).
 \end{equation}
There are similar  asymptotics for   $\zeta \to 0, \infty$, as can be deduced
from formal properties of the supersymmetric Wilson loop combined with duality. Using the
same strategy one can argue for the required
reality properties under $\zeta \to -1/\bar \zeta$.

Next we consider wall-crossing. There is no compelling reason to expect any
phase transition in $\inprod{L_\zeta}$ as a function on $\CB \times \IC^*$,
since there is no phase transition in the vacuum structure of the UV theory,
and we are taking the trace of a nice trace class operator so long as $R>0$.
Therefore, $\inprod{L_\zeta}$
should \emph{not} exhibit any wall-crossing behavior. On the other hand,
as we saw in detail in Section \ref{subsec:Halo-Wall-Crossing}, the framed
BPS degeneracies $\fro(L_\zeta,\gamma)$ do undergo wall-crossing. Now
 note that \eqref{eq:trh-large-r}  looks very much like the formal generating function $F(L)$ we
introduced in \eqref{eq:form-gen}, with the formal
variables $X_\gamma$ replaced by the honest functions $\CY_\gamma$ on $\CM$, and
$y$ specialized to $y=-1$.
As $(u,\zeta)$ crosses a wall $\widehat{W}(\gamma_\BPS)$, the functions $\CY_\gamma$ have
a discontinuity given by a symplectomorphism $\CK_{\gamma_\BPS}^{\Omega(\gamma_\BPS)}$,
where $\CK_{\gamma}$ was defined in \eqref{eq:KS-sympo}.
Therefore, comparing with the discussion from \eqref{eq:gen} to \eqref{eq:y-min-one-gen},
we see that $\inprod{L_\zeta} $  will indeed suffer no wall-crossing discontinuity,
provided that $\CY'_{\gamma}$ transforms exactly the same way as $\CY_{\gamma}$.

Finally,  the Heisenberg relations
\eqref{eq:Heis-Alg} for $y=-1$
imply that the functions $\CY_\gamma$ should satisfy a twisted multiplication rule
$\CY_\gamma \CY_{\gamma'} =(-1)^{\langle \gamma,\gamma'\rangle} \CY_{\gamma+\gamma'}$. This is
beautifully consistent with \eqref{eq:Y-twisted}.

This concludes our argument for \eqref{eq:trh-large-r} and \eqref{eq:Ygam-Xgam}.
The main   gap in the argument is the claim that there are sufficiently many
line operators to invert the equations \eqref{eq:trh-large-r}.

\subsection{Remarks on the Darboux expansion}

We would like to make a number of remarks on the previous subsection.

\begin{enumerate}

\item Since the index vanishes on massive representations, one might ask
why \eqref{eq:Lvev-trace} is not simply a sum over the framed BPS states.
(As does happen, for example, in the heat kernel proof of the index theorem
or in the evaluation of the Witten index in supersymmetric quantum mechanics
on a compact target).
As we noted, evaluating the operator in the trace on framed BPS states gives
the semiflat Darboux coordinates, and not the true Darboux coordinates, giving
the wrong answer.  The reason
for this is very similar to the phenomenon discussed at length in \cite{Cecotti:1992qh}:
the continuum makes a nonzero contribution to the trace.

\item As we describe in Section \ref{sec:Six-Dim-View}, there is a very interesting class of
$d=4, \CN=2$ theories arising from compactification of the $(2,0)$ theory
on a Riemann surface $C$ with punctures. In this case the IR
 abelian gauge theory arises from the self-dual 3-form of the abelian $(2,0)$ theory
 compactified on $\IR^{4}\times \Sigma$, where $\Sigma$ is the Seiberg-Witten curve
 covering $C$. In
 the path integral of the abelian six-dimensional theory the sum over topological sectors
 is weighted by a  quadratic refinement of the mod-two intersection
 form on  $H^3(M_6;\IZ)$ \cite{Witten:1996hc,Freed:2000ta,Hopkins:2002rd,Belov:2006jd}.
 Applying this to $M_6= (\IR^3-\{\vec 0 \}) \times S^1 \times \Sigma$
and using compactly supported cohomology the quadratic refinement of the six-dimensional
theory induces a quadratic
refinement $\sigma$ of the mod-two intersection form on $H^1_{cpt}(\Sigma;\IZ) \cong H_1(\Sigma;\IZ)$.

\item
Given the existence of protected spin characters it is natural to generalize
the RHS of \eqref{eq:Lvev-trace} to include  a variable conjugate
to $\CJ_3$ and define the twisted trace:
\begin{equation}\label{eq:y-dep-trace}
\inprod{L_\zeta}_y := {\rm Tr }_{\CH_{u, L_\zeta}} (-1)^F e^{-2 \pi R H} (-y)^{2 \CJ_3} e^{i \theta \cdot \CQ}\sigma(\CQ)
\end{equation}
The argument of the trace still commutes with a supersymmetry $\epsilon_{\alpha A} \CR_{\alpha}^A$.
Nevertheless, the physical interpretation of \eqref{eq:y-dep-trace}
appears to be rather different from \eqref{eq:Lvev-trace}.
The reason is that if we attempt to give a three-dimensional interpretation to
\eqref{eq:y-dep-trace} then the   modified trace requires us to glue space back to itself under a rotation around the $z$ axis and R-symmetry rotation. This is a reminiscent of an  $\Omega$ deformation of the four-dimensional background which preserves only a single supercharge.
(See \cite{Nekrasov:2010ka} for a recent reference with references to the earlier literature.)
The interpretation of this quantity in terms of
three-dimensional field theory is not straightforward, and might involve a noncommutative target space as a replacement for $\CM$. Indeed, one might guess
that at low energies the theory reduces to a quantum mechanics problem on $\CM$ and the trace
projects to the lowest energy states, where the appropriate expansion coefficients are noncommutative (much the way wavefunctions in the lowest Landau level of the quantum Hall effect may be interpreted in terms of noncommutative geometry). This might be a fruitful context in which to give a physical interpretation of the deformed chiral ring of line operators discussed in Section \ref{subsec:Deformed-Ring}.
This is an interesting subject for further research but we will not pursue it  in this paper, except
for some related remarks in Section   \ref{sec:Quantum-Holonomy}.

\item Finally, we can make a simple observation concerning the
meaning of the non-commutative deformation of the ring of holomorphic functions on $\CM$
given by the product of spin characters of the corresponding line operators. The leading correction away from $y=\pm 1$ is given by a Poisson bracket $\{ X_\gamma,X_{\gamma'}\}  = \langle \gamma,\gamma'\rangle X_\gamma X_{\gamma'}$ which coincides with the natural Poisson bracket of the $\CX_\gamma$ functions (remember that the \hk manifold $\CM$ is complex symplectic in all complex structures). Hence the non-commutative product should coincide with the deformation quantization of the ring of holomorphic functions on $\CM$. We illustrate this remark with
several examples in Section \ref{sec:Examples-of-M}.

\end{enumerate}

\subsection{Multiple moduli spaces and maximal mutually local lattices} \label{sec:moreonmoduli}

Let us now return to the precise definition of the moduli spaces on which
$\inprod{L_\zeta}$ are defined.

Let $\CL_i$ denote the distinct possible maximal mutually local collections of
simple line operators.  For each $i$ there is a charge lattice $\Gamma_{\CL_i}$,
and these fit together in a system of inclusions\footnote{Such a
system of inclusions is also called a ``lattice'' in Boolean algebra. However,
as that term would probably lead to awkwardness if not outright confusion, we
will not use it.}
\begin{equation}
\begin{matrix}
    &  &  \Gamma_{mx} &  &   \\
    & \nearrow &   &  \nwarrow & \\
    \Gamma_{\CL_1} &  & \cdots &  & \Gamma_{\CL_n} \\
    & \nwarrow &   &  \nearrow & \\
    &  &  \Gamma  &  &   \\
\end{matrix}
\end{equation}
where $\Gamma_{mx}$ is the union of the $\Gamma_{\CL_i}$. Note that the antisymmetric
pairing $\inprod{,}$ need not be integer-valued on $\Gamma_{mx}$.  This implies in particular
that $\Gamma_{mx}$ generally does not correspond to a lattice of charges in any
physical $\CN=2$ theory.

Each charge lattice $\CL$ corresponds to a slightly different physical theory.  When we reduce to
three dimensions this difference is important:  the target of the three-dimensional sigma model
is a \hk\ manifold $\CM_\CL$ which depends on $\CL$.  It is
a torus fibration
with fiber above $u\in \CB-\CB^{sing}$ given by
\begin{equation}
(\CM_{\CL})_u = {\rm Hom} ((\Gamma_{\CL})_u, \IR/\IZ).
\end{equation}
We may similarly define spaces $\CM$ and $\CM_{mx}$ corresponding to lattices $\Gamma$ and $\Gamma_{mx}$,
although these do not correspond to any physical
$\CN=2$ theories.
So there is a system of finite coverings
\begin{equation}\label{eq:covering-maps}
\begin{matrix}
    &  &  \CM_{mx} &  &   \\
    & \swarrow &   &  \searrow & \\
    \CM_{\CL_1} &  & \cdots &  & \CM_{\CL_n} \\
    & \searrow &   &  \swarrow & \\
    &  &  \CM &  &   \\
\end{matrix}
\end{equation}

In fact, these manifolds are very closely related to one another:  $\CM$ and $\CM_{mx}$ are \hk (just like
the $\CM_{\CL_i}$), and all the covering maps in \eqref{eq:covering-maps} are local isometries. Indeed,
they are fiberwise covering maps of tori.
This follows easily from the constructions of
\cite{Gaiotto:2008cd}:  the
dependence of the metric on the fiber coordinates $\ttheta$ arises only through factors $e^{i \ttheta_\gamma}$,
where $\gamma \in \Gamma$ is the charge of a vanilla BPS particle.  Hence any shift of $\ttheta$ which leaves
$e^{i \ttheta_\gamma}$ invariant for all $\gamma \in \Gamma$ is an isometry.

We believe the above picture extends over the singular locus $\CB^{sing}$. The
reason for this is the following. At  a generic point  $u^*\in \CB^{sing}$ the singularity
in the fibration is due to a   single BPS particle of charge $\gamma$ becoming  massless.
In this case a single circle in the torus shrinks to a point over $u^*$.
(Using the results of Section 13.7 of \cite{Gaiotto:2009hg} this is the
 circle described by the flow $\frac{d\ttheta_{\tilde \gamma}}{dt} = \langle \gamma, \tilde \gamma \rangle$.)
At least at large   radius $R$ of the spacetime circle
we can understand the behavior of the full \hk metric $u^*$  in terms
of the one loop correction to the naive large radius (semiflat)  metric due to the particle of charge
$\gamma$. Using the results of \cite{Gaiotto:2008cd}   we see that if
 $\gamma$ is irreducible, the circle shrinks at a single location in the fibre,
at $\exp i \ttheta_\gamma =1$.
Otherwise if $\gamma = Q \gamma_0$, then the metric has $Q$ codimension two singularities which
are - generically - families of $A_{Q-1}$ singularities. Using this picture of the degeneration of the
fiber we see that the relation of quotients in \eqref{eq:covering-maps} is preserved.

We will comment on these issues again in Section \ref{subsec:Hitchin-Isogenous}, and we
will illustrate these ideas with concrete examples in Section \ref{sec:Examples-of-M}.

\subsection{A vev for a twisted moduli space} \label{sec:twisted-vev}

Having opened Pandora's box by introducing \eqref{eq:y-dep-trace} we should note that
in addition to the case $y=-1$ discussed above there \emph{is} an interpretation for a closely related
trace:
\begin{equation}\label{eq:y=1-trace}
\inprod{L_\zeta}' := {\rm Tr }_{\CH_{u,L_\zeta}} (-1)^{2I_3} e^{-2 \pi R H}   e^{i \theta \cdot \CQ}
\end{equation}
where we choose $u,\theta$ and use the trace to \emph{define} the LHS.
Once again, all four supercharges are preserved.
The extra factor $(-1)^{2 I_3}$ acts trivially on non-exotic framed BPS states.

As in the $y=-1$ case, $\inprod{L_\zeta}'$ can be understood in terms of
a three-dimensional sigma model, obtained by compactification of our 4-d theory on $S^1$, now with the extra
twist $(-1)^{2 I_3}$.
The resulting theory is, as before, a sigma model into a \hk manifold, which we denote as $\tCM$.

How is $\tCM$ related to $\CM$?  Both of them can be described by the method of
\cite{Gaiotto:2008cd}.  The main nontrivial ingredient in that description is a set of instanton corrections, coming from
BPS states of the 4-dimensional theory, with world-lines wrapping the compactification $S^1$.
Now how does the factor $(-1)^{2 I_3}$ modify this story?  If there were exotic BPS states, then the corresponding instanton
corrections would be affected by this factor.  So in that case $\tCM$ would be distinct from $\CM$.
In the absence of exotic BPS states, though, we expect that $\CM$ and $\tCM$ are isomorphic as \hk manifolds
(though perhaps not canonically so).
At least in the examples we encounter below, this will indeed be the case.

The same reasoning as above allows us to expand
\begin{equation} \label{eq:y1}
\inprod{L_\zeta  }' = \sum_\gamma \fro(L_\zeta, \gamma, y=1) \tCY_\gamma,
\end{equation}
where $\tCY_{\gamma}$ are a collection of functions on $\tCM$, satisfying
properties analogous to the $\CY_{\gamma}$.  An important point is that unlike $\CY_\gamma$, these functions have an
\ti{untwisted} multiplication law,
\begin{equation}
\tCY_{\gamma} \tCY_{\gamma'} = \tCY_{\gamma + \gamma'}.
\end{equation}

Since we have commented that the manifolds $\CM$ and $\tCM$ should coincide in the absence of exotic
BPS states, one may wonder how to see this identification.
The most obvious way to do it is to identify the functions $\tCY_{\gamma}$ on $\tCM$ with
the functions $\sigma(\gamma) \CY_{\gamma}$ on $\CM$, for some choice of the quadratic refinement $\sigma$.
This identification is consistent with the discontinuity properties of $\CY_\gamma$ and $\tCY_\gamma$
if and only if $\sigma(\gamma) = (-1)^{2 \CJ_3}$ acting on all vanilla BPS states.
This property indeed holds for the canonical $\sigma$ defined in
\cite{Gaiotto:2009hg} for the $A_1$ theories.
If strong positivity holds, we should expect a similar simple behavior for more general theories.

\section{The six-dimensional viewpoint}\label{sec:Six-Dim-View}

Our considerations thus far have relied only on $d=4, \CN=2$
supersymmetry. In the remainder of the paper we are going to focus
on theories denoted as the class $\CS$ (for ``six'') in
\cite{Gaiotto:2009hg}.  These theories arise from compactification of the
superconformal $ADE$ $(2,0)$ theories on Riemann surfaces $C$ with
punctures.  They form a rich set of examples, and the six-dimensional
viewpoint will allow us to construct some very interesting examples
of line operators.  For $A_1$ theories we will be able to compute
the vacuum expectation values of the line operators in complete detail.

\subsection{Review of the 6d $(2,0)$ theories}\label{subsec:6d-review}

General considerations in Type IIB string theory or M-theory suggest
the existence of six-dimensional local quantum field theories with superconformal
symmetry \cite{Witten:1995zh,Strominger:1996ac,Witten:1995em}.  Very little
is known about these theories.
For nice reviews of what is known see \cite{Seiberg:1997ax,
Witten:2009at}.  We summarize below a few of the known results relevant for the present paper.
In particular, we focus on the theories with $(2,0)$ superconformal
symmetry. For a summary of our conventions on supersymmetry in five and six dimensions
see Appendix \ref{app:Six-Super}.

Abelian theories of tensor multiplets are examples of $(2,0)$ theories.
These are free field theories and can be described by a Lagrangian (after certain choices are made
\cite{Belov:2006jd}).
We will instead focus on the interacting theories for which no fundamental field-theoretic
degrees of freedom, and certainly no Lagrangian, is
known or even expected to exist. These theories have an ADE classification, and in
this section $\lieg$ denotes the compact real form of a simple and simply-laced Lie algebra.

In the case of the $A_{r}$ theories a partial definition can be given
using DLCQ and quantum mechanics on instanton moduli space
\cite{Aharony:1997th,Aharony:1997an}. What we do know about the ADE theories is
based on two statements which we will treat as axiomatic. Once the theories
are properly defined these should become theorems:

\begin{enumerate}

\item When the theory is compactified on $\IR^{1,4} \times S^1$ where $S^1$ has
radius $R$, with periodic boundary conditions for fermions, the long distance dynamics
is governed by the maximally supersymmetric Yang-Mills theory with a
gauge   Lie algebra $\lieg$ and coupling constant $g^2_{YM}$ proportional to $R$.\footnote{See equation (3.36) of \cite{Gaiotto:2009hg} for the precise normalization in our conventions.}

\item The theory on $\IR^{1,5}$ has a moduli space of vacua given by
\begin{equation}
\CM(\IR^{1,5}) = (\IR^5\otimes \liet)/\CW,
\end{equation}
where $\liet$ is a Cartan subalgebra of $\lieg$ and $\CW$ is the Weyl group.
At smooth points of $\CM(\IR^{1,5})$,
the low energy dynamics is described by a theory of a free $(2,0)$ tensormultiplet
valued in $\liet$.
\end{enumerate}

Of course, these two deformations must be compatible. For example, the moduli
space of vacua of the 5-d supersymmetric Yang-Mills theory can be identified
with $\CM(\IR^{1,5})$, since the vacuum expectation values $\langle Y^I \rangle$ of the
adjoint scalars in the ${\bf 5}$ of $so(5)_R$ must be simultaneously diagonalizable.
A point of the moduli space is thus identified with an ordered $5$-tuple of
simultaneously diagonalizable elements of $\lieg$, up to gauge invariance.
Just the way these vev's spontaneously
break the $\lieg$-symmetry to the Cartan subalgebra $\liet$ in the 5d SYM, we should think of the
``gauge symmetry'' of the 6d theory being spontaneously broken to the abelian gauge group of
the free tensormultiplets valued in $\liet$.  For this reason the moduli space $\CM(\IR^{1,5})$
is often referred to as the ``Coulomb branch.''

Furthermore, in the case of the
$A_r$ theories there is a nice intuitive picture of the theory as the worldvolume
of $(r+1)$ coincident M5 branes in a gravitational decoupling limit,
after the center of mass has been factored out \cite{Strominger:1996ac}.  The $so(5)$
$R$-symmetry is then interpreted in terms of rotations in the space transverse to the
5-branes, and the moduli space $(\IR^5 \otimes \IR^{r+1})/S_{r+1}$ parametrizes the
positions of $r+1$ parallel singly wrapped M5's in the transverse space.
The low energy dynamics of a single M5 brane is described by a $U(1)$ tensormultiplet,
with the scalar fields $Y^I$ representing transverse fluctuations.  A point of moduli
space can thus be specified by the vacuum expectation values of the scalar fields $\langle Y^I_s \rangle$,
$s=1,\dots, r+1$.  The $A_r$ theory is obtained after ``decoupling'' the tensormultiplet
describing the overall center of mass.

The $(2,0)$ theories can be defined on certain Lorentzian six-manifolds $M_6$
and their partition functions can be Wick rotated to Euclidean six-manifolds.
The $3$-form fieldstrength of the gauge potential $B$ in the tensormultiplet is
constrained to be self-dual.  This implies that $M_6$ must be oriented, and in
addition must be equipped with some extra topological data.
\footnote{Technically, a differential integral lift of the fourth Wu class \cite{Hopkins:2002rd}.
For physical discussions of this condition see \cite{Witten:1996hc} and
\cite{Belov:2006jd}.}
Moreover, it must be
equipped with a spin structure in order to define the fermions.

Finally, we mention a very subtle aspect of these  six-dimensional theories which we
do not wish to discuss  in depth here. The six-dimensional theory is not a standard
quantum field theory but is probably best regarded as a six-dimensional theory
valued in   an invertible 7-dimensional
 topological field theory. \footnote{We will not
go into the precise details of this notion here. We thank D. Freed and E. Witten
 for clarifying discussions on this issue.} The term ``vector-valued quantum field theory''
 is also sometimes used because the partition function of these theories on a six-manifold
 is valued in a vector space, and not just a number.  This subtlety reveals itself
  upon compactification on a circle. We said above that the 5d SYM has gauge Lie
  algebra $\lieg$, but to specify the theory one must choose a gauge group $G$
  whose Lie algebra is $\lieg$. There is not a unique choice of this gauge group.
  Extra data must be specified which determine a projection of   the vector space
    of partition functions to a one-dimensional subspace \cite{Witten:2009at}. Physically, such a projection
 might be specified by a choice of 't Hooft flux in the 5-dimensional theory.
This is already visible in the S-duality properties of ${\cal N}=4$ SYM, which is a torus compactification of the six-dimensional theory: at weak coupling, one picks a 5d gauge group when reducing on the ``small'' circle of the torus,
and then the same gauge group appears after reduction on the second circle. S-duality exchanges the two circles, and exchanges the projections corresponding to Langlands dual gauge groups.

\subsubsection{Chiral primary fields}

The DLCQ definition of the $A_r$ theories shows that there are
local chiral primary fields generating short multiplets of the
superconformal algebra \cite{Aharony:1997an,Bhattacharya:2008zy}. After compactification on a circle
these can be identified with gauge
invariant protected operators built out of the Casimirs of the
scalar fields, symmetrized and traceless over the $SO(5)$ indices.
In the $A_r$ case these descend to the operators  ${\Tr} Y^{(I_1 \cdots} Y^{I_k)}$
in the 5d SYM. They have scaling dimension $2k$.
It seems reasonable to assume that such chiral primary multiplets exist not only
for $A_r$ but for all the ADE theories, and we make this assumption in what follows.
A Casimir of rank $d$ gives rise to a dimension $2d$
operator in the symmetric traceless representation of $SO(5)$, with
$d$ indices.

\subsubsection{Dynamical BPS strings}

In the 6d nonabelian theory there are dynamical strings.  In the
M-theory derivation of the $A_r$ theories these arise from
open M2-branes stretched between the M5 branes.  Single BPS
string states are labeled by roots $\alpha \in \Phi(\lieg)$,
where $\Phi(\lieg)$ is the root system of $\lieg$.
A string with unit tangent vector $\hat t^M$ along the string
has a central charge $ Z_M^I = \hat t_M z^I$ in the supersymmetry
algebra. The tension of the string is proportional
to $\parallel \vec z \parallel$, where
$\parallel \cdot \parallel$ is the Euclidean norm on $\IR^5$.
A string excitation of a vacuum on the  Coulomb branch at a point $Y\in \IR^5 \otimes \liet$
has $\vec z$ proportional to $\langle \alpha, Y\rangle$.
These strings should be thought of as analogous to the ``off-diagonal'' gauge bosons
of nonabelian gauge theory.

We can check these assertions by compactifying on $S^1$ and using the expected reduction to 5d SYM.
The BPS strings descend to BPS particles of the
5D SYM theory.  Particles corresponding to singly-wrapped strings
correspond to gauge bosons of charge $\alpha$ and have mass
$\sim R \lVert \langle \alpha, Y\rangle \rVert$. There should
be no other light particles for small $R$ (they would have to be described
by fields not present in the 5d SYM multiplet), so for example there should
not be any strings labeled by other weight vectors of $\lieg$.

The dynamical strings of six dimensions also give BPS string states in
the 5d SYM. On the Coulomb branch these may be described, semiclassically,
as magnetic monopoles in the three dimensions transverse to the worldsheet of
the string. Such solutions are thus labeled by $\Lambda_{G} = {\rm Hom}(U(1),T)$,
the cocharacter lattice of $G$. For \emph{simply laced} Lie algebras $\Lambda_{cr}$
and $\Lambda_{r}$ are isomorphic upon choosing a normalization of
the  Killing form so that roots have length $2$. For this reason the strings in
5 dimensions are most easily understood for the choice of gauge group $\tilde G$,
the simply connected compact group associated to $\lieg$.

For completeness we mention that in the 5D SYM there are also BPS particles
described by instantons in the four transverse dimensions to the worldline.
These have mass proportional to $n/R$, where $n$ is the instanton charge,
and can be interpreted as KK modes arising from compactification on $S^1$.

\subsubsection{Surface operators}

In addition to chiral primary fields and dynamical strings
 there are surface operators in the 6d $(2,0)$ theories and
their 5d SYM compactifications. Our primary goal in this section is to use these to define
the line operators which will be the focus of our studies in the remainder
of the paper.

We begin with a general definition of a surface operator which is closely analogous
 to that of line operators:
A surface operator on a subspace $\IR^{1,1} \subset \IR^{1,5}$ has a transverse space
$\IR^4$. The metric is conformally equivalent to $AdS_3 \times S^3$ and has an isometry
group with Lie algebra $so(2,2)\oplus so(4)$. We define a surface operator on the
embedded $\IR^{1,1}$ to be the result of imposing a conformally invariant boundary
condition on $AdS_3 \times S^3$. Note that $so(2,2)\oplus so(4) \cong (sl(2,\IR) \oplus su(2))
\oplus (sl(2,\IR) \oplus su(2))\cong so(4^*) \oplus so(4^*)$.

As with the line operators we will focus on special surface operators that preserve some
$R$-symmetry and some supersymmetry. To motivate them, let us consider the surface operators in
the low energy tensormultiplet theory on the Coulomb branch. If $\Sigma$ is an oriented surface
in $M_6$
then we can define the holonomy of the self-dual field\footnote{We assume for simplicity in the
following discussion that $B$ is topologically trivial. For general spacetimes and topologically
nontrivial $B$-fields we would use instead Cheeger-Simons differential characters.}
\begin{equation}\label{eq:self-dual-holon}
h_v(\Sigma) = \exp[ 2\pi i \int_{\Sigma} (v, B) ]
\end{equation}
where $v \in \Lambda_{wt}$ is in the character group, i.e. the weight lattice,
 of ${\tilde G}$. We impose this restriction because on
topologically nontrivial manifolds large gauge transformations of the $B$-field take the form
$B \to B + \omega$ where $\omega$ is a globally defined closed 2-form with periods in $\Lambda_{cr}$.
Note that here and in the remainder of Section \ref{sec:Six-Dim-View}
$\Sigma$ has \emph{nothing} to do with the IR Seiberg-Witten curve!

Under some circumstances the holonomy can be upgraded to a supersymmetric surface operator. For
simplicity consider a single tensormultiplet with gauge group $U(1)$. Then consider
\begin{equation}\label{eq:single-tm-susy}
\exp\left[ 2\pi i \int_{\Sigma} B + \kappa  n^I Y^I  \vol(\Sigma)\right]
\end{equation}
where $\vol(\Sigma)$ is the volume form on $\Sigma$ from the induced metric, $\kappa$ is a constant,
 and we can
assume without loss of generality that $\vec n$ is a unit vector in Euclidean $\IR^5$ with
components $n^I$.
Let $\xi^\alpha$, $\alpha = 1,2$ be a local coordinate system on $\Sigma$. Then supersymmetries
$\epsilon_i^r Q_r^i$ will annihilate this operator provided
\begin{equation}\label{eq:surface-susy-cond}
\epsilon_i^r \left( \frac{d\xi^\alpha \wedge d \xi^\beta \p_\alpha X^M \p_\beta X^N}{\vol(\Sigma) }(\gamma_{MN})_r^{~~s} \delta^i_{~~j} + \kappa (n^I \Gamma^I)^i_{~~j} \delta_{r}^{~~s} \right) =0
\end{equation}
where $X^M(\xi)$ denote the embedding of the surface into $M_6$.
In order to preserve supersymmetry this equation must be satisfied for
\emph{constant} unbroken supersymmetries $\epsilon_i^r Q_r^i$. For a flat surface
and constant $n^I$ half the supersymmetries will be preserved with $\kappa=\pm 1$. More generally,
(analogously to super Yang-Mills)
  $\Sigma$ can be a curved  surface and $n^I$ can vary. An example which will be
   important below arises when the surface is $\IR \times \wp $ where $\wp$ is a curve
   in, say, the $12$ the plane. If we decompose the $R$-symmetry space $\IR^5 = \IR^2 \oplus \IR^3$,
   identify the $\IR^2$ summand with the $12$ plane, and take $n^I$ to be the unit tangent
   vector to $\wp$ then one-quarter of the supersymmetry is preserved. In general, the   $R$ symmetry is broken to $so(4)$ by the direction $n^I$.
In the special case of a plane in, say, the $01$ direction we can describe the preserved supersymmetry
in a way closely analogous to our discussion of $osp(4^*\vert 2)_\zeta$ for line operators. We
define an involution of $osp(8^*\vert 4)$ by reflection in the plane of the surface, together with
an $R$-symmetry reflection in the plane orthogonal to $n^I$. The fixed points of this involution
define a superconformal algebra which we can denote $(osp(4^*\vert 2) \oplus osp(4^*\vert 2))_{\vec n}$.

We can now combine \eqref{eq:self-dual-holon} and \eqref{eq:single-tm-susy} to define
supersymmetric surface operators $\IS_{\vec n}(v,\Sigma)$ in the abelian tensormultiplet theory.
It is then natural to conjecture that there are corresponding surface operators
$\IS_{\vec n}(\CR,\Sigma)$ in the nonabelian theory associated with a representation $\CR$ of
$\lieg$ and preserving the superconformal symmetry  $(osp(4^*\vert 2) \oplus osp(4^*\vert 2))_{\vec n}$.
As usual, we can check this assertion with the two available deformations.
When  the gauge symmetry is strongly broken by large expectation values of $Y$ these surface operators
will be described semiclassically by sums over the abelian surface operators
$\IS_{\vec n}(v,\Sigma)$ where $v$ runs over the weights of the representation $\CR$.
As a second check we consider the relation to the 5d SYM which emerges upon
compactification on a circle of radius $R$. If $\Sigma =   S^1\times \wp $ for a path $\wp\subset M_5$
then the IR limit of the nonabelian surface operator should be
\begin{equation}\label{eq:5d-line-op}
L_{\vec n}(\CR, \wp) = {\Tr}_{\CR} P\exp \left[ \int_\wp  A + \kappa  n\cdot R Y ds \right]
\end{equation}
These are indeed   BPS line operators in the 5d SYM.
Finally, the existence of the non-abelian surface operators is
intimately connected to the existence of the dynamical BPS
strings. For example, one could start from an $su(r+1)$ theory,
and move to a partial Coulomb branch, where the theory reduces
to an $su(r) \times u(1)$ theory. There will be dynamical
strings charged under the $u(1)$ abelian factor, whose tension
goes to infinity as the expectation values of the scalars are
sent to infinity, and the $u(1)$ factor decouples. In the
limit, the dynamical strings become surface operators
$\IS_{\vec n}(v,\Sigma)$, where $\vec n$ is the direction of
the $Y$ sent to infinity.

It is well-known that the insertion of line operators such as
\eqref{eq:5d-line-op} into a path integral can be interpreted as coupling the ambient theory to
a $0+1$-dimensional quantum mechanical system (defined by the quantization of
co-adjoint orbits) \cite{Alekseev:1988vx}. Similarly, surface operators can be represented
by coupling  a $1+1$ dimensional quantum field theory defined on $\Sigma$ to the
ambient theory. In this interpretation note that,  as is familiar from studies of the
$AdS_3/CFT_2$ correspondence, the Lie algebra of isometries $so(2,2)\cong sl(2,\IR)_l \oplus sl(2,\IR)_r$
can be interpreted as the conformal symmetry algebra which acts separately on
 the  left- and right-moving degrees of freedom. This
interpretation extends to allow us to interpret the unbroken superconformal symmetry
in the UV as $osp(4^*\vert 2)_l \oplus osp(4^*\vert 2)_r$. Moreover, as we have
seen, in the IR the abelian surface operators descending from  $\IS_{\vec n}(\CR,\Sigma)$
are labeled by weights $v$ of the representation $\CR$. This suggests that the QFT defined
on $\Sigma$ should have vacua labeled by $v$.

Dynamical strings can end on surface operators, and they define
domain walls in the 2D QFT   on $\Sigma$. A string labeled by a
root $\alpha \in \Phi(\lieg)$ will be a domain wall between
vacua $v_1$ and $v_2$ if $v_1 - v_2 = \pm \alpha$.  The
configuration can be $\frac{1}{4}$ BPS, much in the same way as
junctions between dynamical strings can be. A typical example
is a static configuration where all the strings and the surface
operator lie on a plane, the central charge vectors $\langle
\alpha_i,Y\rangle$ or $\vec n$ also lie in a plane in the space
of central charges, and the slopes in the two planes are the
same. Notice that if one is interested in $\frac{1}{4}$ BPS
configurations, the surface operators may be allowed a more
general shape than a straight line in the plane, by allowing
$\vec n$ to vary along the operator, following the variation of
the tangent vector.

\subsection{The twisted theory compactified on a Riemann surface}\label{subsec:Twisted-C}

We want to consider the $(2,0)$ theory wrapped on a Riemann surface $C$ as used in \cite{Gaiotto:2009hg}.
As described in Section 3.1.2 of \cite{Gaiotto:2009hg} in order to preserve $d=4, \CN=2$ supersymmetry
in this compactification we must twist the theory. The compactification $\IR^{1,3} \times C$
breaks the local Lorentz symmetry to   $so(1,3) \oplus so(2)_C$. We then explicitly break the $R$-symmetry $so(2)' \oplus so(3)'\subset so(5)$
and twist the theory so that the spin connection on $C$ is coupled to the diagonal subalgebra
$so(2)_d \subset so(2)_C \oplus so(2)'$.   Eight   supercharges survive, forming an $\CN=2$
susy algebra in four dimensions. The $so(3)'$ is identified with the $R$-symmetry of $d=4, \CN=2$
 and $so(2)_d$ is the anomalous $u(1)_R$ symmetry of $d=4, \CN=2$.
 The multiplet of chiral primary operators of index $d_k$  contains one of
maximal $so(2)_d$ charge, denoted $\CO_k$ so that the vev $\langle \CO_k\rangle$ becomes a
holomorphic $k$-differential on $C$, and the Coulomb branch of the 4d theory
is parameterized by the vevs of these operators
(see \cite{Gaiotto:2009hg} eq. (3.5)):
\begin{equation}
\CM(\IR^{1,3}\times C) =\CB =  \oplus H^0(C, K^{\otimes d_k})
\end{equation}
As in \cite{Gaiotto:2009hg} it is important to introduce codimension $2$ defects placed at punctures of $C$.
These defects do not break any further supersymmetry.  Their effect is to create poles in
$\langle \CO_k \rangle$.
The four-dimensional IR theory is determined by the choice of $C$ and the singularities of
the $\langle \CO_k \rangle$ at the defect points on $C$.
The moduli space of gauge couplings coincides with the moduli space of complex structures of the
punctured Riemann surface $C$.

\subsection{Moduli space}

As explained at length in \cite{Gaiotto:2009hg} the moduli space of vacua
\begin{equation}
\CM = \CM(\IR^{1,2}\times S^1 \times C)
\end{equation}
can be identified with the moduli space of solutions to Hitchin's equations on $C$
with certain singularities at the defects.  These are equations on a connection $A$
and 1-form $\varphi$, which say that the complex connections
\begin{equation} \label{eq:flat-conn}
\CA(\zeta) = R \frac{\varphi}{\zeta} + A + R \zeta \bar\varphi
\end{equation}
are flat for all $\zeta \in \IC^\times$.
In fact, in its complex structure $J^{(\zeta)}$, the moduli space $\CM$ is just identified with a moduli space
of flat $G_\IC$-connections $\CA(\zeta)$ on $C$, having singularities
at the punctures on $C$. The gauge group $G$ has Lie algebra $\lieg$.

The nature of the singularities at the punctures depends on the precise defects we insert on $C$.
The simplest possibility is a ``regular puncture'', which creates simple poles in $A$ and $\varphi$,
hence produces a regular singularity in $\CA(\zeta)$.
Such a puncture determines a ($\zeta$-dependent) conjugacy class in $G_\IC$; the monodromy of $\CA(\zeta)$ is restricted to lie
in that class.  One can also
consider ``irregular punctures,'' which create more intricate irregular singularities.

We can describe $\CM$ more concretely.
If we only consider regular punctures $\CP_i$, then flat connections modulo gauge are completely captured
by their monodromy representation:  so $\CM$ is a space of
representations of $\pi_1(C \setminus \{ \CP_i \})$ into $G_\IC$, with the conjugacy class of the
mondromies around punctures fixed.\footnote{We say ``a space of representations''
instead of ``\ti{the} space of representations'' because of a subtlety involving discrete quotients, to
be discussed momentarily.}  Yet more concretely,
choose a basepoint $x \in C$,
based loops running around $A$ and $B$ cycles on $C$, and based loops running around the punctures
$\CP_i$:  then we are studying tuples
$\{ A_1, \dots, A_g, B_1, \dots, B_g, M_1, \dots, M_n \} \in G_\IC^{2g + n}$ obeying
\begin{equation}
 M_1 \cdots M_n = \prod_{i=1}^g A_i B_i A_i^{-1} B_i^{-1},
\end{equation}
with $M_i$ in fixed conjugacy classes.  These tuples are considered
modulo the action of $G_\IC$ by simultaneous conjugation on all $A_i$, $B_i$, $M_i$.
If we consider irregular punctures the story becomes slightly more complicated, because
at such a singularity there
is more local information than just the monodromy:  the asymptotics of
the flat sections yield additionally a decomposition
of each $M_i$ into Stokes factors and formal monodromy.
See \cite{Witten:2007td} for a more in-depth discussion.

We will see below that the vevs of line operators are natural functions on $\CM$, such as traces of
products of the matrices appearing above.
This description allows us to study their semiring relations and compare with the discussion
we gave in Section \ref{sec:Formal-line-operators}.  On the other hand, to study the relation between line operators and spin characters
we need to identify the functions $\CY_\gamma$ which appeared in Section \ref{sec:linehol}.  These are not given directly
in terms of the matrices above, and in fact we do not know how to construct them at all for general $G$
(except by solving the integral equation of \cite{Gaiotto:2008cd}).  However,
for the $A_1$ theories we do know what $\CY_\gamma$ are \cite{Gaiotto:2009hg}, a fact which will be exploited
in Section \ref{sec:Examples}.

\subsubsection{Isogenous Hitchin moduli spaces}\label{subsec:Hitchin-Isogenous}

Our general discussion from Section \ref{sec:moreonmoduli}
led us to expect not just one moduli space $\CM$ but several closely related $\CM_\CL$,
labeled by families $\CL$ of line operators.
How does this choice show up in the Hitchin system?

In defining the moduli space of solutions to Hitchin equations, one always has to divide out by the group of
gauge transformations.
One possibility is to use gauge transformations valued in the simply connected form of $G$.
This yields some moduli space $\CM_{sc}$.  However, one could also have been more liberal by allowing
gauge transformations valued in the adjoint form of $G$.  The two choices differ by a finite group $\Delta$,
isomorphic to the group of flat bundles on $C$ with structure group $\CZ = Z(\tilde G)$.  $\Delta$ acts by
isometries on $\CM_{sc}$, so we can consider various quotients of $\CM_{sc}$ by subgroups of $\Delta$.
We would like to identify these quotients with the various $\CM_\CL$.

We can state the dictionary between choices of $\CL$ and subgroups of $\Delta$ more precisely in specific examples.
For $A_1$ theories with regular singularities, the relation is particularly simple:  as we will see
momentarily, a choice of $\CL$ will correspond to a choice of a set of closed paths $\wp$ on $C$.
Our proposal will be that $\CM_\CL$ is the quotient of $\CM_{sc}$
by the subgroup of $\Delta$ corresponding to bundles which have trivial holonomy along all paths $\wp$ in $\CL$.
For $A_1$ theories with irregular singularities, the situation is more difficult:  in particular,
$\CM_{sc}$ is not the biggest space which occurs --- one seems to need some further discrete restrictions
on the allowed gauge transformations.  We will not attempt a full analysis in this paper,
but we will present a few examples.

What is the six-dimensional motivation of the relation between a choice of four dimensional gauge group, and of the allowed gauge transformations in the Hitchin system? We do not have a full story
available, but we do have some useful hints.
Although the Lie algebra underlying the Hitchin system is determined by the choice of six-dimensional SCFT, the precise choice of gauge group is rather more subtle. The issues are similar to those discussed for the
reduction along a circle in Section \ref{subsec:6d-review} above. Presumably, when compactifying on a Riemann surface
the choices of 4d gauge group will correspond to certain choices of a projection on the vector space
of partition functions. When we compactify down from 4d to 3d on a further circle, we do not have a further choice of projection available anymore:
whatever choice we made of 4d gauge group of the ${\cal N}=2$ theory will determine the properties of the 5d theory on $C$.

\subsubsection{Twisted local systems} \label{sec:twisted-loc}

There is a second global issue which needs attention.
In Section \ref{sec:twisted-vev} we emphasized that in addition to the moduli space $\CM$, which arises
when we compactify the theory on $S^1$
in the usual way, there is an \ti{a priori} different space $\tCM$ which arises when we
compactify with a twist by $(-1)^{2\CJ_3}$.  What is this space?

We propose that $\tCM$ is a slightly twisted version of $\CM$.
We will not describe the appropriate twisting of Hitchin's equations here; instead we just
describe what $\tCM$ looks like as a complex manifold
in its complex structure $J^{(\zeta)}$, for $\zeta \in \IC^\times$.  It looks very much
like $\CM$, except that we replace flat connections by \ti{twisted} flat connections.
A twisted flat $SL(2,\IC)$-connection on $C$ is a flat $SL(2,\IC)$-connection
on the unit tangent bundle to $C$, which has holonomy $-1$ around each fiber.
(Note that if we replaced $-1$ by $+1$ in the last sentence we would have reduced to
ordinary flat connections.)

It would be desirable to have a direct derivation of this twisting from the physics of the $(2,0)$
theory.  For our purposes in this paper we are content with a consistency argument:
$\tCM$ as just defined will turn out to have all the properties we described
in Section \ref{sec:twisted-vev}.  In particular, the twisting turns out to dispose very
nicely of some tricky sign issues.

Choosing a spin structure on $C$ gives an isomorphism $\tCM \simeq \CM$.  This is because fixing
a spin structure is equivalent to fixing a flat $\IZ_2$-valued local system on the unit tangent
bundle to $C$, which has holonomy $-1$ around each fiber:  tensoring with this local system
gives the desired isomorphism.  The two spaces are
thus isomorphic but not quite canonically isomorphic.  An exception is the case
when $C$ has genus zero, in which case there is only a single spin structure, so the two spaces
are really canonically isomorphic.

The moduli space of twisted local systems was also considered in \cite{MR2233852,Frenkel:2007tx}, for closely related
(but not quite identical) reasons.

\subsection{Line operators from six dimensions}

We can use the nonabelian surface operators $\IS_{\vec
n}(\CR, \Sigma)$ of the interacting $(2,0)$ theory to define
line operators $L_\zeta(\CR, q; \wp )$ in the $d=4, \CN=2$
theory, where $q$ is a path in $\IR^{1,3}$ and $\wp\subset C$ is
a non-self-intersecting path. We  let $\Sigma = \wp \times q$ and work in the
twisted theory of Section \ref{subsec:Twisted-C}.   Now
the twisting has broken the $R$-symmetry $5=2\oplus 3$ and we
want to preserve the $so(3)'$ $R$-symmetry so $\vec n$ should
be a vector in the two-dimensional subspace in the
decomposition $5=2\oplus 3$.  Let us now take $q =
\{ \vec 0 \}\times \IR$ to be a line along the time direction
in $\IR^{1,3}$. If we  write out the supersymmetry condition
\eqref{eq:surface-susy-cond} for the unbroken supersymmetries
preserved by the surface operator we find that they are in fact
four-dimensional supersymmetries of the form in equation
\eqref{eq:Q-zeta-susy}  of Section \ref{sec:Susy-BPS-line}, as
long as we allow $\vec n$ to vary along the path. Actually, we
should be more precise here: after twisting, $\vec n$ is a
vector in the  tangent
bundle to the curve. The best notion of ``constant $\vec n$''
available is that $\vec n$ is transported along $\wp$ so that the
angle $\beta$ between $\vec n$ and the tangent vector to $\wp$ is
constant. This choice is the correct one to preserve four
supersymmetries, labeled by a phase $\zeta = e^{i \beta}$.

Now, as we observed in Section \ref{sec:linehol} if we Wick rotate the time to a Euclidean circle
of radius $R$ then the vevs $\langle L_\zeta(\CR,q;\wp)\rangle$, where $q$ is now the Euclidean time circle,
can be interpreted as holomorphic functions on the hyperkahler moduli space $\CM$. Using the six-dimensional
viewpoint we can give a nice interpretation of these holomorphic functions.

First, consider the
IR limit of the surface operator:
it is written in terms of the surface operators of the tensormultiplet theory valued in $\liet$.
Let
\begin{equation}\label{eq:varph-Y}
\varphi  = \half( Y^1 + i Y^2 )
\end{equation}
(as in equation (3.31) of \cite{Gaiotto:2009hg}). Because of the twisting, this becomes a $(1,0)$ form on $C$.
Then for finite $R$, in the IR limit we must replace fields by their zero-modes along $S^1$ and hence
\begin{equation}
2\pi i \int_{q\times \wp} (n\cdot Y) \vol(\Sigma) \longrightarrow  \pi \int_\wp \left(
\frac{R \varphi}{\zeta} + R \zeta \bar \varphi \right)
\end{equation}
(The overall factor of $i$ is removed by the Euclidean continuation of the time circle.)
Similarly, $2\pi i \int_{q\times \wp} (v, B) \longrightarrow \int_\wp (v, A)$, where $A$ is an abelian
gauge field of the 5d abelian SYM theory and $v$ is a weight vector in $\CR$.\footnote{The
factor of $2\pi$ does not appear on the RHS because we use standard normalization
conventions for the gauge field so that $F/2\pi$ has integral periods.}

Recall that compactification of the nonabelian $(2,0)$ theory on $S^1$ leads to a 5-dimensional
nonabelian super Yang-Mills theory. Now
\eqref{eq:varph-Y} defines an adjoint-valued $(1,0)$-form $\varphi$ on $C$.
In view of the above statements about the abelian theory,
it is natural to conjecture that the vev of the nonabelian surface operator is
\begin{equation}\label{eq:surface-op-vev-to-5d}
\langle \IS_{\vec n}(\CR, q\times \wp) \rangle =  \left \langle {\rm Tr}_{\CR} P \exp \int_\wp \left(
  \frac{\pi R \varphi}{\zeta} + A +  \pi R \zeta \bar \varphi   \right) \right \rangle,
\end{equation}
where now $\varphi$ and $A$ are valued in the nonabelian 5d SYM
multiplet with gauge algebra $\lieg$, and the vev on the
RHS is in the 5d SYM theory. This is nothing else but the
statement that the nonabelian surface operator should descend
to the Wilson loop of the nonabelian 5d theory. The vev on the
LHS is a holomorphic function on $\CM$.  Moreover, thanks to the
twisting it is independent of the scale of $C$, which can
therefore be made arbitrarily large. Finally, the IR limit of
the 5d SYM theory is free. These facts strongly suggest that
the equality in \eqref{eq:surface-op-vev-to-5d} is exact, and,
moreover, the RHS can be replaced by its classical value.
The RHS of \eqref{eq:surface-op-vev-to-5d} is thus the holonomy of $\CA(\zeta)$ on $\wp$
and we finally arrive at the main result of this subsection:
\begin{equation}\label{eq:L-5d-Holon}
\langle L_\zeta(\CR,q;\wp) \rangle = \Tr_\CR \Hol_\wp \CA(\zeta).
\end{equation}
The answer depends on $\wp$ only through its homotopy class.
This is a reflection of the fact that the twisting has eliminated dependence on the
scale of the metric on $C$.

The result \eqref{eq:L-5d-Holon} is only slightly modified if we consider the twisted compactification on $S^1$
as in Section \ref{sec:twisted-loc}.  To any oriented closed non-self-intersecting path $\wp$ in $C$ we can
assign a corresponding oriented closed path $\tilde{\wp}$ in the unit tangent bundle (by parameterizing $\wp$
with unit speed).  Then we can take the holonomies of twisted flat connections $\tilde\CA$ along $\tilde{\wp}$:
\begin{equation}\label{eq:L-5d-Holon-twisted}
\langle L_\zeta(\CR,q;\wp) \rangle' = \Tr_\CR \Hol_{\tilde{\wp}} \tilde\CA(\zeta).
\end{equation}
This gives a holomorphic function on $\tCM$.

\subsection{Cataloging line operators}

Let us make a few more comments about this realization of line operators.

\begin{enumerate}

\item Above we discussed line operators associated to closed non-self-intersecting paths.
What about paths which do have self-intersections?
The expression $\Tr_\CR \Hol_\wp \CA(\zeta)$ clearly makes sense for self-intersecting paths $\wp$,
but we do not understand the physics of self-intersecting surface operators in the 6d theory well enough
to include them in our discussion.  One may also wonder about ``junctions'' on $C$,
where surface operators corresponding to three or more
representations $\CR_i$ come together.  In 5d Yang-Mills the number of such intersections would be
counted by the number of identity representations in $\otimes_i \CR_i$: the ends of the Wilson loops would
be contracted with the corresponding intertwiner. Similar junctions exist in the 2d Toda theories on $C$ \cite{Drukker:2010jp}, which are relevant to $S^4$ compactifications of the 4d gauge theory.
This suggests they should be a feature of the 6d theory as well, and yield line operators in the 4d theories.
We do not pursue this further here.

\item There is some redundancy in our construction.  For example,
\cite{Drukker:2009id} argued that, in the case when the ADE group is $A_1$ and $C$ has only regular punctures,
surface operators along non-self-intersecting paths on $C$ carrying the fundamental
representation of $SU(2)$ suffice to describe all UV gauge theory operators:  no higher representations seem
to be needed.

\item On the other hand, if we consider $C$ with irregular punctures, our discussion so far does not give all the
line operators:  we need to include additional operators involving open paths on $C$.  We will not give a general
construction of such operators here, but in Section \ref{sec:missing} below we will describe what they must look like
in the $A_1$ theories.

\item Finally, not all the line operators we discussed here can be introduced simultaneously, because of the constraints
from mutual locality.  The choice of an allowed set of line operators $\CL$ corresponds in this context to choosing some restrictions
on the representations and closed paths we will allow.  For example, in $A_1$ theories, we may freely allow arbitrary paths carrying
integral-spin representations of $SU(2)$, but we may allow half-integral spins only on some subset $S$ of the paths.
Locality requires that $S$ be chosen so that the intersection number between paths in $S$ is always \ti{even}.
The corresponding moduli space $\CM_\CL$ is the moduli space of $SU(2)$ Hitchin systems, divided by the group of
$\IZ_2$-connections which have trivial holonomy along all paths in $S$.

\end{enumerate}

\section{Some examples of the moduli spaces and holomorphic functions}\label{sec:Examples-of-M}

This section describes the moduli spaces $\CM$, and their relation to the spaces $\CM_{\CL_i}$ and
$\widetilde{\CM}$ introduced in Sections \ref{sec:moreonmoduli}, \ref{sec:twisted-vev},
\ref{subsec:Hitchin-Isogenous}, and \ref{sec:twisted-loc}, in the concrete examples of theories used in this paper.
We will also describe how the algebra of functions on these spaces is related to the noncommutative
algebras of line operators described in Section \ref{sec:Formal-line-operators} and to the
laminations described in Section \ref{sec:Examples}.

\subsection{$U(1)$ gauge theory and periodic Taub-NUT}

We begin with the $U(1)$ theory of Section \ref{subsec:Formal-line-ops-U1}.
The torus fibration structure of $\CM$ is described as follows.
The Coulomb branch $\CB$ of a $U(1)$ gauge theory coupled to an hypermultiplet
of charge $1$ has a singularity at the origin $Z_{0,1} = a = 0$,
where the hypermultiplet is massless.  It also has an unphysical behavior at large distance from the origin,
due to the fact that the theory is not asymptotically free.
Sufficiently close to the origin in $\CB$, the theory makes sense.
Upon dimensional reduction, the 4d gauge fields provide two
circle-valued scalars:
an electric Wilson line $\ttheta_e$ and a magnetic Wilson line $\ttheta_m$.  The magnetic circle
shrinks to a point at the origin of the Coulomb branch.

The metric on $\CM$ is smoothed out by the 3d one-loop corrections to the ``periodic Taub-NUT'' or ``Ooguri-Vafa'' space \cite{Ooguri:1996me,Seiberg:1996ns}.
After these corrections one sees that the
magnetic circle shrinks only in codimension 3 --- at the origin of the 4d Coulomb branch ($a=0$)
and at a specific point in the electric circle ($\ttheta_e=0$).

The subtleties we mentioned about different moduli spaces $\CM_{\CL}$ are absent here:  there are no interesting
choices of $\CL$ to be made in this example.

Now let us turn to the algebra of functions on $\CM$.
In generic complex structure, the manifold can be identified with the complex manifold $V_+ V_- = 1-U$, where $U = \CY_{0,1} = \exp \left( \frac{\pi R}{\zeta} a + i \ttheta_e + \pi R \zeta \bar a \right)$ is valued in $\IC^*$ and $V_\pm$ are more intricate holomorphic functions, valued in $\IC$.
Their asymptotics in the $\zeta$ plane are such that  $V_+ = \CY_{1,0}$ in the sector in the $\zeta$ plane with $\ {\rm
Im}(Z_{0,1}/\zeta)>0$, while $V_- = \CY_{-1,0}$ on the opposite sector. We recognize immediately that
the expansion of $V_+^p U^q$ and of $V_-^p U^q$ in $\CY_\gamma$ and their ring relations
match the ones computed for the formal line operators $F_{p,q}$ and $F_{-p,q}$ discussed in
Section \ref{subsec:Formal-line-ops-U1}, once evaluated at $y=-1$.

It is also instructive to inspect the geometry associated to a $U(1)$ gauge theory with
an hypermultiplet of electric charge $Q$. The equation of the manifold now
 takes the form of  $V_+ V_- = (1+(-U)^Q)^Q$,
and has a set of $Q$ $A_{Q-1}$ singularities, sitting at $U^Q=(-1)^{Q-1}$: the magnetic circle is shrinking
at $Q$ locations on the electric circle, and it is $Q$ times smaller than before, so that it forms $A_{Q-1}$ singularities where it shrinks. Note that this beautifully matches the $y=-1$ specialization of the
ring relations \eqref{eq:Q-U1-algebra}. (To show this it is useful to note that
at $y=-1$ we have $[n]=(-1)^{n-1} n $, $[n]!  = (-1)^{\half n (n-1) } n!$
  and $\qbinom{n }{ j}_{y=-1} = (-1)^{j(n+1)} {n \choose j}$. )

\subsection{$N=3$ Argyres-Douglas theory}\label{subsec:N3-AD-Manfld}

Let us now turn to the AD theory with $N=3$ discussed in Section \ref{sec:formaln3}. We begin by
describing the fibration structure.

The three-dimensional Coulomb branch of this theory has an intuitive structure for large radius $R$ of the compactification circle:
the neighborhood of each of the two singularities in the Coulomb branch looks like the Coulomb branch for the $U(1)$ theory,
coupled to a single particle of charge $1$. The singularities are smoothed out exactly the same way, though of course
 a different circle
shrinks at each. Roughly speaking, the magnetic circle shrinks at one point of the electric circle over one singularity,
and the electric circle shrinks at one point of the magnetic circle over the other singularity.  The shift symmetry of the Wilson lines is completely broken.  As in the
$U(1)$ example, there are no interesting choices for $\CM_{\CL_i}$ in this example.

Now, it follows from the matching to Hitchin systems described in \cite{Gaiotto:2009hg}, Section 9, that
we  can describe $\CM$ as the moduli space of an $SU(2)$ Hitchin system on the sphere,
with a single irregular singularity
of such a degree as to give five Stokes sectors around the singularity. As a complex manifold,
$\CM$ is given by the equation $M=1$, where the monodromy matrix $M$ is decomposed as
\begin{equation}
M  = \begin{pmatrix} 1 & U_1 \\  0  & 1 \\ \end{pmatrix}\begin{pmatrix} 1 & 0 \\  - U_4  & 1 \\ \end{pmatrix}\begin{pmatrix} 1 & U_2 \\  0  & 1 \\ \end{pmatrix}\begin{pmatrix} 1 & 0 \\  - U_5  & 1 \\ \end{pmatrix}\begin{pmatrix} 1 & U_3 \\  0  & 1 \\ \end{pmatrix}\begin{pmatrix} 0 & 1 \\  - 1  & 0 \\ \end{pmatrix}.
\end{equation}
Here the triangular matrices are Stokes matrices, and the last matrix is the formal monodromy.
The $U_i$ are holomorphic functions on the moduli space. They have been labeled for future convenience.
We will also extend the labeling by defining $U_{i+5} = U_i$.
The equations $M=1$, with a bit of massaging,\footnote{One way to prove this is to bring three factors to one side
of the equation $M=1$. This then gives four equations on matrix elements which are easily seen to be equations
\eqref{eq:N3-manifold} for $i= 2,3,4,5$. To get the last equation, multiply the equation
for $i=4$ by $U_2$, and note that if $U_3$ is nonzero then we get the equation for $i=1$.
Similarly, if we multiply the equation for $i=3$ by $U_5$ and divide by $U_4$, we
get the equation for $i=1$. But $U_3$ and $U_4$ cannot simultaneously vanish. }
take an appealing form:
\begin{equation}\label{eq:N3-manifold}
U_{i-1} U_{i+1} = 1+U_i.
\end{equation}

These five equations can be shown to define a smooth two-dimensional complex submanifold of $\IC^5$.
If we associate the five coordinates with vertices of a pentagon, then the coordinates attached to
neighboring vertices never simultaneously
vanish. There are five special divisors $U_i = 0$, intersecting pairwise in a pentagonal configuration
at five special points, namely $U_1 = U_3 =0$, $U_2=U_4=U_5=-1$ and its images under cyclic permutation
of the indices.

Comparing equation \eqref{eq:Deformd-N3-rels} with \eqref{eq:N3-manifold},
we see that we can recover the commutative algebra of
functions by taking the limit $F_i \to U_i$ for $y\to 1$ or $F_i \to - U_i $ for $y\to -1$.
It appears that here $\CM$ and $\tCM$ are identical.
We will match in detail the expansion of the $-U_i$ in $\CY_\gamma$ and the $U_i$ in $\tilde \CY_\gamma$
with the formal line operators in Section \ref{subsec:ADN3-Laminations}.

\subsection{$SU(2)$ and $SO(3)$ $N_f=0$ gauge theories}\label{subsec:SU2-Functions}

Let us now turn to the $SU(2)$ theory discussed in Section \ref{subsec:formal-SU2}. We
begin by describing the fibration structure over the $u$-plane.

The $SU(2)$ and $SO(3)$ $N_f=0$ gauge theories in flat space differ only
by the choice of allowed sets of local line operators. We saw that there are really three possibilities,
and the third possibility can be identified with $SO(3)$ $N_f=0$ upon a shift by $\pi$ of the UV $\theta$ angle.
The 3d Coulomb branches of the three theories, $\CM_1$, $\CM_2$ and $\CM_3$, differ only in
the precise periodicities of the Wilson line parameters.
Let us start by describing the moduli space $\CM_1$ for the $SU(2)$ case.
The 4d Coulomb branch $\CB$ of this theory is remarkably similar to the one for the $N=3$ AD theory,
but the charges of the particles which become massless
at the singularities satisfy $\langle \gamma_1,\gamma_2\rangle=2$.
Conventionally, we can set $\gamma_1 = (1,0)$ and
$\gamma_2=(-1,2)$. All BPS particles have even electric charges, and instanton corrections, and hence $\CM_1$,
are invariant under the $\IZ_2$ shift symmetry $\ttheta_e \to \ttheta_e + \pi$. As the $\gamma_i$ are related by $SL(2,\IZ)$ transformations
to a pure electric charge, the instanton corrections at large radius give a standard periodic Taub-NUT geometry
around each singularity.
For an $SO(3)$ gauge theory we would say instead that  $\gamma_1 = (2,0)$ and
$\gamma_2=(-2,1)$:  magnetic charges are doubled and electric charges halved with respect to $SU(2)$.
Then $\CM_2$ has a different $\tilde \IZ_2$ shift isometry acting on $\ttheta_m$. As $\gamma_1$ is divisible by
$2$, the local geometry near the first singularity of $\CB$ at large radius is expected to have two $A_1$
singularities, permuted by $\tilde \IZ_2$.
For $\CM_3$ the roles of the two singularities are permuted.
In order to relate the $\CM_i$, we can do two things. One possibility is to quotient them by their $\IZ_2$ shift symmetries,
to reduce their torus fibers to a common form, built from the lattice $\Gamma$ instead of the $\Gamma_{\CL_i}$.
The resulting manifold $\CM$ has two $A_1$ singularities, corresponding to the two singularities in the 4d Coulomb branch.
The second possibility is to ``double up'' their torus fibers, to a larger torus modeled on the union of
the $\Gamma_{\CL_i}$ in $\Gamma \otimes \IR$. The resulting manifold $\CM_{mx}$ is smooth and has a $\IZ_2 \times \tilde \IZ_2$ shift isometry. The diagram \eqref{eq:covering-maps} in this case becomes
\begin{equation}\label{eq:A1-covering-maps}
\begin{matrix}
    &  &  \CM_{mx} &  &   \\
    & \swarrow & \downarrow  &  \searrow & \\
    \CM_{1} &  & \CM_{2} &  & \CM_{3} \\
    & \searrow & \downarrow  &  \swarrow & \\
    &  &  \CM &  &   \\
\end{matrix}.
\end{equation}
Then quotient of $\CM_{mx}$ by $\IZ_2$ produces $\CM_1$, by $\tilde \IZ_2$ produces $\CM_2$ and by the diagonal $\IZ_2$ produces $\CM_3$.

How do these manifolds compare with the expected moduli space of solutions of a Hitchin system?
We will see now that $\CM_1$ is the standard moduli space of a
$SU(2)$ Hitchin system, but $\CM_{mx}$ is a natural extension of that.
This example suggests that the story for the case with irregular singularities might be a bit
more complicated than for the case with regular singularities:  even when $C$ has genus zero, there may be choices
of gauge group and of $\CM_\CL$.

The Hitchin system has two irregular singularities
on the sphere, with a single Stokes sector each. The condition $M_1 M_2=1$ becomes
\begin{equation} \label{eq:nf0-matcond}
W \begin{pmatrix} 1 & U_1 \\  0  & 1 \\ \end{pmatrix}\begin{pmatrix} 0 & 1 \\  - 1  & 0 \\ \end{pmatrix} W^{-1} \begin{pmatrix} 1 & U_2 \\  0  & 1 \\ \end{pmatrix}\begin{pmatrix} 0 & 1 \\  - 1  & 0 \\ \end{pmatrix}=1.
\end{equation}
Here $W$ is a matrix of unit determinant. There is a small residual ``gauge freedom''
$W \to -W$. Actual holomorphic functions on the moduli space of this Hitchin system must be invariant under this $\IZ_2$ transformation,
but for the moment we will ignore that.
The matrix elements of $W$ and $U_{1,2}$ can be considered as holomorphic functions on a slightly enlarged moduli space, which will coincide with $\CM_{mx}$.

Massaging \eqref{eq:nf0-matcond} a bit we see it implies $U_1 = U_2 =: U$ and $\Tr W = 0$, i.e.
we can write
\begin{equation}
W = \begin{pmatrix} V_1 & V_0 \\  -V_2  & -V_1 \\ \end{pmatrix}.
\end{equation}
In addition \eqref{eq:nf0-matcond} implies $V_0 + U V_1 + V_2 = 0$; and since $\det W = 1$ we have
$V_0 V_2 = 1 + V_1^2$.
The two-dimensional space $\CM_{mx}$ is coordinatized by $U, V_0, V_1, V_2$ subject to these two equations.

We may extend $V_0$, $V_1$, $V_2$ to an infinite set of useful functions $V_n$ by
\begin{equation}
W \left(\begin{pmatrix} 1 & U \\  0  & 1 \\ \end{pmatrix}\begin{pmatrix} 0 & 1 \\  - 1  & 0 \\ \end{pmatrix} \right)^n=\begin{pmatrix} V_{n+1} & V_n \\  -V_{n+2}  & -V_{n+1} \\ \end{pmatrix}.
\end{equation}
Consistency of these definitions requires that $V_{n-1} + U V_{n} + V_{n+1} = 0$ for all $n$. Moreover,
since $\det W=1$ we have the relations $V_{n-1} V_{n+1} = 1 + V_n^2$ for all $n$.

The trace of the monodromy has the simple form $\Tr M_1 = -U$, and should correspond to the expectation value of a fundamental $SU(2)$ Wilson loop.
We have two natural symmetries: a $\IZ_2$ acting by $V_n \to -V_n$ and a $\tilde \IZ_2$ acting by $U \to -U$, $V_n \to (-1)^n V_n$.
The former is the one which we have to quotient by in order to obtain the moduli space of solutions of the $SU(2)$ Hitchin system; it acts freely, so
the moduli space is smooth.
On the other hand, either $\tilde \IZ_2$ or the diagonal in $\IZ_2 \times \tilde \IZ_2$ does have fixed points:
either $U=0$ and $V_{2n}=0$, which means $V_{2n+1} = \pm (-1)^n i$,
or $U=0$ and $V_{2n+1}=0$, which means $V_{2n} = \pm (-1)^n i$.
Thus the quotient by either $\tilde \IZ_2$ or the diagonal in $\IZ_2 \times \tilde \IZ_2$ gives a space with two $A_1$ singularities exchanged
by the residual $\IZ_2$ symmetry.  The quotient by both symmetries gives a space with two unrelated $A_1$ singularities.
This is exactly what is expected from the gauge theory analysis: the moduli space without any quotients appears to coincide with
$\CM_{mx}$, the $\IZ_2$ quotient with $\CM_1$, and the $\tilde \IZ_2$ quotient and the diagonal quotient
appear to coincide with $\CM_2$ and $\CM_3$ respectively.

It is now of some interest to compare the rings of functions on $\CM_{mx}$ and $\CM_1$ with the results
on formal line operators from Section \ref{subsec:formal-SU2}. To get a commutative ring we
can specialize $y=\pm 1$. If we set $y=1$ then we can compare with the functions $V_n$ on
$\CM$. In this case we have for $y\to 1$ $\hat V_n \to V_n$  and $W \to -U$. Compare the above
relations to equations \eqref{eq:SU2-formring-2} and \eqref{eq:SU2-formring-3}.
When we take the quotient to go to $\CM_1$ then we must
consider even functions generated by the $V_n$. Now we have $W\to -U$, $G_{2n+1} \to V_n^2$,
and $G_{2n} \to V_{n-1} V_n$.   When discussing functions on $\CM_1$
 it is also possible to set $y=-1$ in the algebra generated by
the $G$'s. In this case $G_{2n+1} \to - V_n^2$, $G_{2n+2} \to - V_n V_{n+1}$ and $W \to U$.
Indeed with this identification we can check that the equations \eqref{eq:rec-rels-1},
\eqref{eq:rec-rels-2},
\eqref{eq:SU2-formring-4} and \eqref{eq:SU2-formring-5}, when specialized to $y=-1$ are
satisfied by the corresponding classical functions.

\subsection{$SU(2)$ and $SO(3)$ ${\cal N}=2^*$ theory}\label{subsec:N2-star-moduli}

Finally, let us consider the $\CN=2^*$ theory discussed in Section \ref{subsec:Formal-N2-star}.
As noted above $C$ is a once-punctured torus. The fundamental group of $C$ has generators
$\alpha,\beta, m$ with one relation $\alpha\beta\alpha^{-1}\beta^{-1}=m$.
Hence the moduli space $\CM_{mx}$ of flat $SL(2,\IC)$ connections
with fixed conjugacy class of monodromy around the puncture is given by
the space of $SL(2,\IC)$ matrices $A,B,M$ satisfying
\begin{equation}\label{eq:monod-equa-nstar}
ABA^{-1} B^{-1} = M,
\end{equation}
with a fixed trace $\Tr M = \mu + 1/\mu$, modulo simultaneous overall conjugation.

From the gauge theory point of view it is clear that the algebra of holomorphic functions on
$\CM_{mx}$ is generated by traces of holonomies which are in turn traces of words made out
of $A^{\pm 1}$ and $B^{\pm 1}$. We claim that the algebra of holomorphic functions is in
fact generated by
\begin{equation}
a:= {\Tr} A, \qquad  b:= {\Tr}B, \qquad c:= {\Tr}AB,
\end{equation}
and that $\CM_{mx}$ is simply the space of $(a,b,c)\in \IC^3$ subject to
\begin{equation}\label{eq:moduli-equation}
a^2 + b^2 + c^2 - abc = \mu + 2 + \frac{1}{\mu},
\end{equation}
a result that goes back to  Fricke. For useful information on this
moduli space see \cite{MR2026539,MR1407481}.
The solution space to \eqref{eq:moduli-equation} is smooth so long as $\mu\not=-1$.

To justify these claims note first that since the matrices are $2\times 2$ it suffices to consider
the traces of holonomies in the fundamental representation. Next, recall that if $x$ is a
$2\times 2$ matrix of determinant $1$ then
\begin{equation}\label{eq:CH-thm}
x^2 - {\Tr}(x) x +1 =0.
\end{equation}
Using this for $x=A^{\pm 1}, B^{\pm 1}$ we can clearly reduce an expression
${\Tr}(A^{n_1} B^{m_1} A^{n_2} B^{m_2} \cdots ) $ with $n_i, m_i \in \IZ$ to a
polynomial in traces with $n_i,m_i \in \{0,\pm 1\}$. Then using
\eqref{eq:CH-thm} in the form $x^{-1} = {\Tr}(x) - x$ we can replace $A^{-1}$ by $A$
and $B^{-1}$ by $B$ (changing the polynomial) and finally, applying
\eqref{eq:CH-thm} to $x=AB$ we can reduce ${\Tr}[(AB)^n]$ to a polynomial in
${\Tr}(AB)$. Moreover, using these relations we can write
\begin{equation}
\begin{split}
{\Tr}M & = {\Tr}(ABA^{-1} B^{-1} ) \\
& = ({\Tr} A)^2 - {\Tr}(ABAB^{-1}) \\
& = ({\Tr} A)^2 - {\Tr}(ABA) {\Tr}(B) + {\Tr}(ABAB)  \\
& = ({\Tr} A)^2 + ({\Tr} B)^2 + ({\Tr}AB)^2 - ({\Tr}A )({\Tr} AB) ({\Tr} B) - 2 \\
\end{split}
\end{equation}
from which we get \eqref{eq:moduli-equation}.

In order to reproduce the expected covering moduli spaces
\begin{equation}\label{eq:A1-covering-maps-nstar}
\begin{matrix}
    &  &  \CM_{mx} &  &   \\
    & \swarrow & \downarrow  &  \searrow & \\
    \CM_{1} &  & \CM_{2} &  & \CM_{3} \\
    & \searrow & \downarrow  &  \swarrow & \\
    &  &  \CM &  &   \\
\end{matrix}
\end{equation}
we note that $\CM_{mx}$ defined by \eqref{eq:monod-equa-nstar} has a $\IZ_2 \times \IZ_2$
symmetry group generated by $(A,B) \to (-A, B)$ and $(A,B) \to (A,-B)$ so that the
three nontrivial elements act on $(a,b,c)$ by
\begin{equation}
\begin{split}
g_1: (a,b,c) & \rightarrow (-a,b,-c) \\
g_2: (a,b,c) & \rightarrow (a,-b,-c) \\
g_3: (a,b,c) & \rightarrow (-a,-b,c)\\
\end{split}
\end{equation}
Clearly $\CM_i = \CM_{mx}/\langle g_i \rangle $ and $\CM= \CM_{mx}/(\IZ_2 \times \IZ_2)$.
Note that $\CM$ is the moduli space of flat $PSL(2,\IC)$ connections with
fixed conjugacy class of monodromy around the puncture.

There are correspondingly six fixed points of elements of $\IZ_2 \times \IZ_2$ on $\CM_{mx}$,
e.g. ${\rm Fix}(g_3)$ is $(0,0,\pm (\mu^{1/2}+ \mu^{-1/2}))$.
If we take a quotient by one of the $\IZ_2$ subgroups then there are
two $A_1$ singularities, exchanged by the remaining $\IZ_2$ symmetry group.
For example $\CM_3$ has two $A_1$ singularities
at $[0,0,\pm (\mu^{1/2}+ \mu^{-1/2})]$ exchanged by $g_1$ (or $g_2$).
If we take a quotient by $\IZ_2 \times \IZ_2$ to get $\CM$ there are $3$ distinct
$A_1$ singularities. In the fibration $\CM \to \CB$ these project to the
three singular points in the $u$-plane mentioned in Section \ref{subsec:Formal-N2-star}.

The formal Wilson operators of Section \ref{subsec:Formal-N2-star} correspond neatly
to the functions $a,b,c$. To be precise, the $y\to 1$ limit takes
\begin{equation}
\begin{split}
W_1 & \to a, \\
W_2 & \to b, \\
W_3 & \to c.
\end{split}
\end{equation}
With this understood we recognize that the equation \eqref{eq:quant-skein-rel}
is the quantum skein relation corresponding to
\begin{equation}
{\rm Tr}(A) {\rm Tr}(B) = {\rm Tr}(AB) + {\rm Tr}(AB^{-1}).
\end{equation}
Note that this really is a skein relation of the form familiar from Chern-Simons theory,
since the $A,B$ cycles intersect on the torus and may be resolved in two ways,
with coefficients $y^{\pm 1/2}$. We will comment again on this in Section \ref{sec:Open-Problems}.
Similarly, we can also recognize the equation \eqref{eq:quant-mod-eq} as the
quantum analog of \eqref{eq:moduli-equation}. (One can check that for $y\to 1$ we have
$\sum a_{ijk} \to 1$.)

Finally, note that the $SL(2,\IZ)$ S-duality symmetry is beautifully manifest in the
moduli space since we can write the defining equation as
\begin{equation}
A B = M B A,
\end{equation}
which can be equivalently written as
\begin{equation}
A (B A) = M (B A) A
\end{equation}
\begin{equation}
B A^{-1} = (A^{-1} M A) A^{-1} B.
\end{equation}
Hence we can define actions of the generators of $SL(2,\IZ)$ on the holonomies
$T: A \to A, B\to B A, M\to M$ and $S: A \to B, B\to A^{-1}, M\to A^{-1} M A$.
One can check that $S^4$ is the identity transformation, but the relation
$(ST)^2= S^2$ is not satisfied on the holonomies. Nevertheless
\begin{equation}
STS\cdot (A,B,M) = B^{-1}\left( T^{-1} S T^{-1} \cdot (A,B,M) \right) B
\end{equation}
and hence the action on traces of holonomies does factor through the modular
group.

\section{$A_1$ theories and laminations}\label{sec:A1}

In Sections \ref{sec:A1} and \ref{sec:Examples} we specialize to the $A_1$ theories.  In these theories one can give explicit examples of
the decomposition \eqref{eq:y1} expressing vevs of line operators in terms of the $\tCY_\gamma$,
and hence compute the framed BPS degeneracies.  In this section we give some necessary preliminaries.
Section \ref{sec:Examples} will contain concrete results.

\subsection{A brief technology review}

We begin by recalling a few bits of technology used in \cite{Gaiotto:2009hg}, to which we refer for more
details.

Fix a point $u \in \CB$.
$u$ corresponds to a meromorphic quadratic differential $\phi_2$ on $C$.
$\phi_2$ determines the Seiberg-Witten curve $\Sigma$,
which is of the form $\lambda^2 = \phi_2$.  The charge lattice is $\Gamma = H_1(\Sigma, \IZ)^{odd}$.
The behavior of $\phi_2$ near a regular singularity on $C$, say at $z=0$, is dictated by a boundary condition
\begin{equation}
 \phi_2 \sim m^2 \frac{dz^{\otimes 2}}{z^2}.
\end{equation}
Here $m$ is a mass parameter corresponding to a relevant deformation of the theory.
At irregular singularities $\phi_2$ has a higher-order pole.
At generic $u$, $\phi_2$ has only simple zeroes;
$\CB^{sing} \subset \CB$ is the locus where not all zeroes of $\phi_2$ are simple.

Around any point of $C$ which is neither a pole nor a zero of $\phi_2$, we can define a local coordinate (up
to sign) by $w = \int \sqrt{\phi_2}$.
We define a ``WKB curve of phase $\vartheta$'' on $C$ to be one
which becomes a straight line of inclination $\vartheta$ when mapped to the $w$-plane.
A generic WKB curve parameterized by $t$ is asymptotic to poles of $\phi_2$ as $t \to \pm \infty$.
Near a regular singularity for $m e^{i \vartheta} \notin i\IR$, WKB curves describe exponential spirals falling
into the singularity.
Near an irregular singularity where $\phi_2$ has a pole of order\footnote{In   \cite{Gaiotto:2009hg} the order of the irregular singularity
 was expressed in terms of an integer $L$, but that notation would clash with our notation for line
 operators. } $k+2$, WKB curves are clustered into
$k$ distinct families, each one asymptotic to one of $k$ ``WKB rays.''

In \cite{Gaiotto:2009hg}
we also introduced a triangulation $T_\WKB(u, \vartheta)$ of $C$ --- or more precisely of
a surface obtained from $C$ by cutting out a small disc around each irregular singularity,
and marking boundary points which divide each boundary circle into $k$ arcs.
We abuse notation by calling this marked surface $C$ as
well when no confusion can result.
The edges of $T_\WKB(u, \vartheta)$
are generic WKB curves.  The vertices are the asymptotic ends of the WKB curves:
each regular singularity is a vertex, and each irregular singularity gives a set of $k$
vertices, one on each arc, plus an identification between these vertices and the Stokes rays.
The triangulation $T_\WKB(u, \vartheta)$ is ``decorated'' by some additional discrete data, as follows.
As we have mentioned, in its complex structure $J^\zeta$, $\CM$ is a moduli space of flat connections with
singularities.
Given such a connection, and given a WKB curve on $C$, we obtain a distinguished ``small flat section''
(defined up to rescaling) at each asymptotic infinity:  it is the section which is
exponentially decaying as we follow the WKB curve into the singularity.
At a regular singularity this small flat section is independent of
which particular WKB curve we consider; moreover it is one of the two eigensections of the monodromy
around the singularity, with eigenvalue
\begin{equation}
\mu = \exp \left[ \pm 2 \pi i \left( R \zeta^{-1} m - 2 m^{(3)} - R \zeta \bar{m} \right) \right].
\end{equation}
At an irregular singularity where $\phi_2$ has a pole of order $k+2$,   the situation
is slightly more complicated, because of Stokes phenomenon:  the WKB curves are clustered into $k$ families
as we have mentioned,
each one asymptotic to one of $k$ WKB rays, and each WKB ray carries a different
small flat section. So in either case, we get a distinguished flat section attached to each vertex of $T_\WKB$.

Using the decorated triangulation $T_\WKB(u, \vartheta = \arg \zeta)$,
we defined in \cite{Gaiotto:2009hg} a canonical collection of
coordinate functions $\CX_\gamma$ on $\CM$.  (We review this definition in Appendix \ref{app:fgcoords}.)
We also defined in \cite{Gaiotto:2009hg} a canonical quadratic refinement $\sigma$ on $\Gamma$,
so we can define another collection of coordinate functions by $\CY_\gamma = \sigma(\gamma) \CX_\gamma$.
The arguments of \cite{Gaiotto:2009hg} show that these $\CY_\gamma$ indeed coincide
with the ones we have been using in this paper, for $\gamma \in \Gamma$.
On the other hand, in our discussion of framed BPS degeneracies above, we also needed $\CY_\gamma$ where $\gamma$ belongs to
one of the extended lattices $\Gamma_\CL$.  For these charges, \cite{Gaiotto:2009hg} did not provide a definition of
$\CY_\gamma$.  Since $2 \gamma \in \Gamma$ we do have a definition of $\CY_\gamma^2$, so the difficulty is to
fix the sign of a square root.
We will not give a prescription for fixing these signs here.  Instead we will sidestep
the difficulty by working on the twisted space $\tCM_\CL$.  On this space the sign difficulties are less severe and we
can indeed define the desired functions $\tCY_\gamma$ for all $\gamma \in \Gamma_\CL$.
This is done in Appendix \ref{app:traffic}.

Finally, in Appendix A of \cite{Gaiotto:2009hg} we reviewed a recipe for expanding the holonomy functions
$\Tr \Hol_\wp \CA$ on $\CM$ in terms of the coordinates $\CY_\gamma$, modulo some sign ambiguities which we did not fix
in general.  In Appendix \ref{app:traffic} we slightly improve this recipe, by giving an expansion of the twisted
holonomy functions on $\tCM$ in terms of the $\tCY_\gamma$, with no sign ambiguities.
This is the expansion we will use to extract the framed BPS degeneracies $\fro(u, L_\zeta, \gamma; y = 1)$.

\subsection{The puzzle of the missing line operators} \label{sec:missing}

As we have just described, an $A_1$ theory in class $\CS$ is characterized by a choice of a Riemann surface $C$ carrying some
number of ``regular'' or ``irregular'' punctures.  Much of the literature on these theories so far has focused on regular punctures, but some of the most natural (as well as combinatorially simplest) examples
really require the irregular ones:  for example, asymptotically free theories like the pure $SU(2)$ theory always
require irregular punctures, as do the realizations of Argyres-Douglas CFTs described in \cite{Gaiotto:2009hg}.

How can we construct line operators in such a theory?  We might try to obtain them from
closed paths on $C$, as described above and in \cite{Drukker:2009tz}.  However, consideration
of examples quickly shows that this does not give all of the line operators we would expect
to exist.  For example, the pure $SU(2)$ theory corresponds to $C = \IC\IP^1$
with two irregular punctures \cite{Gaiotto:2009hg}; so up to isotopy there is just a single non-self-intersecting closed path
on $C$.  This one can be identified with the Wilson line operator.  But where are the Wilson-'t Hooft operators?

As we will see shortly, the missing line operators correspond to certain \ti{integral laminations},
combinations of paths which are allowed to have ends on the irregular punctures.

\subsection{Decoupling flavors}\label{subsec:Decouple-Flavor}

Let's consider what happens to the line operators when a conformal theory degenerates to an asymptotically free one.

We begin with an $A_1$ theory $T'$ defined using a curve $C'$.
Suppose that on $C'$ we have a pair of regular singularities, in a patch with local coordinate $z$.
We think of these singularities as sitting close together, say at $z = \pm \eps$, with mass parameters
$m_1, m_2$.  We then have
\begin{equation}
\begin{split}
\phi_2 & = \left( \frac{m_1^2}{(z-\eps)^2} + \frac{m_2^2}{(z+\eps)^2} + \cdots \right) \, dz^{\otimes 2}\\
& = \frac{F(z)}{(z - \eps)^2 (z + \eps)^2} \, dz^{\otimes 2}.\\
\end{split}
\end{equation}
Consider expanding $F(z)$ in a Taylor series around $z = 0$.
By adjusting the residues $m_1$, $m_2$ at the singularities $+\eps$, $-\eps$ we can fix the constant and linear terms of this series
arbitrarily.  Now suppose we take $\eps \to 0$ while holding these two terms finite.  In the limit we can write
\begin{equation} \label{eq:limit-q}
\phi_2 = \left(\frac{\Lambda}{z^2} + \frac{m}{z} + \cdots\right)^2  \, dz^{\otimes 2}
= \left( \frac{\Lambda^2}{z^4} + \frac{2 \Lambda m}{z^3} + \cdots\right)  \, dz^{\otimes 2}
\end{equation}
where $\eps^2(m_1^2 + m_2^2) \to \Lambda^2$, and $\frac{(m_1^2 - m_2^2)^2}{m_1^2 + m_2^2} \to m^2$.  In particular,
both $m_1$ and $m_2$ scale like $1 / \eps$, while either $m_1 + m_2$ or $m_1 - m_2$ is of order $1$.
We could take a further limit $m\to \infty, \Lambda \to 0$ holding $\Lambda m$ fixed.

So in this scaling limit the original theory $T'(m_1, m_2)$ ``degenerates'' into
a theory $T(m, \Lambda)$.
The new theory is defined in terms of a curve $C$ with an irregular singularity:
indeed, \eqref{eq:limit-q} is the form corresponding to an irregular singularity from which two Stokes rays emerge.

Now we are ready to discuss line operators in the limiting theory $T$, by starting with line operators in theory $T'$
and observing what they become in the limit.  For some line operators the answer is rather trivial:  any closed path $\wp'$ which
does not pass near the two colliding singular points on $C'$ simply corresponds to a closed path $\wp$ on $C$, and the holonomy
$\Hol_{\wp'} \CA(\zeta)$ limits to $\Hol_\wp \CA(\zeta)$.  It is thus very natural to extend \cite{Drukker:2009tz} by conjecturing
that $\wp$ corresponds to a line operator $L_\wp$ in theory $T'$, and write
\begin{equation}
 L_{\wp'} \to L_\wp
\end{equation}
in the limit.

A more interesting situation arises when $\wp'$ does pass between the two colliding singular points, so that it gets ``pinched'' in the limit $\eps \to 0$.  In this case we have to work a little harder.
First, a local analysis shows that near the colliding singular points, for small $\eps$ and generic $\vartheta$,
the WKB foliation
looks like the field of a magnetic dipole as depicted in Figure \ref{fig:dipole}.
\insfig{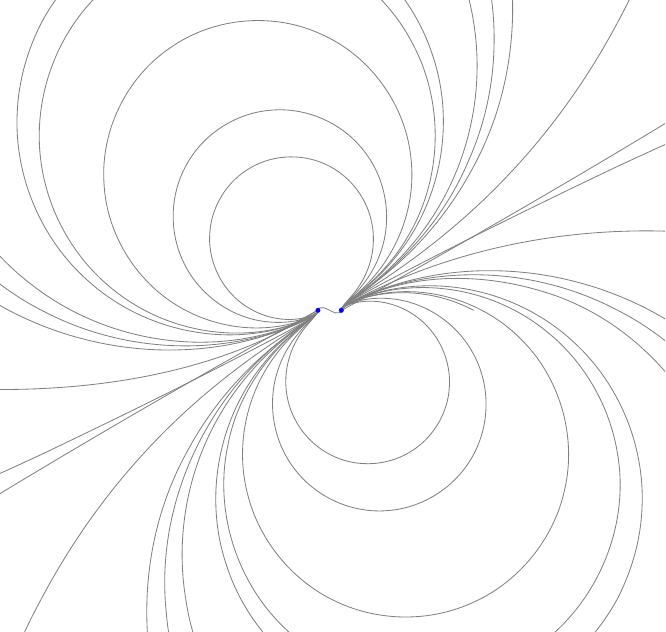}{The WKB foliation in a small neighborhood of a pair of nearby regular singular points on $C'$.}
In particular, we see two distinguished bundles of leaves emerging from the neighborhood of the
colliding singularities.  These two bundles are asymptotically approaching two distinguished rays.  Moreover,
applying the WKB approximation to a vicinity of the irregular singularity after taking $\eps \to 0$, one sees
that these two distinguished rays are exactly the WKB rays emerging from the singularity.
(Compare Figure 42 of \cite{Gaiotto:2009hg}, which depicts an irregular singularity with $4$ WKB rays emerging.)
Similarly the spaces of small flat sections along these WKB curves are smoothly related
in this limit.  It follows that we can choose the normalization of the small flat sections $s_1$, $s_2$ as functions of
$\eps$ in such a way that they limit to the two small flat sections for the two Stokes rays.

Now a convenient trick for studying the lamination $\wp'$
is to chop $\wp'$ into pieces by ``inserting a complete set of states''.  Given a path segment $q$,
and two flat sections $s_1$, $s_2$ along $q$,
a general flat section $s$ along $q$ may be rewritten as a sum of two terms:
\begin{equation}
s = \frac{s_1 \wedge s}{s_1 \wedge s_2} s_2 + \frac{s \wedge s_2}{s_1 \wedge s_2} s_1.
\end{equation}
Hence the operator of parallel transport along $q$ from an initial point $i$ to a final point $f$
is similarly a sum:
\begin{equation}  \label{eq:two-terms}
s(f) = \frac{s_1(i) \wedge s(i)}{s_1 \wedge s_2} s_2(f) + \frac{s(i) \wedge s_2(i)}{s_1 \wedge s_2} s_1(f).
\end{equation}

We are going to apply this taking $q$ to be a short segment of $\wp'$, and choosing $s_1$, $s_2$ to be the small
flat sections around our two singularities, continued to $q$ along the obvious shortest route.  See Figure \ref{fig:path-passage}.
\insfig{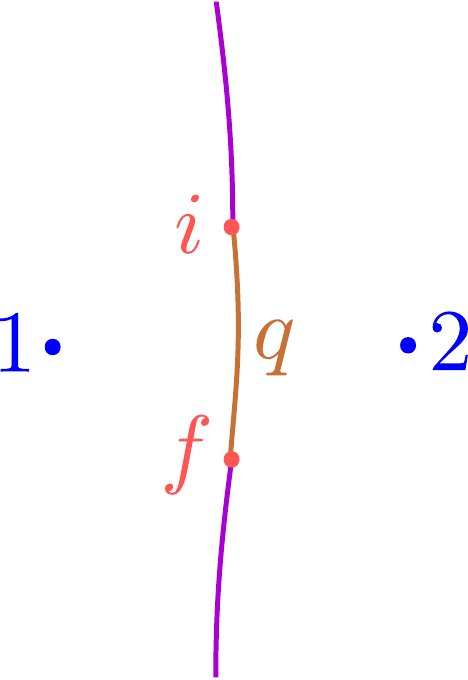}{The subpath $q$ of $\wp'$.}

Our goal is to rewrite the answer in a way that does not involve any
paths running between the singularities.  To do so, we treat the two terms in \eqref{eq:two-terms} separately.  Let us
consider the first term.  We deform $\wp'$ as illustrated in Figure \ref{fig:path-deform}.
\insfig{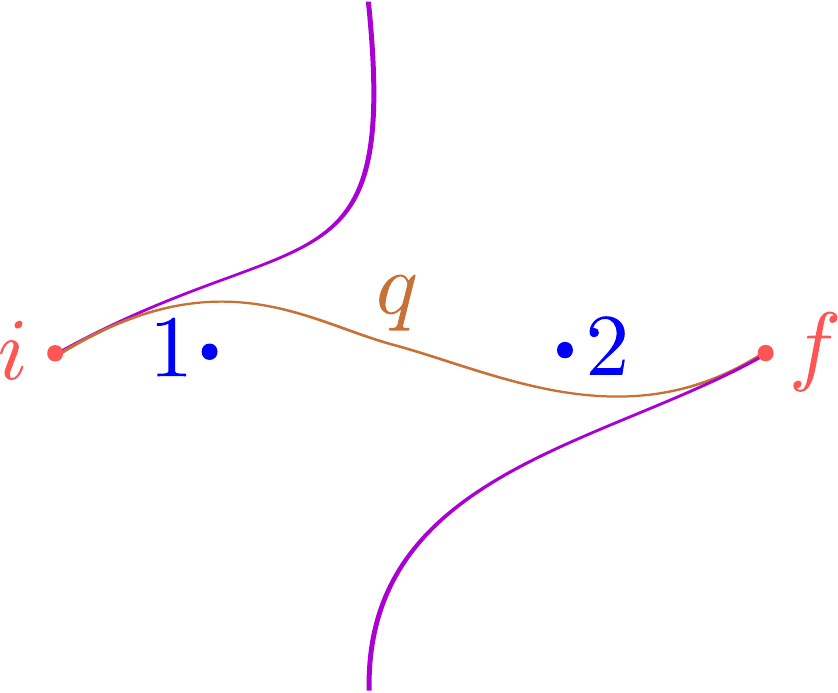}{A deformation of $\wp'$ so that the endpoints of $q$ survive the $\eps \to 0$ limit.}
After this deformation, the quantities $s_1(i)$ and $s_2(f)$ which appear still make sense in the limit where the singularities collide.
On the other hand, when we try to evaluate $s_1 \wedge s_2$ we encounter a difficulty:  the fact that $s_1(i)$ and
$s_2(f)$ are finite does not help us here, because we have to evaluate $s_1$ and $s_2$
at the \ti{same} point.  This involves transporting one or the other along $q$, which runs between the singularities.
To get around this difficulty, we can re-express $q$ as the composition of a small loop around
$z_1$ and a new path $q'$ which does
not get pinched, as shown in Figure \ref{fig:path-deform-2}.
\insfig{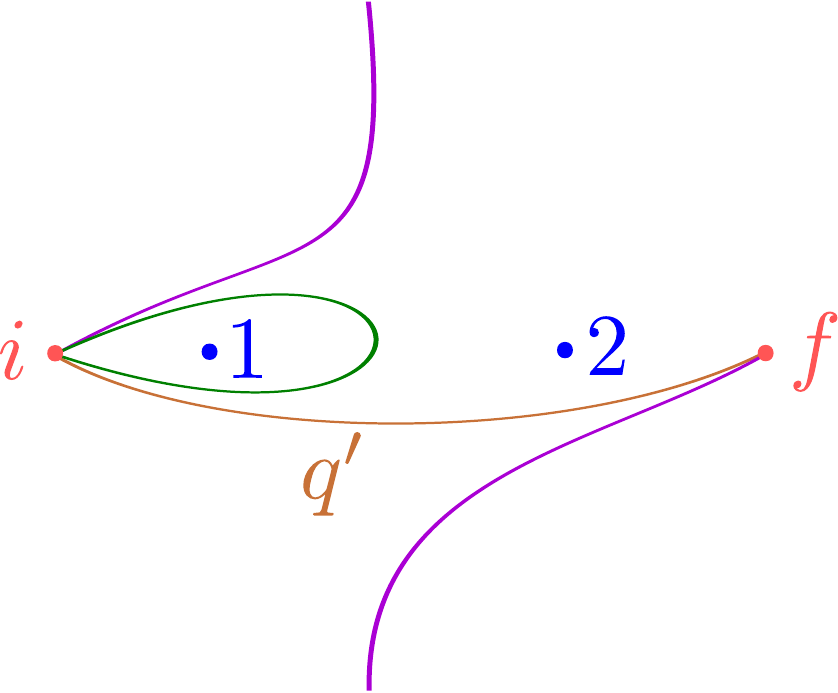}{A further deformation of $\wp'$.}
Evaluating $s_1 \wedge s_2$ along $q$
thus gives a factor $\mu_1^{-1}$ (from the monodromy of $s_1$ around $z_1$ --- recall
we define $\mu_i$ to be the counterclockwise monodromy of $s_i$ around $z_i$),
times a finite piece.  This finite piece is again $s_1 \wedge s_2$, now with the understanding that
they are transported to a common point along $q'$ rather than $q$.
A very similar analysis applies to the second term in \eqref{eq:two-terms}.
So the parallel transport becomes
\begin{equation}
s(f) = \mu_1^{-1} \frac{s_1(i) \wedge s(i)}{s_1 \wedge s_2} s_2(f) + \mu_2 \frac{s(i) \wedge s_2(i)}{s_1 \wedge s_2} s_1(f),
\end{equation}
where all factors are finite and nonzero as $\eps \to 0$ except for $\mu_1$ and $\mu_2$.

Now let us consider the \ti{trace} of the parallel transport around $\wp'$ in the fundamental representation of $SL(2,\IC)$.
We have written the transport along $q$ as a sum of two rank-1 operators.
Correspondingly the trace decomposes as a formal sum of two distinct objects,
\begin{equation} \label{eq:lp-decomp}
 L_{\wp'} = \mu_1^{-1} L_1 + \mu_2 L_2.
\end{equation}
We represent each of $L_{1,2}$ as a union
of oriented path \ti{segments}, with endpoints on the circle $S^1(\CP)$ around the irregular singularity $\CP$, and carrying weights $\pm 1$.
See Figure \ref{fig:limit-laminations}.
\insfig{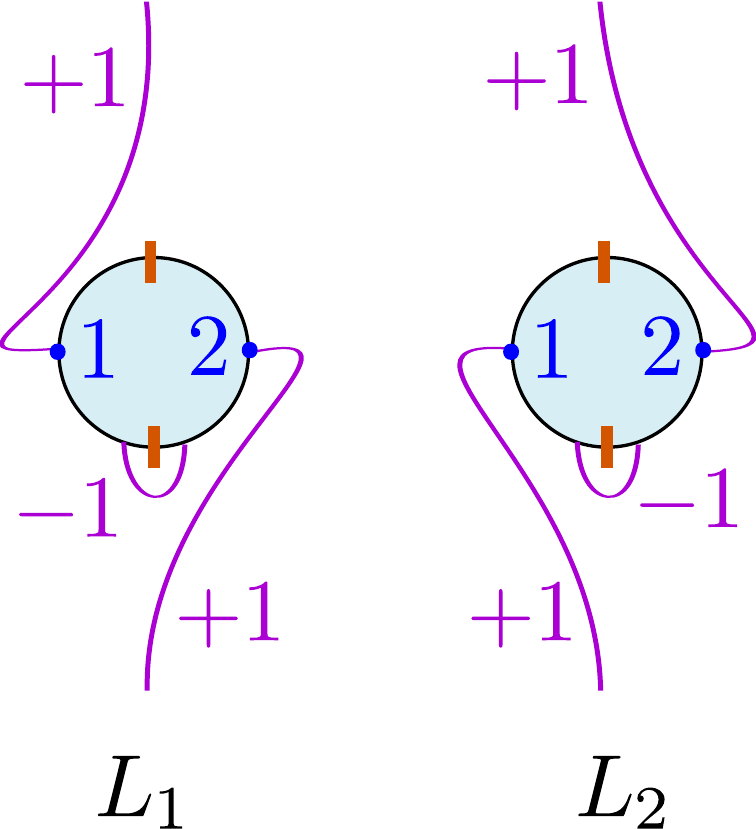}{The limiting objects $L_{1,2}$.}
Each oriented path segment has both ends on Stokes sectors emerging from the irregular singularity, so there are privileged
small flat sections $s_i$, $s_f$ at the two ends $i$, $f$.
The vev of either of $L_{1,2}$ is a product over the individual segments:  a segment running from $i$ to $f$ and carrying weight $k$
contributes $(s_i \wedge s_f)^k$.

In the strict $\eps \to 0$ limit, for generic values of the phases of $m_{1,2}$ and $\zeta$,
both $\mu_1$ and $\mu_2$ go to $0$ or $\infty$.  One of the two terms in \eqref{eq:lp-decomp} will dominate the other.
So we can say that in the $\eps \to 0$ limit $L_{p'}$ approaches one of the two objects $L_1$ or
$L_2$, multiplied by an infinite ``renormalization'' factor.  We regard $L_1$ and $L_2$ as good line operators
for the theory $T'$.

So the effect of the $\eps \to 0$ limit is to ``cut'' the path $\wp'$ in a specific fashion.  More generally, by colliding multiple pairs
of singular points, we could cut $\wp'$ into many different segments.

Above we considered a trace of holonomy in the fundamental representation.  We could similarly consider a higher
spin representation, say the $k$-th symmetric power of the fundamental.  The analog of \eqref{eq:lp-decomp} in that case would be
a decomposition
\begin{equation}
 L_{\wp', k} = \sum_{n=0}^k \mu_1^{-k-n} \mu_2^{n} L_1^{k-n} L_2^n.
\end{equation}
\insfig{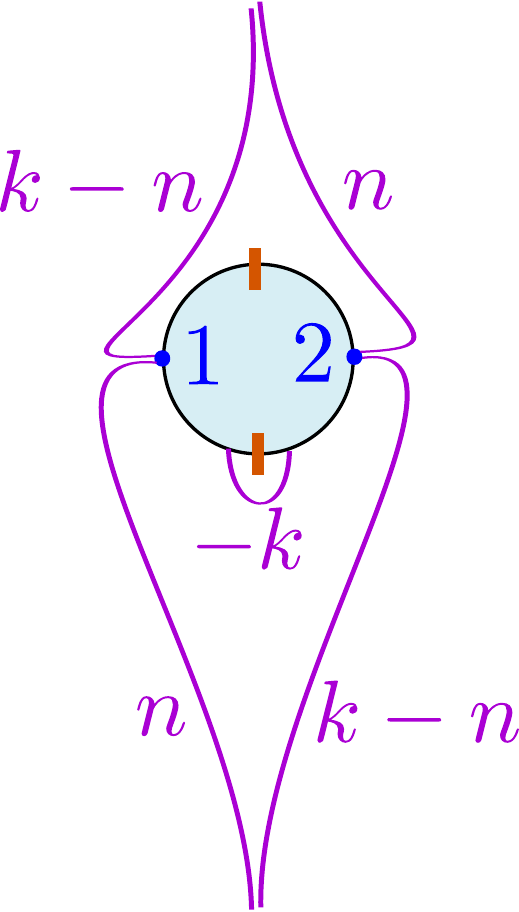}{The limiting object $L_1^{k-n} L_2^n$.}
The object $L_1^{k-n} L_2^n$ is shown in Figure \ref{fig:limit-laminations-higher}; as the notation suggests,
its vev is $\inprod{L_1^{k-n} L_2^n} = \inprod{L_1}^{k-n} \inprod{L_2}^{n}$.  In the $\eps \to 0$ limit, the sum is again
dominated by either the $n=0$ or the $n=k$ term.

Our construction above is closely related to one given in \cite{fgtoappear}.

\subsection{Laminations}\label{subsec:Lams-Def}

What we have found above is that in $A_1$ theories with irregular singularities from which
two Stokes lines emerge, the class of good line operators includes not only closed paths on $C$ carrying representations
of $SL(2,\IC)$, but also open paths which end on Stokes sectors around
the irregular singularities, carrying integer weights.  Our derivation involved a certain limiting procedure,
but it is natural to believe that the line operators attached to open paths make
sense even when the irregular singularities do not arise from this procedure:  after all, the physics defining
the surface operators in the $(2,0)$ theory is local on $C$.

What if $C$ carries more general irregular singularities,
with an arbitrary number of Stokes lines?  There is a natural proposal which
generalizes what we found above for two Stokes lines and which can be derived
by colliding multiple regular singularities. \footnote{We have not carried out
  the full details of this derivation.} It is very similar to a construction used
by Fock-Goncharov in a closely related context \cite{MR2349682}.  Following (but slightly abusing)
their terminology, we will call the objects we consider \ti{laminations} (short for ``integral
$\CA_0$-laminations'').

Recall
that when $C$ has irregular punctures, we cut out
a little disk around each irregular puncture, and
divide each boundary circle into a number of arcs.  A
lamination $L$ on $C$ is a union of curves $\wp_i$ on
$C'$, each either closed or ending on boundary arcs, non-self-intersecting and mutually
non-intersecting, and considered up to isotopy.  Each closed curve carries a nontrivial irreducible
representation of $SL(2,\IC)$, with one exception for technical convenience below:
a closed curve which surrounds a single regular puncture carries an integer weight
(positive or negative) rather than
a representation.  In addition, each open curve $\wp_i$ carries
an integer weight $k_i$, required to be positive,
with one exception:  if an open curve can be retracted to a segment of the
boundary containing exactly one marked point, then its weight is
allowed to be negative.
The sum of the weights of all curves ending on each segment of
the boundary must vanish. There are the following equivalences:  if a
curve is contractible or can be retracted to a segment of the
boundary containing no marked points, then it can be removed; if the
lamination $L$ contains two curves $\wp_i$, $\wp_j$ which are isotopic
and carry weights $k_i$ and $k_j$, it is equivalent to a new
lamination where $\wp_j$ is removed and $\wp_i$ carries weight $k_i + k_j$.

\emph{We conjecture   that for any $A_1$ theory the laminations are in 1-1 correspondence with simple line operators.}

The vacuum expectation value of the line operator attached to
such a lamination is multiplicative on the components
of the lamination.  The factor attached to a closed path $\wp$ is
the trace of the holonomy along $\wp$ as usual --- with the exception of closed paths
of weight $k$ surrounding a single puncture, to which we attach the factor $\mu^k$.
For an open path with weight $k$ we have a factor
\begin{equation} \label{eq:open-vev}
(s_{1} \wedge s_{2})^{k},
\end{equation}
where $s_1$, $s_2$ are the small flat sections at the initial and final vertices
of the path, respectively.
Although the factors associated to the open paths depend on the choice
of normalization of the small flat sections $s_i$, when we take the product
over components in the lamination, this normalization cancels out. This is why
we put the constraints on the weights $k_i$ in the above definition.

We have not been careful to fix the overall sign of the vev above.  Indeed, although there is
a definite prescription for this sign, we will not need it in what follows.  Instead, when we
want to be careful about signs, we are going to work on the space of twisted local systems.
On the latter space there is a natural way to fix the signs, described in Appendix
\ref{app:traffic}.

\section{Examples of $A_1$ theories} \label{sec:Examples}

In this Section we derive the framed BPS degeneracies in some simple
examples of $A_1$ theories.

We will also comment on results for the
protected spin characters of these theories. We will be able to deduce
the PSC's using the following simple remark, which we exalt with the
moniker \emph{Promotion Principle}.  The idea is very simple. If one
knows that the specialization of a spin character to $y=1$ is equal to $1$,
then the representation must be a single copy of the trivial (i.e. spin-zero)
representation of $SU(2)$, so its character is identically equal to $1$
for all $y$. Note that this principle clearly fails if all we know is that
the PSC is in the representation ring of $SU(2)$, and therefore we can only
invoke the promotion principle when the strong positivity conjecture holds.
In Section \ref{sec:Quantum-Holonomy} below we will exploit this principle
a little more systematically.

\subsection{$N=3$ Argyres-Douglas theory}\label{subsec:ADN3-Laminations}

We begin with the $N=3$ Argyres-Douglas theory, which we briefly reviewed in Section \ref{sec:formaln3}.
The realization of $\CM$ in this case in terms of Hitchin systems was described in Section 9.4.4 of \cite{Gaiotto:2009hg}.
In this case one has $C = \IC\IP^1$, with a single irregular singularity, from which $5$ Stokes lines emerge;
correspondingly there are $5$ marked points on the boundary $S^1$.  These marked points divide the boundary into
$5$ sectors, which we number $1$ through $5$, increasing clockwise around the circle.
\insfig{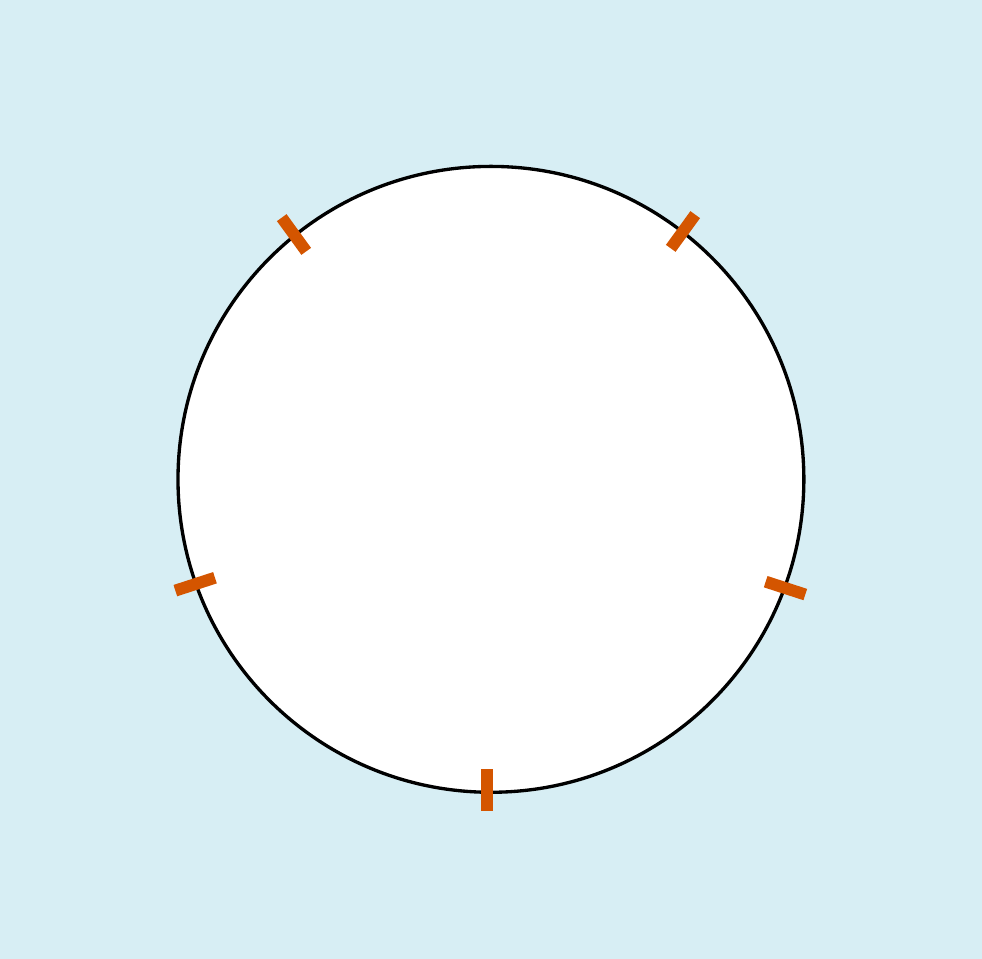}{The curve $C = \IC\IP^1$ with a disc around infinity cut out, and five marked points on the boundary.}

In this theory there is a distinguished set of $5$ laminations.
Begin with $L_1$ as depicted in Figure \ref{fig:pentagon-l1}.
\insfig{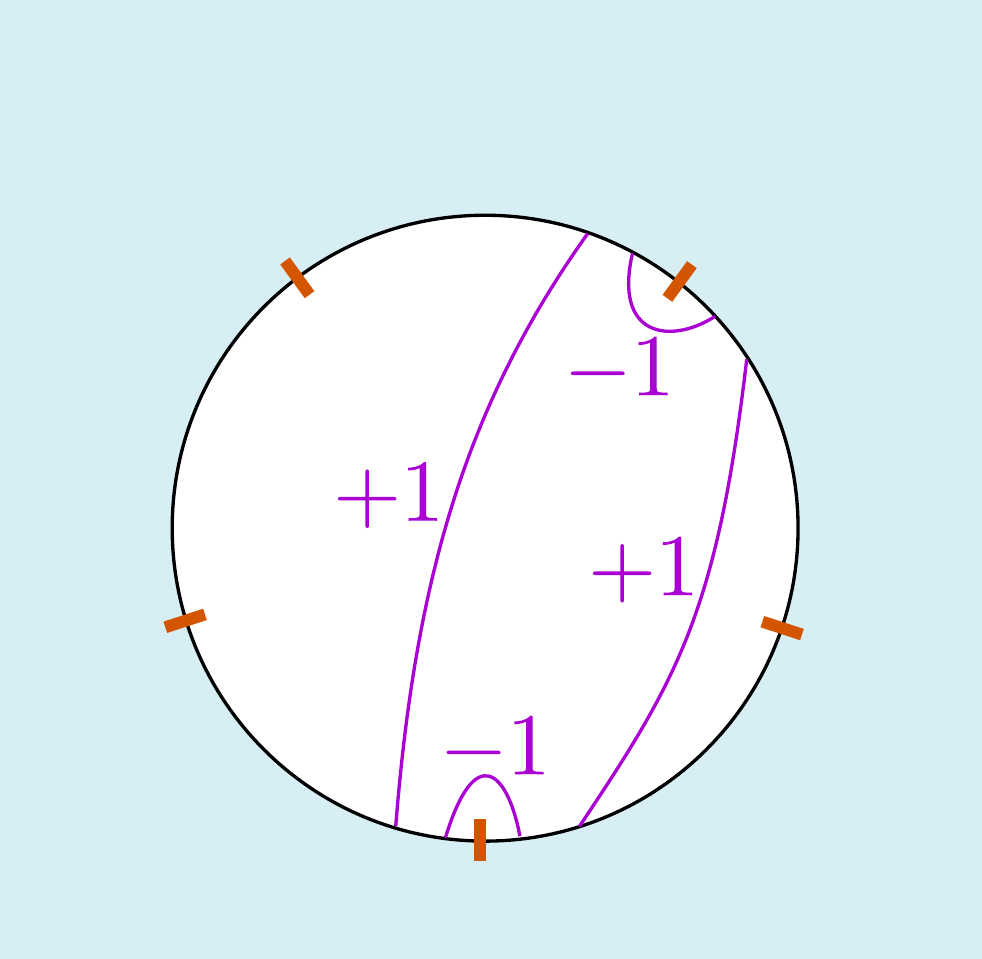}{The lamination $L_1$ in the $N=3$ theory.}
$L_{i+1}$ is obtained from $L_i$ by rotating the circle by an angle $\frac{4 \pi}{5}$ counterclockwise.
Repeating this operation $5$ times brings us back to the original lamination, so $L_{i+5} = L_i$.

Each of the laminations $L_i$ gives rise to a family of line operators $L_{i,\zeta}$ parameterized by $\zeta$.
To calculate the dimensions $\fro(\gamma; u, \zeta, y = 1)$, according to \eqref{eq:y1}, we should expand
the vevs $\inprod{L_{i,\zeta}(u)}'$ in terms of the coordinate functions $\tCY_\gamma$.  This can be done by the
``traffic rule'' algorithm we described in Appendix \ref{app:traffic}.  We begin by drawing the WKB triangulation
$T_\WKB(\zeta, u)$.  The triangulations
of a pentagon all look the same up to a cyclic permutation of the vertices, so even without knowing what $\zeta$ and $u$ are,
we know what $T_\WKB(\zeta, u)$ looks like.  We label the vertices
as shown in Figure \ref{fig:pentagon}.
\insfig{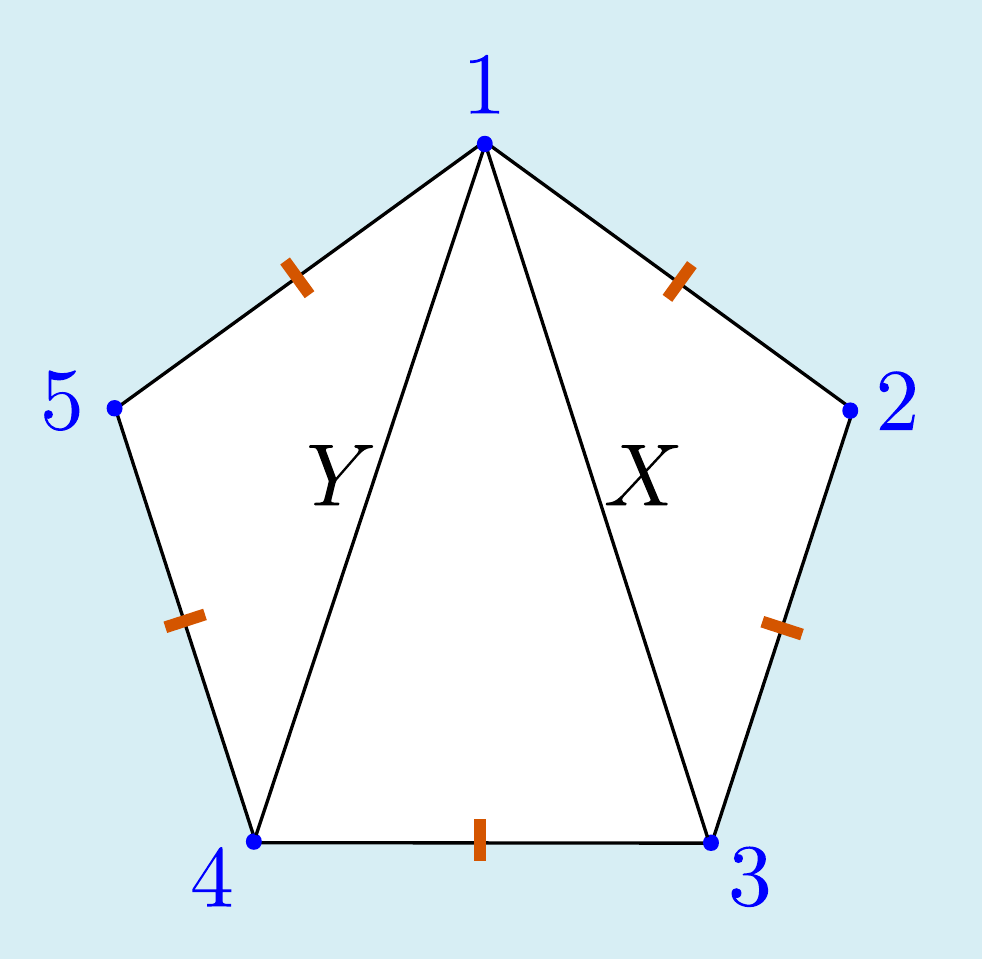}{A triangulation in the $N=3$ theory.  The five edges of the pentagon are boundary edges.
Each orange mark on the boundary is now identified
with a Stokes ray emerging from the irregular singularity.  The blue points are vertices of the triangulation; we identify them with anti-Stokes rays.}

In Appendix \ref{app:traffic} we defined for each edge $E$ a function $\tCY_E(\zeta)$ on the moduli space $\tCM$.
For convenience, let us lighten the notation by writing $X$ for the function $\tCY_X$ and $Y$ for $\tCY_Y$.
Then applying the traffic rules, we quickly obtain the expansion of the vevs in terms of $X$ and $Y$:
\begin{align}
\inprod{L_1}' &= X, \label{eq:pentagon-vevs-1} \\
\inprod{L_2}' &= Y + XY, \\
\inprod{L_3}' &= \frac{1}{X} + \frac{Y}{X} + Y, \\
\inprod{L_4}' &= \frac{1}{X} + \frac{1}{XY}, \\
\inprod{L_5}' &= \frac{1}{Y}. \label{eq:pentagon-vevs-5}
\end{align}
From these vevs we immediately read off the spectrum of framed BPS states.  For example,
the line operator $L_1(\zeta)$ supports a single framed BPS state in the vacuum $u$.
This state then necessarily has spin zero.
On the other hand, $L_3(\zeta)$ supports three framed BPS states, with three different charges.
Again, all these states have spin zero.

A general lamination in this theory is of the form
\begin{equation}\label{eq:ADN3-gen-lam}
L_i^m L_{i+1}^n, \qquad m, n \in \IZ_{\ge 0}.
\end{equation}
(The product of two nonconsecutive $L_i$ does not give a lamination, because the
edges of a lamination are not allowed to intersect one another.)
The corresponding expectation values are
\begin{equation}
\inprod{L_i}^m \inprod{L_{i+1}}^n, \qquad m, n \in \IZ_{\ge 0}.
\end{equation}
Expanding such an expectation value generally leads to a more interesting spectrum of BPS states,
with multiplicities greater than $1$ in some charge sectors.

It is interesting to compute the commutative ring of these operators.  For this purpose it is sufficient to give the ring relations
for the five generators $L_i$.  Inspection of \eqref{eq:pentagon-vevs-1}-\eqref{eq:pentagon-vevs-5}
shows that $L_{i+1} L_{i-1} = 1 + L_i$.  (This can be also understood more directly as a consequence of
``skein relations'' which relate the product of two intersecting laminations to a sum of laminations where the intersection has
been resolved.)

Note that the expressions \eqref{eq:pentagon-vevs-1}-\eqref{eq:pentagon-vevs-5} perfectly match up with those of the
formal line operators $F_i$ evaluated in chamber $c_1$ of Section \ref{sec:formaln3},
if we make the identification $Y \to X_{-\gamma_1}$ and $X \to X_{\gamma_2}$.  Moreover,
\eqref{eq:ADN3-gen-lam} corresponds nicely to the simple formal line operators
$F_{p,q}^{(i)}$ defined in Section \ref{sec:formaln3}.

\subsection{$N = 4$ Argyres-Douglas theory}

Now let us consider the $N=4$ Argyres-Douglas theory.  Its realization in terms of Hitchin systems is very similar
to the $N=3$ case we just discussed:  the only difference is that we have $6$ marked points on the boundary $S^1$ instead of $5$.
See Figure \ref{fig:hexagon-bare}.
\insfig{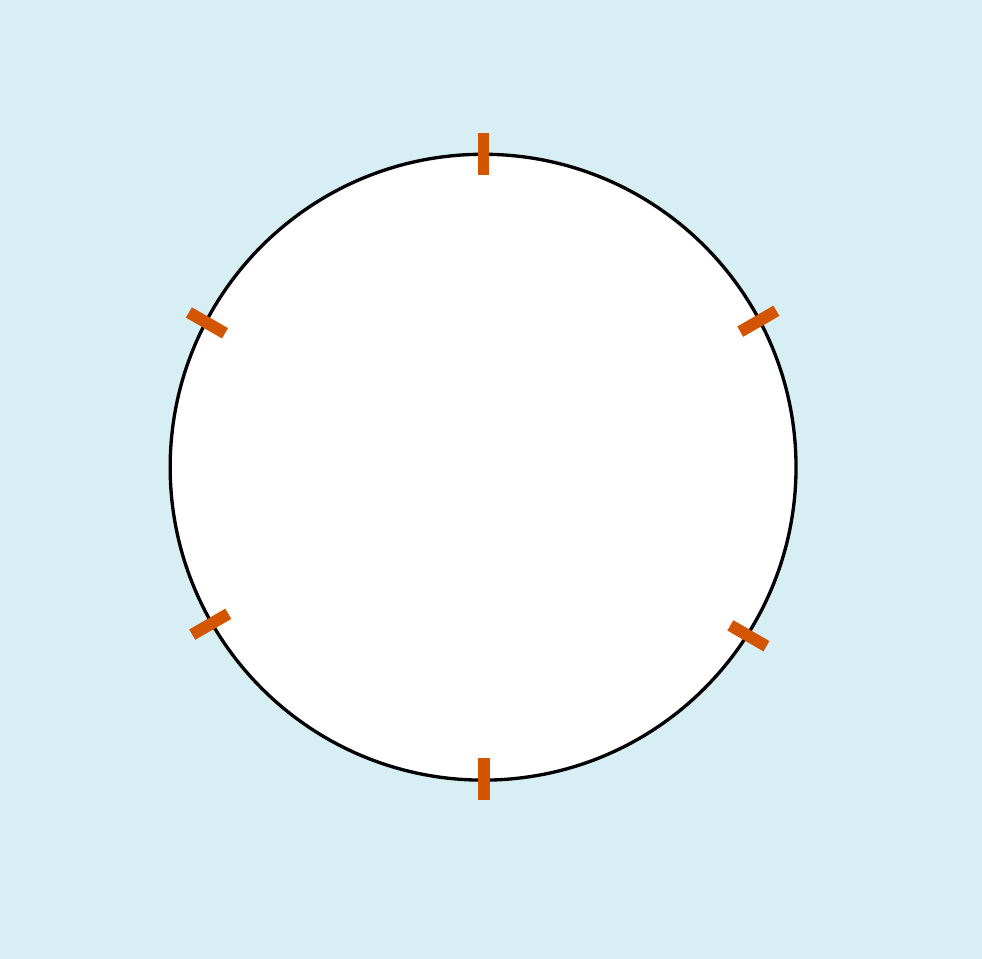}{The curve $C = \IC\IP^1$ with a disc around infinity cut out, and six marked points on the boundary.}

One important difference from the previous case arises because $6$ is even:
there are nontrivial laminations which can be pushed into an arbitrarily small neighborhood of the boundary.
An example is shown in Figure \ref{fig:hexagon-mu}.
\insfig{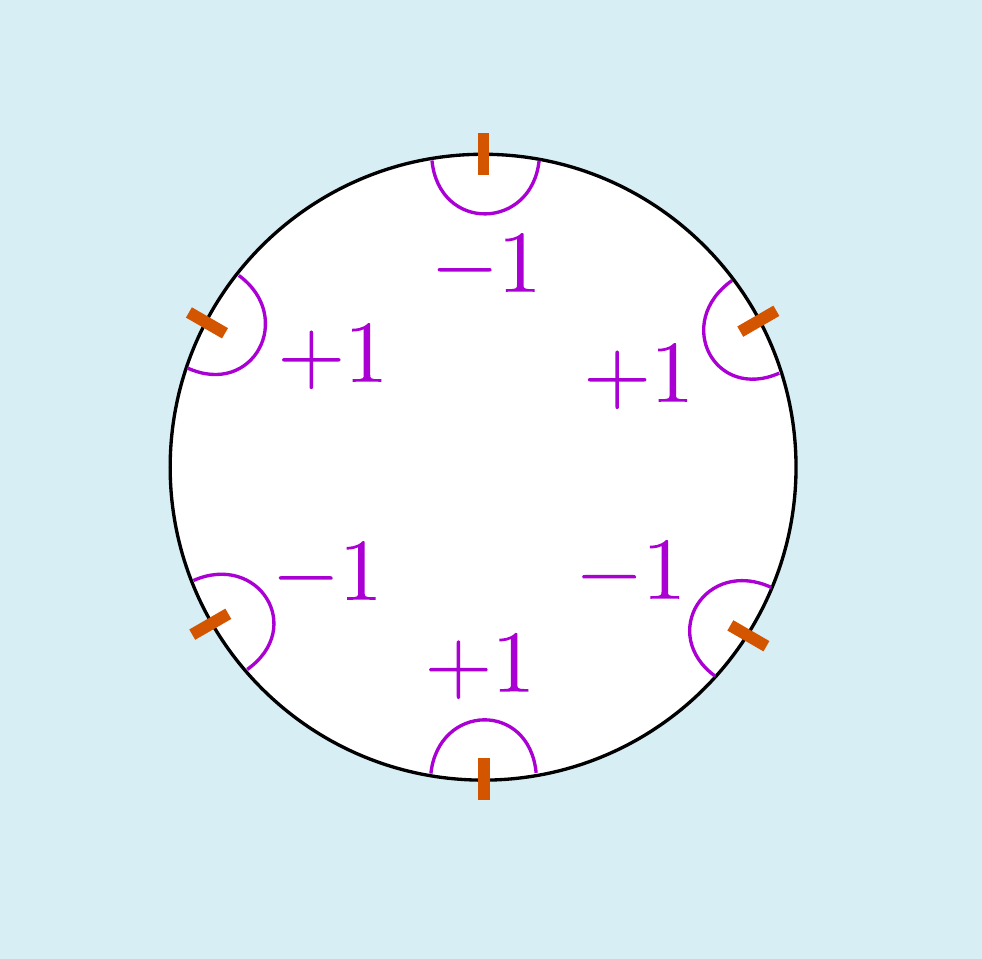}{A nontrivial lamination $\mu$ which can be pushed into an arbitrarily small neighborhood of the boundary.}
$\inprod{\mu}$ is independent of the position on $\CM$:
it can however be changed by a non-normalizable deformation of the theory.

Now define two laminations $L_1$ and $L_2$ as in Figure \ref{fig:hexagon-l1}.
\insfig{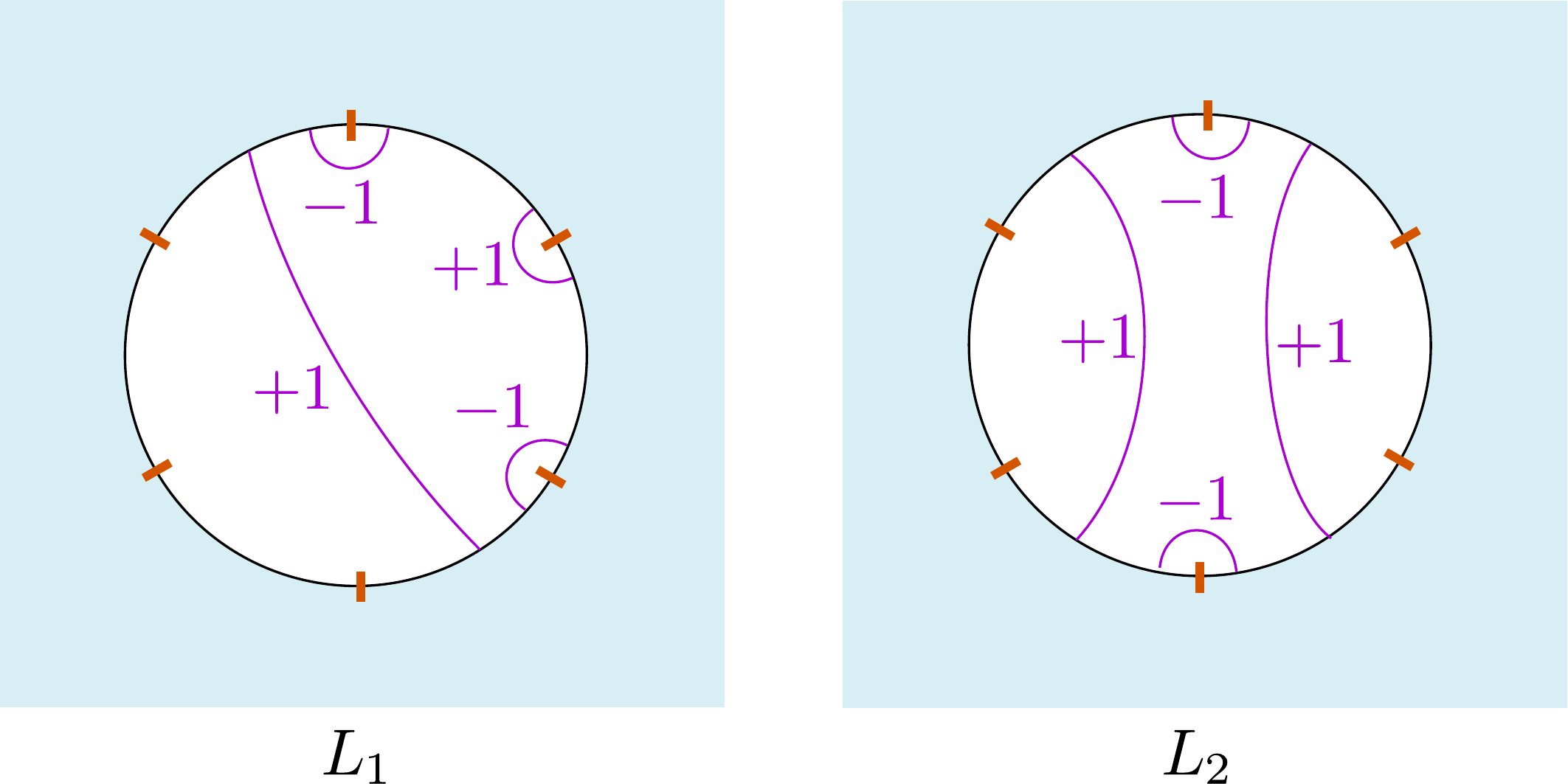}{The laminations $L_1$ and $L_2$.}
Also define $L_{i+2}$ to be the lamination obtained from $L_i$ by rotating the circle by an angle $\frac{2 \pi}{6}$ clockwise.
The sequence $L_i$ is thus periodic with period $12$.  Moreover, $L_{i+6}$ can be expressed in terms of $L_i$ using the relations
\begin{align}
L_7 &= \mu^{-1} L_1, \\
L_8 &= L_2,
\end{align}
and their images under applications of the operation $(L_i \to L_{i+2}, \mu \to \mu^{-1})$.
Any lamination is of the form
\begin{equation}
\mu^r L_i^m L_{i+1}^n, \quad r \in \IZ, \quad m, n \in \IZ_{\ge 0}
\end{equation}
for some $i$.

\insfig{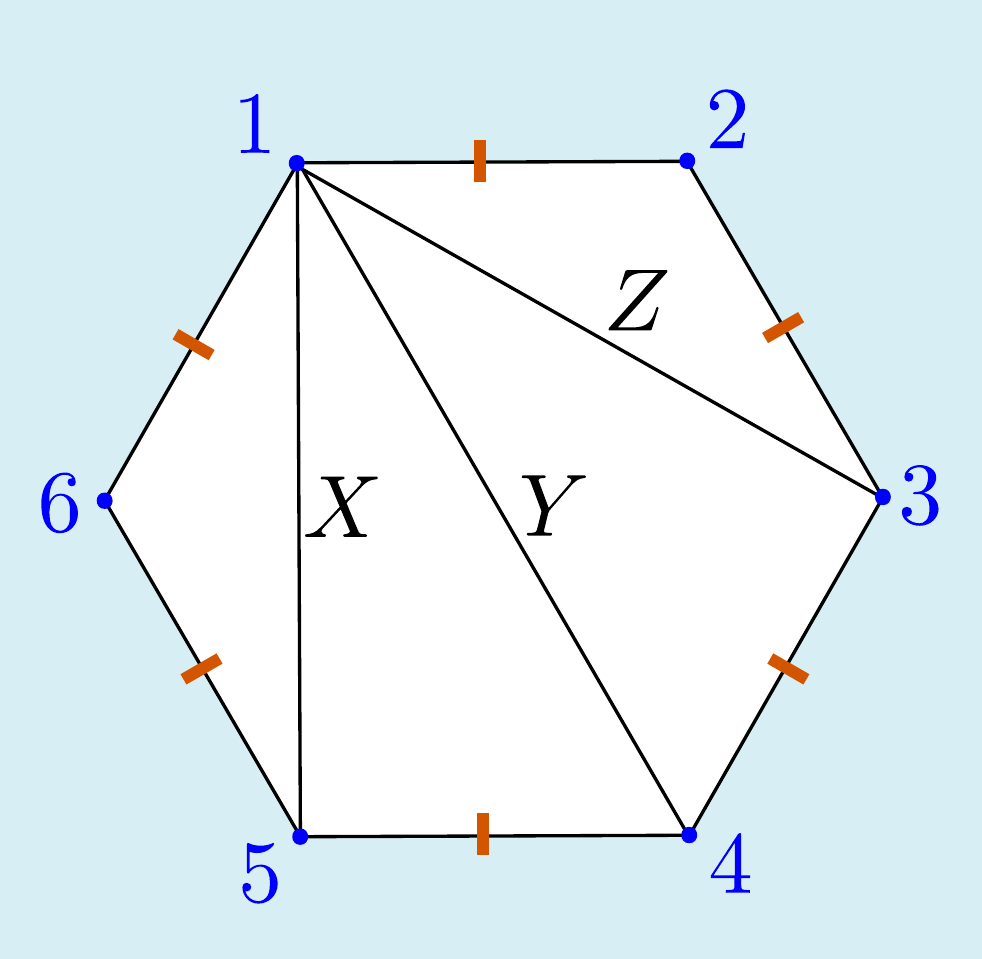}{A sample triangulation in the $N=4$ Argyres-Douglas theory.}

Now consider the triangulation shown in Figure \ref{fig:hexagon}, which could arise as $T_\WKB(u, \zeta)$ for some $(u, \zeta)$.
Expanding our laminations in terms of the coordinates attached to this triangulation, again using the traffic rules, gives
\begin{align}
\inprod{\mu}' &= XZ, \\
\inprod{L_1}' &= Z, \\
\inprod{L_2}' &= Y + YZ, \\
\inprod{L_3}' &= \frac{1}{Z} + \frac{Y}{Z} + Y, \\
\inprod{L_4}' &= \frac{(1+Y)(1 + X + XY + XYZ)}{YZ}, \\
\inprod{L_5}' &= \frac{1}{Y} + \frac{X}{Y} + X, \\
\inprod{L_6}' &= \frac{1}{XY} + \frac{1}{Y}.
\end{align}
Using these explicit formulas we can easily read out the relations among the $L_i$:  we have
\begin{align}
L_1 L_3 &= 1 + L_2, \\
L_2 L_4 &= (1+ L_3)(1 + L_9),\\
L_1 L_4 &= 1+ \mu + L_5 + L_9,
\end{align}
and their images under applications of the operation $(L_i \to L_{i+2}, \mu \to \mu^{-1})$.

\subsection{$SU(2)$, $N_f = 0$}\label{subsec:SU2-Lams}

Next we treat the pure $SU(2)$ gauge theory.  This is the $A_1$ theory where we choose $C$ to be $\IC\IP^1$
with two irregular singularities, at each of which $\phi_2$ has a pole of order $3$.  Topologically, as we have
mentioned, we should cut out a small disc around each of these two singularities, and divide each boundary
component into a single arc, i.e. mark a single point on each --- see Figure \ref{fig:nf0-bare}.

\insfig{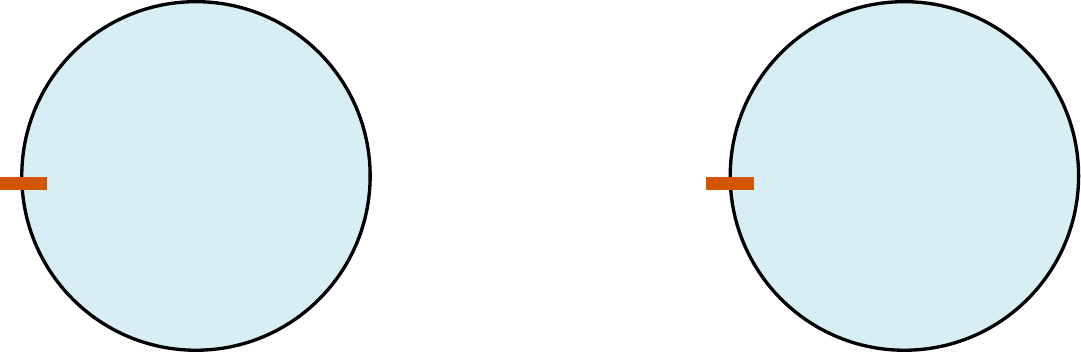}{The curve $C = \IC\IP^1$ with two discs cut out, and one marked point on each boundary component.}

Define a lamination as follows.
First take two small semicircles $\wp_1$, $\wp_2$ passing over
the marked points of the two boundary circles.  Next take two nonintersecting curves $\wp_3$, $\wp_4$
running from one boundary circle to the
other, differing from one another by 1 unit of winding.  These four curves form a ``maximal'' set, in the sense that
we cannot add any more curves which do not intersect and are not homotopic to any of them.  Assign them integer
weights $k_i$, subject to the restrictions
\begin{equation} \label{eq:nf0-cond}
2 k_1 + k_3 + k_4 = 0, \quad 2 k_2 + k_3 + k_4 = 0, \quad k_3 \ge 0, \quad k_4 \ge 0.
\end{equation}
In this way we obtain a family of laminations, parameterized by the set of solutions to \eqref{eq:nf0-cond}.
We get one solution for every $k_3$, $k_4$ with $k_3 + k_4$ even.
See Figure \ref{fig:nf0-maximal}.
\insfig{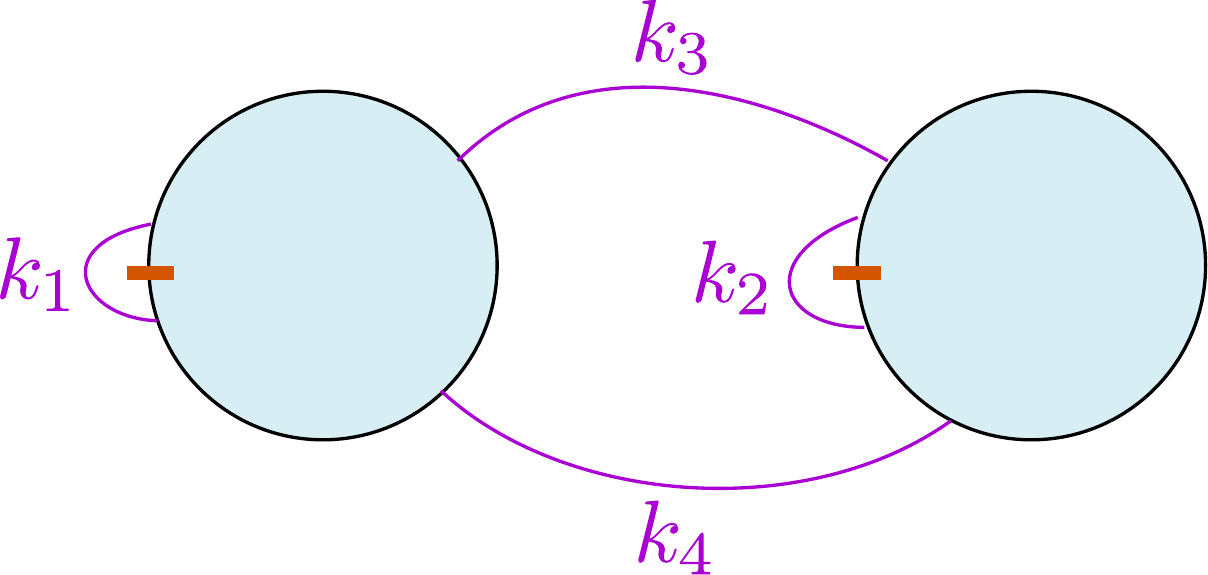}{A family of laminations.}
This is only one in an infinite set of such families, since we can choose the overall winding of
$\wp_3$, $\wp_4$ arbitrarily.

All of these laminations give supersymmetric line operators.  Which ones are they?  By applying the
correspondence given in \cite{Drukker:2009tz} to the $SU(2)$ theory with $N_f = 4$ and then sending masses to
infinity along the lines of Section \ref{subsec:Decouple-Flavor}, one finds that
these laminations correspond to Wilson-'t Hooft operators.  They are the operators with
labeling $[(\frac{p}{2} H_\alpha , \frac{q}{2} \alpha)]$ where the magnetic charge is $p = k_3 + k_4$,
and the electric charge $q$ is equal to the total winding, i.e. $k_3$ times the winding of $\wp_3$ plus $k_4$ times the
winding of $\wp_4$.\footnote{There is an ambiguity in what we mean by the ``total winding'' of a curve running from
one boundary to the other:  all that is really canonically
defined is the relative winding between two different such curves.  This ambiguity here reflects a similar
difficulty in defining the electric charge of a Wilson-'t Hooft operator in the gauge theory.}
Under the monodromy $\zeta \to e^{2 \pi i} \zeta$ both $\wp_3$ and $\wp_4$ gain $2$ units of winding,
so $q$ shifts by $2p$ units.

There are a few more laminations we have not yet described:  namely, we could take a single closed curve separating the
two singularities, carrying the spin-$\frac{q}{2}$ representation of $SL(2,\IC)$.
This lamination corresponds to a pure Wilson loop operator with electric charge $q$, i.e.
in the representation with highest weight $\frac{q}{2} \alpha$.

Now we are ready to consider the IR expansions of the vevs.
As usual, we begin by drawing the WKB triangulation $T_\WKB(u, \zeta)$.  In this case the combinatorics are so simple that
every such triangulation must look like the one in Figure \ref{fig:nf0}.  (In particular, this is the case independent of whether
$u$ is in the weak or strong coupling region!)
There are two internal edges
which we label as $X$ and $Y$, differing by one unit of winding.
\insfig{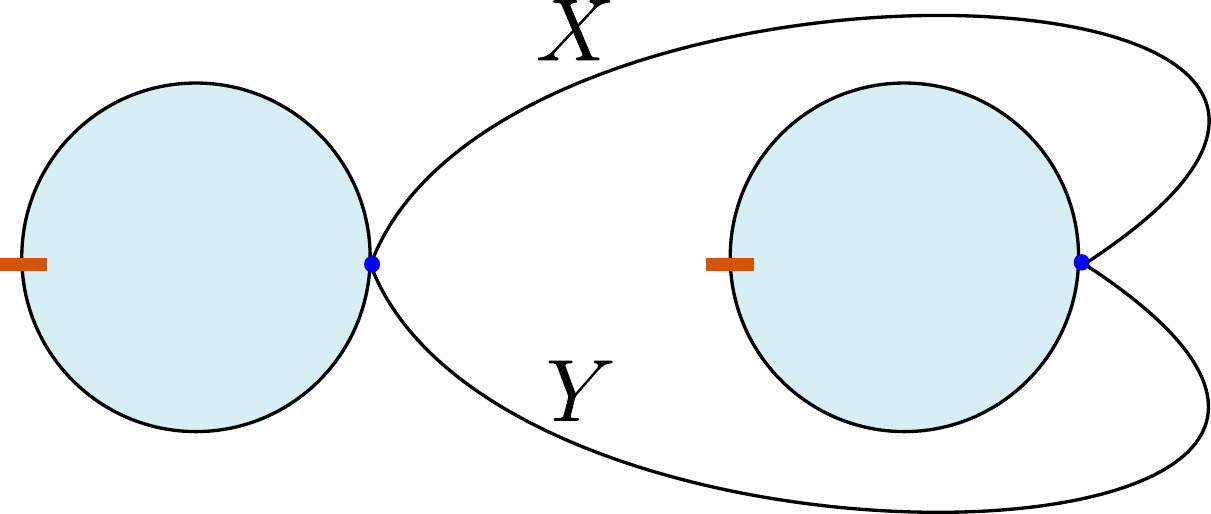}{A triangulation in the $SU(2)$ theory with $N_f = 0$.}
Let $L_0$ and $L_1$ be the laminations shown in Figures \ref{fig:nf0-l0} and \ref{fig:nf0-l1}.
More generally we can define $L_{n+2k}$ by beginning with $L_n$ and applying a Dehn twist $k$ times
around one of the two boundary $S^1$.  In this way we obtain laminations $L_n$ for all $n \in \IZ$.
See for example Figure \ref{fig:nf0-l2} for $L_2$.
We have
\begin{equation}
L_{2n+1}^2 = L_{2n} L_{2n+2}.
\end{equation}
(To check this just note that by Dehn twists we can reduce to the case $n=0$, and that case follows from
Figures \ref{fig:nf0-l0}-\ref{fig:nf0-l2}.)
\insfig{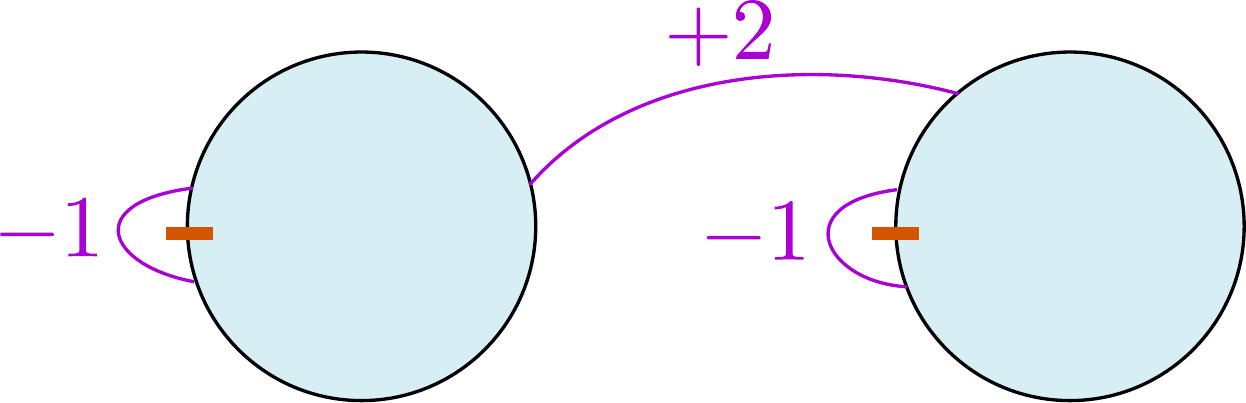}{The lamination $L_0$ in the $SU(2)$ theory with $N_f = 0$.}
\insfig{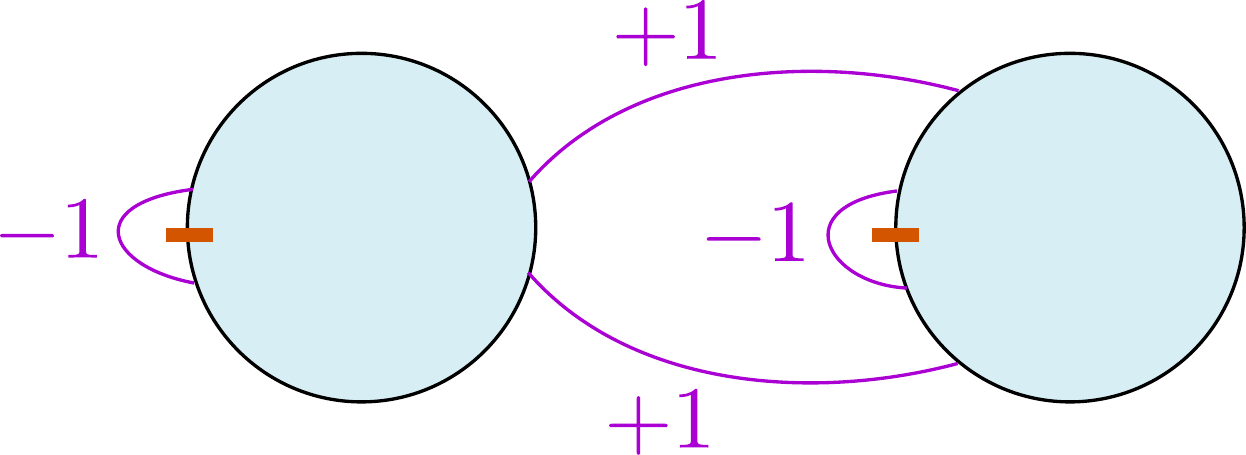}{The lamination $L_1$ in the $SU(2)$ theory with $N_f = 0$.}
\insfig{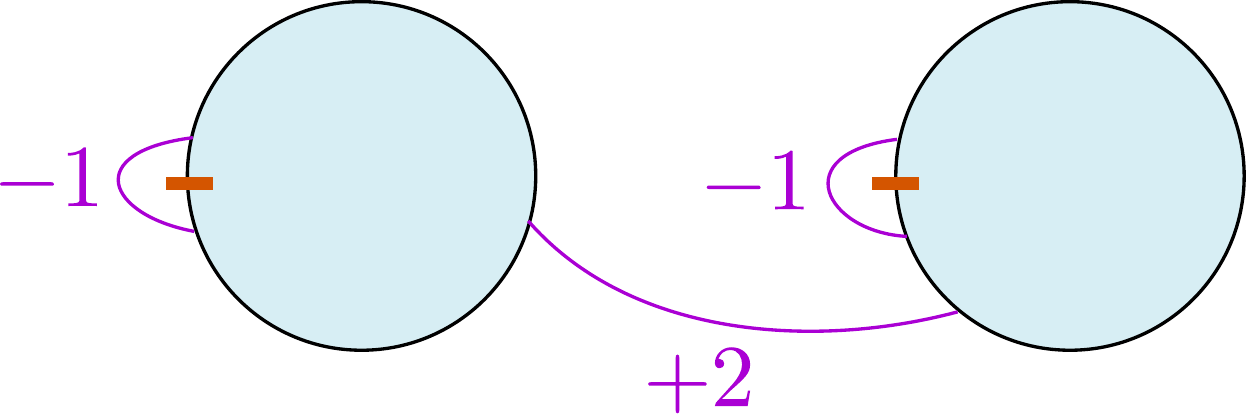}{The lamination $L_2$ in the $SU(2)$ theory with $N_f = 0$.}

The most general lamination is of the form $L_n^a L_{n+1}^b$, or a closed loop carrying some representation
of $SL(2,\IC)$.
We let $L_*$ denote the lamination consisting of a single
closed loop carrying the fundamental representation of $SL(2,\IC)$.
See Figure \ref{fig:nf0-lstar}.
\insfig{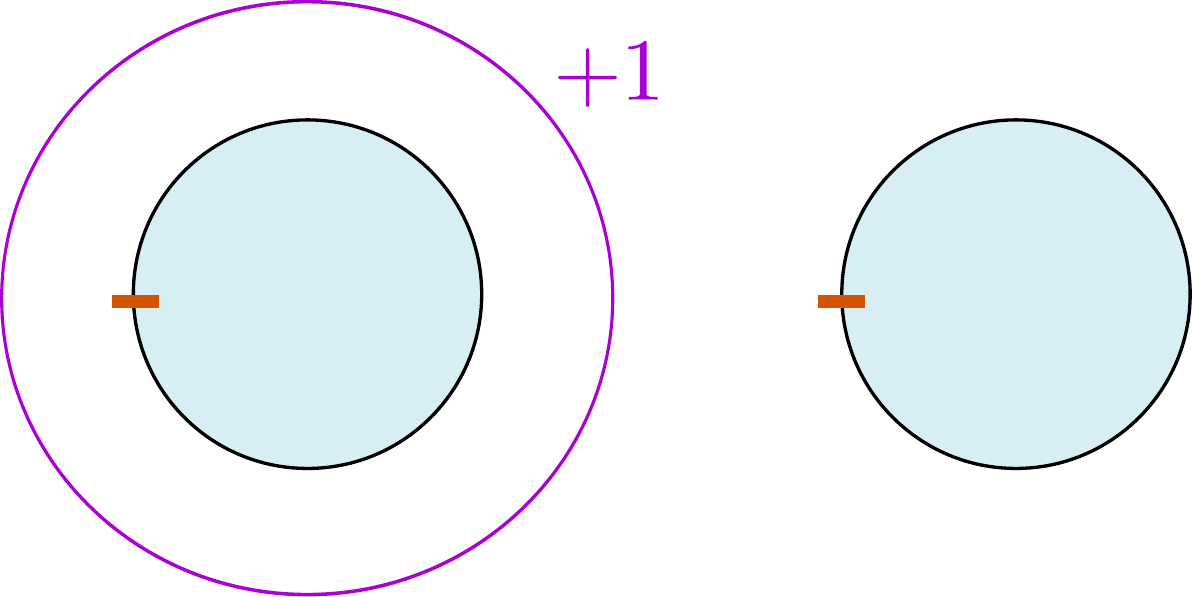}{The lamination $L_*$ in the $SU(2)$ theory with $N_f = 0$.  Note that the figure is drawn on $\IC\IP^1$, so although we draw the circle
going around the irregular singularity on the left, it is homotopic to a circle which goes around the one on the right.
We abuse notation a bit by using $+1$ to stand for the fundamental representation of $SL(2,\IC)$.}

Now we want to compute the expectation values $\inprod{L}'$ of the line operators associated to these
laminations, and to expand them in terms of the functions $\tCY_X$ and $\tCY_Y$ on $\tCM$.
As usual we abbreviate those functions as $X$ and $Y$ below.
The result for any particular lamination
can be obtained straightforwardly from the traffic rules.  In fact, however, there is also a nice uniform formula
for the answer, which can be proven by exploiting the formula \eqref{eq:open-vev} more directly.
We give that derivation in Appendix \ref{app:chebyshev}, and here just report the answer.

Introduce Tchebyshev polynomials defined by $U_n(\cos\theta) := \frac{\sin(n+1)\theta}{\sin\theta}$
and $T_n(\cos\theta) := \cos(n\theta)$.  Defining
\begin{equation}
 \alpha := \frac{1}{2 \sqrt{XY}} \left( XY + Y + 1 \right),
\end{equation}
the formula is
\begin{equation}\label{eq:chebyshev-form}
\inprod{L_{2n}}' = Y^{-1} \left[T_n(\alpha) + \frac{XY- Y - 1}{2 \sqrt{X Y}} U_{n-1}(\alpha) \right]^2.
\end{equation}
In particular, all the expectation values $\langle L_{2n} \rangle'$ can be expressed
as Laurent polynomials in $X$ and $Y$ with positive integer coefficients, as we expected.
For example,
\begin{align}\label{eq:chebyshev-expls}
\inprod{L_{-4}}' & = X^{-2} Y^{-3} (1+ 2 Y + Y^2 + X Y^2)^2 \\
&= \frac{1}{Y^3 X^2}+\frac{4}{Y^2
   X^2}+\frac{Y}{X^2}+\frac{6}{Y X^2}+\frac{2
   Y}{X}+\frac{2}{Y
   X}+Y+\frac{4}{X^2}+\frac{4}{X}, \\ \displaybreak[0]
\inprod{L_{-2}}' &= \frac{1}{X Y^2} (1 + Y)^2=\frac{1}{Y^2 X}+\frac{2}{Y X}+\frac{1}{X}, \\ \displaybreak[0]
\inprod{L_{0}}' &= \frac{1}{Y}, \\ \displaybreak[0]
\inprod{L_{2}}' &= X, \\ \displaybreak[0]
\inprod{L_{4}}' &=Y(1+X)^2 = Y X^2+2 Y X+Y, \\ \displaybreak[0]
\inprod{L_{6}}' & = X^{-1} (1+ (1+X)^2 Y)^2 \\
&= Y^2 X^3+4 Y^2 X^2+6 Y^2 X+\frac{Y^2}{X}+4
   Y^2+2 Y X+\frac{2 Y}{X}+4 Y+\frac{1}{X}, \label{eq:chebyshev-expll}
\end{align}
and so on.
We also have
\begin{equation} \label{eq:wilson-vev-nf0}
\inprod{L_*}' = \frac{1}{\sqrt{XY}} + \sqrt{\frac{Y}{X}} + \sqrt{XY}.
\end{equation}
According to \eqref{eq:y1} these expansions capture the framed BPS degeneracies.  For example, in \eqref{eq:wilson-vev-nf0} we find three framed BPS
states in the Hilbert space with the fundamental UV Wilson loop inserted.
The contributions $(XY)^{\pm \half}$ correspond to electrically charged states, of charges $\pm 1$, as we naively expected.
More surprisingly, there is also the extra contribution $\sqrt{X/Y}$, corresponding to a state which carries no electric charge,
and magnetic charge given by the difference of the windings of the edges $X$ and $Y$.
It would be interesting to reproduce this result from a weakly coupled gauge theory computation.
In addition, the expansions \eqref{eq:chebyshev-expls}-\eqref{eq:chebyshev-expll} imply an intricate pattern of framed BPS states
bound to the UV 'tHooft-Wilson loops, which should also be interesting to discover from the weak coupling point of view.

The commutative ring relation among the line operators also bears examination.  For example, a simple
consequence of \eqref{eq:Trig-Ln} is
\begin{equation} L_{2k} L_* = L_{2k-1} + L_{2k+1}. \end{equation}
This is physically very reasonable: if we bring together a Wilson loop and a 'tHooft-Wilson loop,
the latter breaks $SU(2)$ to $U(1)$, hence the Wilson loop decomposes into the sum of two $U(1)$
contributions of opposite charge.  We thus obtain two 'tHooft-Wilson loops with electric charges
shifted by one unit.
We also have relations $L_{2k} L_{2k+4} = (1+L_{2k+2})^2$, $L_{2k} L_{2k+6} =  1+(1+ L_{2k+2})(1 +  L_{2k+4}) + L_*^2$,
$L_{2k+1} L_{2k+3} = L_{2k+2} (1+L_{2k+2})$, etc.
These relations are rather challenging to understand directly in the gauge theory.

Finally, it is interesting to compare with the formal line operators in
Section \ref{subsec:formal-SU2}.
We recognize that with the identification $X=[0,2]$ and $Y=[-2,0]$ we can
identify $L_{2n}$ with $G_{2n-1}$ and hence $\sqrt{L_{2n}}$ with $\hat V_{n-1}$ evaluated at $y=1$. The above
ring relations become those of equations \eqref{eq:SU2-formring-1} to
\eqref{eq:SU2-formring-5} evaluated at $y=1$. This comparison is
compatible with the discussion at the end of Section \ref{subsec:SU2-Functions} and moreover
shows that we can use the formal line operators to compute explicitly the protected spin characters
in these theories.  Note that in general one does not simply promote an integer $n$ in
\eqref{eq:chebyshev-expls} to $[n]$.

\subsection{$SU(2)$, $N_f = 1$}

Next we treat the $SU(2)$ gauge theory with one fundamental flavor.  As explained in Section
10.2 of \cite{Gaiotto:2009hg}, this is the $A_1$ theory where we choose $C$ to be $\IC\IP^1$
with two irregular singularities $\CP_1$ and $\CP_2$, supporting poles of $\phi_2$ of orders $3$ and $4$
respectively.  Topologically, we cut out a disc around each singularity, and
mark a single point on $S^1(\CP_1)$, two points on $S^1(\CP_2)$.

A sample lamination in this theory is shown in Figure \ref{fig:nf1-maximal}.
The integer weights $k_i$ are subject to the restrictions
\begin{equation} \label{eq:nf1-cond}
2 k_1 + k_4 + k_5 + k_6 = 0, \quad k_2 + k_3 + k_4 = 0, \quad k_2 + k_3 + k_5 + k_6 = 0, \quad k_4 \ge 0, \quad k_5 \ge 0, \quad k_6 \ge 0.
\end{equation}
Solutions to the conditions \eqref{eq:nf1-cond} are
determined by $k_4, k_5, k_6$, all $\ge 0$, such that $k_4 + k_5 + k_6$ is even.
\insfig{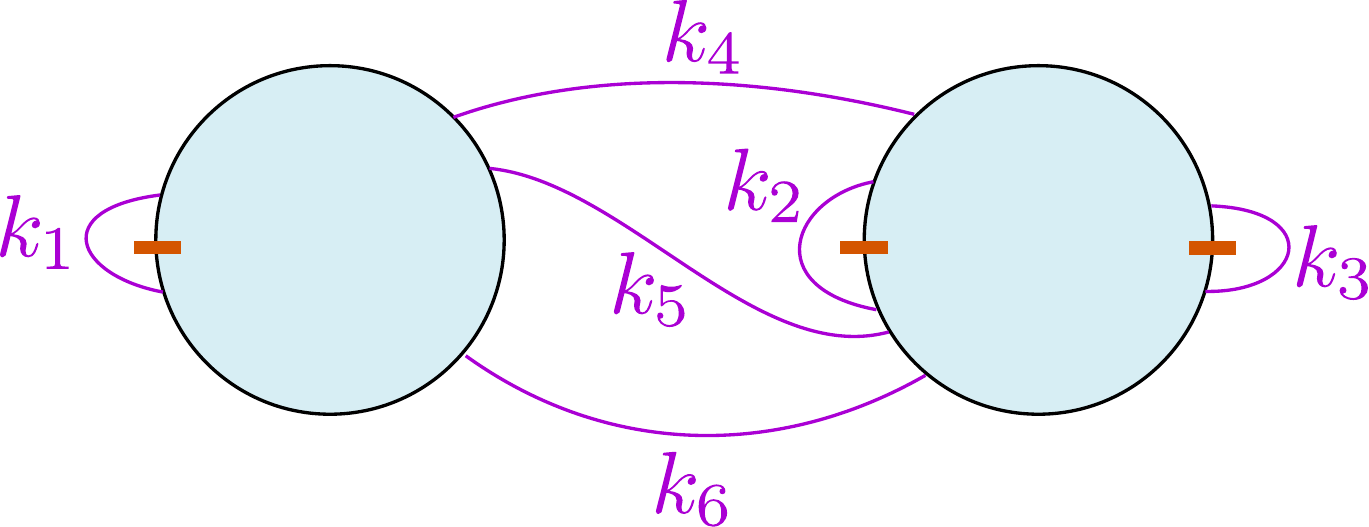}{A family of laminations in the $SU(2)$ theory with $N_f=1$.}
So we have obtained a family of laminations parameterized by such $(k_4, k_5, k_6)$.

These are not all the possible laminations, however, since our choice of curves was not
the most general possible.  As in the $N_f = 0$ theory, we can generate more laminations by a kind of Dehn twist:
continuously rotate a neighborhood of $S^1(\CP_2)$ by an angle of $\pi$ (so that the two marked points are exchanged).
The twist drags the paths $\wp_4$, $\wp_5$, $\wp_6$ into new ones.  By performing repeated clockwise
or counterclockwise twists
on the laminations we constructed above, we can obtain almost all of the possible laminations.
As before, we let $L_*$ denote the lamination consisting of
a single closed curve, which is not obtained by the above construction.

Now choose a triangulation as follows.
The boundary circle $S^1(\CP_2)$ is divided into two segments.  One
segment meets a single edge; label that edge $X$.  The other segment
meets two edges; label them $Y$ and $Z$.  All three edges end on
the boundary circle $S^1(\CP_1)$.  Traveling clockwise around an arc
which begins and ends on $S^1(\CP_1)$ we meet $Z$, $X$, $Y$ in order.
See Figure \ref{fig:nf1}.
\insfig{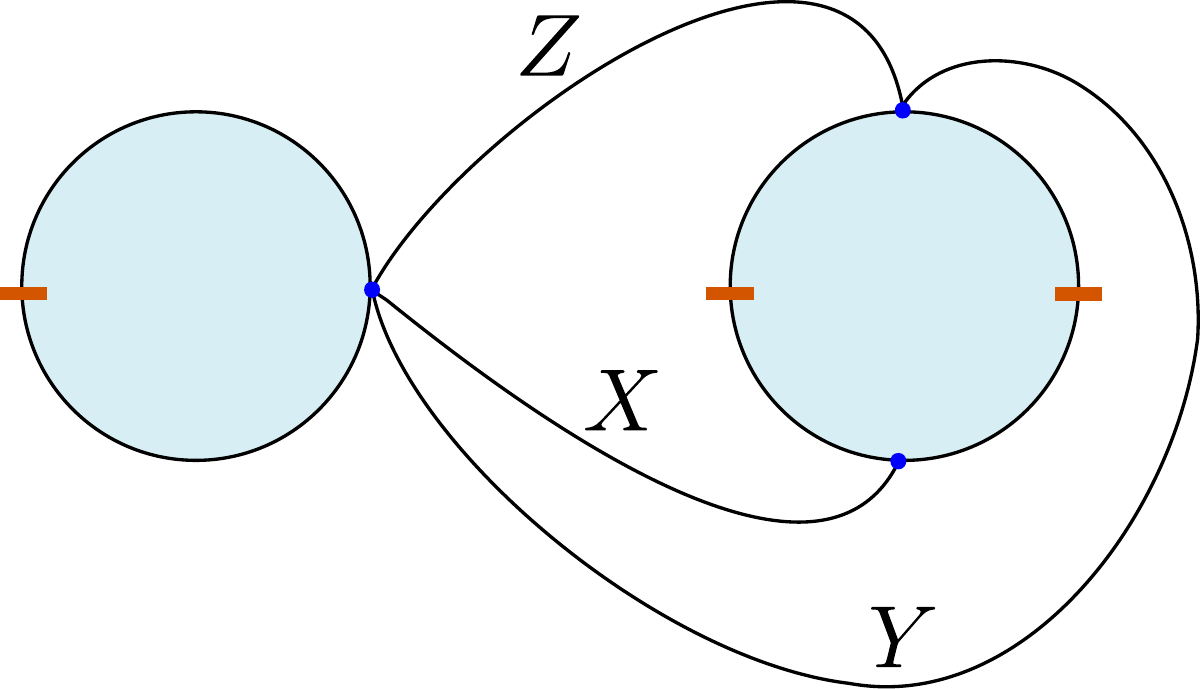}{A triangulation in the $SU(2)$ theory with $N_f = 1$.}

Let $L_0$ be the lamination shown in Figure \ref{fig:nf1-l0}.
\insfig{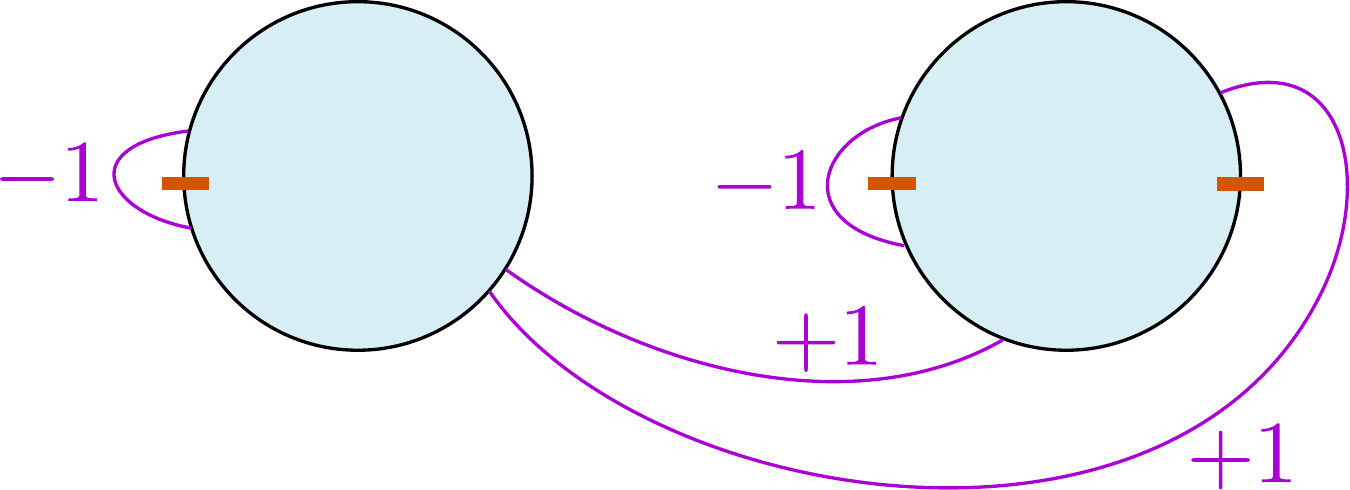}{The lamination $L_0$ in the $SU(2)$ theory with $N_f = 1$.}
Other basic laminations
$L_n$ are obtained from $L_0$ by twisting around $\CP_2$ as described above.
Computing their vevs then gives e.g.
\begin{align}
 \inprod{L_{-2}}' &=  Y + XY  + YZ + X Y Z^2 + 2 XYZ, \\
 \inprod{L_{-1}}' &= X + XZ, \\
 \inprod{L_{0}}' &= Z, \\
 \inprod{L_{1}}' &= 1/Y, \\
 \inprod{L_{2}}' &= 1/X + 1/XY, \\
 \inprod{L_3}' &= 1/Z + 1/XZ + 1/X Y^2 Z + 1/YZ + 2/XYZ.
\end{align}
In general, $\inprod{L_k}'$ is related to $\inprod{L_{1-k}}'$
by the simultaneous exchanges $Z \leftrightarrow 1/Y$, $X \leftrightarrow 1/X$.

We also have
\begin{equation}
\inprod{L_*}' = \frac{1 + X + XY + XYZ}{\sqrt{XYZ}}.
\end{equation}

\subsection{$SU(2)$, other $N_f$}

One can similarly describe the laminations for all the $SU(2)$, $N_f > 0$ theories.  Here we confine ourselves
to a few brief comments.

The theory with $N_f=2$ has two realizations as described in Sections 10.3 and 10.5 of
\cite{Gaiotto:2009hg}.  Let us consider
its ``first realization'', in which $C$ is $\IC\IP^1$ with two irregular singular points, each
supporting two Stokes rays.
A sample lamination is shown in Figure \ref{fig:nf2-l}.  This lamination
is maximal in the usual sense that no more curves can be added to it.  We could get a more general family
of laminations by changing the weights attached to the curves.
\insfig{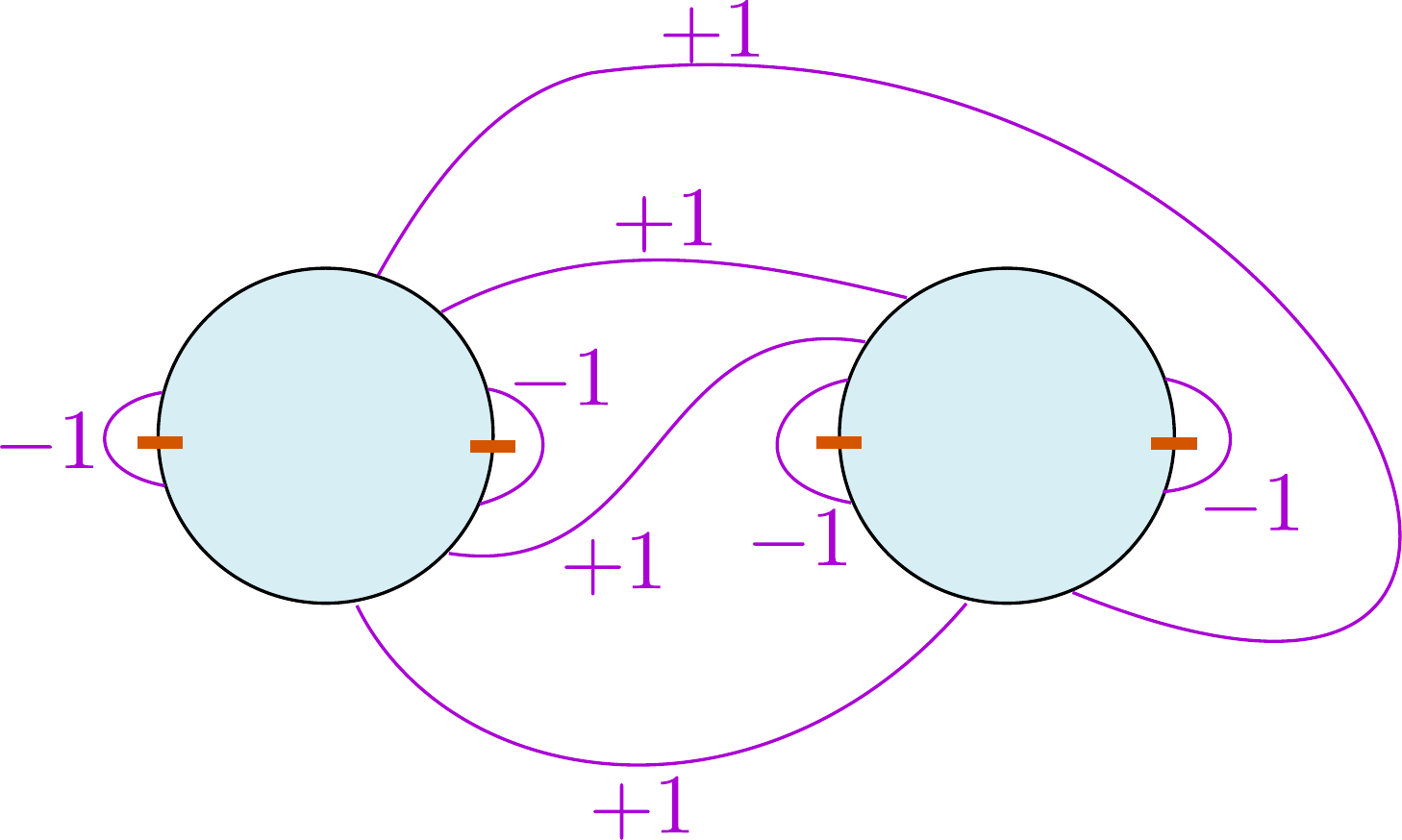}{A sample lamination in the $SU(2)$ theory with $N_f=2$, in its first realization.}

The theory with $N_f = 3$ is obtained by taking $C$ to be $\IC\IP^1$, with
one irregular singular point supporting two Stokes rays, and two regular singular points.
A sample lamination is shown in Figure \ref{fig:nf3-l}.
\insfig{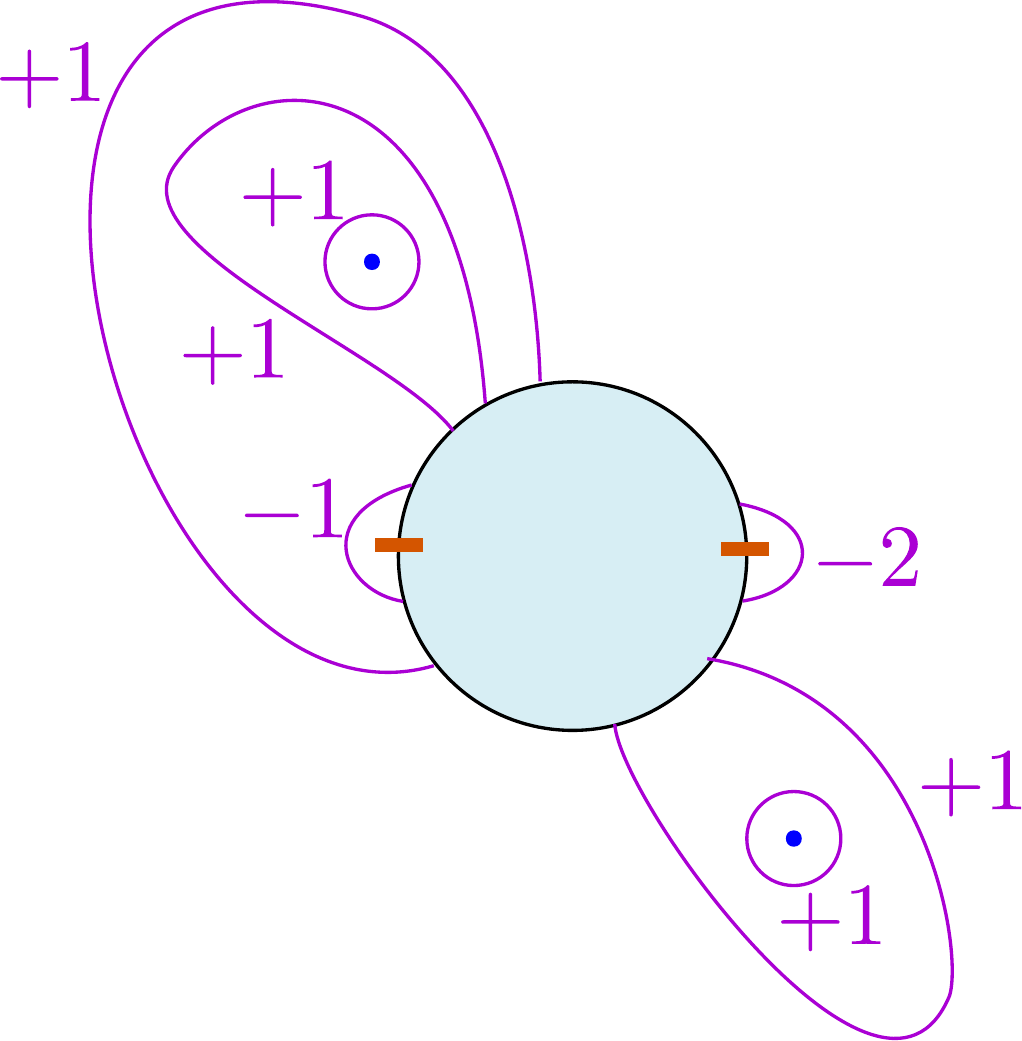}{A sample lamination in the $SU(2)$ theory with $N_f=3$.}

\subsection{$SU(2)$, $N_f = 4$}

Next let us briefly consider the $SU(2)$ theory with $N_f = 4$.
This theory corresponds to $C = \IC\IP^1$ with $4$ regular singularities.
For some choice of $(\zeta, u)$, $T_\WKB(\zeta, u)$ is
the ``tetrahedral'' triangulation pictured in Figure \ref{fig:nf4}.
\insfig{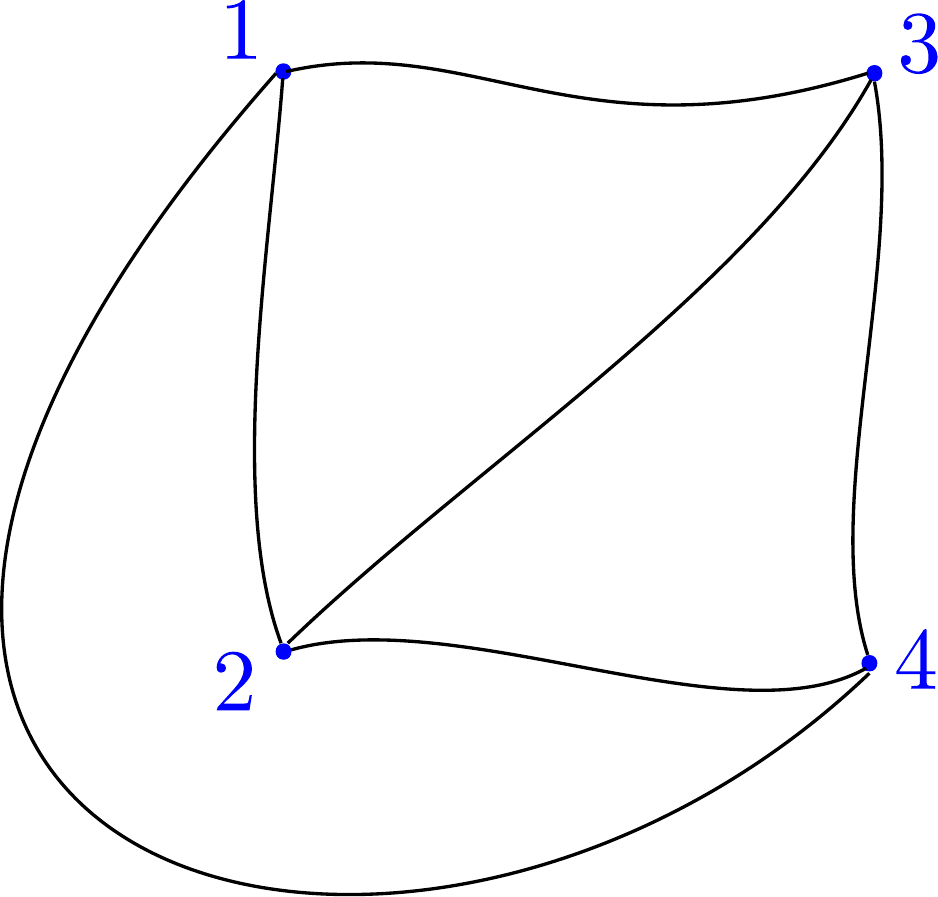}{The ``tetrahedral'' triangulation of $\IC\IP^1$, appearing in the $SU(2)$ theory with $N_f=4$.  The edge between vertex $i$ and vertex
$j$ is labeled $X_{ij}$.}
Then consider the lamination $L_*$ shown in Figure \ref{fig:nf4-lstar}.
Applying the rules of \cite{Drukker:2009tz} we know that the corresponding operator $L_*(\zeta)$ is a Wilson loop in one duality frame.
\insfig{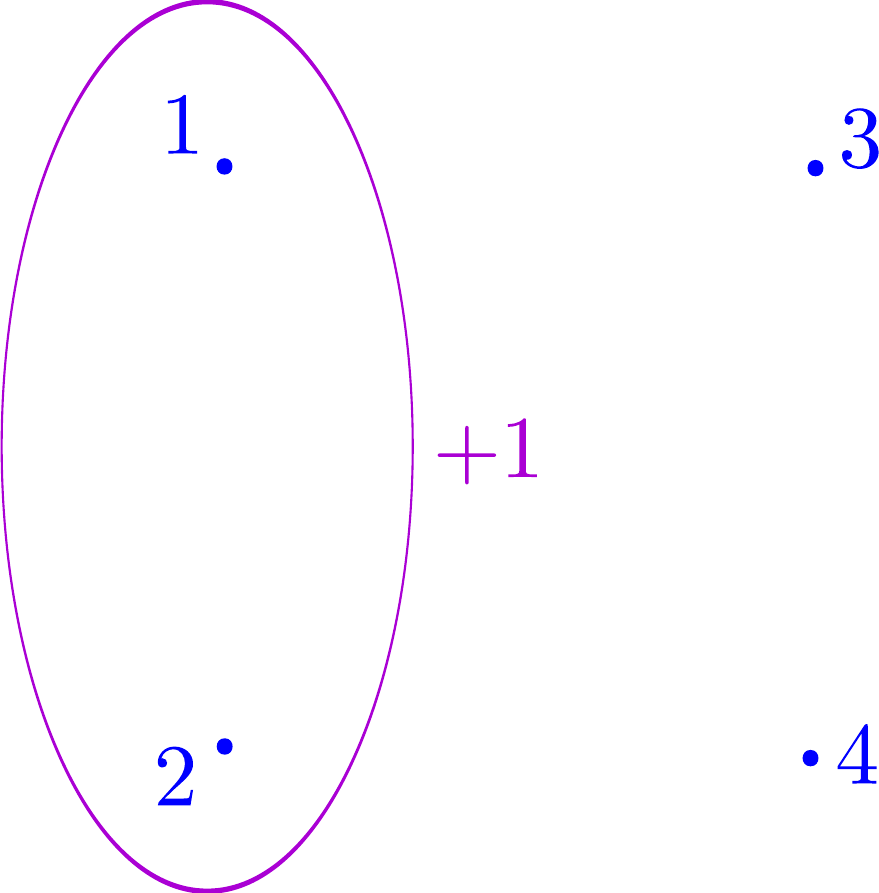}{The lamination $L_*$ in the $SU(2)$ theory with $N_f=4$.}
Using the traffic rules, its vacuum expectation value is

\begin{equation}
\inprod{L_*}= {\Tr} \, L M_{23} R M_{13} L M_{14} R M_{24}
\end{equation}
which works out to be
\begin{equation}\label{eq:Nf4-Wilson}
\inprod{L_*} = \frac{1 + X_{13} + X_{24} + X_{13} X_{24} + X_{13} X_{14} X_{24} + X_{13} X_{23} X_{24} + X_{13} X_{14} X_{23} X_{24}}{\sqrt{X_{13} X_{14} X_{23} X_{24}}}.
\end{equation}
So this line operator supports $7$ BPS states, all with different charges.

\subsection{$SU(2)$, $\CN=2^*$}\label{subsec:SU2-N2star-Examples}

\insfig{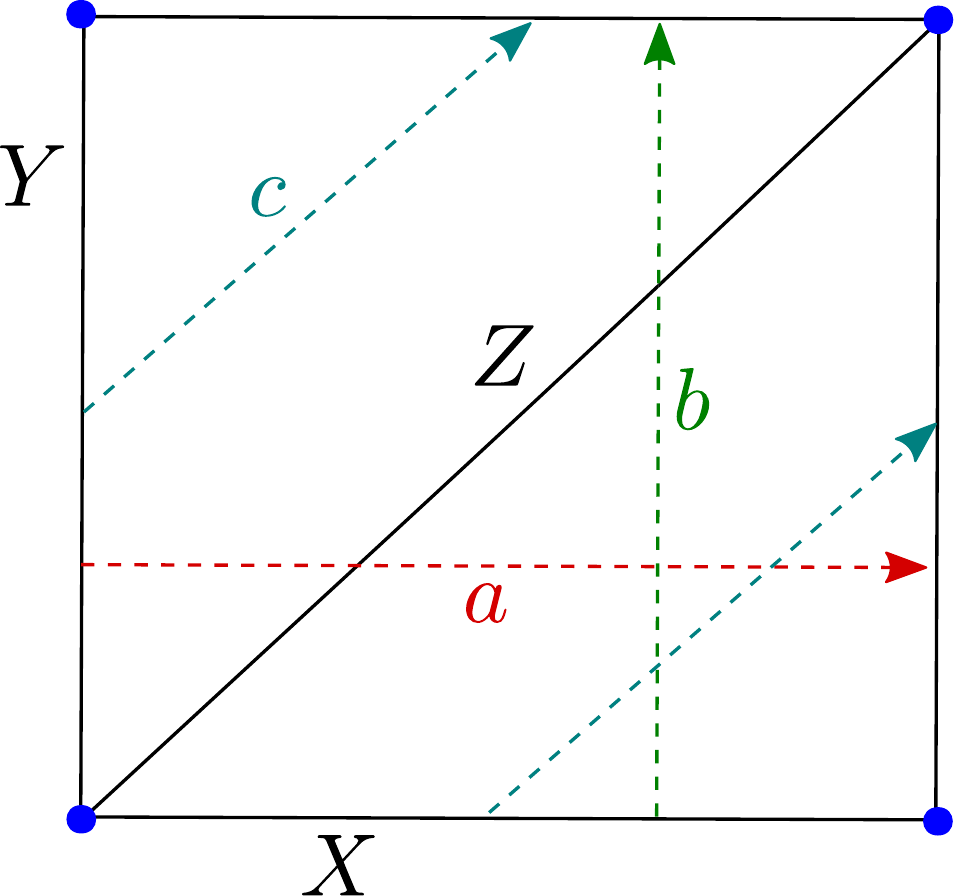}{The generic WKB triangulation for the $\CN=2^*$ theory.}

Finally we consider the most interesting example, the $\CN=2^*$ theory.
As we have reviewed in Section \ref{subsec:Formal-N2-star} the theory
corresponds to taking $C$ to be a once-punctured torus with a regular singular
point.   The relevant quadratic differential is
$\phi_2 = u + m^2 \wp(z\vert \tau)$.  There are two turning points, and all WKB triangulations have
the same topology: two triangles, splitting some fundamental region of the torus in two. Without loss of generality
we can represent the edges by three lines $(0,1)$, $(0,\tau)$, $(0, \tau+1)$, as shown in
 figure \ref{fig:n2star-triang}. We will denote the corresponding cross-ratios as
$X$, $Y$, $Z$. Then the monodromy around the puncture is $\mu = - XYZ  $.
The holonomies ${\rm Tr}(A)$ and ${\rm Tr}(B)$ around the basic cycles $z \to z+1$ and $z \to z+ \tau$ of the torus are easily computed, as the cycles cross two edges each, and hence   have the standard three-term expansion. The same is true of the cycle $z \to z+1+ \tau $, which gives ${\rm Tr}(AB)$. Thus we have the expansions:
\begin{align}\label{eq:N2-star-traffic}
\langle L_{0,1}\rangle' &={\rm Tr}(A) = \sqrt{Z Y} + \sqrt{Z/Y} + 1/\sqrt{Z Y}, \\
\langle L_{1,0}\rangle'  &= {\rm Tr}(B) =\sqrt{X Z} + \sqrt{X/Z} + 1/\sqrt{X Z},  \\
\langle L_{1,1}\rangle' &={\rm Tr}(AB) = \sqrt{Y X} + \sqrt{Y/X} + 1/\sqrt{Y X}.
\end{align}
We can compare these with the formal Wilson line operators \eqref{eq:star-formal-Wilson}
so that $W_1$ corresponds to (the classical limit) of ${\rm Tr}(A)$ etc., exactly
as in Section \ref{subsec:N2-star-moduli}.
As a check, it follows from \eqref{eq:N2-star-traffic} that if $X,Y,Z$ are considered to
be commutative variables then substitution of \eqref{eq:N2-star-traffic} indeed leads to
\begin{equation}
({\rm Tr}(A))^2 + ({\rm Tr}(B))^2 + ({\rm Tr}(AB))^2  -{\rm Tr}(A) {\rm Tr}(B) {\rm Tr}(AB) = \mu + 2 + \frac{1}{\mu}
\end{equation}
where we have used $\mu = - XYZ$. This is
in agreement with the equation \eqref{eq:moduli-equation} and is the classical
limit of \eqref{eq:quant-mod-eq}.

Using the ``promotion principle''
%
%
 it follows
from these expressions that the formal line operators constructed
in Section \ref{subsec:Formal-N2-star} compute the protected spin characters of the
$\CN=2^*$ theory.

\subsection{The millipede expansion}

In the $A_1$ theories there is a simple description of the vanilla BPS states:  they correspond to
BPS strings of the $(2,0)$ theory on $C$ with finite total mass.  This amounts to WKB curves which are either closed or begin
and end on zeroes of the quadratic differential $\phi_2$.
This picture was introduced in \cite{Klemm:1996bj, Shapere:1999xr} and played an important role in \cite{Gaiotto:2009hg}.
In this paper we introduced framed BPS states and it is natural to ask whether they are captured by
a similar geometric picture.
We believe the answer is yes:  in
this section we briefly sketch a proposal to identify the framed BPS degeneracies $\fro(L, \gamma, \zeta)$
directly as counting some objects which we call \ti{millipedes with body $L$ and phase $\arg \zeta$}.

We do not define these objects in general here, but restrict to the case
when the lamination $L$ is just a closed loop, carrying the fundamental representation of $SL(2,\IC)$.
In this case a millipede is a closed oriented curve $\xi$ on $\Sigma$, which can be divided into segments
as follows.  Some of the segments are called \ti{body} segments:  each body segment is a lift to $\Sigma$
of a corresponding segment of $L$.
We require that each point of $L$ lifts to exactly one point on a body segment:  so the body segments make up
a kind of broken lift of $L$ to $\Sigma$.  The rest of the segments are called \ti{legs}.
These are oriented curves along which $\lambda / \zeta$ is real and nonnegatively oriented.  (Such segments are lifts
of WKB curves from $C$.)
The \ti{charge} of the millipede is defined to be the class of $\xi$ in $H_1(\Sigma, \IZ)$.
We propose that the framed BPS degeneracies should be ``counting'' millipedes $\xi$
with body $L$, phase $\arg \zeta$, and charge $\gamma$.

Let us consider the most basic way of constructing a millipede.
Begin with a broken lift of $L$.
Where the lift jumps from one sheet of $\Sigma$ to the other, it has boundary points $p_1$, $p_2$
(which project to the same
point $p$ of $C$.)  So the body of $\xi$ is not closed.  To fix this problem, at each boundary point $p$
we attach a leg segment, which runs from
$p_1$ along $\Sigma$ to a zero of $\lambda$ and then returns along
the other sheet of $\Sigma$ to $p_2$.  (Note that this is only possible if there is a WKB curve which
runs from $p$ to a turning point:  and there
are only finitely many such ``special WKB curves,'' so there are only finitely many possibilities
for attaching legs to any given $L$.)
The projection of this picture to $C$ explains the terminology --- see Figure \ref{fig:millipede}.
\insfig{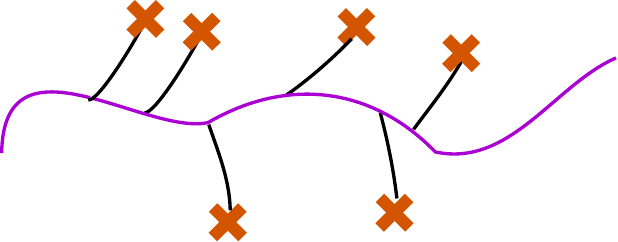}{A portion of the projection of a millipede to $C$.  The body is shown in purple
while the legs are black.  This particular millipede is of the simplest kind, where all legs end on
turning points, shown as orange crosses.}

As in the case of vanilla BPS states, the precise rules for counting these objects are a bit subtle.
The simplest case is the one we just discussed, where all legs are running from $L$
to turning points.  Such a millipede is isolated (has no moduli).
If these are the only millipedes, and if all millipedes have just a single leg ending on each turning point,
then it is relatively straightforward to see using the traffic rules
that the number of millipedes with body $L$ and charge $\gamma$ indeed coincides with the number
$\fro(L, \gamma)$.

More generally though, we may have multiple legs ending on a single turning point; in these cases
the millipede sometimes contributes $-1$ instead of $+1$.
In addition, there can also be millipedes where some legs both begin and end on $L$.
Such a millipede has moduli (the legs can slide back and forth along $L$).  We conjecture
that the isolated millipedes contribute spin zero framed BPS states and non-isolated millipedes
contribute higher spin framed BPS multiplets; this is the case at least
in some simple examples, but we have not studied the question systematically.

\section{Quantum holonomy} \label{sec:Quantum-Holonomy}

In Section  \ref{sec:Examples}
we have shown how to compute the framed BPS degeneracies
for $A_1$ theories. In this section we sketch how
one can --- in principle --- go further and compute the full Protected Spin Characters
of the $A_1$ theories, thereby computing the deformed
algebras of functions in these examples. The key idea is to combine
wall-crossing formulae with the Promotion Principle explained at the
beginning of Section \ref{sec:Examples}.

\insfig{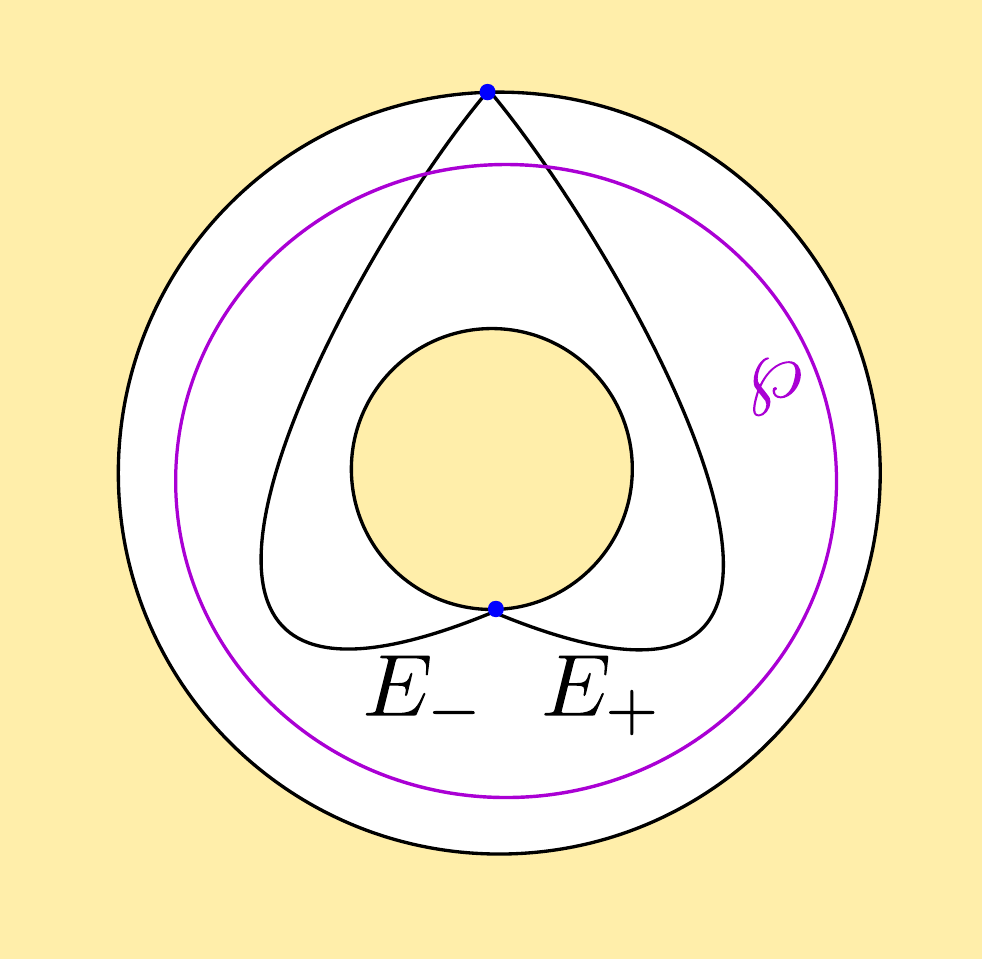}{In regions of parameter space for $(u,\zeta)$ (and the gauge couplings)
 where $T_\WKB(u, \zeta)$ looks like this picture, we can use the
 promotion principle to compute the PSC's for the line operator associated to the
 curve $\wp$.}

We can apply the promotion principle very straightforwardly in some special situations
(as was already done in passing in Section \ref{sec:Examples}.)
Suppose we have a curve $\wp$ and a triangulation such that the local neighborhood
of $\wp$ is of the form of Figure \ref{fig:cylinder}.
Suppose we wish to compute the specialization to $y=1$ of the PSC of the line operator $L_\zeta(\wp)$
(where the representation $\CR=\bf{2}$ is understood. )  That is, we wish to compute the
expectation value $\langle L_\zeta(\wp) \rangle'$. Applying the traffic rules of
Appendix \ref{app:traffic}, we find
\begin{equation}\label{eq:Prime-Wilson-Spec}
\langle L_\zeta(\wp) \rangle' =  \sqrt{\tCY_+ \tCY_- } + \frac{1}{\sqrt{\tCY_+ \tCY_-}} + \sqrt{\frac{\tilde
\CY_+ }{\tCY_-}}.
\end{equation}
We can clearly apply the promotion principle to this expression and conclude that
\begin{equation}\label{eq:PromPrinc-1}
F(L_\zeta(\wp)) =   X_{\half(\gamma_ + + \gamma_-)}+X_{-\half(\gamma_ + + \gamma_-)} +
X_{\half(\gamma_ + - \gamma_-)}.
\end{equation}

Now suppose we wish to compute $F(L_\zeta(\wp))$ for some generic $(u,\zeta)$.
The data $(u,\zeta)$ (and the coupling constants)
determines a Seiberg-Witten differential and an associated
WKB triangulation $T_\WKB(\vartheta,u)$ with
$\vartheta = \arg \zeta$, as described in \cite{Gaiotto:2009hg}, Section 6.
By varying $u,\zeta$ and the coupling constants of the theory we attempt to
find a morphism of the WKB triangulation to one of the form of Figure \ref{fig:cylinder}.
If we can do so, then by successively applying the transformation rules
\eqref{eq:Spin-KS-tmn-S}, \eqref{eq:Spin-KS-ii}
for the series of flips connecting these two
triangulations, we can begin with \eqref{eq:PromPrinc-1} and produce the Protected
Spin Characters at $(u,\zeta)$.

There are two important loopholes in the above algorithm for computing the PSC's.
First, the algorithm can only work if $C$ admits a triangulation
of the form of Figure \ref{fig:cylinder}.  In reality, there are situations when  there is no such
triangulation. Fortunately, we can address this case by invoking a procedure used by J. Teschner
in the quantization of Teichm\"uller space \cite{Teschner:2005bz}.  Briefly,
we choose a trinion decomposition of $C$ such that $\wp$ is one of the cutting
cycles in the decomposition.
Boundaries of the trinions are labeled ``ingoing''
and ``outgoing.''
There is a triangulation compatible with the trinion decomposition
used to define Fock coordinates on Teichm\"uller space (as described in
Sections 13 and 14 of \cite{Teschner:2005bz}). The algorithm we have described above
applies to the case of equation (15.4) of \cite{Teschner:2005bz}. The other case
occurs in equation (15.5) of \cite{Teschner:2005bz}.  However, as described there,
one can invoke a recursive procedure for computing $\langle L_\zeta(\wp) \rangle'$,
and the promotion principle will apply to the result obtained from this recursive
procedure.

The second loophole is that even when a triangulation of the form of Figure \ref{fig:cylinder}
exists it is not immediately evident that a WKB triangulation of this form exists.
We do not think this is a serious loophole, since one can always choose a suitable
weak coupling region so that a tubular neighborhood of $\wp$ is a long thin tube, and
$L_\zeta(\wp)$ corresponds to the elementary Wilson loop of the corresponding $SU(2)$
factor in the gauge group. It seems to us quite reasonable that in such a region of parameter
space the WKB triangulation
will be of the appropriate type, but we leave the detailed demonstration of this claim
undone in this paper.

Let us make a few further remarks:

\begin{enumerate}
\item We have seen in Section \ref{subsec:formal-SU2}
for the case of
$SU(2)$ with $N_f=0$ that some strongly positive formal line operators can
be readily computed purely algebraically and the relation to cluster
algebras described in Section \ref{sec:Cluster-Algebras} shows that
one can proceed with local rules for joining chambers in moduli space.
Thus, using the promotion principle we see that one can compute some PSC's fairly efficiently.

\item The above rules for computing the PSC's are in a sense nonlocal on $C$.
It would be desirable to  have \emph{local} rules which compute framed protected spin characters,
analogous to the traffic rules which we have at $y = +1$.
We expect such local rules to follow from
the detailed 2d-4d wall-crossing formula whose existence was suggested in
\cite{Gaiotto:2009fs}. This work is in progress.\footnote{We
have been informed by A. Goncharov that there are some
known, but unpublished, rules for computing analogous quantities in the theory of
cluster ensembles \cite{MR2567745}.}

\item It follows from the relation to Teschner's work that the deformed
algebra of functions on $\CM$ in the $A_1$ theories is naturally
isomorphic to the algebra of quantum geodesic operators (quantizing the geodesic lengths
in the metric of constant curvature $-1$) in quantum
Teichm\"uller theory for $C$. We would like to stress that
our framework incorporates irregular
singularities and laminations, and thus provides interesting generalizations of what has been
done in the Teichm\"uller context.

\item It is possible that one can go further and show that
the formal expansion \eqref{eq:form-gen} can be given a
concrete meaning in terms of some quantization of an appropriate real slice of
$\CM$ as
\begin{equation}\label{eq:q-proposal}
 \hat L_\wp = \sum_{\gamma} \fro(u, L,\zeta,\gamma;y) X_\gamma
 \end{equation}
where $X_\gamma$ are now concrete quantum operators acting on
a definite Hilbert space.  Indeed, concrete operator interpretations of the
$X_{\gamma}$ were constructed in \cite{Teschner:2005bz} and in \cite{MR2567745,clqd2,qd-cluster}.
However, \eqref{eq:q-proposal} should have a wider
range of applicability, and should not depend on a choice of real slice.
Upon choosing certain real slices and specific quantization schemes we might
expect to make some contact with the
work of Nekrasov and Shatashvili \cite{Nekrasov:2009ui,Nekrasov:2009rc} and Nekrasov and
Witten \cite{Nekrasov:2010ka}.

\end{enumerate}

\section{Tropical labels}\label{sec:Tropic}

In this section we will sketch a way to label the simple line operators of
an $\CN=2$ theory in terms of IR data.

There are three motivations for finding such a labeling.   First,
such a labeling is desirable since so far we were
only able to give labels in the case of theories which have Lagrangians (see Section
\ref{subsec:LabelsForLines}.)
An IR labeling could be applied equally well to non-Lagrangian theories.
Second, it is of interest to specify the  large $R$ or $\zeta$ asymptotics of
$\langle L \rangle $.  This is motivated from the math
viewpoint because we know that traces of holonomies are interesting functions on
Hitchin moduli space; it thus interesting to think about the $\zeta \to 0, \infty$
asymptotics, which turn out to be very subtle.  There also appear to be some interesting applications
to physics.  Third, we would like to fill in a gap in
Section \ref{sec:linehol}:  we would like to show that there are sufficiently many distinct
line operators $L_i$ to
``invert'' the expansion \eqref{eq:trh-large-r} and give $\CY_\gamma$ in
terms of $L_i$.

It turns out that the subject of ``tropical varieties'' is useful for addressing
these questions.  We will make use of an aspect of the work of Fock and Goncharov \cite{MR2233852}
which so far has not played a role in our story.

The key idea will be to consider the leading asymptotics of a vev $\langle L \rangle$ for a
simple line operator, say, as $\zeta \to 0$. The Darboux expansion \eqref{eq:trh-large-r}
expresses  $\langle L \rangle$  as a sum of  $\CY_\gamma$.
As $\zeta \to 0$ along some ray, or $R\to \infty$ for some fixed $u,\zeta$
one term in the sum will dominate. If we consider a path  $(u_s,\zeta_s)$ in $\widehat \CB \times \widehat \IC^*$
then two terms $\fro(L_1,\gamma_1)\CY_{\gamma_1}$
and $\fro(L_2,\gamma_2)\CY_{\gamma_2}$ can only exchange dominance at a point on the path where
 ${\Re}(Z_{\gamma_1}/\zeta) = {\Re}(Z_{\gamma_2}/\zeta)$. For simple line operators it is
natural from the halo picture    to expect that two such terms must have
$\gamma_1 - \gamma_2 = \gamma$ where $\gamma$
supports a vanilla BPS state. Therefore we define an  \emph{anti-BPS wall} $\check W(\gamma)$
for $\gamma \in \Gamma$ to be a wall \footnote{In our only example, that of the $A_1$ theories, this definition
will prove to be correct. However, we note that while it is natural, it does not follow rigorously
from the halo picture that the only walls where there is an exchange of dominance are these walls.
It could in principle happen that $\gamma_1 - \gamma_2$ does not support a vanilla BPS state. If
there are examples of such a kind then the definition of a tropical theory below will have to be
modified.}
\begin{equation}
\check W(\gamma):=  \{(u,\zeta): Z_\gamma/\zeta \in - i \IR_+  \qquad {\rm and} \qquad  \Omega(\gamma;u)\not=0\}.
\end{equation}
The reason we have chosen the sign $-i$ will become evident below.
The  complement of these walls have connected components which are defined to be the
\emph{anti-chambers} and will be denoted $\check c$. In general we expect that within
a given anti-chamber a single term in the Darboux expansion \eqref{eq:trh-large-r}
 will dominate the asymptotics and we can write
 \begin{equation}
 \langle L\rangle  \sim \fro(L;\gamma_t) \exp ( \pi R Z_{\gamma_t}/\zeta) \left( 1 + \cdots\right)
 \end{equation}
for a vector $\gamma_t \in \Gamma_\CL$. We will call this the \emph{tropical label}. In
general it depends on $L$ and $\check c$ so we write $\gamma_t(L, \check c)$.
Note that we have the ``tropical formulae'':
\begin{equation}\label{eq:tropic-mult-line}
\gamma_t(LL',\check c) = \gamma_t(L,\check c) + \gamma_t(L', \check c)
\end{equation}
and
\begin{equation}
{\rm Re} (Z_{\gamma_t(L+L',\check c)}/\zeta)  = {\rm Max}[{\rm Re} (Z_{\gamma_t(L,\check c)} / \zeta), {\rm Re} (Z_{\gamma_t(L', \check c)} / \zeta)].
\end{equation}
(Regarding \eqref{eq:tropic-mult-line}, the product $LL'$ in general will not be a simple
line operator, but it can be decomposed in terms of simple line operators, and one of the
resulting ones will have the tropical label given in  \eqref{eq:tropic-mult-line}.)

More formally, we will define an $\CN=2$ theory to be \emph{tropical} if there is a
(anti-chamber-dependent) basis
$\{ \gamma_a \} $ for $\Gamma$ so that \footnote{We could also work with
$\langle L_\zeta  \rangle' $  and $\widetilde \CY_{\gamma}$, which has the advantage
that all expansion coefficients are positive and there can be no cancelations.}

\begin{enumerate}

\item  In any anti-chamber $\check c$  the $R\to \infty$ and $\zeta\to 0 $ asymptotics
for all $a$ satisfy $\CY_a \to \infty$. Moreover,  all simple line
operators have  the form:
\begin{equation}
\langle L \rangle =  \fro(L,\gamma_t) \CY_{\gamma_t} P(1/\CY_a)
\end{equation}
where $P$ is a polynomial whose lowest order term is $1$ and $\gamma_t\in \Gamma_\CL$ is
constant within the antichamber $\check c$. (The polynomial $P$ can change within the anti-chamber. For
example ordinary chambers will divide up the anti-chamber into more than one component.)

\item Across anti-BPS-walls $\check W(\gamma_0) $ we have the transformation law
\begin{equation}\label{eq:Tropical-Transformation}
\gamma_t(L_i, \check c') = \gamma_t(L_i, \check c) -  \Omega(\gamma_0;c) \langle \gamma_t(L_i, \check c),\gamma_0 \rangle_+  \gamma_0
\end{equation}
(the anti-wall sits inside the chamber $c$).

\item  For all $\gamma \in \Gamma_{\CL} $ and any anti-chamber $\check c$
there is a simple line operator $L_\gamma$ so that
$\gamma = \gamma_t(L_\gamma, \check c)$.

\item  If $\gamma_t(L , \check c)= \gamma_t(L', \check c)$ then $L=L'$.

\end{enumerate}

In physical terms this says that a simple line operator, which usually is given a
UV label (such as a vector in $\CL$) can also be uniquely labeled by an IR label
in the (extended) charge lattice $\Gamma_{\CL}$ of the theory. However, unlike the UV
label, the IR label depends on the IR parameters (through the $\check c$-dependence)
and undergoes wall-crossing.

The work of Fock and Goncharov \cite{MR2233852} suggests the conjecture that all theories in the
class $\CS$ are tropical. We will now sketch why the $A_1$ theories are indeed tropical.
(The following argument relies heavily on the technology
developed in \cite{Gaiotto:2009hg}, and we assume the reader is familiar with that paper.)

For simplicity, we restrict attention in this argument to the case of only regular singular
points on $C$.  We consider a simple line operator.  As explained in \cite{Drukker:2009tz}
and Section \ref{subsubsec:A1-labels} the UV label of the operator is an isotopy class
of a non-self-intersecting closed curve $\wp\subset C$.
The leading asymptotics of $\langle L_\zeta(\wp) \rangle$  will be extracted using the
traffic rule algorithm.
 We choose $(u,\zeta)$ which is not on an anti-BPS wall  and set
$\zeta = \vert \zeta \vert e^{i \vartheta} $   from which
we extract the WKB triangulation $T_{\WKB}(\chi)$ for $\chi = \vartheta- \frac{\pi}{2}$.
(See Section 6 of \cite{Gaiotto:2009hg}.)
We can assume that a closed path $\wp$ does not backtrack through the
triangles of $T_{\WKB}(\chi)$. Now we consider the
traffic rule algorithm for this triangulation. It was shown in Section 7.8
of \cite{Gaiotto:2009hg} that for any WKB triangulation the corresponding
vectors $\gamma^{\chi}_E$, where $E$ runs over the edges of the triangulation,
form a basis of simple roots. In particular ${\rm Im} ( e^{-i \chi}Z_{\gamma_E^\chi}(u)) > 0 $
for all the edges $E$ and therefore
\begin{equation}
{\rm Re} ( e^{-i \vartheta} Z_{\gamma_E^\chi}(u)) > 0
\end{equation}
and hence the $\CY_{\gamma_E^{\chi} }(u,\theta; \zeta' ) $ have asymptotics going to infinity, for
all $\zeta' \to 0 $ in the half-plane $\IH_{\chi}$. This includes the $\zeta = e^{i \vartheta}$ on its boundary
so we still have $\CY_{\gamma_E^{\chi}}(u,\theta; \zeta) \to \infty$.
It now follows from the traffic rules that the dominant term in $\langle L_\zeta(\wp) \rangle$
is simply given by
\begin{equation}
 \prod_{\wp \cap E \not=\emptyset }  \sqrt{\CY_{\gamma_E^{\chi}} }
\end{equation}
and hence the tropical charge is the sum of the charges associated with the edges
crossed by $\wp$:
\begin{equation}
\gamma_t(L(\wp), \check c) = \half \sum_{\wp \cap E \not=\emptyset } \gamma_E^{\chi}
\end{equation}

Now, let us ask how this tropical label can change. Such a change can only happen
when there is a flip of the WKB triangulation.\footnote{We explicitly exclude the
 ``juggle'' transformation at this point, which would lead to a much more complicated
 analysis.}  But the results of \cite{Gaiotto:2009hg}
show that $T_{\WKB}(\chi)$ will only change when there is a hypermultiplet of charge $\gamma$
such that $\chi = {\rm arg} Z_\gamma(u)$. In the present situation $\chi = \vartheta-\frac{\pi}{2}$
 and hence this means that $Z_{\gamma}/\zeta \in - i \IR_+$. That is, the tropical vector $\gamma_t$
 for $L_\zeta(\wp)$ can only change across the anti-BPS-walls, which are the boundaries of
 $\check c$. This completes the proof of property 1 in our definition of a tropical theory.

\insfig{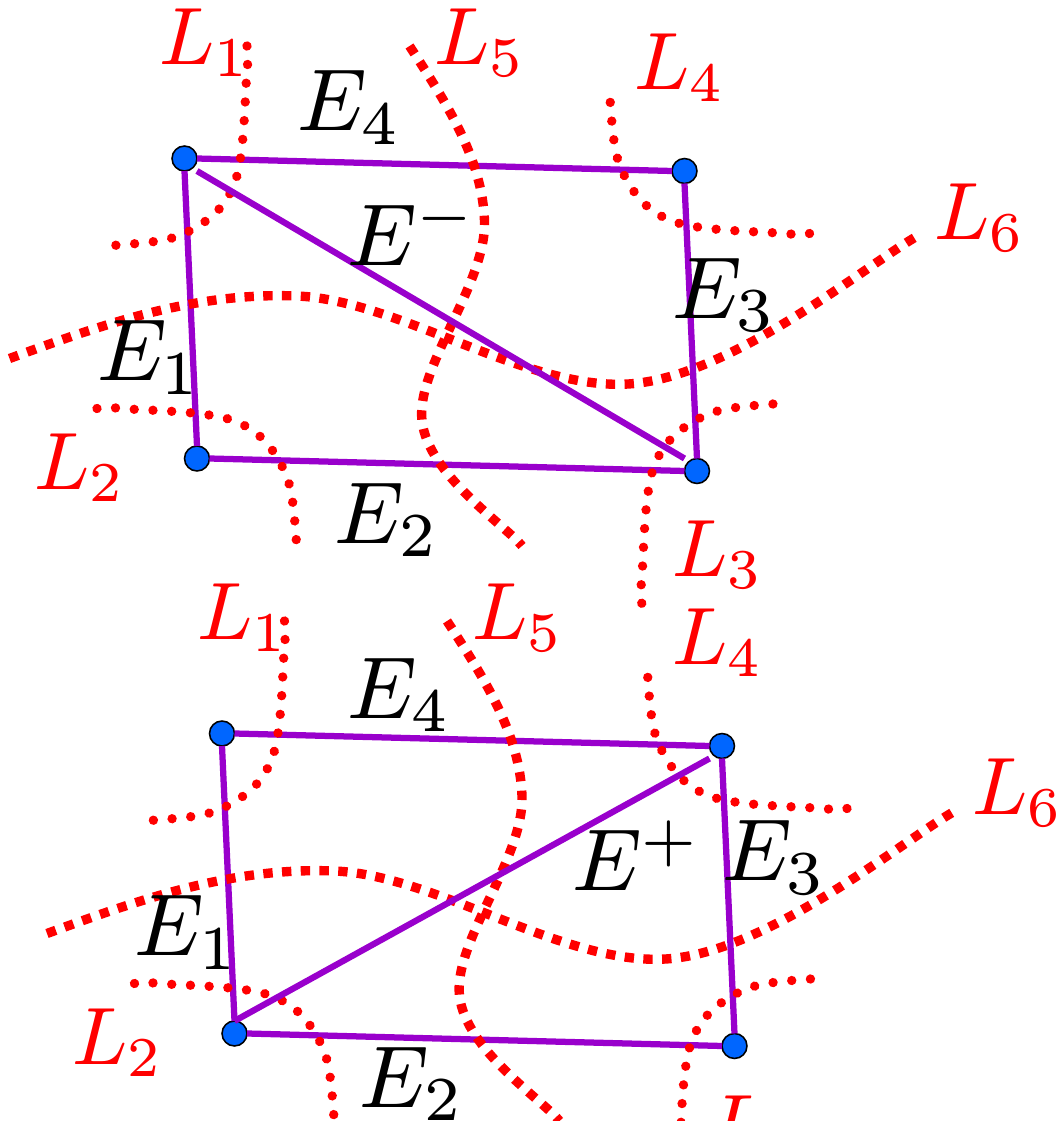}{Using the traffic rules and the mutations of the simple roots under a flip
one can check the tropical transformation rule for $\gamma_t$ across chambers related
by a flip. Here the red curves are pieces of six possible line operators that intersect the relevant quadrilateral
without backtracking.}

Now, to check property 2, consider the flip in a quadrilateral and how it affects the
laminations $L_1, \dots, L_6$ which have segments intersecting the lamination as in
Figure \ref{fig:Flip-Traffic}. Using equations (7.23)-(7.24) of \cite{Gaiotto:2009hg} we see
that the simple roots change by a mutation as $\chi$ moves in the counterclockwise direction
through a critical phase:
\begin{equation}
\begin{split}
\gamma_{E^+}^+ & =  - \gamma_{E^-}^- \\
\gamma_{E_i^+}^+ & =  \gamma_{E_i^-}^- + \langle \gamma_{E_i^-}^-, \gamma_{E^-}^- \rangle_+ \gamma_{E^-}^- \\
\end{split}
\end{equation}
Using this one easily checks the tropical transformation rule \eqref{eq:Tropical-Transformation}.

\insfig{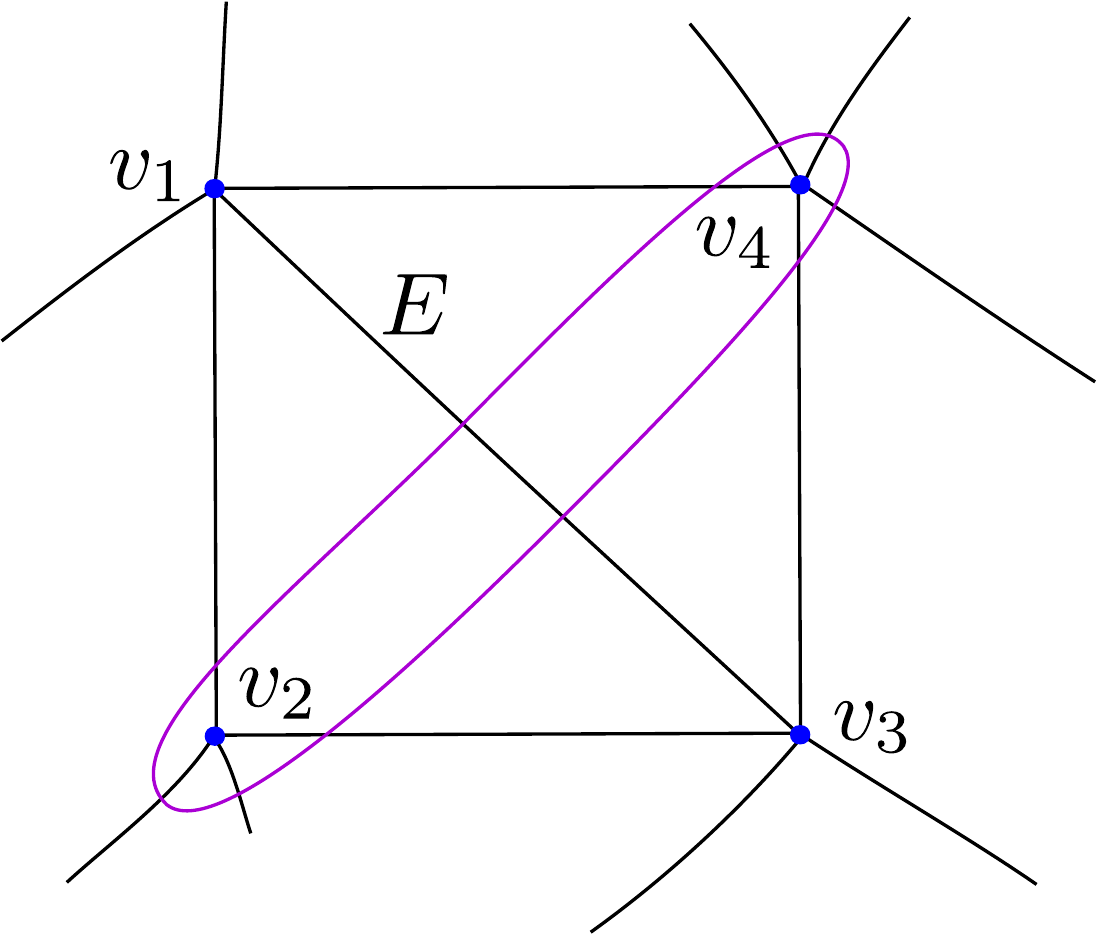}{After multiplying by suitable line operators with purely flavor charge,
the above line operator has tropical label $\gamma_E$. }

\insfig{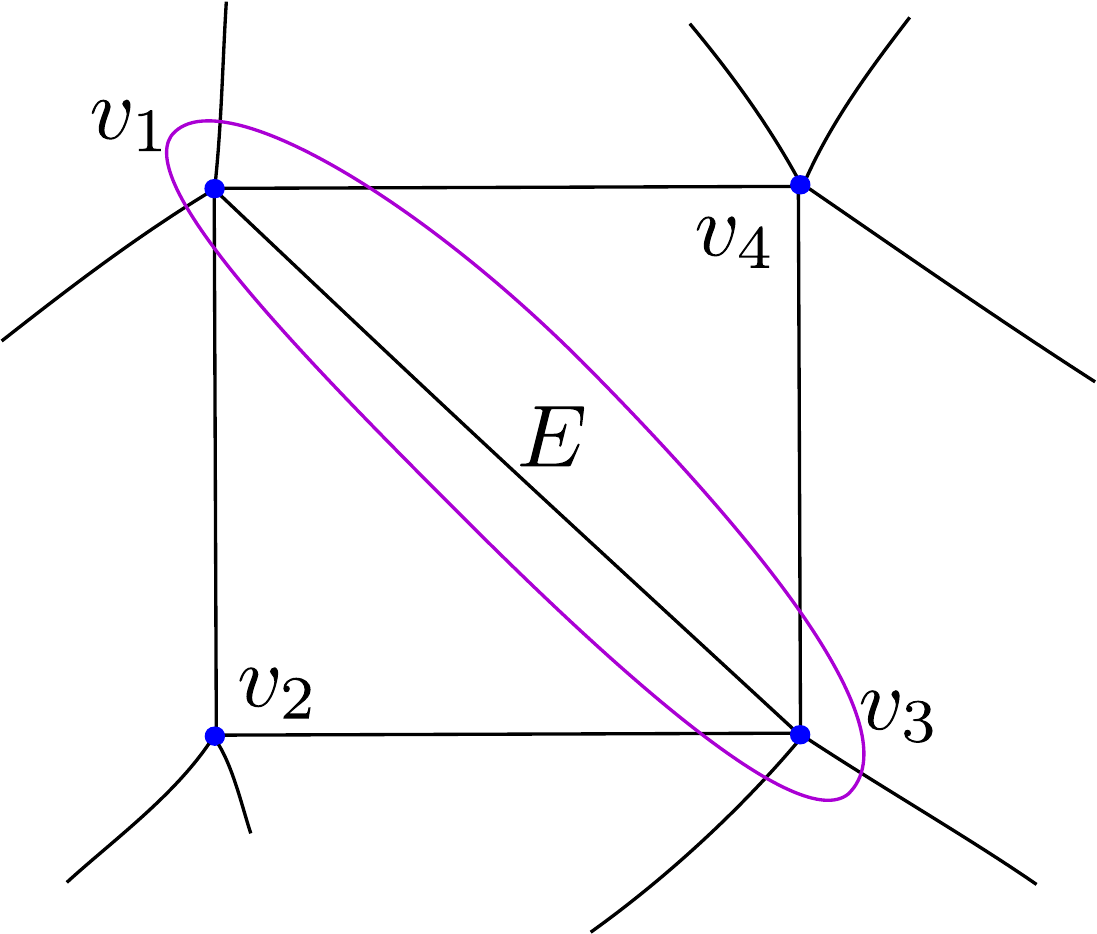}{After multiplying by suitable line operators with purely flavor charge,
the above line operator has tropical label $-\gamma_E$. }

In order to establish property 3 it suffices to exhibit simple line operators whose tropical labels
give $\pm \gamma_E^\chi$. Then, products of such line operators will give any desired vector
thanks to \eqref{eq:tropic-mult-line}. Now, referring to Figure \ref{fig:tropic-onto-A}, we see that
the path $\wp$ has tropical label
\begin{equation}
\gamma_t(L(\wp)) = \left( \half \sum_{v_2 \in E_i} \gamma_{E_i} + \half \gamma_E \right)
+ \left( \half \sum_{v_4 \in E_i} \gamma_{E_i} + \half \gamma_E \right)
\end{equation}
Next, note that a lamination which is a small circle $C_v$ around the vertex $v$ with weight $-1$ has
$L(C_v) = \mu_v^{-1}$ and hence tropical label $-  \half \sum_{v \in E_i} \gamma_{E_i} $.
Therefore $L(C_{v_1}) L(C_{v_2}) L(\wp)$ has tropical label $\gamma_E$. Similarly, using the path
$\wp'$ in Figure \ref{fig:tropic-onto-B} we find that $L(C_{v_1}) L(C_{v_2}) L(\wp')$ has
tropical label $-\gamma_E$.

Finally, to establish property 4, suppose we are presented with a tropical vector $\gamma_t$.
 Since the $\gamma_{E_i}$ are simple roots this has a unique decomposition into a sum of
 $\gamma_{E_i}$ and hence we know which edges are crossed. Now consider the dual
   cell-decomposition to the  WKB triangulation. To each edge $E_i$ occuring in the tropical
   vector (counted with multiplicity) we associate an edge in the dual cell-decomposition.
   These piece together in a unique way to give a lamination, thus completing the proof that
   $A_1$ theories (with only regular singular points) are tropical. We expect that a similar
   but perhaps more elaborate argument will cover the case with irregular singular points.

We end this section with a few remarks:

\begin{enumerate}

\item The transformation \eqref{eq:Tropical-Transformation} is a ``tropical'' version of the
Kontsevich-Soibelman transformation $\CK_{\gamma_0}^{\Omega(\gamma_0)}$.
We may denote it by $\CK_{t,\gamma_0}^{\Omega(\gamma_0)}$.  There is a
tropical wall-crossing formula in which $\CK$ is replaced everywhere by $\CK_t$.
Note that in the above we have only considered
flips, associated with BPS hypermultiplets;
we have not considered the more difficult ``juggle'' transformations associated with vectormultiplets
(see \cite{Gaiotto:2009hg} Section 6.6.3.).

\item There is a close analogy here with Stokes theory. This is not an accident, given the
$\zeta$-differential equations of \cite{Gaiotto:2008cd}.

\item In the $A_1$ theories the tropical vector $\gamma_t$ of $L(\wp)$ has a very beautiful physical
interpretation: It is the homology class of the WKB path  in the Seiberg-Witten
curve $\Sigma$ which dominates the asymptotics.

\end{enumerate}

 \subsection{Example: Tropical labels for the $N=3$ Argyres-Douglas theory}
In equation \eqref{eq:ADN3-gen-lam} we wrote the general simple lamination of the $N=3$ AD theory.
This decomposes the set of simple   laminations into five ``quadrants'' coordinatized by $m, n \ge 0$.
 It is actually natural to glue these five quadrants together into a single copy of $\IZ^2$, as follows.
 Given any simple line operator $L$ and any triangulation with associated coordinates $X$, $Y$,
 we expand $\inprod{L}'$ in terms of $X$ and $Y$.  The result always has the form
 \begin{equation}
 \inprod{L}' = X^a Y^b P(1/X, 1/Y)
 \end{equation}
 for some polynomial $P = 1 + \cdots$.  The pair $(a,b)$ is the tropical vector for the  operator $L$ (in the basis $\gamma_X$, $\gamma_Y$ for $\Gamma$).

 For example, using the triangulation we introduced above, the operators $L_i^m L_{i+1}^n$ in each of the
 five quadrants are mapped respectively to $(n+m, n)$, $(n, m+n)$, $(-n, m)$, $(-m, -n)$, $(n, -m)$; note that
 taken together these five regions fill up $\IZ^2$, so every possible
IR charge has a corresponding simple line operator.
 This really just follows from the fact that the leading terms in our
 five canonical $L_i$, cyclically ordered, are $X$, $XY$, $Y$, $1/X$, $1/Y$, which are cyclically ordered in the plane of monomials.

 For each
 triangulation, we get in this way a natural coordinatization of the space of all simple
line operators, identifying it with $\IZ^2$.
Flipping the triangulation gives a simple piecewise-linear transformation of
 these coordinates.  This structure is described very nicely in \cite{qd-pentagon}.

\section{Open problems}\label{sec:Open-Problems}

There are a number of interesting directions for future research involving both
the physical and mathematical aspects of framed BPS states and their protected spin
characters. We list some of them here.

\begin{enumerate}

\item The (motivic) Kontsevich-Soibelman wall-crossing formula is expected
to apply not only to the BPS degeneracies of $d=4, \CN=2$ field theories
but also to those of Type II string compactifications on Calabi-Yau manifolds.
Unfortunately, a clear physical derivation of the formula in the supergravity
case remains elusive, although, ironically, the semiprimitive wall crossing
formula was first derived in the supergravity context \cite{Denef:2007vg}.
It would therefore be very interesting to find some analog of the line operators used
in this paper in the supergravity context.

\item The framed PSC's $\fro(L,\gamma;y)$ and framed BPS indices should have a
more direct mathematical definition. It seems clear that it should be possible
to give a rigorous definition in terms of zero modes of certain  Dirac operators on
moduli spaces of (singular) monopoles coupled to appropriate
vector bundles along the lines of \cite{Sen:1994yi,Gauntlett:1993sh, Weinberg:2006rq,Hitchin-L2}.
Such a definition could give a mathematical framework
for testing our gauge theory predictions. In particular, it would be
interesting to do this to check our strong positivity conjecture. \footnote{We would like to thank Edward Witten for an illuminating discussion
about this.}     It might well be that there are other definitions of framed BPS indices and
their protected spin characters,
closer to Donaldson-Thomas theory on Calabi-Yau manifolds. \footnote{We thank E. Diaconescu
for suggesting this possibility.}

\item Do all theories in class $\CS$ have the cluster property of  Section \ref{subsec:Seeds-from-N=2}?
Can this be used to determine the BPS spectrum of such theories? In particular, can it be
used to solve the difficult problem of finding the BPS spectrum of $A_r$ theories for $r>1$?

\item Is there a one-one correspondence between formal laminations and physical
line operators? All all $\CN=2$ theories tropical in the sense of Section \ref{sec:Tropic}?

\item
We believe that  another physical justification for the insertion of
$\sigma(\CQ)$ in \eqref{eq:Lvev-trace} could possibly be given along the
following lines. We would like to have a physical understanding
of the role of the $U(1)$ valued function $\psi_{\gamma}: = e^{i \ttheta_{\gamma}}$
and understand in particular why it is  a \emph{twisted} homomorphism $T_u \to U(1)$ in the sense that
$\psi_{\gamma} \psi_{\gamma'}  = (-1)^{\langle \gamma, \gamma'\rangle}
\psi_{\gamma + \gamma'}$.  Suppose one of the directions in $\IR^3$
is considered to be the time direction. Then we could adiabatically transport
a dyon of charge $\gamma$ around the circle $S^1$. The Aharonov-Bohm phase
picked up by this particle, which we view as measuring the electric and
 magnetic Wilson lines, will be $\psi_{\gamma}$.  Now comparing
  the product $\psi_{\gamma} \psi_{\gamma'}$ to $\psi_{\gamma+\gamma'}$ the
  main difference is that the naive $(-1)^F$ parity
  of the pair of   $\gamma$ and $\gamma'$ particles differs from that of the
  boundstate by a factor of $(-1)^{\langle \gamma,\gamma'\rangle}$ due to the
   the spin degrees of freedom of the electromagnetic field of the pair of dyons.  Thinking this remark
   through one encounters a number of subtleties  which we will not try to sort out. We think it would
   be very nice if a physical derivation of the twisted homomorphism could be given along the above lines.

\item As we have mentioned, it would be interesting to understand whether
 the noncommutative deformations
of functions on $\CM$ can be related to a twisted trace of the form
\eqref{eq:y-dep-trace}.  Related to this, as we have mentioned above,
it would be very interesting to
understand the noncommutative generating functions such as \eqref{eq:form-gen}
in terms of concrete operators acting on explicit Hilbert spaces. This will surely
involve the study of various real sections of the spaces $\CM_{\CL}$.

\item As we have remarked, the relations \eqref{eq:quant-skein-rel} are
quantum skein relations closely analogous to those familiar from Chern-Simons
theory. Moreover, the
3-manifold invariants for noncompact Chern-Simons  involve the quantum
dilogarithm \cite{Dimofte:2009yn}.   These two facts suggest
that there is a potential connection of our
results to Chern-Simons theory.

\item   In \cite{Zamolodchikov:1991et} Alyosha Zamolodchikov stated a recurrence conjecture for solutions
of certain $Y$-systems. For some references and background see \cite{keller-periodicity}
and the talk \cite{volkov-talk}.
Remarkably, several of the relevant equations
coincide with ring relations for Argyres-Douglas theories.  We would like to suggest that Zamolodchikov's
conjecture can be understood and extended in terms of the discrete symmetries of Argyres-Douglas theories.
This proposal was also made by Sergio Cecotti and Cumrun Vafa, and
is developed in more detail in \cite{CNV}.

\item As we indicated in the introduction, many things remain to be elucidated
concerning the relation of our work to cluster algebras.  Let us mention but two examples.
First, some of the formulae in \cite{Lampe} are closely related to laminations and
line operators in $SU(2)$ gauge theories.  In another direction,
in   \cite{MR2567745}, Fock and Goncharov have stated a very interesting duality
conjecture pairing two different kinds of cluster varieties, which they call
$\CX$-varieties and $\CA$-varieties.  Their conjecture seems to be
related to the labeling of asymptotic behaviors of line operator vevs by tropical
labels.  We hope to give a physical interpretation of the Fock-Goncharov duality conjecture
on some other occasion.

\end{enumerate}

\section*{Acknowledgements}

We thank   Dan Freed, Emanuel Diaconescu,
Alexander Goncharov,  Anton Kapustin,  Juan Maldacena, David Morrison,
 Hiraku Nakajima, Nikita Nekrasov, Tony Pantev, Nathan Seiberg, Yan Soibelman,
Valerio Toledano Laredo,  and Edward Witten for discussions.

We would like
to thank the Simons Center for Geometry and Physics for hosting several
excellent workshops related to this project and for hospitality at those
workshops.

The work of GM is supported by the DOE under grant
DE-FG02-96ER40959.   The work of AN is supported in part by the NSF under grant numbers
PHY-0503584 and PHY-0804450.  DG is supported in part by the NSF grant PHY-0503584.
DG is supported in part by the Roger Dashen membership in the Institute for Advanced
Study.

\appendix

\section{Some technical details on  $d=4, \CN=2$
supersymmetry}\label{app:Susy-Conventions}

We follow the conventions of Bagger and Wess  for $d=4, \CN=1$
supersymmetry \cite{Wess:1992cp}. In particular $SU(2)$ indices are raised/lowered with
$\epsilon^{12}= \epsilon_{21}=1$. Components of tensors in the
irreducible spin representations of $so(1,3)$ are denoted by
$\alpha,\dot\alpha$ running over $1,2$. The rules for conjugation
are that $(\CO_1 \CO_2)^\dagger = \CO_2^\dagger \CO_1^\dagger$ and
$(\psi_\alpha)^\dagger = \bar \psi_{\dot \alpha}$.

The $\CN=2$ supersymmetry operators are $(Q_\alpha^{~A}, \bar
Q_{\dot \alpha B})$ where $A,B$ are $SU(2)_R$ indices running from
$1$ to $2$. They  satisfy the Hermiticity conditions
\begin{equation}
(Q_\alpha^{~A})^\dagger = \bar Q_{\dot \alpha A}
\end{equation}
and the $\CN=2$ algebra
\begin{equation}
\begin{split}
\{ Q_{\alpha}^{~A}, \bar Q_{\dot \beta B} \} & =    2
\sigma^m_{\alpha\dot\beta}P_m \delta^{A}_{~B} \\
\{ Q_{\alpha}^{~A}, Q_{  \beta}^{~ B} \} & =    2 \epsilon_{\alpha\beta}\epsilon^{AB} \bar Z \\
\{ \bar Q_{\dot\alpha A}, \bar Q_{  \dot \beta  B} \} & =   -2
\epsilon_{\dot\alpha\dot\beta}\epsilon_{AB}   Z\\
\end{split}
\end{equation}
where $Z$ is the central charge and $P_m$ is the Hermitian
energy-momentum vector with $P^0 \geq 0$.

A line operator inserted at an origin $x^i =0$ of spatial
coordinates preserves an $so(3) \oplus su(2)_R$ symmetry as well as
the supersymmetries:
\begin{equation}
\CR_{\alpha}^{~A} = \xi^{-1} Q_{\alpha}^{~A} + \xi
\sigma^0_{\alpha\dot\beta} \bar Q^{\dot \beta A}
\end{equation}
Here $\xi$ is a phase: $\vert \xi \vert =1 $. These operators
satisfy the Hermiticity conditions
\begin{equation}
\begin{split}
(\CR_{1}^{~1})^\dagger & = - \CR_{2}^{~2} \\
(\CR_{1}^{~2})^\dagger & =   \CR_{2}^{~1} \\
\end{split}
\end{equation}
and the algebra
\begin{equation}
\{ \CR_{\alpha}^{~A} , \CR_{\beta}^{~B} \} = 4 \left( E + {\rm
Re}(Z/\zeta) \right) \epsilon_{\alpha\beta} \epsilon^{AB}
\end{equation}
where $E=P^0$ is the energy operator and $\zeta = \xi^{-2}$. This
algebra implies that
\begin{equation}
\left( \CR_{1}^{~1} + (\CR_{1}^{~1})^\dagger \right)^2 = \left(
\CR_{1}^{~2} + (\CR_{1}^{~2})^\dagger \right)^2 =  4 (E + {\rm
Re}(Z/\zeta) )
\end{equation}
from which we obtain the BPS bound
\begin{equation}
E + {\rm Re}(Z/\zeta) \geq 0.
\end{equation}

\section{Holomorphy of line operator vevs}\label{app:Holomorphy}

In this appendix we would like to explain in detail how the correlation
functions $\inprod{L_\zeta }$ of operators annihilated by the $\CR_{\alpha}^{~A}$
may be regarded as holomorphic functions on $\CM$ in complex structure $\zeta$.
We will also comment briefly about the relation with Rozansky-Witten twists of the 3d theory.

The crucial step of the identification is to match the 4d supercharges
and the supercharges of the low energy 3d sigma model with \hk target space $\CM$.
After 4d Lorentz invariance is broken by the circle compactification,
the 4d supercharges can be collected into a doublet
$\left(Q_{\alpha}^{~A}, \sigma^0_{\alpha\dot\beta} \bar Q^{\dot \beta A}\right)$ which we can denote as
$Q^{~A}_{a~\alpha}$, where the new index $a$ takes values $1,2$.
The 4d spinor index $\alpha$ can be identified with a 3d spinor index,
and $Q^{~A}_{a~\alpha}$ can be identified with the supercharges of a 3d ${\cal N}=4$ theory.
Notice that whereas the $A$ index is acted upon by the 4d $SU(2)_R$ R-symmetry,
which remains a symmetry of the 3d theory, in general there is no symmetry of the 4d theory which
rotates the $a=1,2$ supercharges into each other. The $a$ index is \emph{not} an index
 for any $SU(2)$ symmetry. The 4d $U(1)_R$ symmetry, if not broken by masses or
gauge coupling scales, of course rotates the $a=1,2$ components in opposite directions.

These facts agree neatly with the general properties of 3d sigma models with a generic \hk target space.
As we will detail shortly, these models always have an $SU(2)_R$ R-symmetry rotating the
fermionic fields, but a second R-symmetry group must take the form of
an isometry of the bosonic target space, which rotates the \hk forms among themselves.
Hence a second R-symmetry group may be present only if the \hk target manifold has special properties.
It can take the form of a $U(1)_R$ isometry which rotates two of the three \hk forms among themselves,
or of an $SU(2)_{R}'$ isometry group rotating all three \hk forms.

In the target space of the sigma
model $\CM$ the Riemannian structure group $SO(4n)$ is reduced to
$(SU(2)\times USp(2n))/\IZ_2 $, and the tangent space $TX\otimes\IC
\cong S \otimes V$ where $S$ is a trivial rank 2 complex vector bundle,
while $V$ is a rank $2n$ complex bundle with structure
group $USp(2n)$. A choice of line in $S$ determines a complex structure on $\CM$.
These are parametrized by $\IC P^1$ and we let $\zeta $ be an
inhomogeneous coordinate on this twistor sphere. Locally we can introduce vector fields $W_{a
i}$, with $i=1,\dots, 2n$ spanning $V$ and $a=1,2$ spanning $S$.
The fermionic fields $\psi_{\alpha}^{A i}$ are an $SU(2)_R$ doublet of space-time spinors,
which are sections of $V$.
The general form of the SUSY transformations acting on a  function $F$ on $\CM$
is
\begin{equation}
[Q^{~A}_{a~\alpha} , F] = \psi_{\alpha}^{A i} W_{a i} F
\end{equation}
In particular, the supercharges are also sections of the trivial bundle $S$.

When we reduce the 4d vector multiplets to 3d, we easily recover this structure.
The 4d fermions indeed transform as doublets of $SU(2)_R$. The $n$ 4d fermions
$\psi_\alpha^{A}$ join with the $n$ conjugate $\sigma^0_{\alpha\dot\beta} \bar \psi^{\dot \beta A}$
into the $2n$ components of the symplectic bundle $V$.
One can check that   $Q_\alpha^A$ acts as a holomorphic
differential operator on functions on $\CM$ in complex structure
$\zeta=0$: it obviously annihilates the anti-holomorphic 4d scalars, and less obviously
annihilates the anti-holomorphic combinations of electric and magnetic Wilson lines.
Similarly $\bar Q^{\dot \beta A}$ acts as an anti-holomorphic
differential operator in complex structure
$\zeta=0$. This completes the identification of the index $a$
of the 4d supercharges with the index $a$ in the 3d sigma model
transformation laws. The identification is unaffected by the quantum corrections,
which leave the target space \hk, and do not modify the complex structure at $\zeta=0$.

In particular, the choice of a set  $\CR_{\alpha}^{~A}$ of 4d supercharges
coincides with the choice of a line in $S$, and a complex structure
in $\CM$. The vector fields which appear in
\begin{equation}
[\CR_{\alpha}^{~A} , F] = \psi_{\alpha}^{A i} \left( \xi^{-1} W_{1 i}+ \xi W_{2 i} \right) F
\end{equation}
are anti-holomorphic in complex structure $\zeta = \xi^2$.
Therefore $\langle L(\zeta,\dots )
\rangle$  are holomorphic functions on $\CM$ in complex structure
$\zeta$.

It may be interesting to draw a comparison with the
topologically twisted version of the 3d sigma model, a la Rozansky-Witten
\cite{Rozansky:1996bq}. The topological twist replaces the 3d Lorentz group with the diagonal
combination of the old 3d Lorentz group and $SU(2)_R$. There is a $\IC P^1$ worth
of topological charges, which coincide with $\CR_{\alpha}^{~A} \delta^{\alpha}_{~A}$.
Clearly the 4d line operators we consider here give rise to point-like
topological observables in the 3d theory. This might prove to be a useful point of view.

\section{Fixed point equations}\label{app:Fixed-Point-Equations}

An $\CN=2$ vectormultiplet has a  scalar $\varphi$, fermions
$\psi_{\alpha A}$, in the $({\bf 2};{\bf 1})\otimes 2$ of
$so(1,3)\oplus su(2)_R$, (and their complex conjugates $\bar
\psi_{\dot \alpha A} := (\psi_{\alpha}^{~A})^\dagger$), an Hermitian
gauge field $A_m$ and an auxiliary field $D_{AB} = D_{BA}$
satisfying the reality condition $(D_{AB})^* = - D^{AB}$. After
multiplication by $i$ all these fields are valued in the adjoint.
\footnote{Our convention is that generators of $u(N)$ are $N\times
N$ anti-hermitian matrices. In ``geometric'' conventions where the
covariant derivative is $d +A$, the present gauge field is related
by $A^{\rm geometric} = i g A^{\rm hermitian}$.}  The supersymmetry
transformation laws are \footnote{Here we deviate slightly from
Bagger-Wess conventions. Our scalar field $\varphi=\sqrt{2}A^{\rm
BW} $ where $A^{\rm BW} $ is the scalar component of a BW chiral
multiplet.   }
\begin{equation}
\begin{split}
[Q_{\alpha A}, \varphi] & = - 2 \psi_{\alpha A} \\
[\bar Q_{\dot\alpha A}, \varphi] & = 0 \\
[Q_{\alpha A}, A_m] & = i \bar \psi_{\dot \beta A} (\bar
\sigma_m)^{\dot \beta}_{~~\alpha} \\
[\bar Q_{\dot \alpha A}, A_m] & = -i (\bar
\sigma_m)_{\dot \alpha}^{~~\beta}  \psi_{ \beta A}  \\
 [Q_{\alpha A}, \psi_{\beta B} ] & =
\sigma^{mn}_{\beta\alpha} F_{mn} \epsilon_{AB} + i D_{AB}
\epsilon_{\beta\alpha} + \frac{i}{2} g
\epsilon_{\beta\alpha} \epsilon_{AB} [\varphi^\dagger, \varphi] \\
[\bar Q_{\dot \alpha A}, \psi_{\beta B} ] & = - i   \epsilon_{AB} \sigma^m_{\beta\dot\alpha} D_m\varphi \\
[Q_{\alpha A}, D_{BC}  ] & = \left( \epsilon_{AB}
\sigma_{\alpha}^{m\dot \beta} D_m \bar\psi_{\dot \beta C} +
B\leftrightarrow C \right) \\
& +   g \left(\epsilon_{AB} [ \varphi^\dagger, \psi_{\alpha C}] +
B\leftrightarrow C   \right)\\
\end{split}
\end{equation}

The fixed point equations for the supersymmetries
$\CR_{\alpha}^{~A}$ are
\begin{equation}\label{eq:Q-zeta-fix}
\begin{split}
F_{0\ell} - \frac{i}{2} \epsilon_{jk\ell} F_{jk} - i   D_\ell
( \varphi/\zeta) & = 0\\
D_0(  \varphi/\zeta) - \frac{g}{2} [  \varphi^\dagger,
  \varphi ] & = 0 \\
\end{split}
\end{equation}

If the structure group $\lieg$ is reduced to the Cartan subalgebra
then we can write a static solution of the form
\begin{equation}\label{eq:Embed-fix-sol}
\begin{split}
F & = \frac{1}{2} \omega_S \otimes \rho^M \\
\varphi & = \zeta \frac{\rho^M}{r} + \varphi_\infty \\
\end{split}
\end{equation}
where $\rho^M\in \liet$ and $\omega_S = \sin\theta d\theta d\phi$ in
terms of standard angular coordinates around an origin. If $\rho^M
\in \Lambda_G$ then $A$ will be properly quantized. This solution
defines UV boundary conditions defining an 't Hooft operator labeled
by an element of $\Gamma_G$.

(There is an interesting subtlety that comes up in these equations.
If we take $\rho^M, \rho^E\in \liet$ to be constant and
$\varphi = \zeta(\rho^M-i \rho^E)/\sqrt{2} + \varphi_\infty$
and $F = \half \omega_S \otimes \rho^M + \half * \omega_S \otimes \rho^E$
then the fixed-point equations and Bianchi identities  are satisfied but the equation of
motion \emph{is not}! )

\section{Fixed-Point Equations in the Low Energy Effective Theory}\label{app:Low-Energy-FxdPoint}

The   fixed-point equations \eqref{eq:Q-zeta-fix} apply in the infrared theory, but
their interpretation is a little different.

Let us recall that the charge lattice in the IR theory, $\hat
\Gamma$ has an antisymmetric form $\langle \cdot, \cdot \rangle $
and fits in a sequence
\begin{equation}
0 \rightarrow \Gamma_f \rightarrow \Gamma \rightarrow
\Gamma_{\rm gauge} \rightarrow 0
\end{equation}
Here the lattice of flavor charges is in the annihilator of
 $\langle \cdot, \cdot \rangle $, and the quotient $\Gamma_{\rm gauge} $ is
 symplectic. We denote the projection map $\gamma \to \bar
 \gamma$.

The low energy Seiberg-Witten effective IR theory is a  self-dual
abelian gauge theory with an invariant fieldstrength $\CF \in
\Omega^2(\IR^{1,3}) \otimes V$, where  $V= \Gamma_{\rm gauge} \otimes \IR$
is a symplectic vector space. The values of the moduli determine a
compatible complex structure $\CI$ on $V$. The fieldstrength
satisfies $d \CF =0$ and the anti-self-duality constraint:
\begin{equation}
(*\otimes \CI)\CF = - \CF
\end{equation}
Moreover, in the sector of the Hilbert space labeled by $\gamma_c$,
there is a quantization condition:
\begin{equation}
\int_{S^2_{\infty}} \frac{\CF}{2\pi}  =  \overline{\gamma_c}.
\end{equation}

A solution to the fixed point equations \eqref{eq:Q-zeta-fix} in the
low energy effective theory can be obtained by taking
\begin{equation}
\CF =\frac{1}{2} \left( \omega_S \otimes \gamma_c - \omega_H \otimes
\CI(\gamma_c) \right)
\end{equation}
where $\omega_H = *_4 \omega_S = {dr \wedge \frac{dt}{r^2}}$. The
fixed point equations will be solved if the vectormultiplet moduli
become $r$-dependent according to the  equation
\begin{equation}\label{eq:Mod-ATT}
2 {\rm Im} \left[ \zeta^{-1} Z(\gamma; u(r))\right] = -
\frac{\langle \gamma, \gamma_c\rangle }{r} +2 {\rm Im} \left[
\zeta^{-1} Z(\gamma; u)\right] \qquad\qquad \forall \gamma \in \hat
\Gamma
\end{equation}
This is obtained by considering the imaginary part of equation
\eqref{eq:Q-zeta-fix}. That equation is written in a fixed duality
frame. Making the equation   duality invariant leads to
\eqref{eq:Mod-ATT}.

The equation \eqref{eq:Mod-ATT} is a modification of the standard
attractor equation. The usual attractor equation is written in
$\CN=2$ supergravity. The field theoretic limit of that equation
gives an equation of the form \eqref{eq:Mod-ATT} with the important
exception that $\zeta = e^{i \alpha}$ is the (constant) phase of the
central charge $Z(\gamma_c;u)$ and $\Gamma_f=0$. This generalization
of the attractor equation is not entirely new. The equations for
``orientiholes'' similarly replace the phase $e^{i \alpha}$ in the
attractor equations by $-e^{i \alpha}$ \cite{Denef:2009ja}.

The dynamics of a probe BPS particle of charge $\gamma_h\in \hat
\Gamma $ moving in one of the above field configurations is governed
by the action
\begin{equation}
\int \vert Z(\gamma_h;u(r))\vert ds + \int \langle
\overline{\gamma_h}, \CA\rangle
\end{equation}
where we integrate along the worldline of the probe particle. The
energy of such a particle at rest is therefore
\begin{equation}
\begin{split}
E  & =  \vert Z(\gamma_h;u(r))\vert - \langle \overline{\gamma_h},
\CA_0 \rangle\\
& =\vert Z(\gamma_h;u(r))\vert  + \frac{ (
\overline{\gamma_h},\overline{\gamma_c} )}{r}
 \\
\end{split}
\end{equation}
The second term, $\frac{ ( \overline{\gamma_h},\overline{\gamma_c}
)}{r}$  is the Coulomb energy and it is expressed in terms of the
positive definite symmetric metric on $V$ formed using the complex
structure: $(v_1, v_2) := \langle v_1, \CI v_2 \rangle$. Note this
expression is symmetric in the charges and positive definite, as is
physically reasonable.

Now, by taking the real part of   the fixed point equations
\eqref{eq:Q-zeta-fix} and writing the duality invariant extension we
find
\begin{equation}
{\rm Re}\left[ \zeta^{-1} Z(\gamma_h;u(r))\right] =\frac{ (
\overline{\gamma_h},\overline{\gamma_c} )}{r} +{\rm Re} \left[
\zeta^{-1} Z(\gamma_h;u)\right]
\end{equation}
In this way we derive the formula \eqref{eq:Halo-Particle-Energy}
for the energy of a halo particle in the $\zeta$-attractor
background. The halo radius \eqref{eq:halo-rad} is obtained by
setting the right hand side of \eqref{eq:Mod-ATT} to zero for
$\gamma=\gamma_h$.

\section{Six-dimensional supersymmetry}\label{app:Six-Super}

The $(2,0)$ superconformal algebra in $\IR^{1,5}$, denoted here as $osp(8^*\vert 4)$,
has even subalgebra
$osp(8^*\vert 4)^0 = so(6,2) \oplus so(5)$. To form the odd subspace we choose one of the chiral spinor
representations $\Delta_\pm$ of $so(6,2)$. $\Delta_\pm$  are quaternionic spaces, each isomorphic to $\IH^4$.
Let us choose $\Delta_+$ for definiteness. Next let   $\Delta'\cong \IH^2$ be the
 irreducible spinor rep of $so(5)$. The odd subspace is $osp(8^*\vert 4)^1 = \Delta_+ \otimes_{\IR} \Delta'$.  It is $32$-real-dimensional.

 The sub-superalgebra of Poincar\'e supersymmetry  has
 even subalgebra $so(5,1) \oplus so(5)$ and now there is a chiral spinor
 $\Delta_+'\cong \IH^2$ of $so(5,1)$ which is used to define the odd subspace $\Delta_+'\otimes \Delta'$
 of $16$ Poincar\'e supercharges.

 It is useful to
 introduce a representation of the Clifford algebra
 $Cl(1_+,5_-)$ in terms of $8\times 8$ complex-valued matrices of the form
 \begin{equation}
 \Gamma^M = \begin{pmatrix} 0 & \bar\gamma^{M~\dot s}_{r}   \\    \gamma^{M~ s}_{\dot r} & 0 \\ \end{pmatrix}
 \end{equation}
 with $r=1, \dots,4 $ indexing the chiral spin rep while $\dot r = 1, \dots, 4$ indexes the antichiral rep.
 Of course $M=0,1,\dots 5$ run over spacetime dimensions.
 Then $(\Gamma^M)^*  = B \Gamma^M B^{-1} $ where
 \begin{equation}
 B = \begin{pmatrix} B_{r}^{~~s} & 0 \\ 0 & B_{\dot r}^{~~ \dot s} \\ \end{pmatrix}
 \end{equation}
The chiral representations $\Delta_\pm'$ are pseudoreal. There is also
 \begin{equation}
 C = \begin{pmatrix} 0 & c^{r \dot s} \\ \bar c^{\dot r s} & 0 \\ \end{pmatrix}
 \end{equation}
 so that $(C \Gamma^M)^{tr} = - (C \Gamma^M)$ and $(\Gamma^M C^{-1})^{tr} = - (\Gamma^M C^{-1})$.
 The tensor $c^{r \dot  s}$ gives the invariant contraction of $\Delta_+'$ with $\Delta_-'$
 to the singlet.

 For the $R$-symmetry $so(5)\cong usp(4)$ identify the spinor $\Delta'$ with the $4$ of $USp(4)$. If we
  wish to introduce indices we let   $i,j =1,\dots, 4$.
  Gamma matrices have index structure $\Gamma^{Ii}_{~~j}$ with $I=1, \dots 5$. Indices are
  raised and lowered with the symplectic matrix $J^{ij}$ and its inverse.
  Thus $(\Gamma^I)_{i}^{~j} = J_{ii'} J^{jj'} \Gamma^{Ii'}_{~~j'}$. We have
 $(\Gamma^{Ii}_{~~j})^* = - \Gamma^{I~j}_{i}$ and $\Gamma^I_{ij} := J_{ii'} \Gamma^{Ii'}_{~~j}$
 is antisymmetric.

 If we denote Poincar\'e supersymmetries by $Q_{r}^i$ then we  impose
 \begin{equation}
 (Q_{r i})^\dagger = B_r^{~~s} J^{ij} Q_{s j}
 \end{equation}
 so there are $16$ Hermitian supercharges. The  basic  supersymmetry algebra is
 \begin{equation}
 [Q_{r}^i, Q_{s}^j ] =  2i  (\bar\gamma^M\bar c)_{rs} J^{ij} P_M
 \end{equation}
 where $P_M$ is the Hermitian translation operator.
 The coefficient $2i$ is determined by dimensional reduction to the $d=4, \CN=2$ algebra
 described above. Acting on field multiplets
 there can be extra terms on the RHS due to gauge transformations. Acting on
 states in solitonic sectors there can be extra terms from ``central extensions.''
Indeed, the   Poincar\'e superalgebra is extended in the presence of string-excitations to
 \begin{equation}
 [Q_{ri}, Q_{sj} ] =  2i  (\bar\gamma^M\bar c)_{rs} J^{ij} P_M +2  (\bar\gamma^M\bar c)_{rs} \Gamma^{I ij} Z_M^I
 \end{equation}
where the $Z_M^I$ correspond to the ``central charges'' of  a BPS string.

 The superconformal algebra has a field representation known as the ``tensormultiplet''
 usually denoted $(B, \psi, Y)$ where $B$ is a locally-defined real two-form potential on $\IR^{1,5}$
 with anti-self-dual fieldstrength (with orientation $\epsilon^{012345}=+1$),
 $\psi \in (\Delta_+'\otimes \Delta')$ satisfies a reality constraint,
 and $Y$ is a  real scalar field in the $5$ of $so(5)$ \cite{Howe:1983fr}.
 If we write out indices then we denote the fields in the tensormultiplet by
   $(B_{MN}, \psi_{r}^i, Y^I)$. The supersymmetry transformations under the Poincar\'e supersymmetries are:
 \begin{equation}
 \begin{split}
 [Q_{r}^i, Y^I] &=   i \Gamma^{Ii}_{~~ j} \psi_{r}^j \\
 [Q_{r}^i, \psi_{sj} ]  & =   i (\bar \gamma^M\bar c)_{rs} \Gamma^{Iij} \p_M Y^I  - \frac{i}{12} J^{ij} (\bar\gamma\bar c)^{MNP}_{rs} H_{MNP} \\
 [Q_{r}^i, B_{MN} ] & =  i (\gamma_{MN})_r^{~~s} \psi_{s}^{i} \\
 \end{split}
 \end{equation}
The  scaling dimension of  $Y^I$ is $2$, that of $\psi$ is $5/2$ and $B_{MN}$ has scaling dimension $2$
 so that $B = \half B_{MN} dx^M \wedge dx^N$ and $H:=dB$ are dimensionless. The fieldstrength $H$ must be
  anti-self-dual and in our  normalization
 for a $U(1)$ tensormultiplet $H$ has integral periods on manifolds of nontrivial topology.

The supersymmetry transformations only close onshell and the closure of  $ [Q_{ri}, Q_{sj} ]$
on $B_{MN}$ is only up
to a gauge transformation by $\Lambda_N = -2 (\bar \gamma_N\bar c)_{sr} \Gamma^{Iij} Y^I + 2 J^{ij}
(\bar \gamma_P\bar c)_{sr}B_{NP}$. Acting on a string state the induced gauge transformation
acts by measuring the charge of the string. This leads to the central charge $Z_{M}^I= \hat t_M Y^I$,
from which one computes the tension of the string $\sqrt{Y^I Y^I}$.

The superpoincare algebra in 5 dimensions has bosonic subalgebra $so(1,4) \oplus so(5)$.
The spin representation $\Delta''$ of $so(1,4)$ is again isomorphic to $\IH^2$ and
the   odd subspace is  $\Delta''\otimes \Delta'$ for $16$ real supersymmetries.
The super-Yang-Mills multiplet is $(A,\chi,Y)$ where $A$ is a connection, $\chi \in
\Delta''\otimes \Delta'$ and $Y$ is a scalar, all valued in a Lie algebra $\fg$.
We have deliberately used the same notation $Y$ for the scalars since the dimensional reduction
of a tensormultiplet gives a $u(1)$ vectormultiplet. However, in this case it is better
to normalize the scalar to be  $\Phi^I = R Y^I$ of scaling dimension $1$.

\section{Fock-Goncharov coordinates and traffic rules}

In \eqref{eq:trh-large-r}  we identified certain line operator vevs
$\inprod{L_\zeta}$ as the traces of holonomies of flat connections around loops on $C$.
In \eqref{eq:y1} we identified the vevs
$\inprod{L_\zeta}'$ similarly as traces of holonomies of twisted flat connections.
In this appendix we explain how to expand these holonomies in terms of the corresponding
Fock-Goncharov-like coordinates.  The basic idea of the calculation is not new and can be
found in many places; the only part which may be novel is that in the twisted case
we are able to define positive expansions even for holonomies in the fundamental representation of $SL(2,\IC)$.

\subsection{The coordinates} \label{app:fgcoords}

We first briefly recall the definition of the Fock-Goncharov coordinates.  These were introduced in \cite{MR2233852}; a slightly adapted version,
convenient for our purposes, was described in \cite{Gaiotto:2009hg}.
We suppose given a ``decorated triangulation'' $T$ of $C$.  This means a triangulation of $C$ such that each vertex
is a regular singularity, or a marked point on a small disc cut out around an irregular singularity.  The ``decoration'' means a certain discrete
choice
associated to each vertex.  For regular singularities we choose one of the two eigenspaces of the monodromy;
for irregular singularities we choose an identification between the marked points and the Stokes lines emerging from the singularity.
In either case, given a vertex and a flat connection $\CA$, the choice of decoration gives us a way to pick out
a flat section $s$ (solution of $(d + \CA) s = 0$) up to scalar multiple.  For regular singularities $s$ is chosen to be
an eigenvector of the monodromy.  For irregular singularities it is a section which is exponentially decaying along the anti-Stokes (aka WKB )  ray.
$s$ might not exist globally on $C$ because of monodromies, but at least we can take it
to exist on any simply connected domain.

Given any decorated triangulation $T$ of $C$ and any $E \in \Edges(T)$, we consider the quadrilateral $Q_E$
which has $E$ as its diagonal.  Number the four vertices $1$ through $4$, counterclockwise from one of the ends of $E$.
Then the decoration of $T$ provides four flat sections $s_i$.  We can parallel transport them to any common point in $Q_E$
and then evaluate their $SL(2,\IC)$ invariant cross ratio:
\begin{equation} \label{eq:def-Y}
\CY_E = \frac{(s_1 \wedge s_2) (s_3 \wedge s_4)}{(s_2 \wedge s_3)(s_4 \wedge s_1)}.
\end{equation}

We also want a version of the above for the case of twisted local systems.  In this case the flat sections $\ts_i$ are defined over
the punctured tangent bundle of $C$, or more precisely over some simply connected subsets thereof.
So we must be somewhat more careful
about how we transport the flat sections $\ts_i$.  We use non-self-intersecting and mutually non-intersecting paths from the vertices to a common
point $\CP_*$ in the interior of $Q_E$.  Having done so we consider the circle of tangent vectors over $\CP_*$:  we have a rank 2 flat connection over this circle, with holonomy $-1$, and four vectors $\ts_i$ at points
$t_i$ cyclically ordered around the circle.  We want to define a ``twisted cross ratio'' of these four vectors.
We adopt the convention that wedge-products $\ts_i \wedge \ts_{i+1}$ are defined by parallel-transporting the two vectors
to a common point along the arc between $t_i$ and $t_{i+1}$ which does not contain the other two $t_j$
(i.e. taking the ``short way'' around the circle), taking their wedge product, and then
dividing by the $SL(2,\IC)$ invariant volume form.  With this convention we can write
\begin{equation} \label{eq:def-tY}
\tCY_E = \frac{(\ts_1 \wedge \ts_2) (\ts_3 \wedge \ts_4)}{(\ts_2 \wedge \ts_3)(\ts_4 \wedge \ts_1)}.
\end{equation}

\subsection{Traffic rules} \label{app:traffic}

In Appendix A of \cite{Gaiotto:2009hg} we reviewed a well-known algorithm for
computing holonomies and their traces
in terms of Fock-Goncharov coordinates associated
to a triangulation.  We call this algorithm the ``traffic rule algorithm.''
Here we describe the extensions of the traffic rules needed to cover the situations of twisted local systems
and laminations.

So fix a triangulation $T$ and a twisted local system, i.e. a point of $\tCM$.
Also fix an edge $E$, and a co-orientation $v$ of $E$.  The pair $(E,v)$ determine a simply connected
domain $D(E,v)$ in the unit tangent bundle, fibered over $E$ (consisting of all tangent vectors which point to one
side of $E$; by abuse of notation we also use $v$ to denote any of these tangent vectors).
Let $l$ and $r$ denote the two vertices of $E$, chosen so that $v$ is on the left of $E$ when it is traversed
from $l$ to $r$.  We will be interested in flat sections of our twisted local system over $D(E,v)$.
These form a two-dimensional space $S(E,v)$, with two distinguished vectors $\ts_l$ and $\ts_r$ determined
by the decorations at the vertices (up to overall scale, which we fix arbitrarily as usual).

Then we can define a quantity $N(E)$ by $N(E) = \frac{\ts_l \wedge \ts_r}{\vol}$, where
$\vol$ is the $SL(2,\IC)$-invariant volume form.
Crucially, $N(E)$ so defined does not depend on the choice of
co-orientation $v$.  Indeed, reversing $v$ exchanges $\ts_l$ and $\ts_r$, which introduces a minus
sign; but reversing $v$ also
changes the domain in which we evaluate the wedge product, and as shown in Figure \ref{fig:twisted-sign}, this change
introduces a second minus sign.  (This is the moment where working with \ti{twisted} local systems helps us.)
\insfig{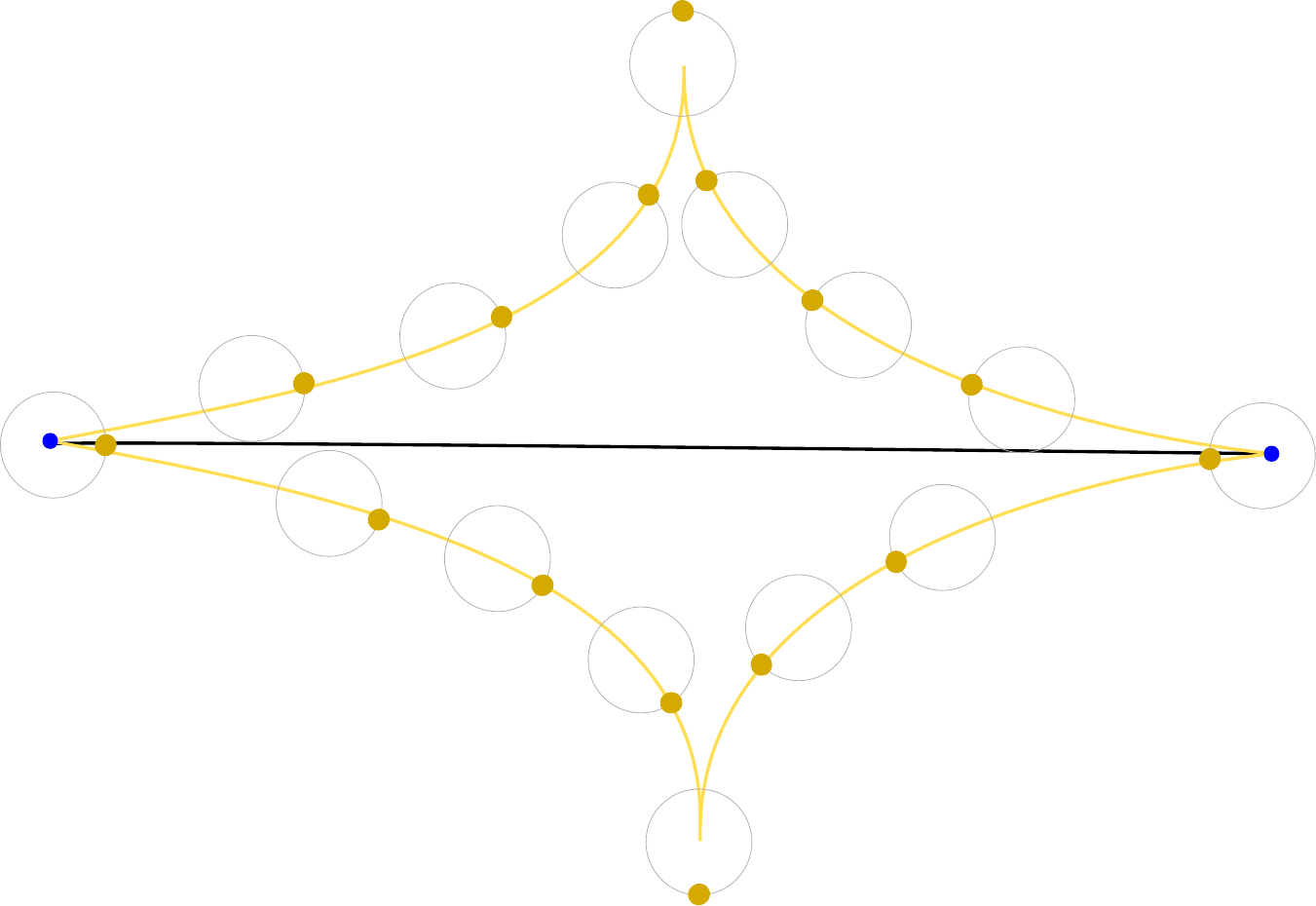}{The sections $\ts_l$ and $\ts_r$ are initially defined in contractible regions of the unit
tangent bundle over patches near the two vertices.  To evaluate
the wedge product $\ts_l \wedge \ts_r$, they must be transported to a common
point $x \in C$ and a common tangent vector at $x$.
The figure shows two different ways of doing so, corresponding to two choices of co-orientation of the edge:
either transport both sections north or south, along the indicated paths.
The gray circles indicate the fibers of the unit tangent bundle, to which the paths are lifted.
The small dots on the gray circles indicate the lifted tangent vectors.
These two ways of defining $\ts_l \wedge \ts_r$ differ by a factor $-1$:
to see this, note that as we go around the loop formed by the four paths,
the tangent vector winds once around the fiber, and the holonomy of our local system
around the fiber is $-1 \in SL(2,\IC)$.}

Define
\begin{equation}
\tau(E) = \sqrt{N(E)}
\end{equation}
where we just choose once and for all one of the two possible square roots; this is an independent
choice for each $E$.  Having made these choices, we
also obtain a definition of $\sqrt{\tCY_E}$:  with notation as in the previous section, we fix
\begin{equation}
 \sqrt{\tCY_E} = \frac{\tau(E_{12}) \tau(E_{34})}{\tau(E_{23)} \tau(E_{41})}.
\end{equation}

Now consider an edge $E$ lying on a face $F$.  $F$ determines two co-orientations of $E$, namely
$v^{\inn}$ pointing into $F$ and $v^\out$ pointing out of $F$.
Let $l$ and $r$ denote the two vertices of $E$, chosen so that $v^\inn$ is on the left of $E$ when it is traversed
from $l$ to $r$; let $E_l$ and $E_r$ be the other two edges of $F$, containing $l$ and $r$ respectively.
Now define two bases of the spaces of flat sections:
\begin{align}
 B^{\inn}(E,F) &= \left( \ts_l \frac{\tau(E_r)}{\tau(E_l) \tau(E)}, \ts_r \frac{\tau(E_l)}{\tau(E_r) \tau(E)} \right), \\
 B^{\out}(E,F) &= \left( \ts_r \frac{\tau(E_l)}{\tau(E_r) \tau(E)}, \ts_l \frac{\tau(E_r)}{\tau(E_l) \tau(E)} \right),
\end{align}
defined over $D(E, v^\inn)$ and $D(E, v^\out)$ respectively.
The normalization factors have been chosen so that both of these bases have determinant $1$.

We want to study parallel transport along paths $p$ on $C$.  Chopping $p$ into pieces, we can write this parallel
transport concretely as a product of matrices relative to the bases $B^{\inn,\out}(E,F)$ for various $E$, $F$
which $p$ encounters.

A path segment running from edge $E$ to $E'$ on a face $F$ lifts to the unit tangent bundle to give
a path segment from $D(E, v^\inn)$ to $D(E', v'^\out)$.
The corresponding parallel transport, relative to the bases $B^\inn(E, F)$ and $B^\out(E', F)$, depends on whether
the path turns right or left:  a short direct computation using the Pl\"ucker relations gives
\begin{equation}
 L = \begin{pmatrix} 1 & 1 \\ 0 & 1 \end{pmatrix}, \qquad R = \begin{pmatrix} 1 & 0 \\ 1 & 1 \end{pmatrix}.
\end{equation}
Similarly, parallel transport along
a path segment crossing an edge $E$ from face $F$ to $F'$ is given by the matrix
\begin{equation}
 M_E = \begin{pmatrix} \sqrt{\tCY_E} & 0 \\ 0 & 1 / \sqrt{\tCY_E} \end{pmatrix}
\end{equation}
That is, $B^\inn(E,F') = B^\out(E,F) M_E $.   See Figure \ref{fig:turns}.

\insfig{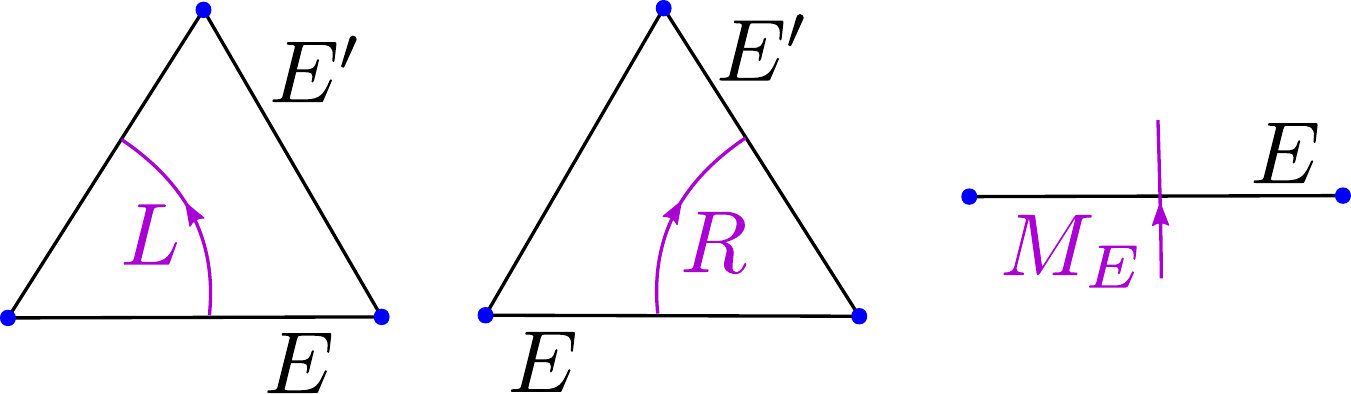}{Parallel transport matrices associated to path segments.}

Define $\tCY_p = \prod_E \tCY_E$ where the product runs over all edges crossed by $p$.  Using the definition of
$\tCY_E$, we see that $\tCY_p$ is the product of various $N(E)$, all raised to \ti{even} powers.
This means that $\sqrt{\tCY_p}$ can be canonically defined independent of any choices.
The holonomy of the twisted local system
along a path $p$ is a product of matrices $R$, $L$ and $M_E$ over all the edges crossed by $p$. Hence this
holonomy is $1 / \sqrt{\tCY_p}$ times a matrix with entries polynomials in the $\tCY_E$, with \ti{positive} integer coefficients.  In particular this means that the \ti{trace} of the holonomy is is $1 / \sqrt{\tCY_p}$ times a
polynomial in the $\tCY_E$, with positive integer coefficients.

We can also calculate the expectation values of line operators corresponding to laminations.
Recall that a lamination can contain, in addition to closed curves, open curves ending on boundary arcs,
carrying integer weights.
This requires us to  augment slightly our traffic rules to specify what to do at the open ends.
In fact, we cannot give simple rules for a single open curve:  rather we have to consider combinations of
curves such that the total weight ending on each boundary arc is zero.  Without loss of generality we can restrict
to the simplest such combination, shown in Figure \ref{fig:lamination-basic-unit}.
\insfig{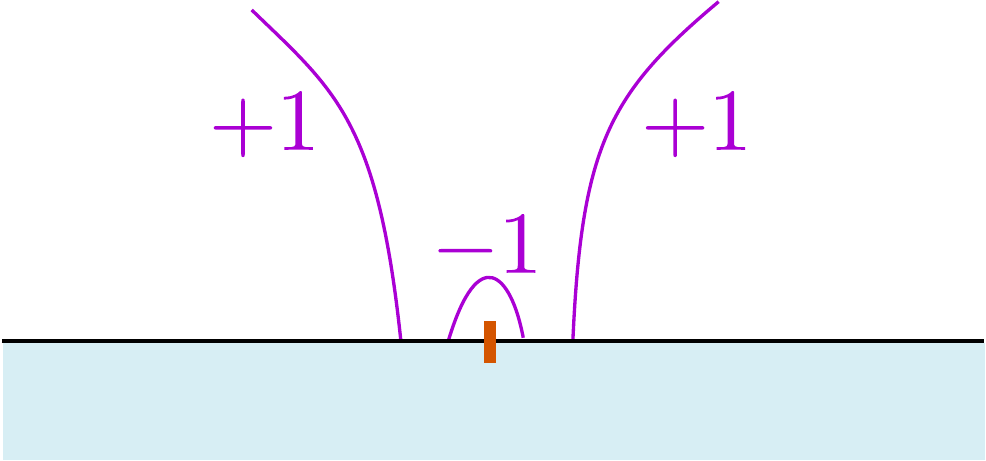}{The basic combination of open curves with net weight zero on each boundary arc.}
By an isotopy we can always arrange that all four of the ends appearing in that figure
lie on the same boundary edge.  After doing so, we simply drop the curve segment carrying
weight $-1$ (we will choose normalized sections such that this segment
contributes $1$).  As for the other two curve segments, we first choose
an orientation on each segment, and then assign them row or column vectors (representing basis
elements in $B(E, v^\inn)$ or $B(E, v^\out)^*$), given by the rules in
Figure \ref{fig:lamination-ends} and
\begin{gather}
 B^L = \begin{pmatrix} 0 & 1 \end{pmatrix}, \quad B^R = \begin{pmatrix} 1 & 0 \end{pmatrix}, \\
 E^L = \begin{pmatrix} 1 \\ 0 \end{pmatrix}, \quad E^R = \begin{pmatrix} 0 \\ 1 \end{pmatrix}.
\end{gather}

\insfig{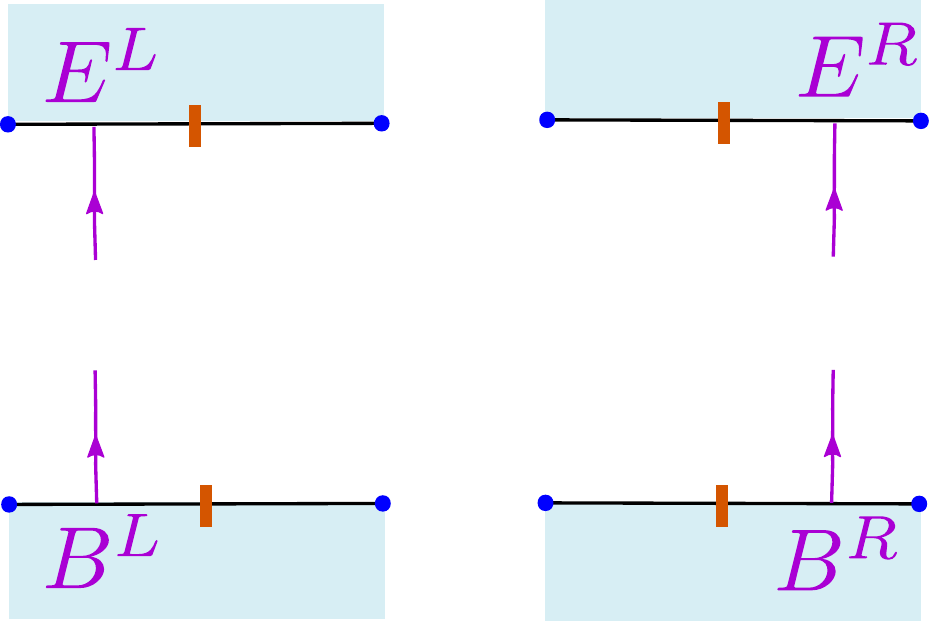}{Four ways an open edge can begin or end, and the corresponding basis vectors in the space
of flat sections.}

These formulas follow directly from the definition of the lamination vev (which we recall says we parallel transport
the flat sections associated to the vertices at the two ends of the path to a common point and take their
wedge product):  here we are transporting the normalized section from the end of the path to the beginning.

\section{$SU(2)$ $N_f = 0$ and Tchebyshev polynomials} \label{app:chebyshev}

In this appendix we establish the formula \eqref{eq:chebyshev-form}, giving the
expansion of $\inprod{L_{2n}}'$ in the $SU(2)$ theory with $N_f = 0$.  We use freely
the notation from Section \ref{subsec:SU2-Lams}.

As we have mentioned in the main text, we simplify our lives and avoid the
words ``twisted local system'' by working on $\CM$ instead of $\tCM$.  As a result it
will be difficult to fix some signs and choices of square root; we simply fix them at the
end by demanding positivity, since we have already shown that on $\tCM$ the desired expansions
indeed have positive coefficients.

Map Figure \ref{fig:nf0-bare} to an annulus so there is an inner and outer circle.
Introduce small flat sections $s_1$ attached to the WKB ray on the outer circle
and $s_2$ attached to the WKB ray on the inner circle.  Let $M$ denote the clockwise
monodromy operator.  Then combining the factors \eqref{eq:open-vev} gives
\begin{equation}
\langle L_{2n} \rangle = - \frac{ (s_2 \wedge M^n s_1)^2 }{(s_1 \wedge M s_1 )(s_2 \wedge M s_2) }.
\end{equation}
In particular, $\langle L_0 \rangle = 1/Y$ and $\langle L_2 \rangle = X$.  It follows that
$\inprod{L_1} = \sqrt{\frac{X}{Y}}$.

Also introduce $s_\pm$ of $M$ so that $M s_\pm = \lambda^{\pm 1} s_{\pm}$,
for some $\lambda \in \IC^*$. We can expand $s_2 = \alpha s_+ + \beta s_- $, $s_1 = \gamma s_+ + \delta s_-$,
and a small computation shows that
\begin{equation}
\langle L_{2n} \rangle = - \frac{ \frac{\beta \gamma}{\alpha\delta} \lambda^{2n} - 2 + \frac{\alpha \delta }{\beta \gamma} \lambda^{-2n} }{(\lambda - \lambda^{-1})^2}.
\end{equation}
(We assume $s_1, s_2$ are in general position.) Then, defining
\begin{equation}
e^{ \theta}:= \sqrt{\frac{\beta \gamma}{\alpha\delta} }, \qquad e^{\varphi}:=\lambda,
\end{equation}
we can express
\begin{align}\label{eq:Trig-Ln}
\begin{split}
\sqrt{\langle L_{2n} \rangle } & = i \frac{\sinh(\theta + n \varphi)}{\sinh\varphi} \\
& = i \left( \frac{\sinh\theta}{\sinh\varphi} T_n(\cosh\varphi) + \cosh\theta \, U_{n-1}(\cosh\varphi)\right),
\end{split}
\end{align}
where we have introduced the Tchebyshev polynomials $U_n(\cos\theta) := \frac{\sin(n+1)\theta}{\sin\theta}$
and $T_n(\cos\theta) := \cos(n\theta)$.   In particular
\begin{equation}\label{eq:Trig-X-Y}
\frac{1}{\sqrt{Y}} = i \frac{\sinh\theta}{\cosh\varphi}, \qquad \sqrt{X} = i \frac{\sinh(\theta+\varphi)}{\sinh\varphi}.
\end{equation}
Inverting \eqref{eq:Trig-X-Y} gives
\begin{align}\label{eq:solve-coshphi}
2\cosh \varphi &= \frac{1}{\sqrt{XY}} \left(   XY + Y + 1 \right), \\
2\cosh \theta &= \frac{i}{  \sqrt{X}Y}(XY-Y-1),
\end{align}
so altogether we obtain the desired result
\begin{equation}
\inprod{L_{2n}} = Y^{-1} \left[T_n(\cosh\varphi) + \frac{XY- Y - 1}{2 \sqrt{X Y}} U_{n-1}(\cosh\varphi) \right]^2.
\end{equation}

\bibliography{Lamination-Paper}

\end{document}